\documentclass[preprint,pre,floats,aps,amsmath,amssymb, standalone]{revtex4}
\usepackage{braket}
\usepackage{graphicx}
\usepackage{bm}
\usepackage{color}
\begin{document}

\title{Wigner distributions and GTMDs in a proton using light-front quark-diquark model}
\author{Satvir Kaur and Harleen Dahiya}
\affiliation{Department of Physics,\\ Dr. B.R. Ambedkar National
	Institute of Technology,\\ Jalandhar, 144011, India}

\begin{abstract}
We investigate the Wigner distributions and generalized transverse momentum-dependent distributions (GTMDs) for $u$ and $d$ quarks in the proton by using light-front quark-diquark model. We consider the contribution of scalar and axial-vector diquark having spin-0 and spin-1 respectively. We take different polarization configurations of quark and proton to calculate the Wigner distributions. The Wigner distributions are studied in the impact-parameter space, momentum space and mixed space for $u$ and $d$ quarks in the proton. We also study the relation of GTMDs with longitudinal momentum fraction carried by the active quark $x$ for different values of $\zeta$ (skewness) which is defined as the longitudinal momentum transferred to the proton. Further, we study the GTMDs in the relation with $x$ for zero skewness $(\zeta=0)$ at different values of quark transverse momentum $\textbf{p}_\perp$ as well as at different values of total momentum transferred to the proton ${\bf \Delta}_\perp$.
\end{abstract}

\maketitle

\section{Introduction}
\label{sec:intro}
The major aim of the hadron physics is to expose the relationship between partons (basic degrees of freedom of QCD) and hadrons. The parton distribution functions (PDFs) $f(x)$ provide the spread of the parton carrying a longitudinal momentum fraction $(x)$ in the hadron. On the other hand, the distribution located in the direction transverse to the motion of the hadron is explained through the generalized parton distributions (GPDs) \cite{gpd, gpda, gpdb} as a function of longitudinal momentum fraction carried by the parton, the longitudinal momentum transferred to the hadron and the total momentum transferred ($x$, $\zeta$ and $t$ respectively). Further, to describe the structure in momentum space, transverse momentum-dependent parton distributions (TMDs) were introduced \cite{tmda, tmd, tmd1, tmd2}, which depend on the transverse momentum carried by the parton ($\textbf{p}_\perp$). The GPDs and TMDs explain well the three-dimensional picture of internal structure of hadron, however, to understand the hadron structure more precisely, joint position and momentum distributions: Wigner distributions were instigated \cite{wigner}. Wigner distributions are quasi-probabilistic distributions which on application of certain limits provide the probabilistic distributions.  Wigner distributions are associated with the generalized parton correlation functions (GPCFs)  which when integrated over the light-cone energy of the parton reduce to the generalized transverse momentum-dependent parton distributions (GTMDs) \cite{gpcf}: \textit{mother distributions}. After suitable integrations, GTMDs can further be reduced to GPDs and TMDs.

The GPDs can be experimentally obtained via hard exclusive processes namely deeply virtual Compton scattering (DVCS) \cite{dvcs, dvcs1, dvcs2, dvcs3, dvcs4, dvcsa} where the interaction of the virtual photon with the parton of the nucleon leads to the radiation of a real photon from that parton and deeply virual meson production (DVMP) \cite{dvmpa, dvmp, dvmp1} where the interaction of the virtual photon with the parton of the nucleon leads to the emission of light-vector meson from that parton. GPDs are also accessible via $\rho$-meson photoproduction \cite{rho, rhoa}, timelike Compton scattering \cite{tcs}, heavy charmonia photoproduction for the production of gluon GPDs \cite{hcp, hcpa} and exclusive pion or photon-induced lepton pair-production \cite{epa, epb, ep}. The information on GPDs and nucleon structure can be extracted from the measurements of ongoing and upcoming experiments at Hall-A and Hall-B of JLab with CLAS collaboration \cite{jlaba, jlabb, jlabc, jlab, jlab1}, J-PARC \cite{jparc, jparca, jparc1} and COMPASS  \cite{compass}. The TMDs are obtainable via semi-inclusive deep inelastic scattering (SIDIS) \cite{sidisa, sidisb, sidis} and Drell-Yan processes \cite{eetmd, drell}. The SIDIS data is accessible from the upgraded experiments at JLab \cite{jlab2}, electron ion collider (EIC) \cite{eic}, DESY etc. \cite{desy}. The rich data of Drell-Yan process is accessed via experiments at FNAL, BNL, J-PARC etc. \cite{etmd, bnl}. 

Even though Wigner distributions have been executed in many fields of physics like heavy ion collision, quantum information, quantum molecular dynamics, signal analysis, non-linear dynamics  \cite{signal, signal1, signal2} and have been studied in some experiments \cite{exp, exp1} but no experiments have been done so far been done to extract Wigner distributions describing the multi-dimensional picture of the proton. Theoretical studies to understand Wigner distribution and GTMDs have however been attempted widely using light-cone spectator model \cite{spectator}, AdS/QCD quark-diquark model \cite{ads, ads1, maji-AdS}, light-front dressed quark model \cite{dressed1, dressed}, light-cone constituent quark model, chiral soliton model \cite{constituent1, constituent, soliton} etc.. Recently, Wigner distributions and GTMDs for electron have also been studied \cite{electron}. The spin-orbital angular momentum and spin-spin correlation between the polarized nucleon and quark can be determined by applying the phase space average to the Wigner distributions \cite{ads, dressed}. Other possible versions of phase-space distributions are: Husimi distribution (smeared version of Wigner distribution) and Kirkwood distribution where the former one is real and positive definite and later is complex \cite{husimi}.  It has been introduced in recent times that the gluon GTMDs can be accessible through diffractive di-jet production in deep-inelastic lepton-nucleon scattering \cite{exp_gtmd, exp_gtmd1, exp_gtmd2} and also in virtual photon-nucleus quasi-elastic scattering \cite{exp_gtmd3}. It has also been identified that the GTMDs of gluons can be measured in proton-nucleus collisions \cite{exp_gtmd4}. The quark GTMDs were recently measured by considering the exclusive double Drell-Yan process \cite{gtmd_DY}.

The dynamical front-form framework was introduced \cite{lc} to describe the constituent picture of hadron. A remarkable advantage of light-front dynamics is the simple light-front vacuum in QCD where the massive fluctuations are completely absent in the ground state. The absence of square root in the Hamiltonian simplifies the dynamical structure. The boost invariant light-front wavefunctions provide the inherent information about the structure of hadron \cite{lc1, lc2, lc3}. 
One of the important model which finds application in non-perturbative regime of QCD is the light-front quark-diquark model \cite{tmd1}. This model is a phenomenological approach to the work done in Ref. \cite{model}. In light-front quark-diquark model, the proton can be considered as a bound state of a quark and a diquark $(p=\Ket{u(ud)}+\Ket{d(uu)})$ with a diquark spin to be 0 (scalar diquark) or 1 (axial-vector diquark).  Using this model, all the T-even and T-odd TMDs of the proton are calculated using scalar and axial-vector diquarks and at nucleon-quark-diquark vertex where different choices of the form factors are considered \cite{tmd1}. The standard parton distribution functions and quasi-parton distribution functions are successfully explained in Ref. \cite{quasi-pdf}. In this model, the proton wavefunction does not exhibit $SU(4)=SU(2)\otimes SU(2)$ spin-isospin symmetry because of the non-vanishing relative orbital angular momentum of the quark-diquark system in its ground state. Using light-front quark-diquark model, GPDs,  TMDs  and spin transverse asymmetries for electron and hadron have already been evaluated \cite{tmd1,di-quark, di-quark-nkumar,bacchetta}. However, the Wigner distributions and GTMDs which provide the maximum information of internal structure of proton have not been evaluated so far.

Considering the above developments of the light-front quark-diquark model in studying
internal structure of the hadrons through GPDs and TMDs, it becomes desirable to extend this model to investigate the Wigner distributions and GTMDs of the hadrons.
In the present work, we have investigated the Wigner distributions in light-front quark-diquark model by considering different polarization configurations of quark and proton. We have also studied the GTMDs of quark contained by the proton for the case with different values of the longitudinal momentum transferred to the proton $\zeta \neq 0$ (non-zero skewness) as well as for the case where the longitudinal momentum transferred to the proton is zero $\zeta=0$ (zero skewness). For the case with $\zeta=0$, we have further studied the variation of GTMDs with longitudinal momentum fraction $x$. The implications of different values of quark transverse momentum $\textbf{p}_\perp$ and different values of the momentum transferred ${\bf \Delta}_\perp$ have also been discussed.

The paper is organized as follows. In Section II, we have given the essential details of the light-front framework and light-front quark-diquark model. The basic introduction to the Wigner distributions with the different polarization considerations and the explicit expressions of the quark Wigner distributions have been given in Section III. We have then presented the relation between the quark-quark correlators and the GTMDs and presented the analytical results for the 16 quark GTMDs for non-zero skewness $(\zeta \neq 0)$ in Section IV. In Section V, we have given the graphical interpretations to the Wigner distributions corresponding to the analytical results discussed in Section IV for the unpolarized proton, longitudinal-polarized proton and transversely-polarized proton. The results of the 16 quark GTMDs have also been presented in this section. Finally, the results have been summarized in  Section VI.

\section{Light-front quark-diquark model}
\subsection{General framework}
In the light-cone frame \cite{lc2}, we consider two light-like four-vectors $n_{\pm}$  which disintegrate a general four-vector $a=\left[ a^{+},a^{-},\textbf{\textit{a}}_{\perp}\right] $ to
\begin{eqnarray}
a=a^{+}n_{-}+a^{-}n_{+}+\textbf{\textit{a}}_{\perp},
\end{eqnarray}
satisfying $n_{\pm}^{2}=0$, $n_{+}.n_{-}=1$. The transverse tensor can be expressed as
\begin{equation}
\epsilon_{\perp}^{ij}=\epsilon^{\mu\nu ij} n_{+\mu} n_{-\nu},
\end{equation} with $\epsilon_{\perp}^{12}=-\epsilon_{\perp}^{21}=1$. The co-ordinates of a general four-vector $a^{\mu}$ are defined as 
\begin{equation}
a^{+}=\frac{1}{\sqrt{2}}\left( a^{0}+a^{3}\right),~~~~~ a^{-}=\frac{1}{\sqrt{2}}\left( a^{0}-a^{3}\right), ~~~~~ {\rm and} ~~~~~\textbf{\textit{a}}_{\perp}=\left( a^{1},a^{2}\right).
\end{equation}
For convenience, we take a frame where the four-momenta are defined as 
\begin{eqnarray}
P&=&\bigg[ P^{+},\frac{\vec{\Delta}_{\perp}^{2}+4M^{2}}{8(1-\zeta^{2})P^{+}},{\textbf{0}}_{\perp}\bigg] ,\nonumber\\
p_q&=&\bigg[xP^{+},p^{-},{\textbf{p}}_{\perp}\bigg],\nonumber\\
P_{D}&=&\bigg[(1-x)P^{+},P_{D}^{-},-{\textbf{p}}_{\perp}\bigg], \nonumber\\
\Delta &=& \bigg[-2\zeta P^{+},\frac{\zeta {\Delta}_{\perp}^{2}+4\zeta M^{2}}{4(1-\zeta^{2})P^{+}},{\bf{\Delta}_{\perp}} \bigg],
\label{eq1}
\end{eqnarray}
where $P=\frac{P'+P''}{2}$, $p_q$, $P_{D}$ and $\Delta=P'-P''$ are the average momentum of hadron, momenta of active quark, momenta of diquark and the four-vector momentum transferred to the proton respectively. Here $P'$ and $P''$ are the initial and final momenta of the proton and $M$ is the mass of proton. The momentum transferred to the proton and momentum fraction carried by the active quark in longitudinal direction are denoted as $\zeta=-\frac{\Delta^{+}}{2P^{+}}$ (skewness) and $x=\frac{p^{+}}{P^{+}}$ respectively. The quark transverse momentum and diquark transverse momentum are expressed in terms of the relative transverse momentum of quark $\textbf{p}_{\perp}$ and proton transverse momentum $\textbf{P}_{\perp}$ as follows
\begin{equation}
\textbf{p}_{q \perp}=x \textbf{P}_{\perp} + \textbf{p}_{\perp}, ~~~~~{\rm and} ~~~~~\textbf{P}_{D \perp}=(1-x) \textbf{P}_{\perp}-\textbf{p}_{\perp}.
\end{equation}
The total momentum transferred to the proton when the proton transverse momentum is not zero ($\textbf{P}_{\perp} \neq 0$) can be expressed as 
\begin{equation}
\textbf{D}_\perp={\bf \Delta}_\perp +2\zeta \textbf{P}_\perp.
\end{equation} 

\subsection{Light-cone wave functions}
The proton state can be defined as a superposition of the quark-diquark states and we have
\begin{eqnarray}
|P; \pm\rangle = c_s|u~ s^0\rangle^\pm + c_a|u~ a^0\rangle^\pm + c'_{a}|d~ a'^1\rangle^\pm, \label{PS_state}
\end{eqnarray}
where $|u~ s^0\rangle$, $|u~ a^0\rangle$ and $|d~ a'^1\rangle$ are defined as scalar isoscalar diquark, vector isoscalar diquark and vector isovector states respectively.
 
For $J_z=+\frac{1}{2}$ proton spin component, the two-particle Fock state expansion with spin-0 diquark can be expressed in terms of light-cone wave functions (LCWFs) $\psi^{\lambda_N}_{\lambda_q}$ with $\lambda_N$ and $\lambda_q$ denoting the helicities of proton and quark respectively. We have
\begin{eqnarray} 
|u~ s\rangle^\pm & =& \int \frac{dx~ d^2\textbf{p}_\perp}{2(2\pi)^3\sqrt{x(1-x)}} \bigg[ \psi^{\pm(u)}_{+}(x,\textbf{b}_\perp)|+\frac{1}{2}~s; xP^+,\textbf{p}_\perp\rangle \nonumber \\
 &+& \psi^{\pm(u)}_{-}(x,\textbf{p}_\perp)|-\frac{1}{2}~s; xP^+,\textbf{p}_\perp\rangle\bigg].\label{fock_PS}
 \end{eqnarray}
 The LCWFs emerging in above equation are defined as 
\begin{eqnarray} 
\psi^{+}_{+} (x,\textbf{p}_{\perp})&=&(m+ x M)\,\varphi/x ,  \nonumber\\
\psi^{+}_{-} (x,\textbf{p}_{\perp}) &=& -(p_x+ i p_y)\,\varphi/x, \nonumber\\
\psi^{-}_{+} (x,\textbf{p}_{\perp}) &=& -\left[\psi^{+}_{-} (x,\textbf{p}_{\perp})\right]^*,\nonumber \\
\psi^{-}_{-} (x,\textbf{p}_{\perp}) &=& \psi^{+}_{+} (x,\textbf{p}_{\perp}),
\end{eqnarray}
with
\begin{eqnarray}
\varphi(x,\textbf{p}_{\perp}) &=& -\frac{g_s}{\sqrt{1-x}}\, 
    \frac{x (1-x)}{\textbf{p}_{\perp}^2+[xM^{2}_{s}+(1-x)m^{2}-x(1-x)M^{2}]} \;,
    \label{scalar:var}
\label{eq:lcwf-s2}
\end{eqnarray}
where $m$, $M$ and $M_{s}$ are masses of quark, proton and spin-0 diquark respectively and $g_s$ is the coupling constant. 

The expansion of two-particle Fock state $|\mu~ A\rangle$ where $\mu$ corresponds to the flavor index $u$, $d$ and $A$ describes the axial-vector diquark with isospin-0 or 1 in the frame defined in Eq. (\ref{eq1}) for proton spin component $J_{z}=+\frac{1}{2}$ with spin-1 diquark, can be expressed  in terms of LCWFs $\psi^{\lambda_N}_{\lambda_q \lambda_a}$. Here $\lambda_N$, $\lambda_q$ and $\lambda_a$ denote the helicities of proton, quark and axial-vector diquark respectively. We have
\begin{eqnarray}
|\mu~ A\rangle^+ &=& \int \frac{dx\ d^2 {\bf p_\perp}}{2(2\pi)^{3}\sqrt{x(1-x)}} \Bigg[\psi^{+}_{+\frac{1}{2}+1}(x,{\bf p_\perp})\Ket{ +\frac{1}{2}+1; x\ P^+, {\bf p_\perp}}+\nonumber\\
&& \psi^{+}_{+\frac{1}{2}-1}(x,{\bf p_\perp})\Ket{ +\frac{1}{2}-1; x\ P^+, {\bf p_\perp}}+ \psi^{+}_{-\frac{1}{2}+1}(x,{\bf p_\perp})\Ket{ -\frac{1}{2}+1; x\ P^+, {\bf p_\perp}}+\nonumber\\
&& \psi^{+}_{-\frac{1}{2}-1}(x,{\bf p_\perp})\Ket{- \frac{1}{2}-1; x\ P^+, {\bf p_\perp}}\Bigg].
\label{+1/2}
\end{eqnarray}
The LCWFs appearing in the above equation are further expressed as \cite{tmd1}
\begin{eqnarray}
\psi^{+}_{+\frac{1}{2}+1}(x,{\bf p_\perp})&=& \frac{(p_{x}-i p_y)}{x(1-x)}\phi,\nonumber\\
\psi^{+}_{+\frac{1}{2}-1}(x,{\bf p_\perp})&=& -x\frac{(p_x+i p_y)}{x(1-x)}\phi,\nonumber\\
\psi^{+}_{-\frac{1}{2}+1}(x,{\bf p_\perp})&=& \frac{(m+xM)}{x} \phi,\nonumber\\
\psi^{+}_{-\frac{1}{2}-1}(x,{\bf p_\perp})&=& 0,
\end{eqnarray}
with 
\begin{eqnarray}
\phi(x,\textbf{p}_{\perp})=-\frac{g_{a}}{\sqrt{1-x}}\frac{x(1-x)}{\textbf{p}^{2}_{\perp}+[xM^{2}_{a}+(1-x)m^{2}-x(1-x)M^{2}]}. \label{phi}
\end{eqnarray}
where $m$, $M$ and $M_{a}$ are masses of quark, proton and spin-1 diquark respectively and $g_a$ is the coupling constant. 

Similarly, for $J_{z}=-\frac{1}{2}$, the two particle Fock state expansion with spin-1 diquark can be written as 
\begin{eqnarray}
|\mu~ A\rangle^- &=& \int \frac{dx\ d^2 {\bf p_\perp}}{2(2\pi)^{3}\sqrt{x(1-x)}} \Bigg[\psi^{-}_{+\frac{1}{2}+1}(x,{\bf p_\perp})\Ket{ +\frac{1}{2}+1; x\ P^+, {\bf p_\perp}}+\nonumber\\
&& \psi^{-}_{+\frac{1}{2}-1}(x,{\bf p_\perp})\Ket{ +\frac{1}{2}-1; x\ P^+, {\bf p_\perp}}+ \psi^{-}_{-\frac{1}{2}+1}(x,{\bf p_\perp})\Ket{ -\frac{1}{2}+1; x\ P^+, {\bf p_\perp}}+\nonumber\\
&& \psi^{-}_{-\frac{1}{2}-1}(x,{\bf p_\perp})\Ket{- \frac{1}{2}-1; x\ P^+, {\bf p_\perp}}\Bigg].
\label{-1/2}
\end{eqnarray}
The LCWFs in this case are given as
\begin{eqnarray}
\psi^{-}_{+\frac{1}{2}+1}(x,{\bf p_\perp})&=& 0,\nonumber\\
\psi^{-}_{+\frac{1}{2}-1}(x,{\bf p_\perp})&=& -\psi^{+}_{-\frac{1}{2}+1}(x,{\bf p_\perp}),\nonumber\\
\psi^{-}_{-\frac{1}{2}+1}(x,{\bf p_\perp})&=& \big[\psi^{+}_{+\frac{1}{2}-1}(x,{\bf p_\perp})\big]^{*},\nonumber\\
\psi^{-}_{-\frac{1}{2}-1}(x,{\bf p_\perp})&=& \big[\psi^{+}_{+\frac{1}{2}+1}(x,{\bf p_\perp})\big]^{*}.
\end{eqnarray}
 The nucleon tree-level cut amplitude $N \rightarrow q+X$ is used for the calculation of quark-quark correlation function, where $X$ is considered as either scalar or axial-vector diquark. In the case of the other spectator diquark models used in literature, the proton wavefunction assumes an $SU(4)$ spin-isospin symmetry leading to the probabilistic weights of the scalar isoscalar ($u-$quark with scalar-diquark), vector isoscalar ($u-$quark with axial-vector diquark) and vector isovector ($d-$quark with axial-vector diquark) to be 3:1:2. The overall size of the couplings are balanced such that the total number of quarks become three.
In the present work, because of the non-vanishing relative orbital angular momentum of the quark-diquark system in its ground state, the proton wave-function does not exhibit $SU(4)$ spin-isospin symmetry and the coefficients become 3 times smaller. Here the total number of quarks ``seen'' explicitly is only one, whereas the other two are hidden inside the diquark. It would be important to mention here that the total number of quarks in this case is also three. Also, the diquark which is not an elementary particle and is composed of two quarks  can be probed by a photon.
\section{Quark Wigner distribution}
Wigner distributions of the quark $\rho^{[\Gamma]}$ relate to GTMDs through the quark-quark correlator  or the Wigner operator $W^{[\Gamma]}$ as \cite{dressed, soliton, ads}
\begin{eqnarray}
\rho^{[\Gamma]}({\bf b_\perp},{\bf p_\perp},x;S)= \int \frac{d^2 \bf \Delta_\perp}{(2 \pi)^2} e^{-i {\bf{\Delta}_\perp} \cdot {\bf b_\perp}} W^{[\Gamma]}({\bf \Delta_\perp}, {\bf p_\perp},x;S),
\label{rho}
\end{eqnarray}
where the Wigner operator $W^{[\Gamma]}(\Delta_\perp, {\bf p_\perp},x;S)$ at fixed light-cone time $z^{+}=0$ is defined as
\begin{eqnarray}
W^{[\Gamma]}(\Delta_\perp, {\bf p_\perp},x;S)=\frac{1}{2} \int \frac{dz^- d^2 z_\perp}{(2 \pi)^3} e^{i p \cdot z} \Bra{P'';S} \bar{\psi}(-z/2) \Gamma \mathcal{W}_{[-\frac{z}{2},\frac{z}{2}]}\psi(z/2)\Ket{P';S}\Bigm\vert_{z^{+}=0}.
\label{w}
\end{eqnarray}
Here $P^{''}$ and $P^{'}$ are initial and final momenta of proton state, $S$ is the spin of proton and $\Gamma$ refers to the specific Dirac $\gamma$-matrices $\gamma^{+}$, $\gamma^{+}\gamma_{5}$, $i\sigma^{j+}\gamma_{5}$, where $j$ = 1 or 2.

On integrating over the impact-parameter space $\textbf{b}_{\perp}$, the Wigner distributions reduce to TMD correlators $\Phi^{[\Gamma]}$ in the absence of total momentum transfer to the hadron $\vec{{\Delta}}_{\perp}=\textbf{0}_{\perp}$ \cite{ads, soliton}. We have
\begin{eqnarray}
\int d^{2}b_{\perp}\rho^{[\Gamma]}(\textbf{b}_{\perp},\textbf{p}_{\perp},x;S)&=& W^{[\Gamma]}(\textbf{0}_\perp, {\textbf p_\perp},x;S) \nonumber\\
&\equiv & \Phi^{[\Gamma]}(\textbf{p}_{\perp},x;S).
\end{eqnarray}
On the other hand, the Wigner distributions reduce to two-dimensional Fourier transformations of GPD correlators $F^{[\Gamma]}$, when integrated over the transverse momentum component $\textbf{p}_{\perp}$. In the absence of light-front transverse co-ordinates i.e. at $\textbf{z}_{\perp}=\textbf{0}_{\perp}$, we have
\begin{eqnarray}
\int d^{2}p_{\perp}\rho^{[\Gamma]}(\textbf{b}_{\perp},\textbf{p}_{\perp},x;S) &=& \int \frac{d^{2}\Delta_{\perp}}{(2\pi)^{2}} e^{{-i\bf{\Delta}_{\perp} }\cdot \textbf{b}_{\perp}}F^{[\Gamma]}({\bf{\Delta}_{\perp}},x;S),
\end{eqnarray}
with 
\begin{eqnarray}
F^{[\Gamma]}({\bf{\Delta}_{\perp}},x;S)\equiv \frac{1}{2} \int \frac{ dz^{-}}{2 \pi} e^{i p \cdot z} \Bra{P'';S} \bar{\psi}(-z/2) \Gamma \mathcal{W}_{[-\frac{z}{2},\frac{z}{2}]}\psi(z/2)\Ket{P';S}\Bigm\vert_{z^{+}=\textbf{z}_{\perp}=0}.
\end{eqnarray}
Further, the Wigner distribution reduce to three-dimensional quark densities by integrating over the two orthogonal directions of transverse plane, i.e. over $b_{x}$ and $p_{y}$ or $b_{y}$ and $p_{x}$. At $\Delta_{x}=z_{y}=0$, we have
\begin{eqnarray}
\int db_{x} dk_{y} \rho^{[\Gamma]}(\textbf{b}_{\perp},\textbf{k}_{\perp},x;S)\equiv \bar{\rho}^{[\Gamma]}(b_{y},p_{x},x;S),
\end{eqnarray}
and at $\Delta_{y}=z_{x}=0$, we have
\begin{eqnarray}
\int db_{y} dk_{x} \rho^{[\Gamma]}(\textbf{b}_{\perp},\textbf{k}_{\perp},x;S)\equiv \tilde{\rho}^{[\Gamma]}(b_{x},p_{y},x;S).
\end{eqnarray}

The combinations of different polarization configurations of the proton and the quark describe 16 independent twist-2 quark Wigner distributions \cite{ads1, electron}. We have the Wigner distribution for an unpolarized quark in the unpolarized proton as 
\begin{eqnarray}
\rho_{UU}({\bf b_\perp},{\bf p_\perp},x)&=& \frac{1}{2} \Big[\rho^{[\gamma^+]}({\bf b_\perp},{\bf p_\perp},x;+\hat{S}_z) + \rho^{[\gamma^+]}({\bf b_\perp},{\bf p_\perp},x;-\hat{S}_z)\Big],
\end{eqnarray}
for a longitudinally-polarized quark in the unpolarized proton as
\begin{eqnarray}
\rho_{UL}({\bf b_\perp},{\bf p_\perp},x)&=& \frac{1}{2} \Big[\rho^{[\gamma^+ \gamma_5]}({\bf b_\perp},{\bf p_\perp},x;+\hat{S}_z) + \rho^{[\gamma^+ \gamma_5]}({\bf b_\perp},{\bf p_\perp},x;-\hat{S}_z)\Big],
\label{ul}
\end{eqnarray}
for a transversely-polarized quark in the unpolarized proton as
\begin{eqnarray}
\rho^{j}_{UT}({\bf b_\perp},{\bf p_\perp},x)&=& \frac{1}{2} \Big[\rho^{[i\sigma^{+j} \gamma_5]} ({\bf b_\perp},{\bf p_\perp},x;+\hat{S}_z) + \rho^{[i\sigma^{+j} \gamma_5]}({\bf b_\perp},{\bf p_\perp},x;-\hat{S}_z)\Big],
\label{ut}
\end{eqnarray}
for an unpolarized quark in the longitudinally-polarized proton as
\begin{eqnarray}
\rho_{LU}({\bf b_\perp},{\bf p_\perp},x)&=& \frac{1}{2}\Big[\rho^{[\gamma^+]} ({\bf b_\perp},{\bf p_\perp},x;+\hat{S}_z) - \rho^{[\gamma^+]} ({\bf b_\perp},{\bf p_\perp},x;-\hat{S}_z)\Big],
\label{lu}
\end{eqnarray}
for a longitudinally-polarized quark in the longitudinally-polarized proton as
\begin{eqnarray}
\rho_{LL}({\bf b_\perp},{\bf p_\perp},x)&=&\frac{1}{2} \Big[\rho^{[\gamma^+ \gamma_5]} ({\bf b_\perp},{\bf p_\perp},x;+\hat{S}_z) - \rho^{[\gamma^+ \gamma_5]} ({\bf b_\perp},{\bf p_\perp},x;-\hat{S}_z)\Big], 
\label{ll}
\end{eqnarray}
for a transversely-polarized quark in the longitudinally-polarized proton as
\begin{eqnarray}
\rho^{j}_{LT}({\bf b_\perp},{\bf p_\perp},x)&=& \frac{1}{2}\Big[\rho^{[i \sigma^{+j} \gamma_5]} ({\bf b_\perp},{\bf p_\perp},x;+\hat{S}_z) - \rho^{[i \sigma^{+j} \gamma_5]} ({\bf b_\perp},{\bf p_\perp},x;-\hat{S}_z)\Big],
\label{lt}
\end{eqnarray}
for an unpolarized quark in the transversely-polarized proton as
\begin{eqnarray}
\rho^i_{TU}({\bf b_\perp},{\bf p_\perp},x)&=& \frac{1}{2} \Big[\rho^{[\gamma^{+} ]} ({\bf b_\perp},{\bf p_\perp},x;+\hat{S}_i) - \rho^{[\gamma^{+}]} ({\bf b_\perp},{\bf p_\perp},x;-\hat{S}_i)\Big],
\label{tu}
\end{eqnarray}
for a longitudinally-polarized quark in the transversely-polarized proton as
\begin{eqnarray}
\rho^i_{TL}({\bf b_\perp},{\bf p_\perp},x)&=& \frac{1}{2} \Big[\rho^{[\gamma^{+} \gamma_5 ]} ({\bf b_\perp},{\bf p_\perp},x;+\hat{S}_i) - \rho^{[\gamma^{+} \gamma_5]} ({\bf b_\perp},{\bf p_\perp},x;-\hat{S}_i)\Big],
\label{tl}
\end{eqnarray}
for  a transversely-polarized quark in the transversely-polarized proton as
\begin{eqnarray}
\rho_{TT}({\bf b_\perp},{\bf p_\perp},x)&=& \frac{1}{2} \delta_{ij} \Big[\rho^{[i \sigma^{+j} \gamma_5]}({\bf b_\perp},{\bf p_\perp},x;+\hat{S}_i) - \rho^{[i \sigma^{+j} \gamma_5]}({\bf b_\perp},{\bf p_\perp},x;-\hat{S}_i)\Big],
\label{tt}
\end{eqnarray}
and finally the pretzelous  Wigner distribution as
\begin{eqnarray}
\rho^{\perp}_{TT}({\bf b_\perp},{\bf p_\perp},x)&=& \frac{1}{2} \epsilon_{ij} \Big[\rho^{[i \sigma^{+j} \gamma_5]}({\bf b_\perp},{\bf p_\perp},x;+\hat{S}_i) - \rho^{[i \sigma^{+j} \gamma_5]}({\bf b_\perp},{\bf p_\perp},x;-\hat{S}_i)\Big].
\label{pret}
\end{eqnarray}
The Wigner distributions for the scalar diquarks using Eqs. (\ref{rho}) and (\ref{w}) are
\begin{eqnarray}
\rho^{q(s)}_{UU}({\bf b_\perp},{\bf p_\perp},x)&=&\frac{1}{(2\pi)^2 16 \pi^3} \int d\Delta_x d\Delta_y \ \cos(\Delta_x b_x+ \Delta_y b_y)\nonumber\\
&\times& \frac{1}{x^2}\bigg[ \bigg(\textbf{p}^{2}_\perp-\frac{(1-x)^2}{4} {\bf \Delta}^{2}_\perp \bigg)+(m+xM)^2\bigg]
 \varphi^{\dagger}(\textbf{p}''_{\perp})\varphi(\textbf{p}'_{\perp}),\\
 \rho^{q(s)}_{UL}({\bf b_\perp},{\bf p_\perp},x) &=& \frac{1}{(2\pi)^2 16 \pi^3} \int d\Delta_x d\Delta_y\ \sin(\Delta_x b_x+ \Delta_y b_y)\nonumber\\
&\times& \frac{(\Delta_y p_x- \Delta_x p_y)(1-x)}{x^2}\varphi^{\dagger}(\textbf{p}''_{\perp})\varphi(\textbf{p}'_{\perp}),\\
\rho^{qj(s)}_{UT}({\bf b_\perp},{\bf p_\perp},x)&=& \frac{1}{(2\pi)^2 16 \pi^3} \int d\Delta_x d\Delta_y\ \sin(\Delta_x b_x+ \Delta_y b_y) \nonumber\\
&\times&  \Delta_{y}\ \frac{(1-x)(m + x M)}{x^2}\
 \varphi^{\dagger}(\textbf{p}''_{\perp})\varphi(\textbf{p}'_{\perp}),\\
  \rho^{q(s)}_{LU}({\bf b_\perp},{\bf p_\perp},x)&=&-\frac{1}{(2\pi)^2 16 \pi^3} \int d\Delta_x d\Delta_y \ \sin(\Delta_x b_x+ \Delta_y b_y)\nonumber\\
&\times& \frac{(\Delta_y p_x- \Delta_x p_y)(1-x)}{x^2}\varphi^{\dagger}(\textbf{p}''_{\perp})\varphi(\textbf{p}'_{\perp}),\\
\rho^{q(s)}_{LL}({\bf b_\perp},{\bf p_\perp},x)&=&-\frac{1}{(2\pi)^2 16 \pi^3} \int d\Delta_x d\Delta_y \ \cos(\Delta_x b_x+ \Delta_y b_y)\nonumber\\
&\times& \frac{1}{x^2} \bigg[ \bigg(\textbf{p}^{2}_\perp-\frac{(1-x)^2}{4} {\bf \Delta}^{2}_\perp \bigg) 
 -(m+xM)^2\bigg]
 \varphi^{\dagger}(\textbf{p}''_{\perp})\varphi(\textbf{p}'_{\perp}), \\
  \rho^{qj(s)}_{LT}({\bf b_\perp},{\bf p_\perp},x)&=& -\frac{2 }{(2\pi)^2 16 \pi^3} \int d\Delta_x d\Delta_y \ \cos(\Delta_x b_x+ \Delta_y b_y)\ \nonumber\\
&\times& p_{x}\ \frac{(m + x M)}{x^2}\ \varphi^{\dagger}(\textbf{p}''_{\perp})\varphi(\textbf{p}'_{\perp}),\label{lts}\\
\rho^{iq(s)}_{TU}({\bf b_\perp},{\bf p_\perp},x)&=& -\frac{1}{(2\pi)^2 16 \pi^3} \int d\Delta_x d\Delta_y \ \cos(\Delta_x b_x+ \Delta_y b_y)\ \nonumber\\
&\times& \Delta_{x}\ \frac{(1-x)(m + x M)}{x^2}\ \varphi^{\dagger}(\textbf{p}''_{\perp})\varphi(\textbf{p}'_{\perp}),\label{tus}\\
 \rho^{iq(s)}_{TL}({\bf b_\perp},{\bf p_\perp},x)&=& -\frac{2}{(2\pi)^2 16 \pi^3} \int d\Delta_x d\Delta_y \ \sin(\Delta_x b_x+ \Delta_y b_y)\ \nonumber\\
&\times& p_{y}\ \frac{(m + x M)}{x^2}\ \varphi^{\dagger}(\textbf{p}''_{\perp})\varphi(\textbf{p}'_{\perp}),\label{tls}\\
\rho^{q(s)}_{TT}({\bf b_\perp},{\bf p_\perp},x)&=& \frac{1}{(2\pi)^2 16 \pi^3} \int d\Delta_x d\Delta_y \ \cos(\Delta_x b_x+ \Delta_y b_y)\ \nonumber\\
&\times& \bigg[\bigg(\textbf{p}^2_\perp-\frac{(1-x)^2}{4}{\bf \Delta}^2_\perp \bigg)\bigg]\frac{1}{x^2} \ \varphi^{\dagger}(\textbf{p}''_{\perp})\varphi(\textbf{p}'_{\perp}),\\
\rho^{\perp q(s)}_{TT}({\bf b_\perp},{\bf p_\perp},x)&=& \frac{1}{(2\pi)^2 16 \pi^3} \int d\Delta_x d\Delta_y \ \sin(\Delta_x b_x+ \Delta_y b_y)\ \nonumber\\
&\times& \epsilon_{ij}\bigg[\bigg(p_i p_j-\frac{(1-x)^2}{4} \Delta_i \Delta_j \bigg)\bigg]\frac{1}{x^2} \ \varphi^{\dagger}(\textbf{p}''_{\perp})\varphi(\textbf{p}'_{\perp}).
\end{eqnarray}

For the axial-vector diquarks we have
\begin{eqnarray}
\rho^{q(a)}_{UU}({\bf b_\perp},{\bf p_\perp},x)&=&\frac{1}{(2\pi)^2 16 \pi^3} \int d\Delta_x d\Delta_y \ \cos(\Delta_x b_x+ \Delta_y b_y)\nonumber\\
&\times& \bigg[ \bigg(\textbf{p}^{2}_\perp -\frac{(1-x)^2}{4} {\bf \Delta}^{2}_\perp \bigg) 
 \frac{(1+x)^2}{x^2 (1-x)^2}+\frac{(m+xM)^2}{x^2}\bigg]
 \phi^{\dagger}(\textbf{p}''_{\perp})\phi(\textbf{p}'_{\perp}),\\
 \rho^{q(a)}_{UL}({\bf b_\perp},{\bf p_\perp},x) &=& \frac{1}{(2\pi)^2 16 \pi^3} \int d\Delta_x d\Delta_y\ \sin(\Delta_x b_x+ \Delta_y b_y)\nonumber\\
&\times& \frac{(\Delta_y p_x- \Delta_x p_y)(1-x){(1-x^2)}}{x^2(1-x)^2}\phi^{\dagger}(\textbf{p}''_{\perp})\phi(\textbf{p}'_{\perp}),\\
\rho^{qj(a)}_{UT}({\bf b_\perp},{\bf p_\perp},x)&=& \frac{1}{(2\pi)^2 16 \pi^3} \int d\Delta_x d\Delta_y\ \sin(\Delta_x b_x+ \Delta_y b_y) \nonumber\\
&\times&  \Delta_{y}\ \frac{(m + x M)}{x^2}\
 \phi^{\dagger}(\textbf{p}''_{\perp})\phi(\textbf{p}'_{\perp}),\\
 \rho^{q(a)}_{LU}({\bf b_\perp},{\bf p_\perp},x)&=&\frac{ 1}{(2\pi)^2 16 \pi^3} \int d\Delta_x d\Delta_y \ \sin(\Delta_x b_x+ \Delta_y b_y)\nonumber\\
&\times& \frac{(\Delta_y p_x- \Delta_x p_y)(1-x){(1-x^2)}}{x^2(1-x)^2}\phi^{\dagger}(\textbf{p}''_{\perp})\phi(\textbf{p}'_{\perp}),\\
\rho^{q(a)}_{LL}({\bf b_\perp},{\bf p_\perp},x)&=&\frac{1}{(2\pi)^2 16 \pi^3} \int d\Delta_x d\Delta_y \ \cos(\Delta_x b_x+ \Delta_y b_y)\nonumber\\
&\times& \bigg[ \bigg(\textbf{p}^{2}_\perp-\frac{(1-x)^2}{4} {\bf\Delta}^{2}_\perp \bigg) 
 \frac{(1+x)^2}{x^2 (1-x)^2}-\frac{(m+xM)^2}{x^2}\bigg]
 \phi^{\dagger}(\textbf{p}''_{\perp})\phi(\textbf{p}'_{\perp}), \\
 \rho^{qj(a)}_{LT}({\bf b_\perp},{\bf p_\perp},x)&=& \frac{2 }{(2\pi)^2 16 \pi^3} \int d\Delta_x d\Delta_y \ \cos(\Delta_x b_x+ \Delta_y b_y)\ \nonumber\\
&\times& p_{x}\ \frac{(m + x M)}{x^2(1-x)}\ \phi^{\dagger}(\textbf{p}''_{\perp})\phi(\textbf{p}'_{\perp}),\label{lt1}\\
\rho^{iq(a)}_{TU}({\bf b_\perp},{\bf p_\perp},x)&=& \frac{1}{(2\pi)^2 16 \pi^3} \int d\Delta_x d\Delta_y \ \cos(\Delta_x b_x+ \Delta_y b_y)\ \nonumber\\
&\times& \Delta_{x}\ \frac{(m + x M)}{x}\ \phi^{\dagger}(\textbf{p}''_{\perp})\phi(\textbf{p}'_{\perp}),\label{tu1}\\
 \rho^{iq(a)}_{TL}({\bf b_\perp},{\bf p_\perp},x)&=& -\frac{2}{(2\pi)^2 16 \pi^3} \int d\Delta_x d\Delta_y \ \sin(\Delta_x b_x+ \Delta_y b_y)\ \nonumber\\
&\times& p_{y}\ \frac{(m + x M)}{x(1-x)}\ \phi^{\dagger}(\textbf{p}''_{\perp})\phi(\textbf{p}'_{\perp}),\\
\rho^{q(a)}_{TT}({\bf b_\perp},{\bf p_\perp},x)&=& -\frac{2 }{(2\pi)^2 16 \pi^3} \int d\Delta_x d\Delta_y \ \cos(\Delta_x b_x+ \Delta_y b_y)\ \nonumber\\
&\times& \bigg[\bigg(\textbf{p}^2_\perp-\frac{(1-x)^2}{4}{\bf\Delta}^2_\perp \bigg)\bigg]\frac{1}{x(1-x)^2} \ \phi^{\dagger}(\textbf{p}''_{\perp})\phi(\textbf{p}'_{\perp}),\\
\rho^{\perp q(a)}_{TT}({\bf b_\perp},{\bf p_\perp},x)&=& 0.
\end{eqnarray}
Here $\varphi^{\dagger}(\textbf{p}''_{\perp})\varphi(\textbf{p}'_{\perp})$ and $\phi^{\dagger}(\textbf{p}''_{\perp})\phi(\textbf{p}'_{\perp})$ can be calculated using Eqs. (\ref{scalar:var}) and (\ref{phi}) respectively for  the initial and final momentum of the active quark being
\begin{eqnarray}
{\bf p'_\perp}={\bf p_\perp}- (1-x) \frac{{\bf \Delta_\perp}}{2},\ \ \ \ \ \ \ \ \ \ \ \ {\bf p''_\perp}={\bf p_\perp}+ (1-x) \frac{{\bf \Delta_\perp}}{2}.
 \end{eqnarray}
 respectively.

\section{Generalized transverse momentum dependent parton distributions (GTMDs) }
The GTMDs, under certain kinematic limits, reduce to GPDs and TMDs. For the twist-2 case, the GTMDs relate to the Wigner correlator as follows \cite{gpcf, gtmd1, gtmd2}
\begin{eqnarray}
 W_{\lambda \lambda'}^{[\gamma^+]}
 &=& \frac{1}{2M} \, \bar{u}(p', \lambda') \, \bigg[
      F_{1,1}
      + \frac{i\sigma^{i+} k_\perp^i}{P^+} \, F_{1,2}
      + \frac{i\sigma^{i+} \Delta_\perp^i}{P^+} \, F_{1,3} \nonumber\\*
 & &  + \frac{i\sigma^{ij} k_\perp^i \Delta_\perp^j}{M^2} \, F_{1,4}
     \bigg] \, u(p, \lambda)
     \,, \label{e:gtmd_1}\\
   W_{\lambda \lambda'}^{[\gamma^+\gamma_5]}
 &=& \frac{1}{2M} \, \bar{u}(p', \lambda') \, \bigg[
      - \frac{i\varepsilon_\perp^{ij} k_\perp^i \Delta_\perp^j}{M^2} \, G_{1,1}
      + \frac{i\sigma^{i+}\gamma_5 k_\perp^i}{P^+} \, G_{1,2}
      + \frac{i\sigma^{i+}\gamma_5 \Delta_\perp^i}{P^+} \, G_{1,3} \nonumber\\*
 & &  + i\sigma^{+-}\gamma_5 \, G_{1,4}
     \bigg] \, u(p, \lambda)
     \,, \label{e:gtmd_2}\\
 W_{\lambda \lambda'}^{[i\sigma^{j+}\gamma_5]}
 &=& \frac{1}{2M} \, \bar{u}(p', \lambda') \, \bigg[
      - \frac{i\varepsilon_\perp^{ij} k_\perp^i}{M} \, H_{1,1}
      - \frac{i\varepsilon_\perp^{ij} \Delta_\perp^i}{M} \, H_{1,2}
      + \frac{M \, i\sigma^{j+}\gamma_5}{P^+} \, H_{1,3} \nonumber\\*
 & &  + \frac{k_\perp^j \, i\sigma^{k+}\gamma_5 k_\perp^k}{M \, P^+} \, H_{1,4}
      + \frac{\Delta_\perp^j \, i\sigma^{k+}\gamma_5 k_\perp^k}{M \, P^+} \, H_{1,5}
      + \frac{\Delta_\perp^j \, i\sigma^{k+}\gamma_5 \Delta_\perp^k}{M \, P^+} \, H_{1,6} \nonumber\\*
 & &  + \frac{k_\perp^j \, i\sigma^{+-}\gamma_5}{M} \, H_{1,7}
      + \frac{\Delta_\perp^j \, i\sigma^{+-}\gamma_5}{M} \, H_{1,8}
     \bigg] \, u(p, \lambda),
     \, \label{e:gtmd_3}
\end{eqnarray} 
where the antisymmetric tensor is $\epsilon_{\perp}^{ij}=\epsilon^{-+ij}$ and $\epsilon^{0123}=1$. 

There are 16 complex-valued twist-2 GTMDs which depend upon the variables $(x,\zeta,\vec{k}_{\perp}^{2}, \vec{k}_{\perp}.\vec{\Delta}_{\perp},\vec{\Delta}_{\perp}^{2} ) $. For zero skewness i.e. $\zeta=0$, one is left with 10 non-zero twist-2 GTMDs. 
One can write all GTMDs $X$ in terms of the real  $X^{e}$  and imaginary $X^{o}$ parts as follows
\begin{eqnarray}
X(x,\zeta,\vec{k}_{\perp}^{2}, \vec{k}_{\perp}.\vec{\Delta}_{\perp},\vec{\Delta}_{\perp}^{2} )=X^{e}(x,\zeta,\vec{k}_{\perp}^{2}, \vec{k}_{\perp}.\vec{\Delta}_{\perp},\vec{\Delta}_{\perp}^{2} )+i X^{o}(x,\zeta,\vec{k}_{\perp}^{2}, \vec{k}_{\perp}.\vec{\Delta}_{\perp},\vec{\Delta}_{\perp}^{2} ).
\end{eqnarray}

The explicit expressions of GTMDs for scalar and axial-vector diquarks at $\zeta \neq 0$ (non-zero skewness) are calculated by using Eqs. (\ref{w}), (\ref{e:gtmd_1}), (\ref{e:gtmd_2}) and (\ref{e:gtmd_3}). For the scalar (spin-0) case we have
\begin{eqnarray}
F^{(s)}_{1,1}(x, \zeta, \bf{\Delta}_\perp, \bf{p}_{\perp})&=& \frac{1}{16 \pi^3}\bigg[ \bigg({\bf{p}}^2_\perp-\frac{(1-x')(1-x'')}{4}  {\bf \Delta} ^2_\perp+(p_x \Delta_x + p_y \Delta_y)\frac{(x'-x'')}{2} \bigg) \nonumber\\
&+& (m+x' M)(m+x'' M)\bigg]\frac{1}{x' x''}\  \varphi^{\dagger}(x'',\textbf{p}''_{\perp}) \ \varphi(x',\textbf{p}'_{\perp}),\label{f11s}\\
F^{(s)}_{1,2}(x, \zeta, \bf{\Delta}_\perp, \bf{p}_{\perp}) &=& \frac{1}{16 \pi^3}\ \frac{M}{x' x''} \big[(m+x'' M)-(m+x' M)\big]\ \varphi^{\dagger}(x'',\textbf{p}''_{\perp}) \ \varphi(x',\textbf{p}'_{\perp}), \\
F^{(s)}_{1,3} (x, \zeta, \bf{\Delta}_\perp, \bf{p}_{\perp})&=& \frac{F_{1,1}}{2}-\frac{1}{(2)16 \pi^3} \ \frac{M}{x' x''} \bigg[ (m+x'' M)(1-x')+ (m+x' M)(1-x'')\bigg]\nonumber\\
&\times &  \varphi^{\dagger}(x'',\textbf{p}''_{\perp})  \varphi(x',\textbf{p}'_{\perp}), \\
F^{(s)}_{1,4}(x, \zeta, \bf{\Delta}_\perp, \bf{p}_{\perp}) &=& \frac{1}{2(16 \pi^3)} \ M^2  \ \frac{(2-x'-x'')}{x' x''} \ \varphi^{\dagger}(x'',\textbf{p}''_{\perp}) \ \varphi(x',\textbf{p}'_{\perp}),\label{f14s} \\
G^{(s)}_{1,1}(x, \zeta, \bf{\Delta}_\perp, \bf{p}_{\perp})&=& - \frac{1}{16 \pi^3}\ M^2 \ \frac{(2-x'-x'')}{x' x''} \ \varphi^{\dagger}(x'',\textbf{p}''_{\perp}) \ \varphi(x',\textbf{p}'_{\perp}),\label{g11s} \\
G^{(s)}_{1,2}(x, \zeta, \bf{\Delta}_\perp, \bf{p}_{\perp}) &=& \frac{1}{16 \pi^3} \ \frac{M}{2}\frac{1}{x' x''} \big[(m+x' M)+(m+x'' M)\big] \ \varphi^{\dagger}(x'',\textbf{p}''_{\perp}) \ \varphi(x',\textbf{p}'_{\perp}), \\
G^{(s)}_{1,3}(x, \zeta, \bf{\Delta}_\perp, \bf{p}_{\perp})&=& \frac{1}{16 \pi^3}\ \frac{ M}{2} \big[(m+x'M)(1-x'')-(m+x'' M)(1-x')\big]\nonumber\\
& \times & \varphi^{\dagger}(x'',\textbf{p}''_{\perp}) \ \varphi(x',\textbf{p}'_{\perp}), \\
G^{(s)}_{1,4}(x, \zeta, \bf{\Delta}_\perp, \bf{p}_{\perp}) &=&  -\frac{1}{16 \pi^3}\bigg[ \bigg(\textbf{p}^2_\perp -\frac{(1-x')(1-x'')}{4}  {\bf \Delta} ^2_\perp+(p_x \Delta_x + p_y \Delta_y)\frac{(x'-x'')}{2} \bigg) \nonumber\\
&-&(m+x' M)(m+x'' M)\bigg]\frac{1}{x' x''}\ \varphi^{\dagger}(x'',\textbf{p}''_{\perp}) \ \varphi(x',\textbf{p}'_{\perp}), \\
H^{(s)}_{1,1}(x, \zeta, \bf{\Delta}_\perp, \bf{p}_{\perp}) &=& \frac{1}{16 \pi^3}\ \frac{M^2}{x' x''} \big[(m+x' M)-(m+x'' M)\big]\ \varphi^{\dagger}(x'',\textbf{p}''_{\perp}) \ \varphi(x',\textbf{p}'_{\perp}), \\
H^{(s)}_{1,2}(x, \zeta, \bf{\Delta}_\perp, \bf{p}_{\perp})&=& \frac{1}{16 \pi^3}\ \frac{M^2}{2}\frac{1}{x' x''}\big[(m+x' M)(1-x'')+(m+x''M)(1-x')\big]\nonumber\\ 
&\times & \varphi^{\dagger}(x'',\textbf{p}''_{\perp}) \ \varphi(x',\textbf{p}'_{\perp}), \\
H^{(s)}_{1,3}(x, \zeta, \bf{\Delta}_\perp, \bf{p}_{\perp})&=& -\frac{1}{16 \pi^3}\big[\textbf{p}_{\perp}^2+(p_x \Delta_x + p_y \Delta_y )(x'-x'')\big]\frac{1}{x' x''}\ \varphi^{\dagger}(x'',\textbf{p}''_{\perp})\ \varphi(x',\textbf{p}'_{\perp}), \nonumber\\
\\
H^{(s)}_{1,4}(x, \zeta, \bf{\Delta}_\perp, \bf{p}_{\perp})&=& \frac{2}{16 \pi^3}\ \frac{ M^2} {x' x''} \ \phi^{\dagger}(x'',\textbf{p}''_{\perp}) \ \phi(x',\textbf{p}'_{\perp}), \\
H^{(s)}_{1,5}(x, \zeta, \bf{\Delta}_\perp, \bf{p}_{\perp})&=& \frac{1}{2(16 \pi^3)}\ \frac{M^2}{2 } \frac{(x''-x')}{x' x''}\ \varphi^{\dagger}(x'',\textbf{p}''_{\perp}) \ \varphi(x',\textbf{p}'_{\perp}), \\
H^{(s)}_{1,6}(x, \zeta, \bf{\Delta}_\perp, \bf{p}_{\perp})&=& \frac{1}{16 \pi^3} \ \frac{M^2}{2} \frac{(1-x')(1-x'')}{2 x' x''}\ \varphi^{\dagger}(x'',\textbf{p}''_{\perp}) \ \varphi(x',\textbf{p}'_{\perp}), \\
H^{(s)}_{1,7}(x, \zeta, \bf{\Delta}_\perp, \bf{p}_{\perp}) &=& -\frac{1}{2(16 \pi^3)} \ \frac{M}{x' x''} \big[(m+ x' M)+(m+x'' M)\big]\ \varphi^{\dagger}(x'',\textbf{p}''_{\perp}) \ \varphi(x',\textbf{p}'_{\perp}),\nonumber\\
\\
H^{(s)}_{1,8}(x, \zeta, \bf{\Delta}_\perp, \bf{p}_{\perp}) &=& \frac{1}{2(16 \pi^3)} \ \frac{M}{2\ x' x''} \big[(m+ x'' M)(1-x')+(m+x' M)(1-x'')\big]\nonumber\\
&\times & \varphi^{\dagger}(x'',\textbf{p}''_{\perp}) \ \varphi(x',\textbf{p}'_{\perp}),
\end{eqnarray}
and for the axial-vector (spin-1) case we have
\begin{eqnarray}
F^{(a)}_{1,1}(x, \zeta, \bf{\Delta}_\perp, \bf{p}_{\perp})&=& \frac{1}{16 \pi^3}\bigg[ \bigg({\bf{p}}^2_\perp-\frac{(1-x')(1-x'')}{4}  {\bf \Delta} ^2_\perp+(p_x \Delta_x + p_y \Delta_y)\frac{(x'-x'')}{2} \bigg) \nonumber\\
&\times &\frac{(1+x' x'')}{x' x''(1-x')(1-x'')}
+\frac{(m+x' M)(m+x'' M)}{x' x''}\bigg]\nonumber\\
&\times & \phi^{\dagger}(x'',\textbf{p}''_{\perp}) \ \phi(x',\textbf{p}'_{\perp}), \label{f11}\\
F^{(a)}_{1,2}(x, \zeta, \bf{\Delta}_\perp, \bf{p}_{\perp}) &=& \frac{1}{16 \pi^3}\ M \bigg[\frac{m+x' M}{x'(1-x'')}-\frac{m+x'' M}{x''(1-x')}\bigg]\ \phi^{\dagger}(x'',\textbf{p}''_{\perp}) \ \phi(x',\textbf{p}'_{\perp}), \\
F^{(a)}_{1,3} (x, \zeta, \bf{\Delta}_\perp, \bf{p}_{\perp})&=& \frac{F_{1,1}}{2}+\frac{1}{16 \pi^3} \ M \bigg[ \frac{m+x' M}{x'}+ \frac{m+x'' M}{x''}\bigg] \ \phi^{\dagger}(x'',\textbf{p}''_{\perp}) \ \phi(x',\textbf{p}'_{\perp}), \\
F^{(a)}_{1,4}(x, \zeta, \bf{\Delta}_\perp, \bf{p}_{\perp}) &=& \frac{1}{2(16 \pi^3)} \ M^2  \ \frac{(1-x' x'')(2-x'-x'')}{x' x'' (1-x')(1-x'')} \ \phi^{\dagger}(x'',\textbf{p}''_{\perp}) \ \phi(x',\textbf{p}'_{\perp}), 
\label{f14}\\
G^{(a)}_{1,1}(x, \zeta, \bf{\Delta}_\perp, \bf{p}_{\perp})&=& - \frac{1}{16 \pi^3}\ M^2 \ \frac{(1-x' x'')(2-x'-x'')}{x' x'' (1-x')(1-x'')} \ \phi^{\dagger}(x'',\textbf{p}''_{\perp}) \ \phi(x',\textbf{p}'_{\perp}), 
\label{g11}\\
G^{(a)}_{1,2}(x, \zeta, \bf{\Delta}_\perp, \bf{p}_{\perp}) &=& -\frac{1}{16 \pi^3} \ M \bigg[\frac{m+x' M}{x' (1-x'')}+\frac{m+x'' M}{x''(1-x')}\bigg] \ \phi^{\dagger}(x'',\textbf{p}''_{\perp}) \ \phi(x',\textbf{p}'_{\perp}), \\
G^{(a)}_{1,3}(x, \zeta, \bf{\Delta}_\perp, \bf{p}_{\perp})&=&- \frac{1}{16 \pi^3}\ \frac{ M}{2} \bigg[\frac{m+x'M}{x'}-\frac{m+x'' M}{x''}\bigg]\ \phi^{\dagger}(x'',\textbf{p}''_{\perp}) \ \phi(x',\textbf{p}'_{\perp}), \\
G^{(a)}_{1,4}(x, \zeta, \bf{\Delta}_\perp, \bf{p}_{\perp}) &=&  \frac{1}{16 \pi^3}\bigg[ \bigg(\textbf{p}^2_\perp-\frac{(1-x')(1-x'')}{4}  {\bf \Delta} ^2_\perp+(p_x \Delta_x + p_y \Delta_y)\frac{(x'-x'')}{2} \bigg) \nonumber\\
& \times &\frac{(1+x' x'')}{x' x''(1-x')(1-x'')}
-\frac{(m+x' M)(m+x'' M)}{x' x''}\bigg]\nonumber\\
&\times & \phi^{\dagger}(x'',\textbf{p}''_{\perp}) \ \phi(x',\textbf{p}'_{\perp}), \\
H^{(a)}_{1,1}(x, \zeta, \bf{\Delta}_\perp, \bf{p}_{\perp}) &=& \frac{1}{16 \pi^3}\ M^2 \bigg[\frac{m+x' M}{x' x'' (1-x'')}-\frac{m+x'' M}{x' x''(1-x')}\bigg]\ \phi^{\dagger}(x'',\textbf{p}''_{\perp}) \ \phi(x',\textbf{p}'_{\perp}), \\
H^{(a)}_{1,2}(x, \zeta, \bf{\Delta}_\perp, \bf{p}_{\perp})&=& \frac{1}{16 \pi^3}\ M^2\bigg[\frac{m+x' M}{x' x''}+\frac{m+x''M}{x' x''}\bigg]\ \phi^{\dagger}(x'',\textbf{p}''_{\perp}) \ \phi(x',\textbf{p}'_{\perp}), \\
H^{(a)}_{1,3}(x, \zeta, \bf{\Delta}_\perp, \bf{p}_{\perp})&=& -\frac{1}{16 \pi^3}\big[\textbf{p}_{\perp}^2+(p_x \Delta_x + p_y \Delta_y )(x'-x'')\big]\frac{x'+x''}{x' x'' (1-x')(1-x'')}\nonumber\\
&\times & \phi^{\dagger}(x'',\textbf{p}''_{\perp})\ \phi(x',\textbf{p}'_{\perp}), \\
H^{(a)}_{1,4}(x, \zeta, \bf{\Delta}_\perp, \bf{p}_{\perp})&=& \frac{1}{16 \pi^3}\ M^2 \frac{(x'-x'')}{x' x'' (1-x')(1-x'')}\ \phi^{\dagger}(x'',\textbf{p}''_{\perp}) \ \phi(x',\textbf{p}'_{\perp}), \\
H^{(a)}_{1,5}(x, \zeta, \bf{\Delta}_\perp, \bf{p}_{\perp})&=& \frac{1}{2(16 \pi^3)}\ M^2 \frac{(x'-x'')^2}{x' x'' (1-x')(1-x'')}\ \phi^{\dagger}(x'',\textbf{p}''_{\perp}) \ \phi(x',\textbf{p}'_{\perp}), \\
H^{(a)}_{1,6}(x, \zeta, \bf{\Delta}_\perp, \bf{p}_{\perp})&=& -\frac{1}{16 \pi^3} \ M^2 \frac{(x'+x'')}{x' x''}\ \phi^{\dagger}(x'',\textbf{p}''_{\perp}) \ \phi(x',\textbf{p}'_{\perp}), \\
H^{(a)}_{1,7}(x, \zeta, \bf{\Delta}_\perp, \bf{p}_{\perp}) &=& \frac{1}{2(16 \pi^3)} \ M \bigg[\frac{m+ x' M}{x' x''(1-x'')}+\frac{m+x'' M}{x' x'' (1-x')}\bigg]\ \phi^{\dagger}(x'',\textbf{p}''_{\perp}) \ \phi(x',\textbf{p}'_{\perp}), \\
H^{(a)}_{1,8}(x, \zeta, \bf{\Delta}_\perp, \bf{p}_{\perp})&=& \frac{1}{2(16 \pi^3)} M \bigg[\frac{m+x'M}{x' x''}-\frac{m+x'' M}{x' x''}\bigg]\ \phi^{\dagger}(x'',\textbf{p}''_{\perp}) \ \phi(x',\textbf{p}'_{\perp}).
  \end{eqnarray}
 Here the initial and final momenta of the active quark are expressed as
  \begin{eqnarray}
  {\bf p'_\perp}={\bf p_\perp}- (1-x') \frac{{\bf \Delta_\perp}}{2}, \ \ \ \ \ \ \ \ \ \ {\rm with}\ \ \ \ \ \ \ \ \ \ x'=\frac{x+\zeta}{1+\zeta};
 \\
 {\bf p''_\perp}={\bf p_\perp}+ (1-x'') \frac{{\bf \Delta_\perp}}{2}, \ \ \ \ \ \ \ \ \ \ {\rm with}\ \ \ \ \ \ \ \ \ \ x''=\frac{x-\zeta}{1-\zeta}.
  \end{eqnarray}

\section{Results and discussion}
The calculations of the Wigner distributions and GTMDs involve the inputs for the diquark mass ($M_s$) and the coupling $(c_s)$ for the scalar isoscalar diquark, the axial-vector diquark mass ($M_a$) and the coupling corresponding to $ud$-system with $I_z=0$ ($g_a$) as well as  the axial-vector diquark mass ($M'_a$) and coupling corresponding to $uu$-system with $I_z=1$ ($g'_a$).  The diquark mass and the couplings have been summarized in Table \ref{table1}. The constituent quark mass is taken to be $m=0.33$ GeV in the present calculations. We choose $\zeta < x<1$ region to plot GTMDs and for Wigner distributions, we choose the frame where $\zeta=0$ i.e. $0 < x<1$.  It is important to mention here that this model is not applicable in the  $x \longrightarrow 1$ regime \cite{high-x}. 
\begin{table}[ht]
\begin{center}
\begin{tabular}{@{\hspace{18pt}} c @{\hspace{12pt}} ||
@{\hspace{12pt}} c @{\hspace{12pt}} || @{\hspace{12pt}} c
@{\hspace{12pt}} }

\hline\hline
$Diquark$ & {$M_X$ in $GeV$ {\hspace{9pt}} } & {$c_X$ {\hspace{9pt}}} \\ \hline
$ud$ (Scalar $s$) & 0.822 $\pm$ 0.053 & 0.847 $\pm$ 0.111 \\
$ud$ (Axial-vector $a$) & 1.492 $\pm$ 0.173 & 1.061 $\pm$ 0.085 \\
$uu$ (Axial-vector $a'$) & 0.890 $\pm$ 0.008 & 0.880 $\pm$ 0.008 \\
\hline\hline
\end{tabular}
\caption{The diquark masses $M_X$ and couplings $c_X$ for $X=s,a, a'$ where $s$ corresponds to the scalar isoscalar, $a$ corresponds to the vector  isoscalar diquark and $a'$ corresponds to the vector isovector diquark \cite{tmd1}.}
\label{table1}
\end{center}
\end{table}

The quark flavors are related to the scalar and vector diquarks as \cite{tmd1}
\begin{eqnarray}
\omega^u &=&c_s^2\ \omega^{u(s)}+c^2_a\ \omega^{u(a)},\\
\omega^d &=&c'^2_{a}\ \omega^{d(a')},
\end{eqnarray} 
where $\omega$ can correspond to the Wigner distributions $\rho$ defined in Section III or the GTMDs $F$, $G$ and $H$ defined in Section IV.

Using the above mentioned parameters, we now plot the purely transverse Wigner distributions in impact-parameter space ($\textbf{b}_{\perp}$) and in transverse momentum space ($\textbf{p}_\perp$). 
\begin{equation}
\rho(\textbf{b}_{\perp}, \textbf{p}_{\perp}) \equiv \int dx \ \rho(\textbf{b}_{\perp}, \textbf{p}_{\perp}, x). 
\end{equation}
The Wigner distributions are the quasi-probabilistic distributions due to Heisenberg's uncertainty principle. To get the actual probabilistic results, one has to integrate Wigner distributions over $\textbf{b}_\perp$ and $\textbf{p}_\perp$ leading  to TMDs and GPDs respectively. By integrating Wigner distributions over $b_x$ and $p_y $, we can calculate the mixed probability densities $\rho({b_y, p_x})$ as follows
\begin{equation} 
\int db_{x} dp_{y} \rho(\textbf{b}_{\perp},\textbf{p}_{\perp})\equiv \bar{\rho}(b_{y},p_{x}).
\end{equation}
In Figs. (\ref{unpol})-(\ref{transt}), we plot the Wigner distributions for the unpolarized, longitudinally-polarized and transversely polarized proton with the different polarization combinations of quark ($u$ or $d$). We take the fixed quark transverse momentum and impact-parameter co-ordinate along $\hat{y}$ i.e. $p_y=0.5$ $GeV$ and $b_y=0.4$ $GeV$ to present the Wigner distribution in impact-parameter space and in momentum space respectively. We have presented the results of Wigner distributions under three broad categories: Wigner distributions for unpolarized, longitudinally-polarized and transversely-polarized proton. The quarks in the proton can further be unpolarized, longitudinally-polarized and transversely-polarized.

\subsection{Wigner distributions for unpolarized proton }
\begin{figure}
\centering
\begin{minipage}[c]{0.98\textwidth}
(a)\includegraphics[width=.3\textwidth]{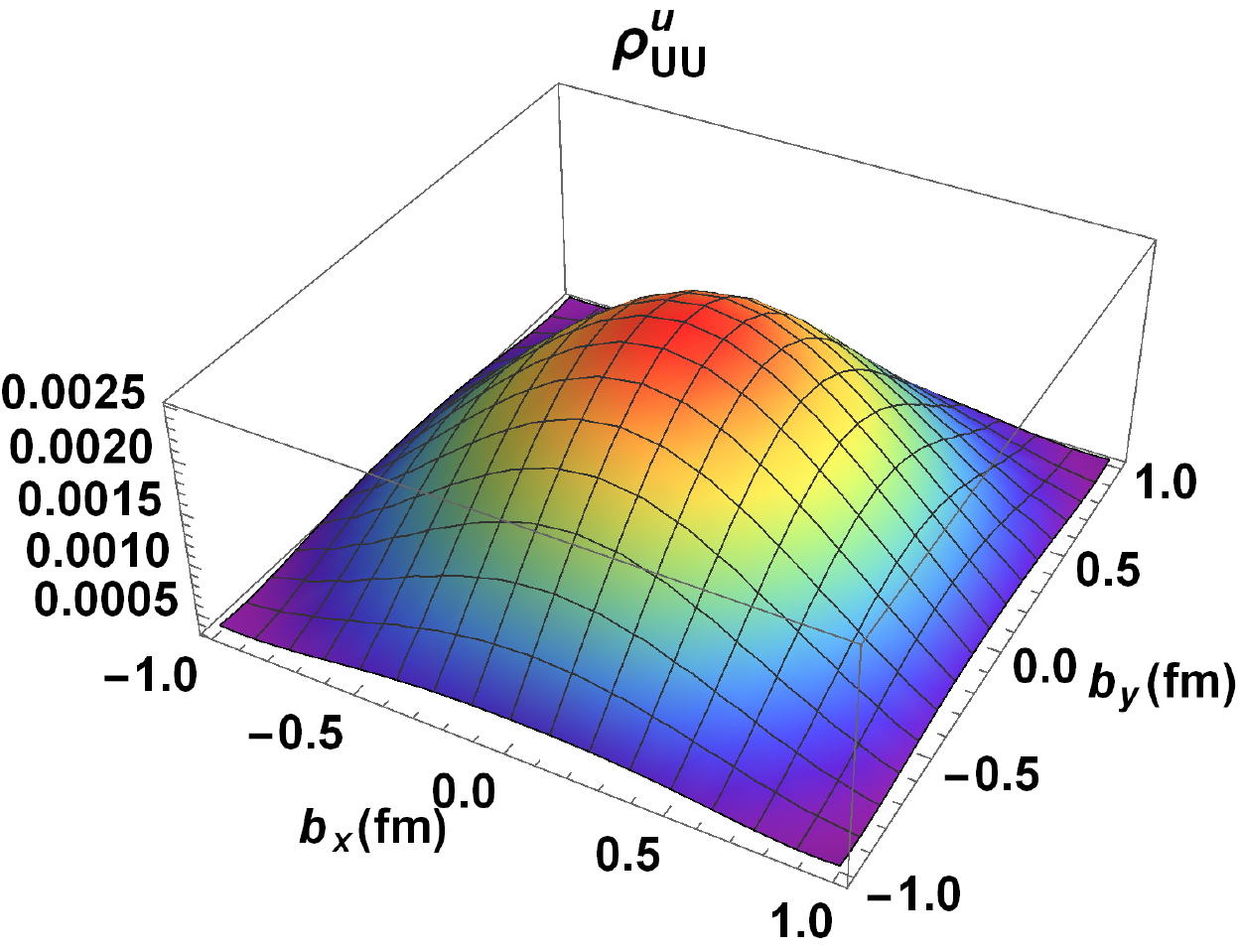}\hfill
(b)\includegraphics[width=.3\textwidth]{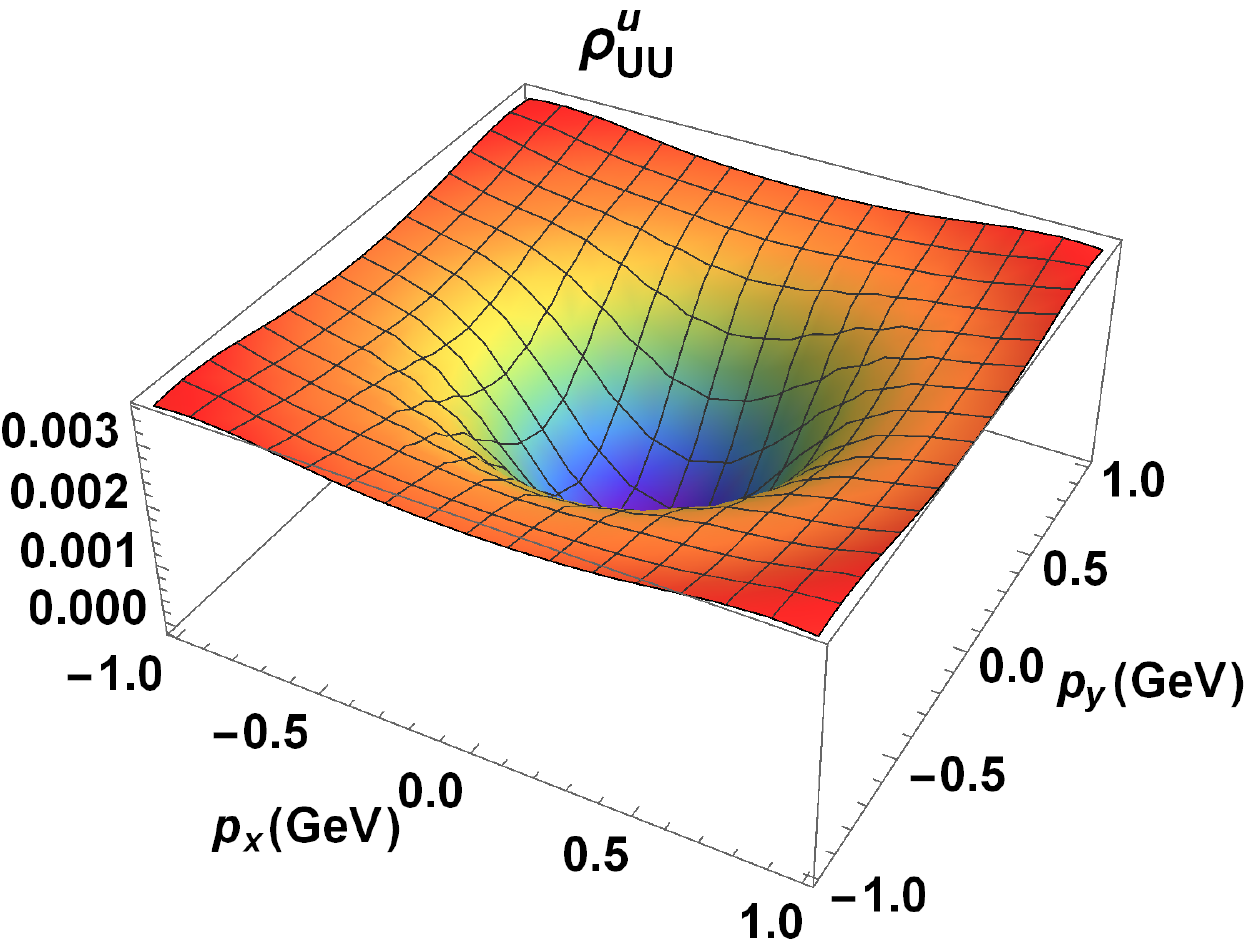}\hfill
(c)\includegraphics[width=.3\textwidth]{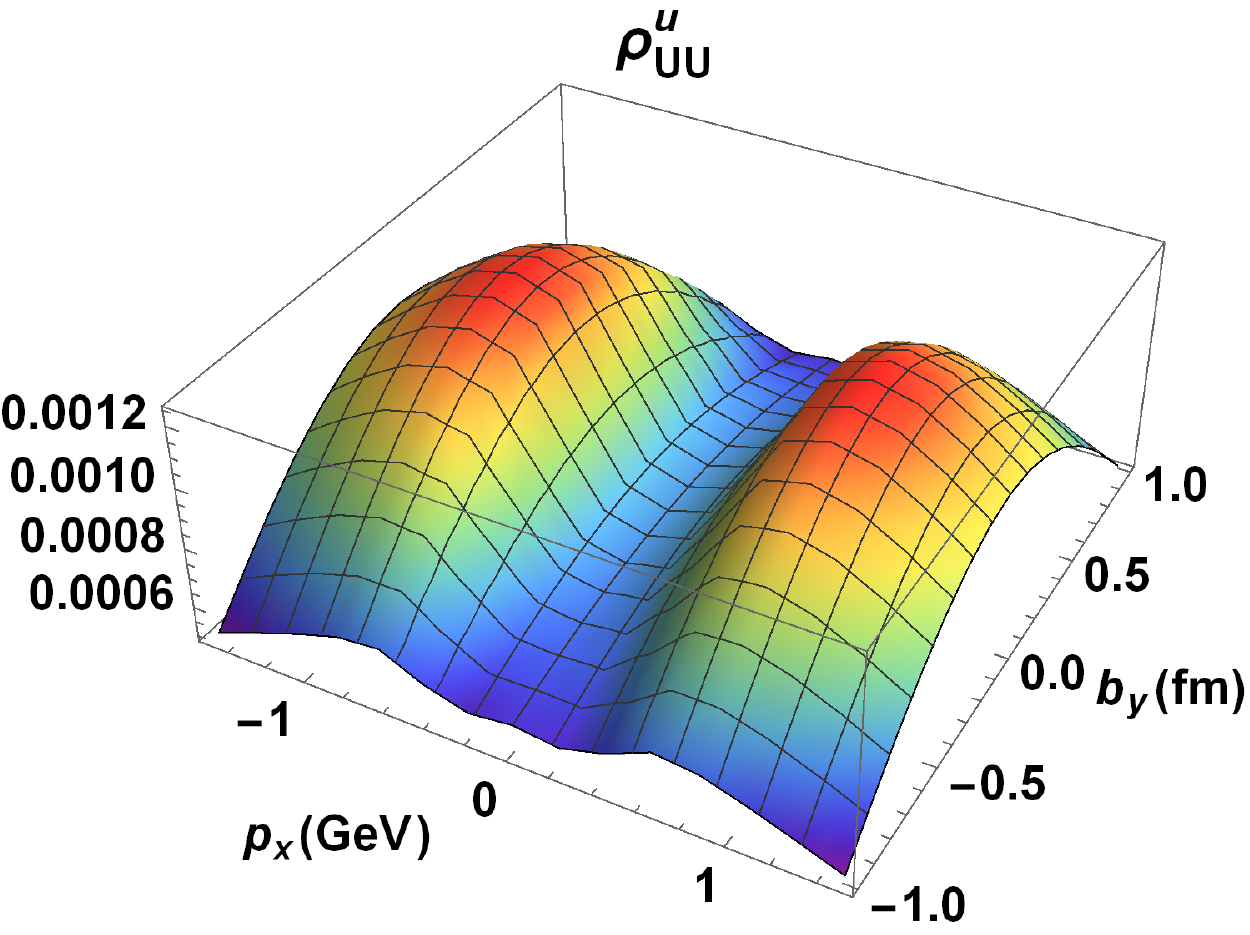}
\end{minipage}
\begin{minipage}[c]{0.98\textwidth}
(d)\includegraphics[width=.3\textwidth]{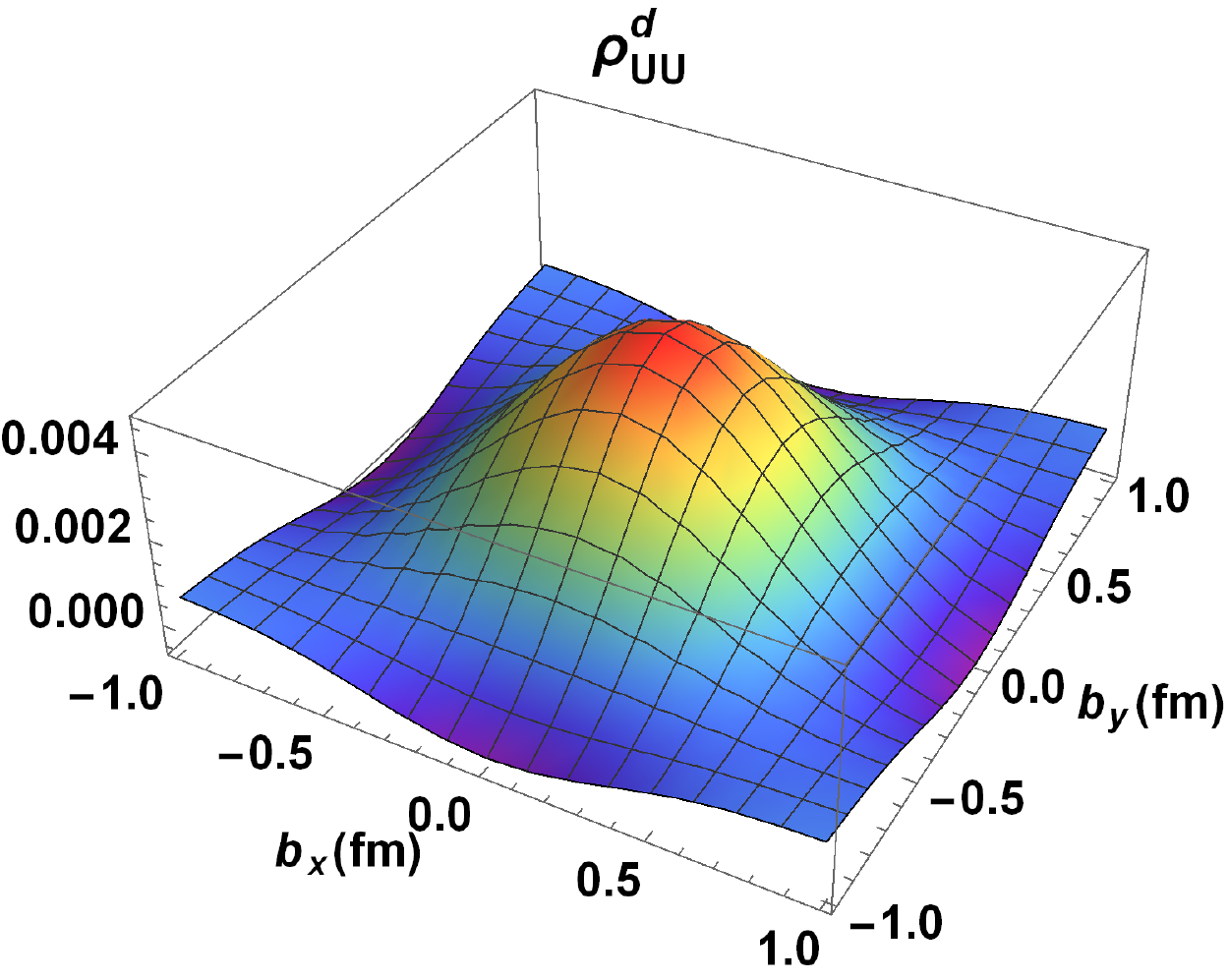}\hfill
(e)\includegraphics[width=.3\textwidth]{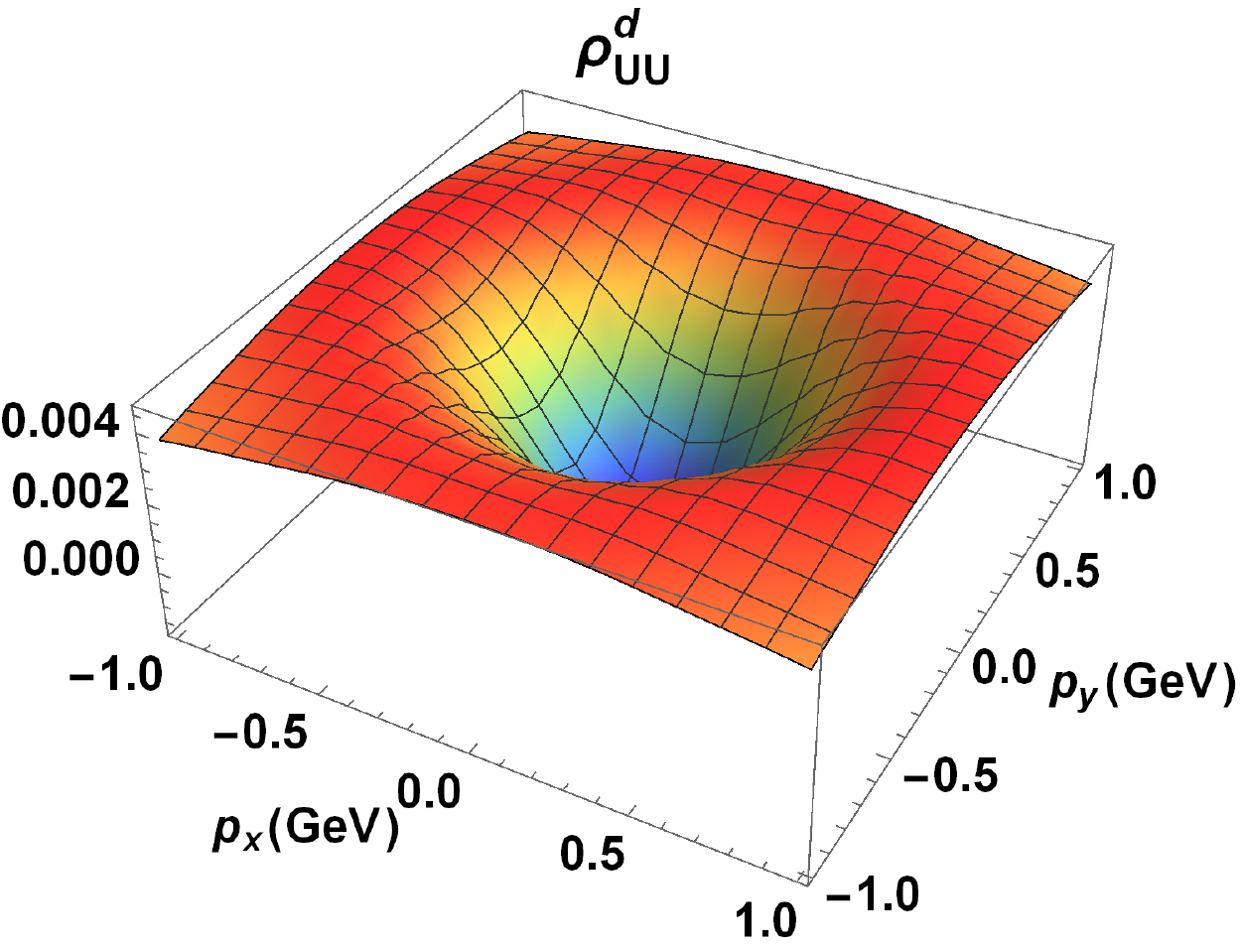}\hfill
(f)\includegraphics[width=.3\textwidth]{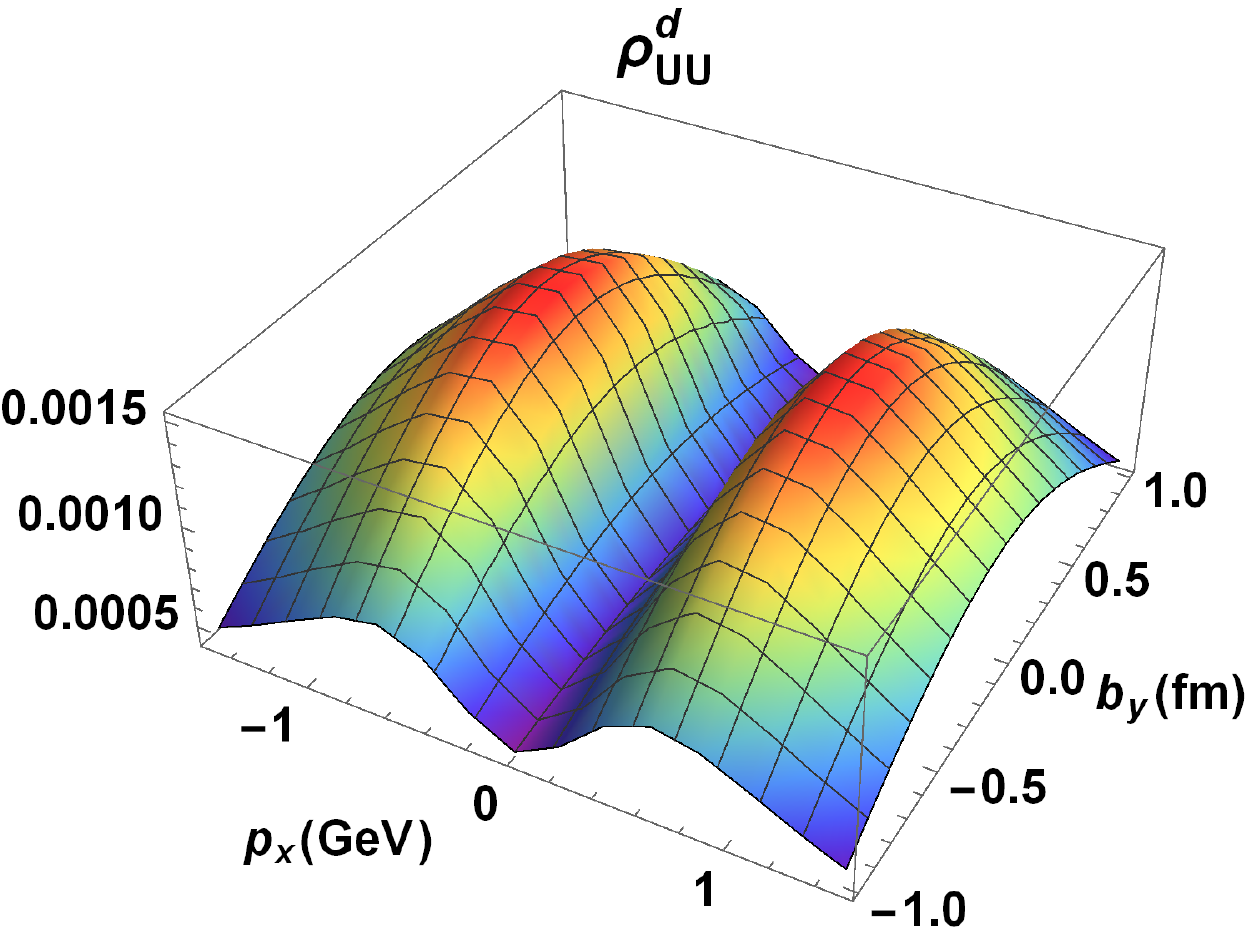}
\end{minipage} 
\caption{The unpolarized Wigner distribution $\rho_{UU}$ in impact-parameter space $(b_x,b_y)$, in momentum space $(p_x,p_y)$ and in mixed space $(p_x,b_y)$ for $u(d)$ quark  presented in  upper (lower) panel.}
\label{unpol}
\end{figure}
 \begin{figure}
 \centering 
 \begin{minipage}[c]{0.98\textwidth}
(a)\includegraphics[width=.3\textwidth]{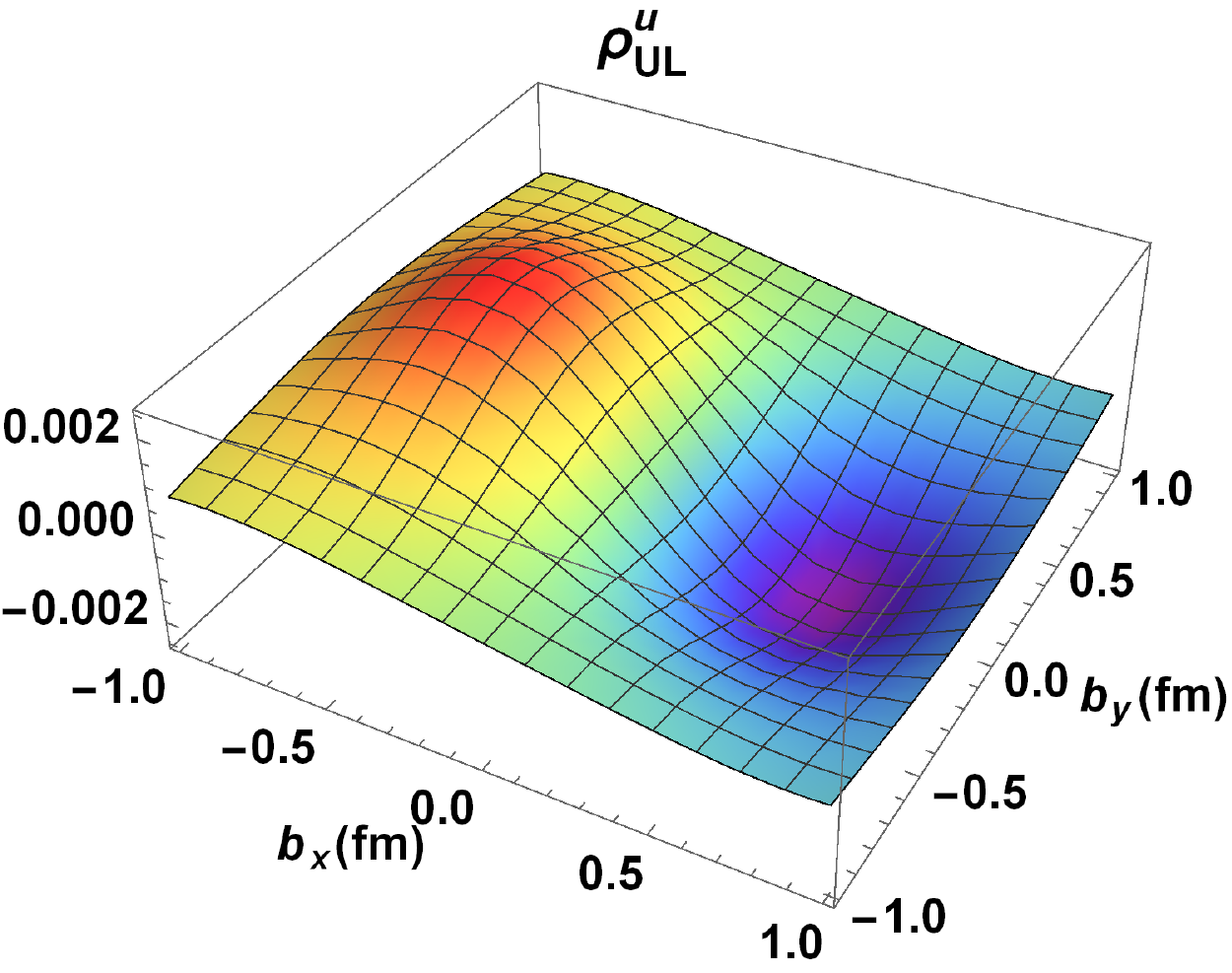}\hfill
(b)\includegraphics[width=.3\textwidth]{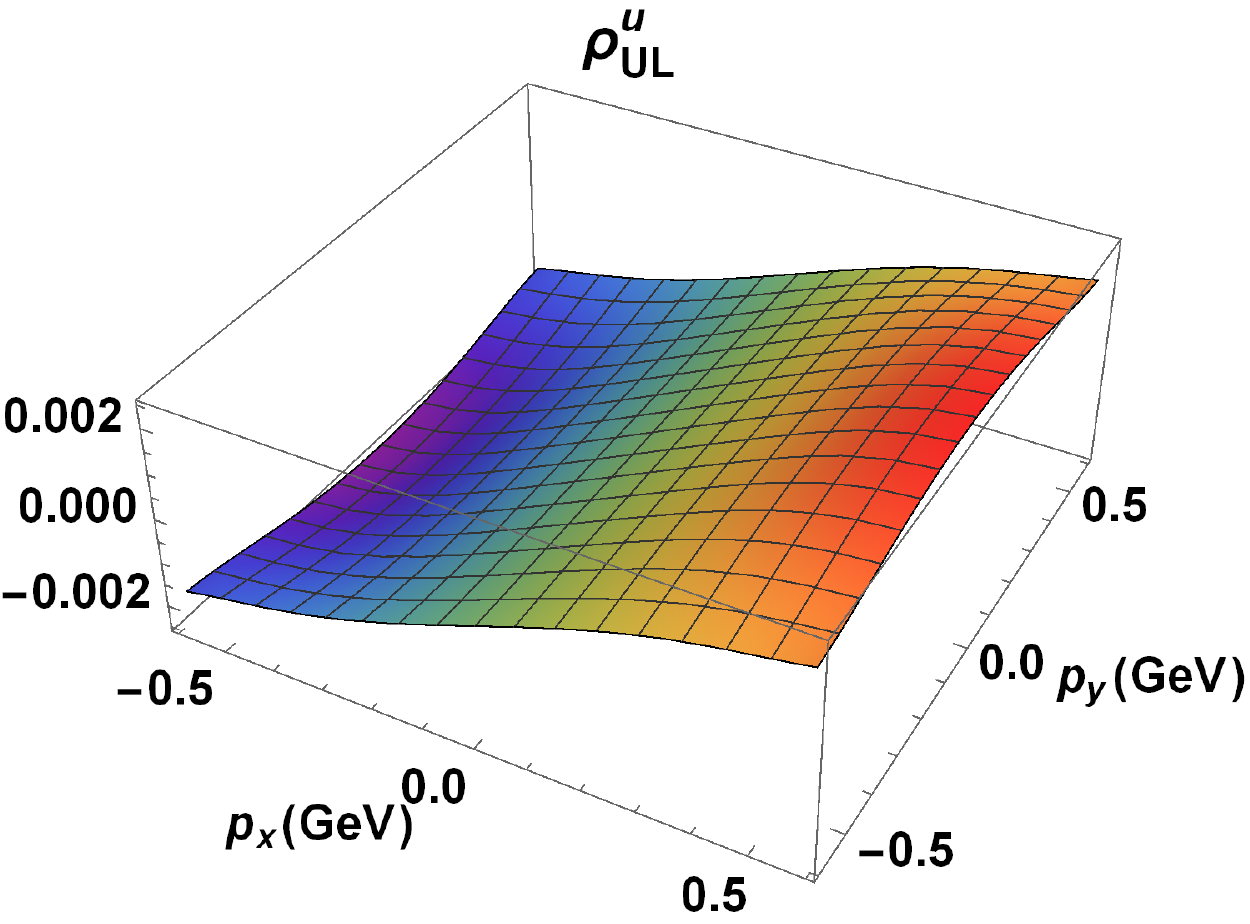}\hfill
(c)\includegraphics[width=.3\textwidth]{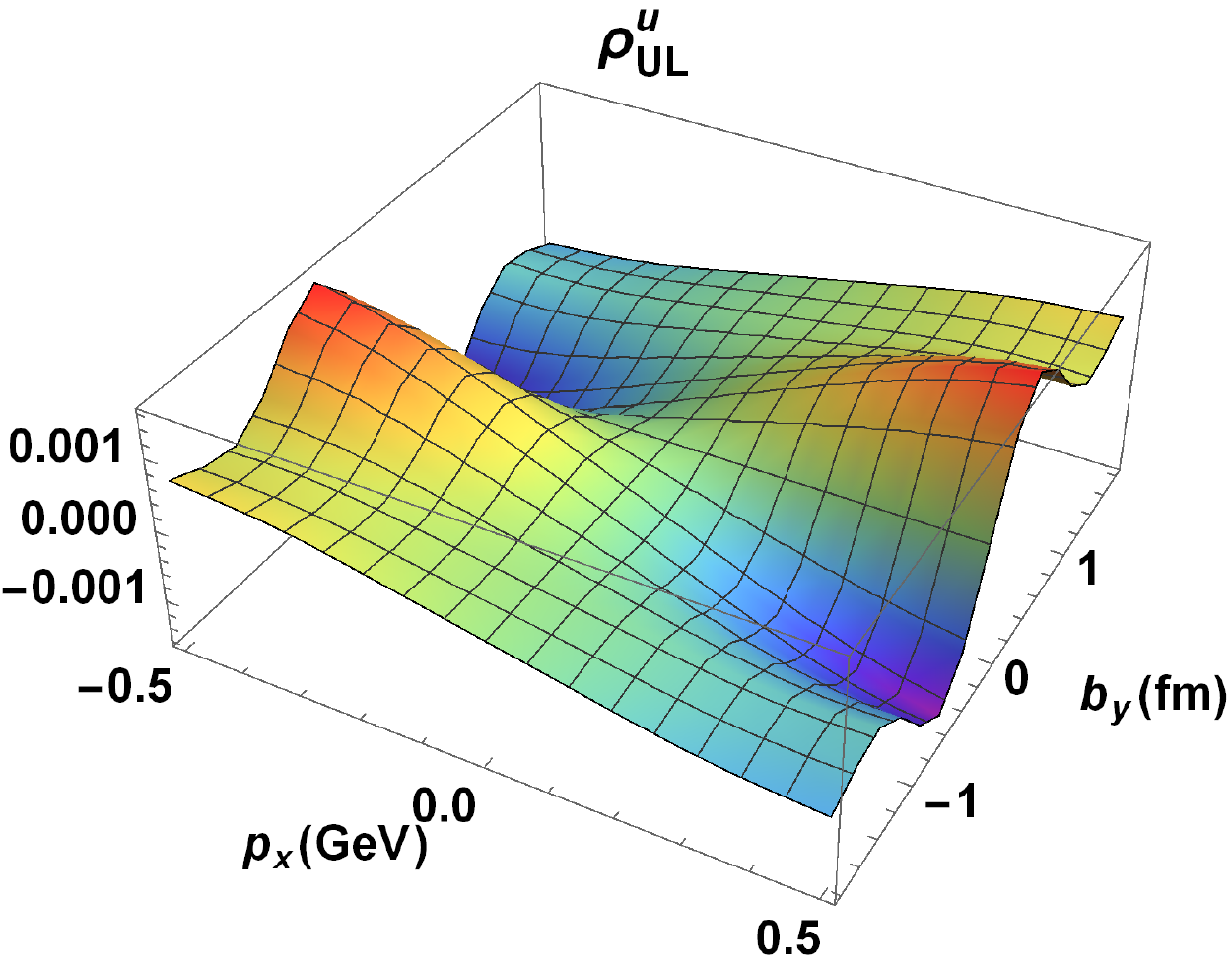}
\end{minipage}
\begin{minipage}[c]{0.98\textwidth}
(d)\includegraphics[width=.3\textwidth]{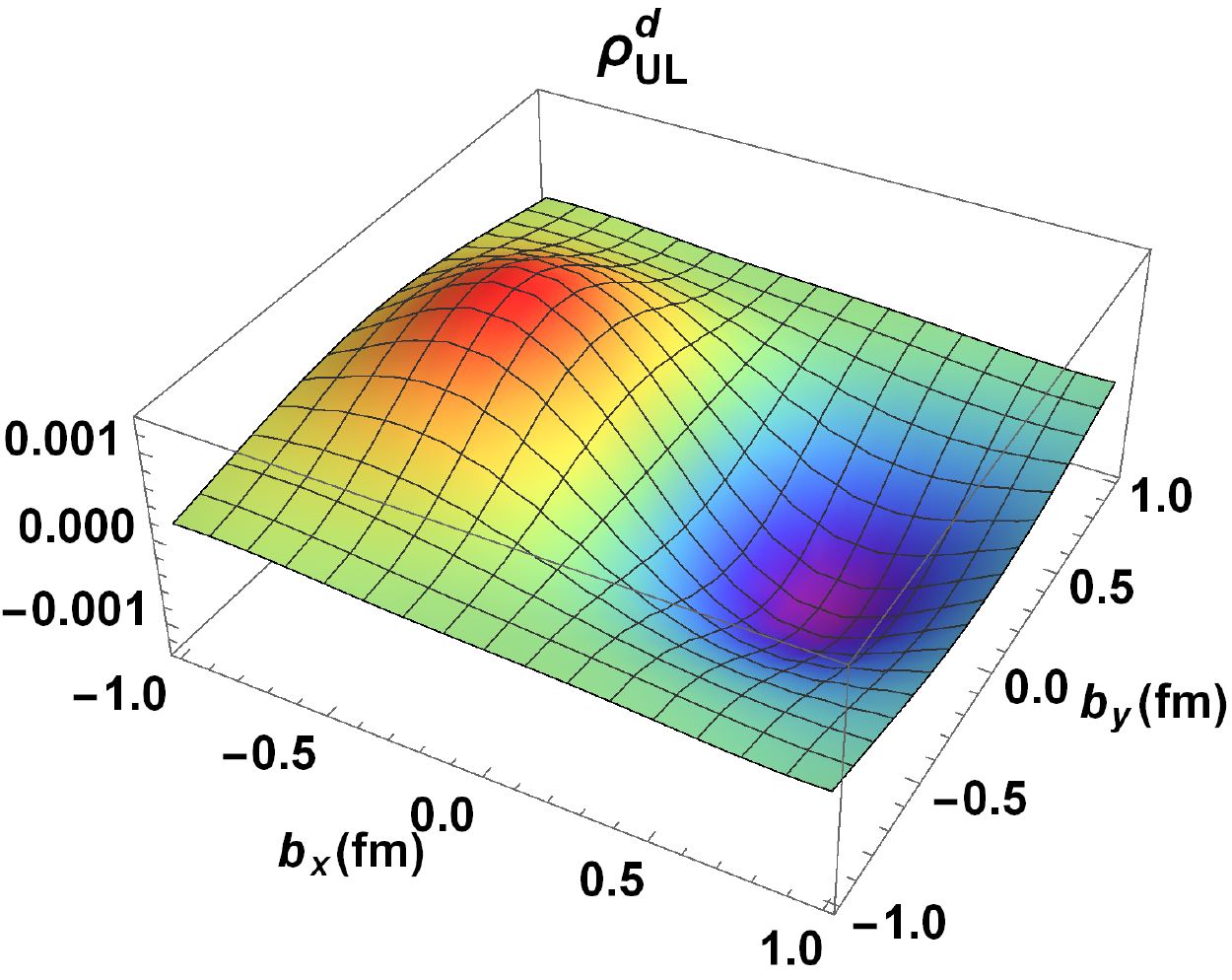}\hfill
(e)\includegraphics[width=.3\textwidth]{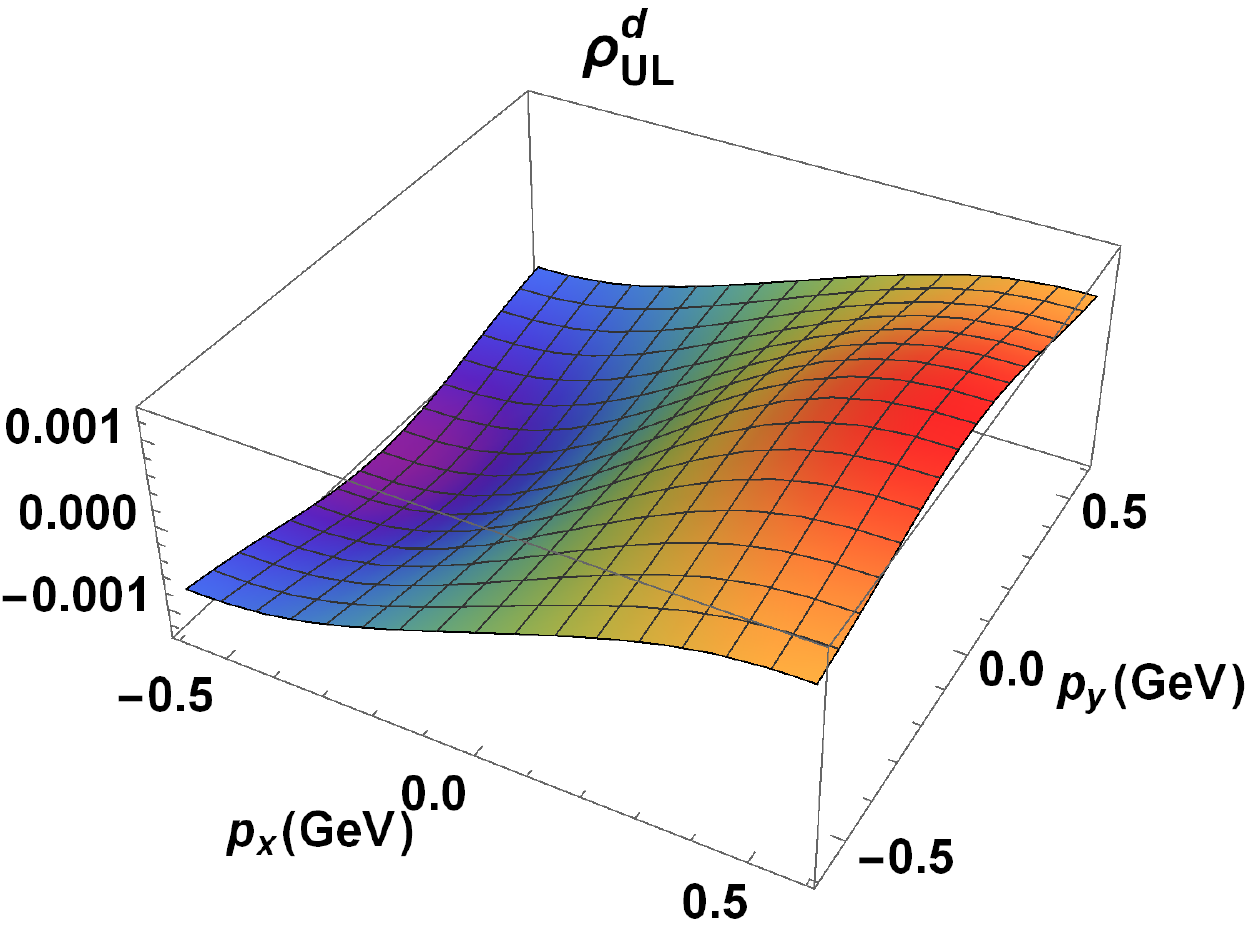}\hfill
(f)\includegraphics[width=.3\textwidth]{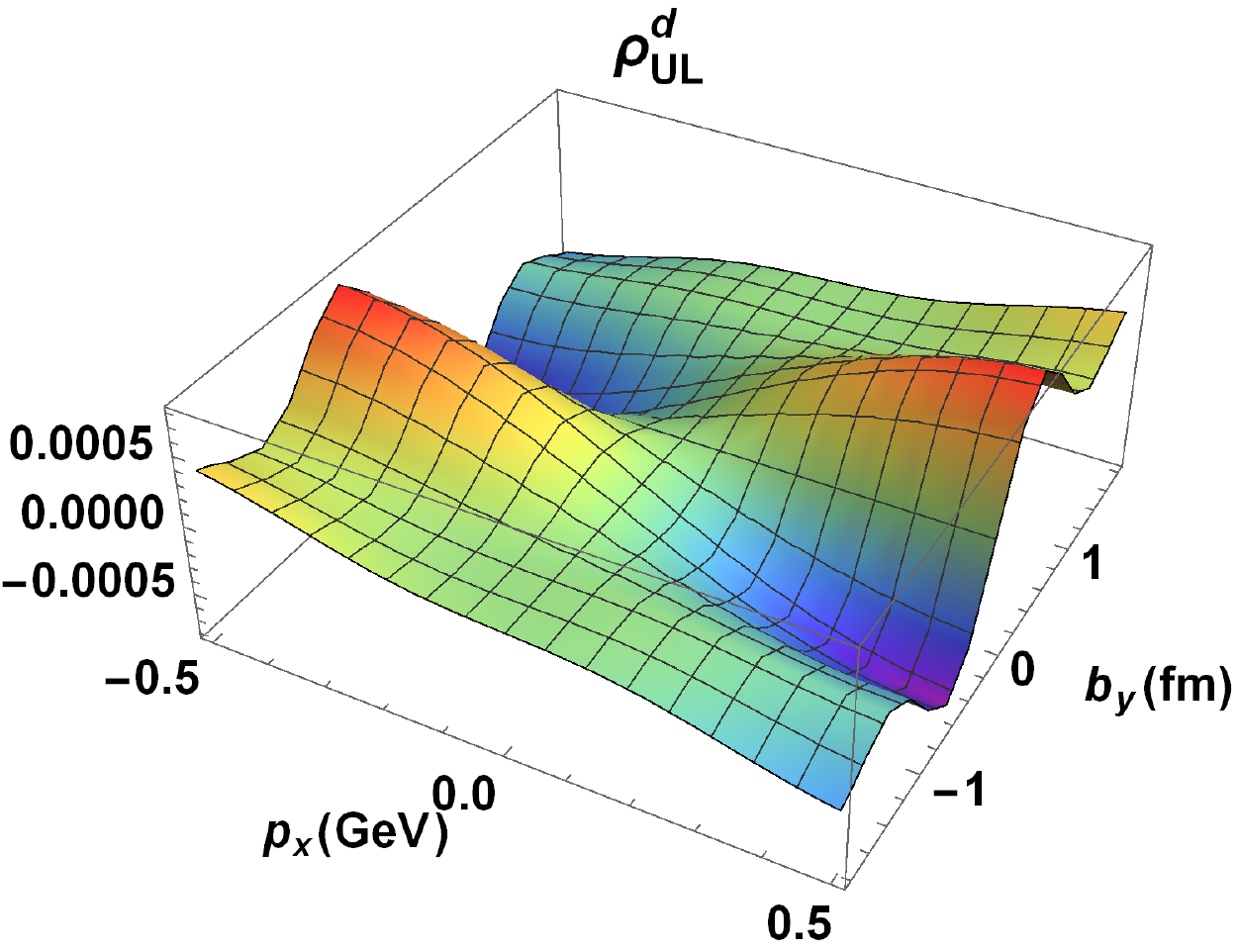}
\end{minipage}
\caption{The unpolarized-longitudinal Wigner distribution $\rho_{UL}$ in impact-parameter space $(b_x,b_y)$, in momentum space $(p_x,p_y)$ and in mixed space $(p_x,b_y)$ for $u(d)$ quark  presented in upper (lower) panel.}
\label{unpoll}
\end{figure}
 \begin{figure}
\centering 
\begin{minipage}[c]{0.98\textwidth}
(a)\includegraphics[width=.3\textwidth]{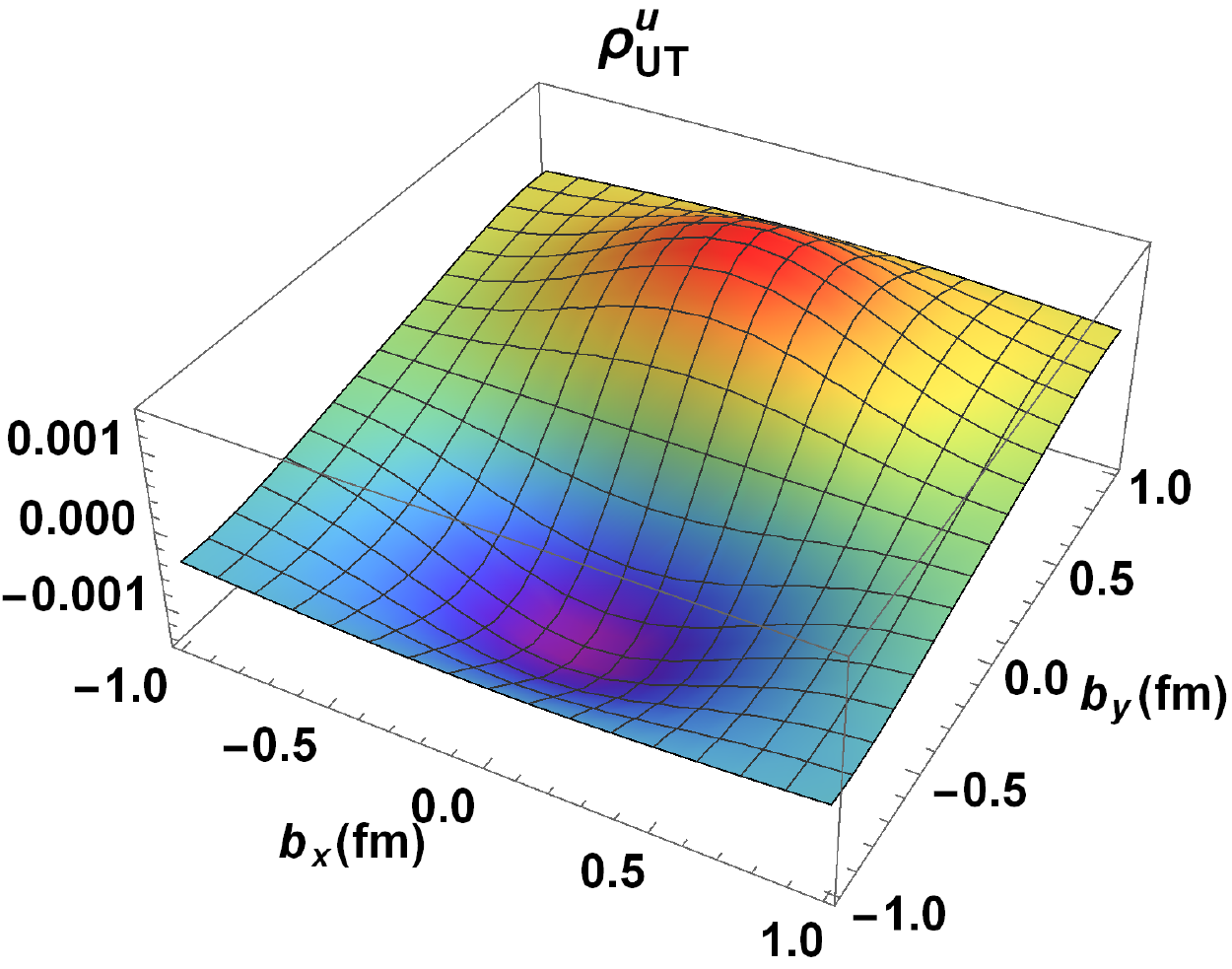}\hfill
(b)\includegraphics[width=.3\textwidth]{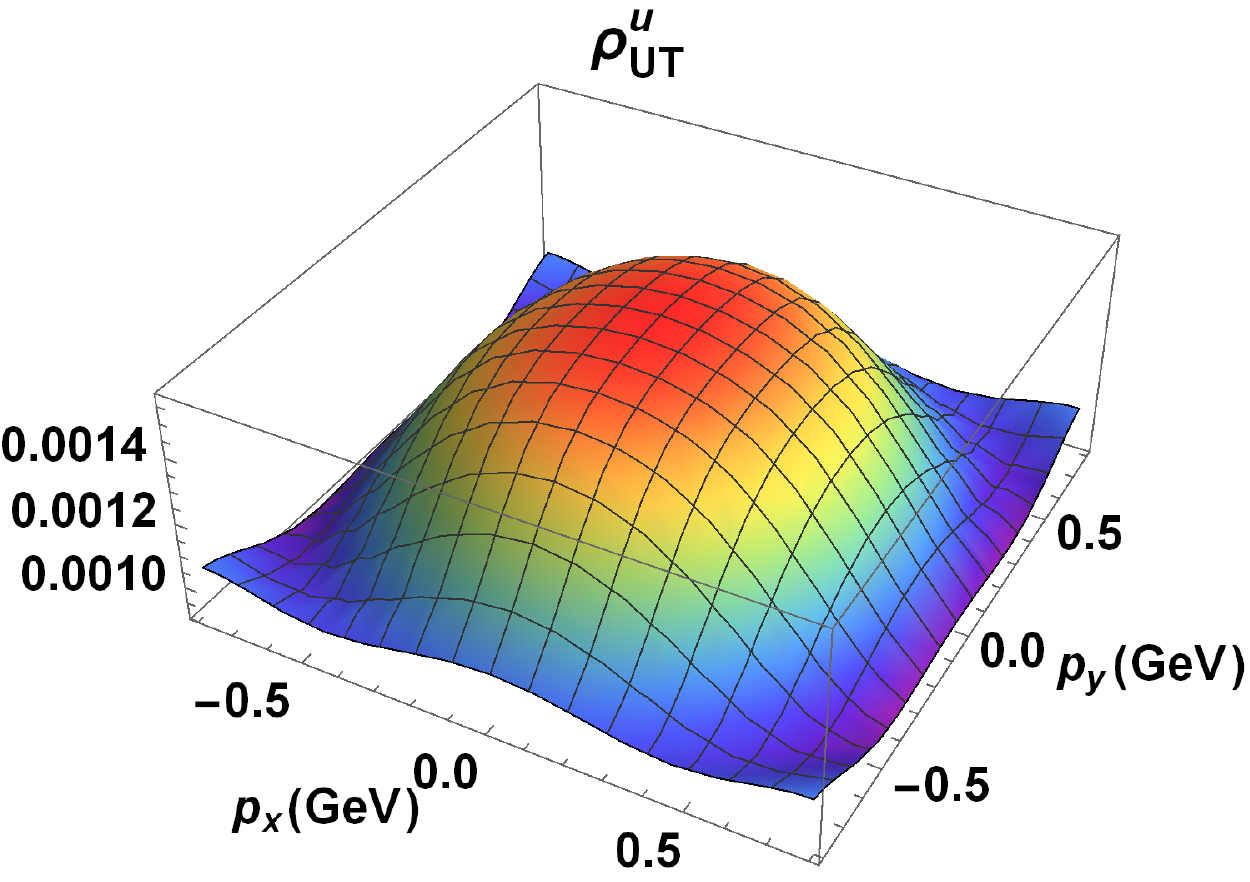}\hfill
(c)\includegraphics[width=.3\textwidth]{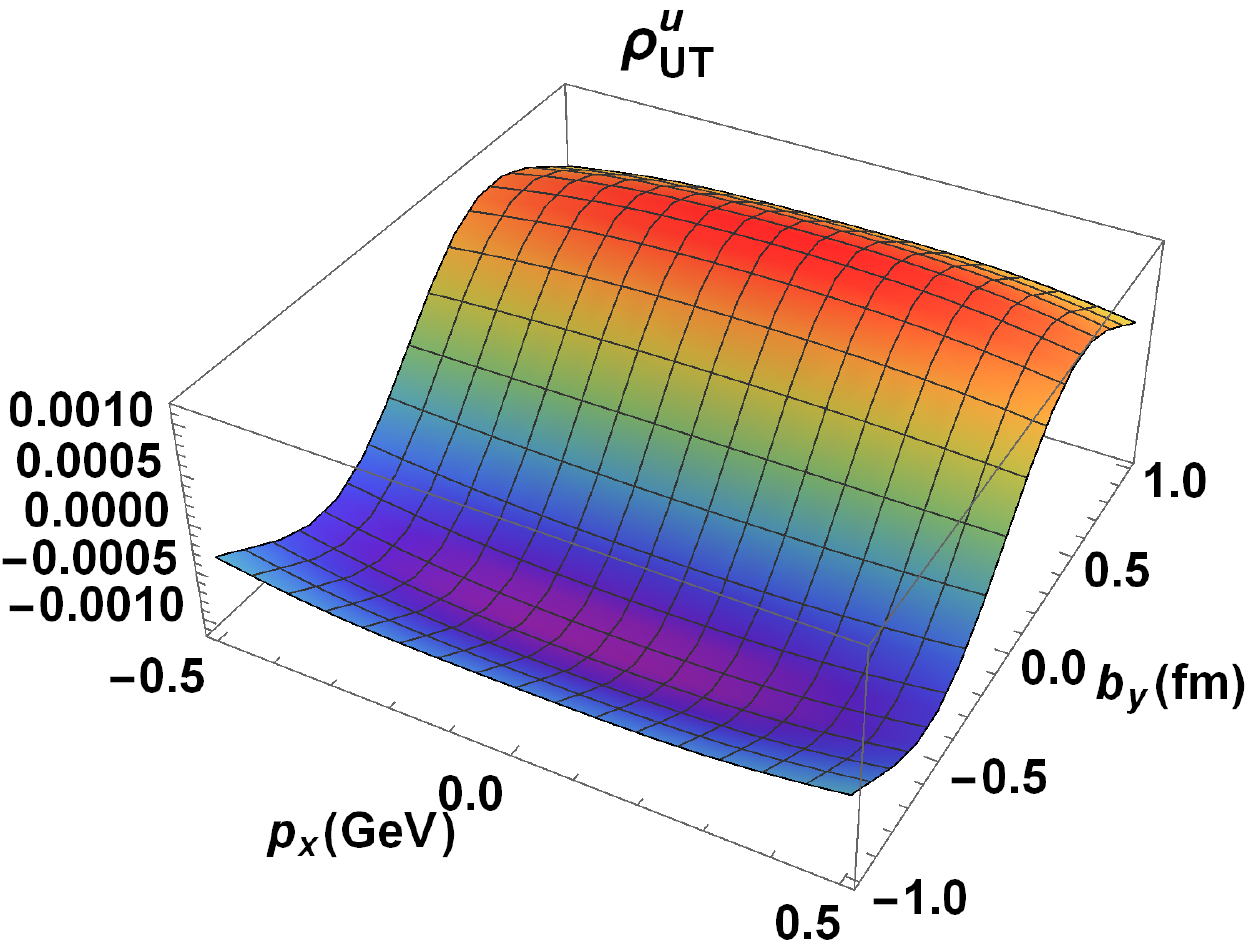}
\end{minipage}
\begin{minipage}[c]{0.98\textwidth}
(d)\includegraphics[width=.3\textwidth]{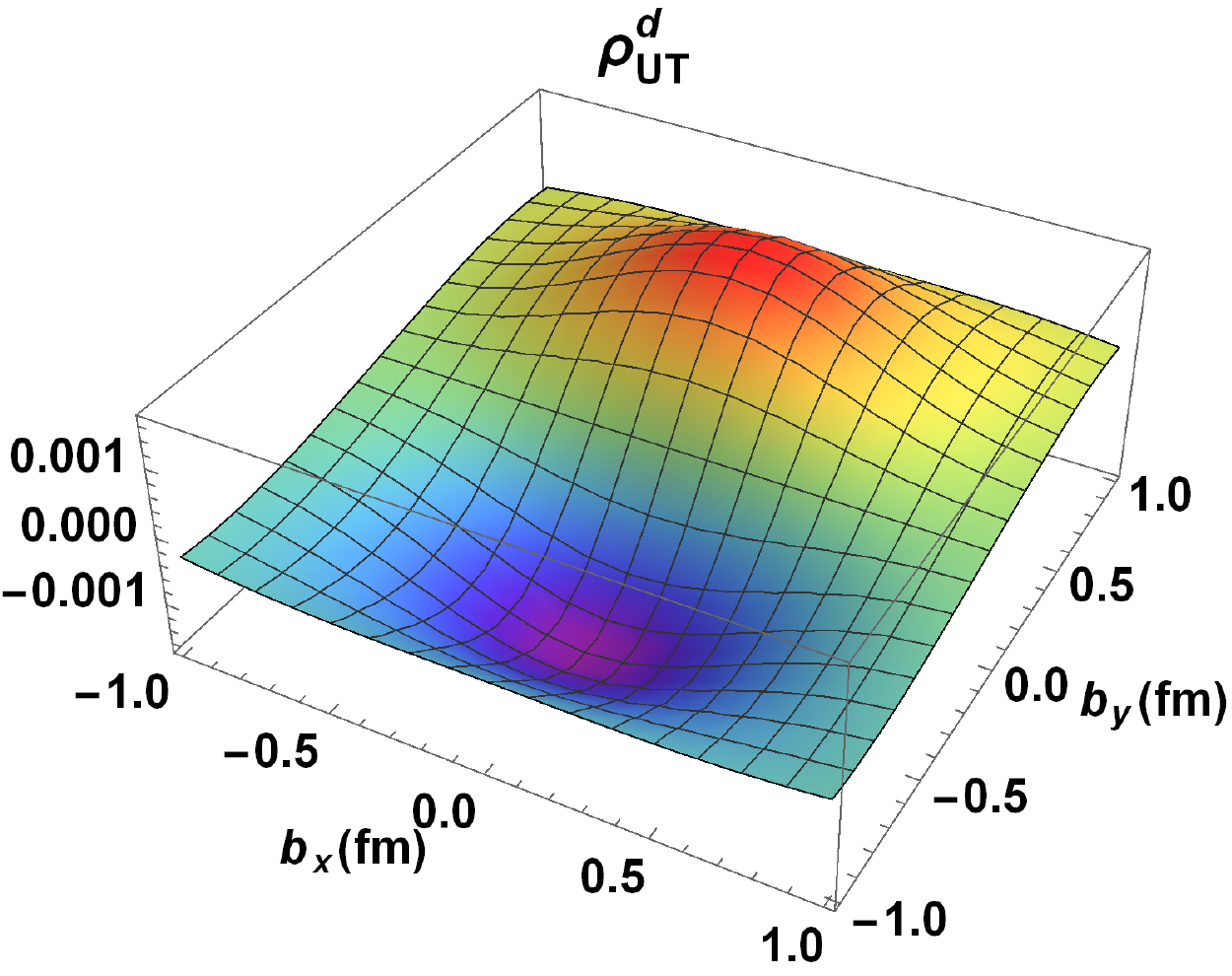}\hfill
(e)\includegraphics[width=.3\textwidth]{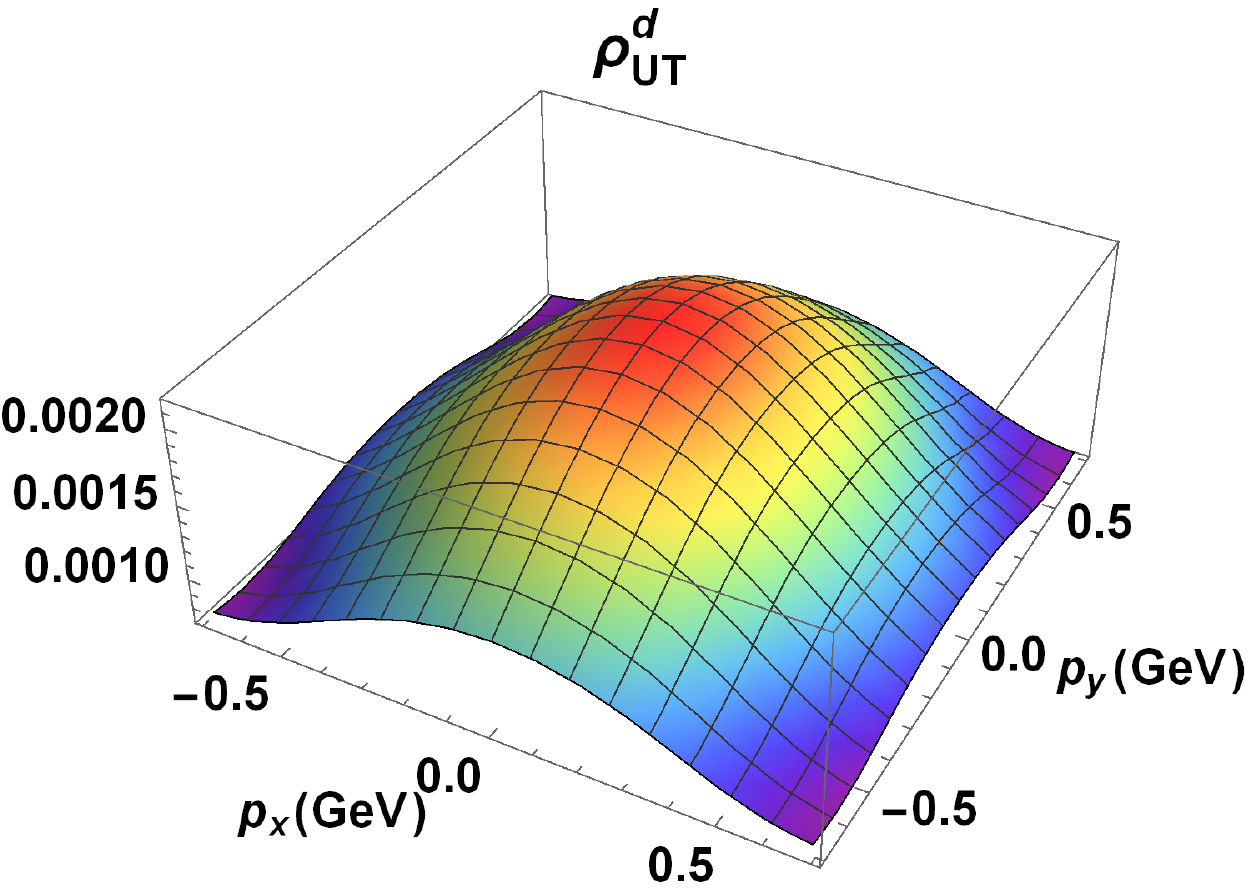}\hfill
(f)\includegraphics[width=.3\textwidth]{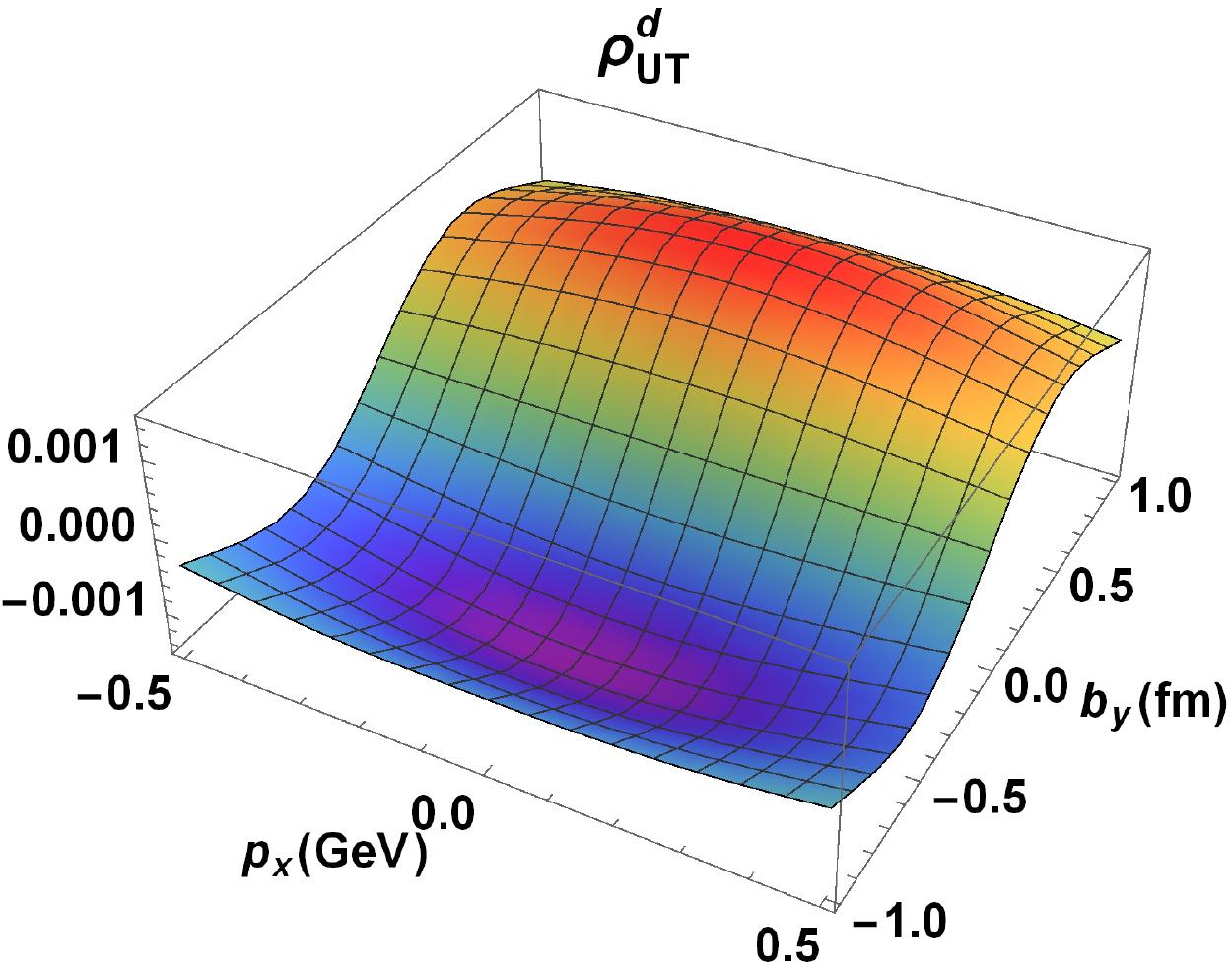}
\end{minipage}
\caption{The unpolarized-transverse Wigner distribution $\rho^j_{UT}$ in impact-parameter space $(b_x,b_y)$, in momentum space $(p_x,p_y)$ and in mixed space $(p_x,b_y)$ for $u(d)$ quark presented in upper (lower) panel.}
\label{unpolt}
\end{figure}
The unpolarized Wigner distributions in Fig. \ref{unpol} throw light on the distributions of a unpolarized quark in the unpolarized proton. Figs. \ref{unpol}(a) and \ref{unpol}(d), presenting the unpolarized Wigner distribution $\rho_{UU}$ for $u$ and $d$ quark respectively in the impact-parameter space, observe left-right symmetry which leads to the fact that the proton is same as viewed from any direction. In other words, the quark has equal probability to move in either clockwise or in anti-clockwise direction. There is also the top-bottom symmetry in this case.  In Figs. \ref{unpol}(b) and \ref{unpol}(e), for $\rho_{UU}$ in momentum space for $u$ quark and $d$ quark respectively, a circular symmetry is observed similar to the distribution in impact-parameter space but in the opposite direction. It would be important to mention here that the distribution $\rho_{UU}$ is connected to the chiral-even GPDs $H$ and $E$ together in the impact parameter space and to the T-even TMD $f_{1}$ in the momentum space. The unpolarized Wigner distributions in the mixed space shows a axially symmetric behavior as shown in Figs. \ref{unpol}(c) and \ref{unpol}(f). The distributions for $d$ quark are  concentrated at 
the center and fall off quickly at low values of impact parameter or momentum. The distributions of $u$ quark, on the other hand, are  concentrated at 
the center but spread out till the higher  values of impact parameter or momentum. 
When compared with similar work carried out in different models like the light-cone spectator model \cite{spectator}, AdS/QCD quark-diquark model \cite{ads}, and light-front dressed quark model \cite{dressed1}, we observe that the distributions in impact-parameter space are consistent with our results. However, the distributions of all the models in the
momentum space are in opposite direction when compared to our results. This leads to a very important observation that even through the the quark and the proton are unpolarized, the distributions have implications on the transverse momentum carried by the quark as well as on the impact-parameter co-ordinate.

In Fig. \ref{unpoll}, we plot the Wigner distribution describing a longitudinally-polarized quark in the  unpolarized proton viz. $\rho_{UL}$ in three spaces namely impact-parameter space, momentum space and mixed space. It shows a dipole behavior in impact-parameter space as well as in momentum space for $u$ and $d$ quark as shown in Figs. \ref{unpoll}(a), \ref{unpoll}(b), \ref{unpoll}(d) and \ref{unpoll}(e). The dipolar structure polarity in the case of impact-parameter space is however inverse as compared to the momentum space. The distribution $\rho_{UL}(p_x,b_y)$ in the mixed space describes quadrupole structures for both $u$ and $d$ quarks with same polarities as shown in Fig. \ref{unpoll}(c) and \ref{unpoll}(e). The dipole structure is due to the advantaged direction by the quark polarization. There are no TMDs and GPDs corresponding to $\rho_{UL}$ but the physical significance of this distribution is related to spin-orbital correlations. This correlation can be conformed with the quadrupole structure of the distribution in mixed space. Our results are in agreement with the light-cone spectator model \cite{spectator}, the AdS/QCD quark-diquark model \cite{ads}, and the light-front dressed quark model \cite{dressed1} for the distributions in impact-parameter space, momentum space as well as mixed space.

In Fig. \ref{unpolt}, we plot unpolarized-transverse Wigner distribution $\rho^j_{UT}(\textbf{b}_\perp, \textbf{p}_\perp)$ describing a transversely polarized quark (polarized along $x$-axis i.e. $j=1$) in the unpolarized proton. The distributions for $u$ and $d$ quark in the impact-parameter space  show a dipolar structure (Figs. \ref{unpolt}(a) and \ref{unpolt}(d)) and a circularly symmetric behavior in momentum space (Figs. \ref{unpolt}(b) and \ref{unpolt}(e)).  We observe that Wigner distribution $\rho^j_{UT}$ become zero, if the direction chosen for quark transverse co-ordinate is parallel to the quark polarization. In other words, there is a strong correlation between quark spin direction and the perpendicular transverse co-ordinate whereas no correlation exists between quark polarization and the parallel transverse co-ordinates. The mixed distribution $\rho^j_{UT}(p_x,b_y)$ relates to a dipolar structure with the extended peaks towards impact-parameter co-ordinate $b_y$ shown in Figs. \ref{unpolt}(c) and  \ref{unpolt}(f). The unpolarized-transverse Wigner distribution is correlated with T-odd TMD $h_{1}^{\perp}$ entitled as Boer-Mulder's function and the combination of $E_{T}$ and $\tilde{H}_{T}$ at TMD limit and IPD (impact-parameter distribution) limit respectively. Since, in the present calculations, we are not taking the gluon contribution into consideration, so no T-odd TMD will be generated from $\rho^j_{UT}$. This may lead to the possibility of building some relation between the TMD $h_{1}^{\perp}$ and combination of $E_{T}$ and $\tilde{H}_{T}$ at TMD limit and IPD (impact-parameter distribution) limit respectively. In this case our results agree with the light-cone spectator model \cite{spectator} and the AdS/QCD quark-diquark model \cite{ads} for the distributions in impact-parameter space, momentum space as well as mixed space. The light-front dressed quark model \cite{dressed1} however show a dipolar structure for the distributions in momentum space as compared to other models. This may not be taken as a contradiction because of different model assumptions. The light-front dressed quark model calculates the distributions for the composite spin-1/2 system whereas other models present the explicit distributions corresponding to $u$ and $d$ quarks.

\subsection{Wigner distributions for longitudinally-polarized proton}
 \begin{figure}
\centering
\begin{minipage}[c]{0.98\textwidth}
(a)\includegraphics[width=.3\textwidth]{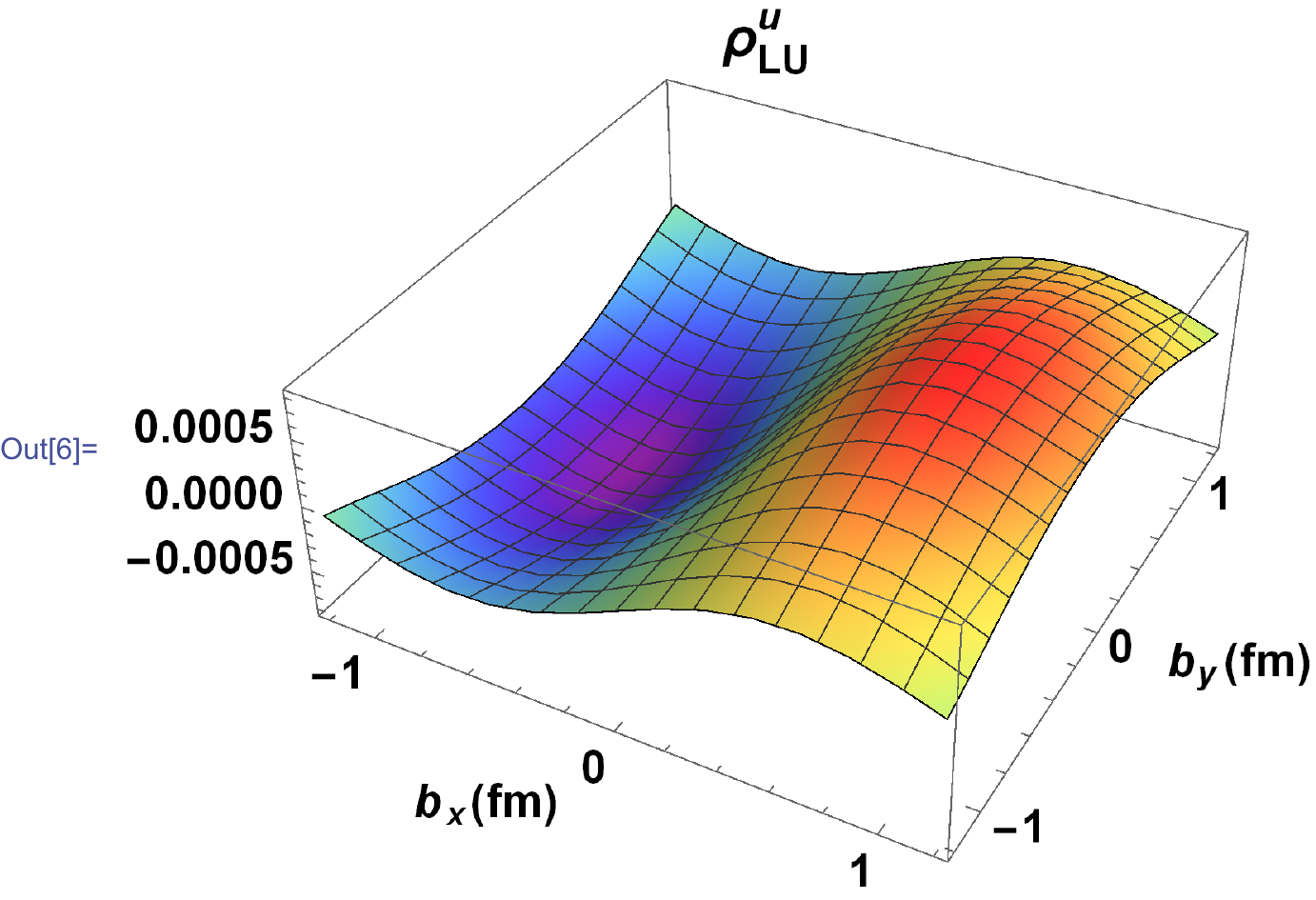}\hfill
(b)\includegraphics[width=.3\textwidth]{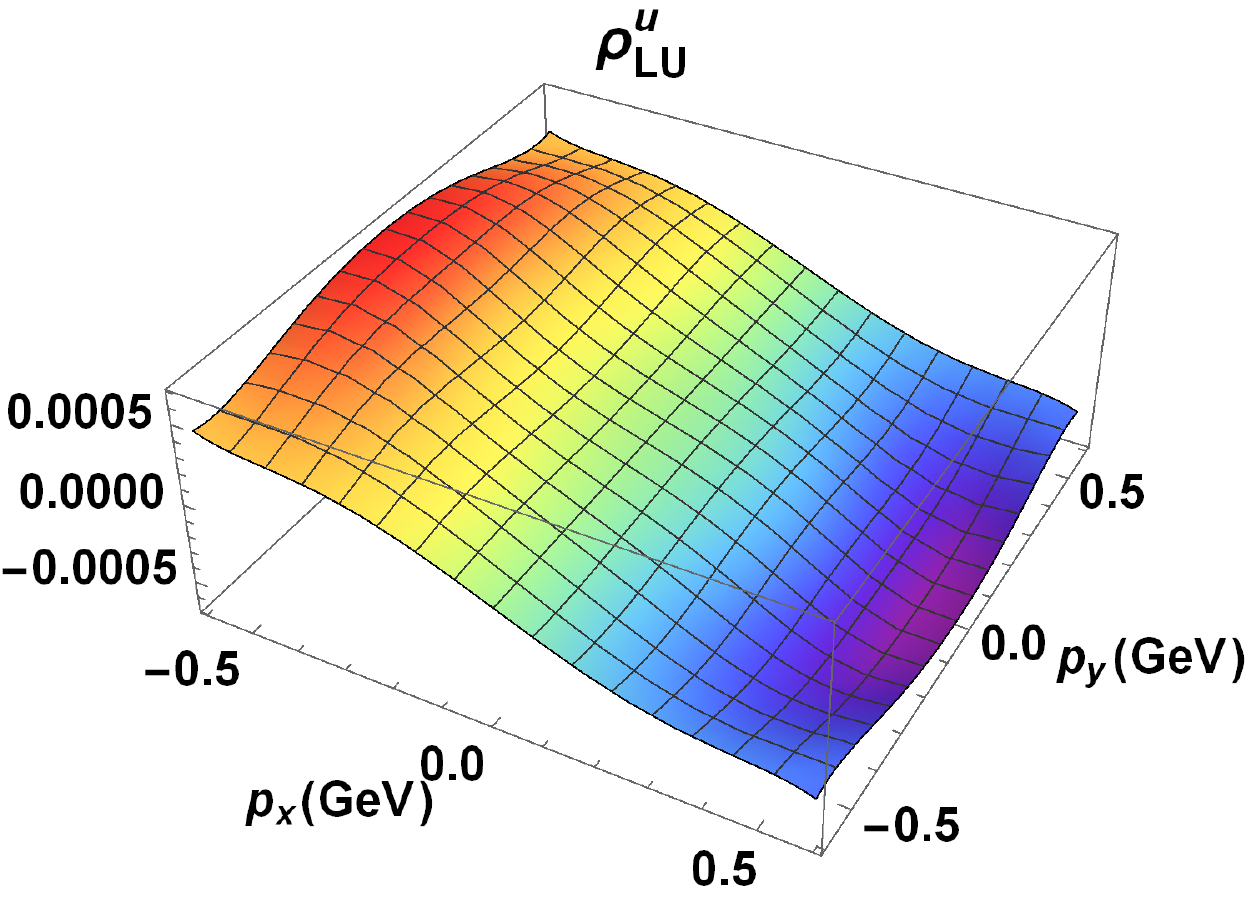}\hfill
(c)\includegraphics[width=.3\textwidth]{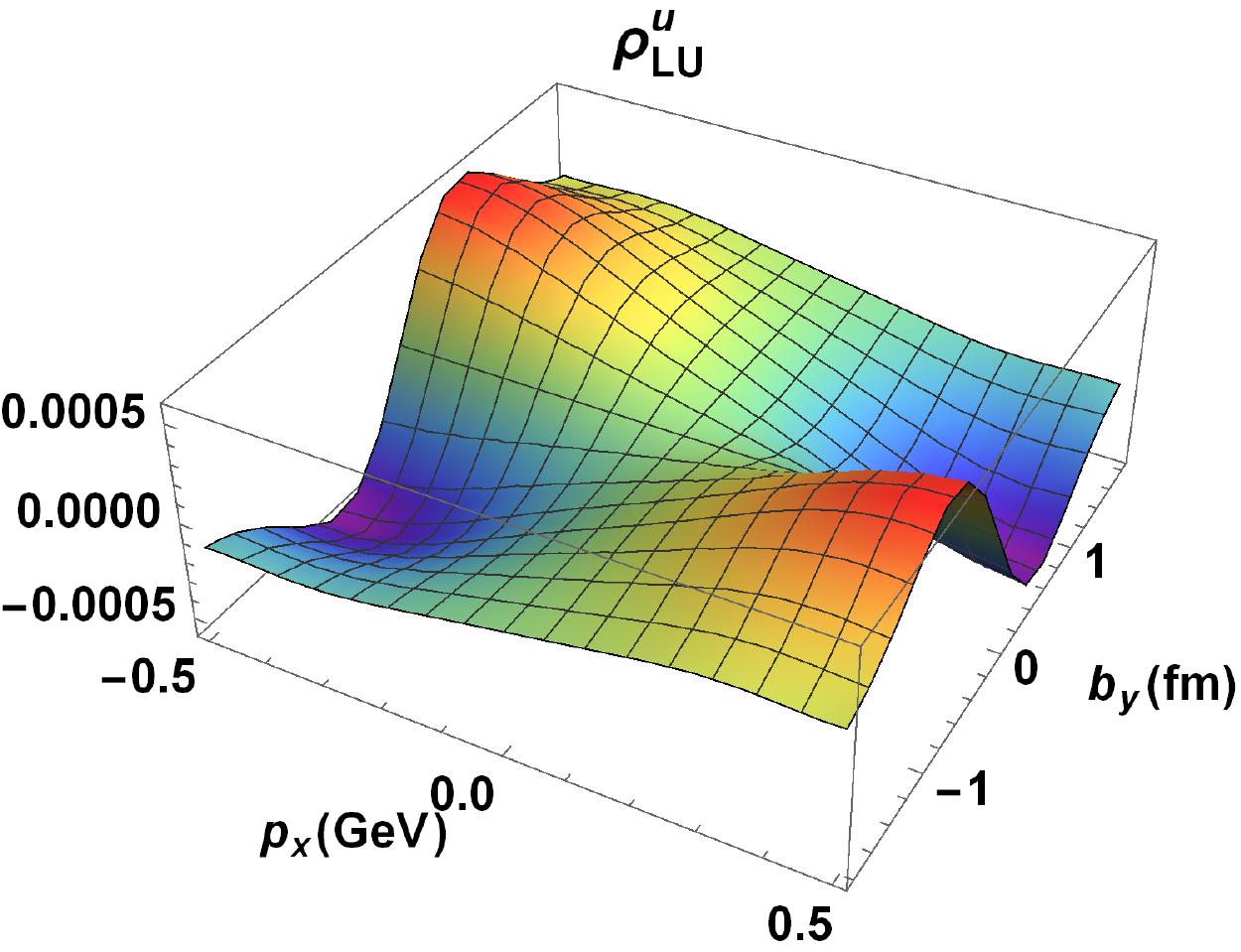}
\end{minipage}
\begin{minipage}[c]{0.98\textwidth}
(d)\includegraphics[width=.3\textwidth]{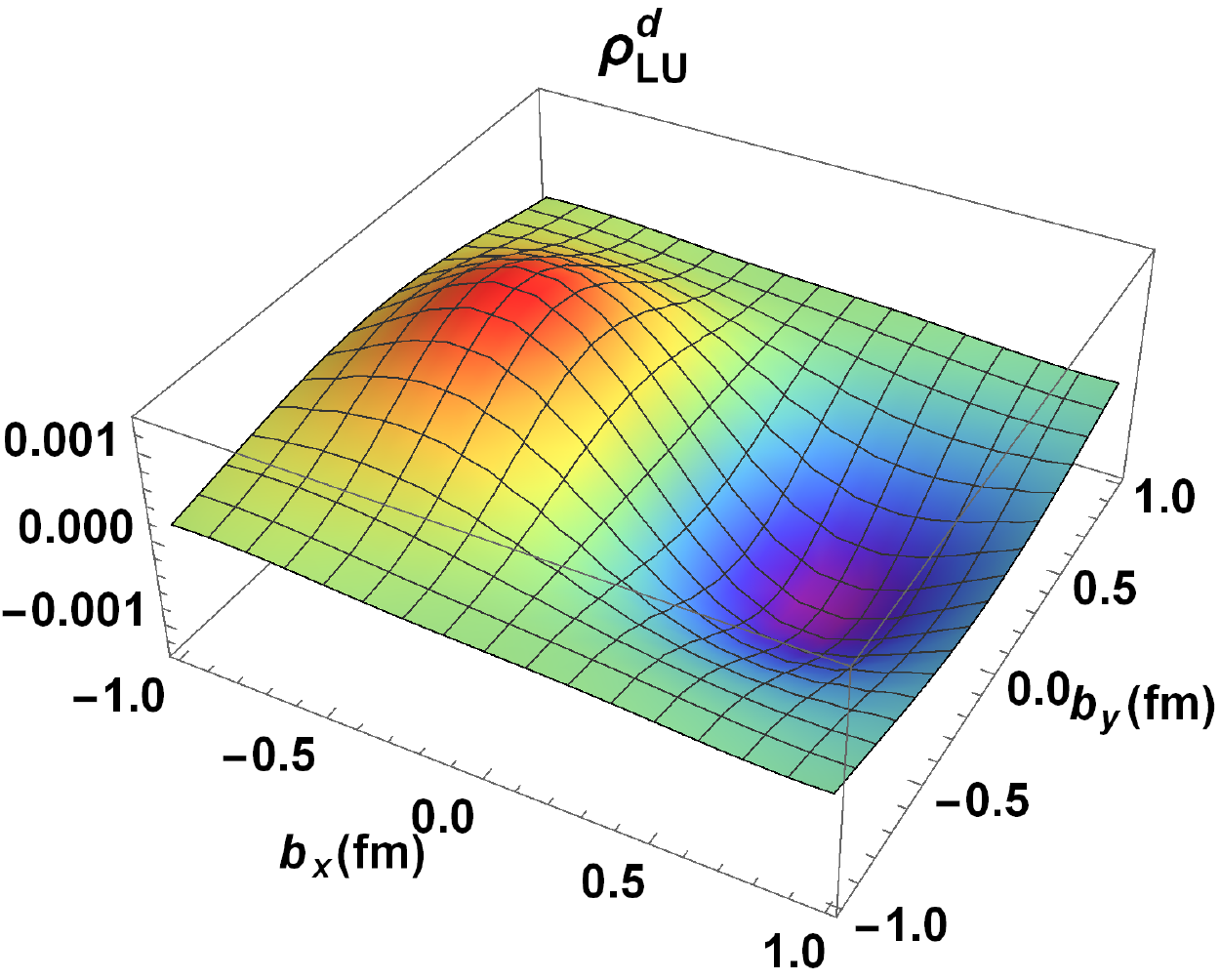}\hfill
(e)\includegraphics[width=.3\textwidth]{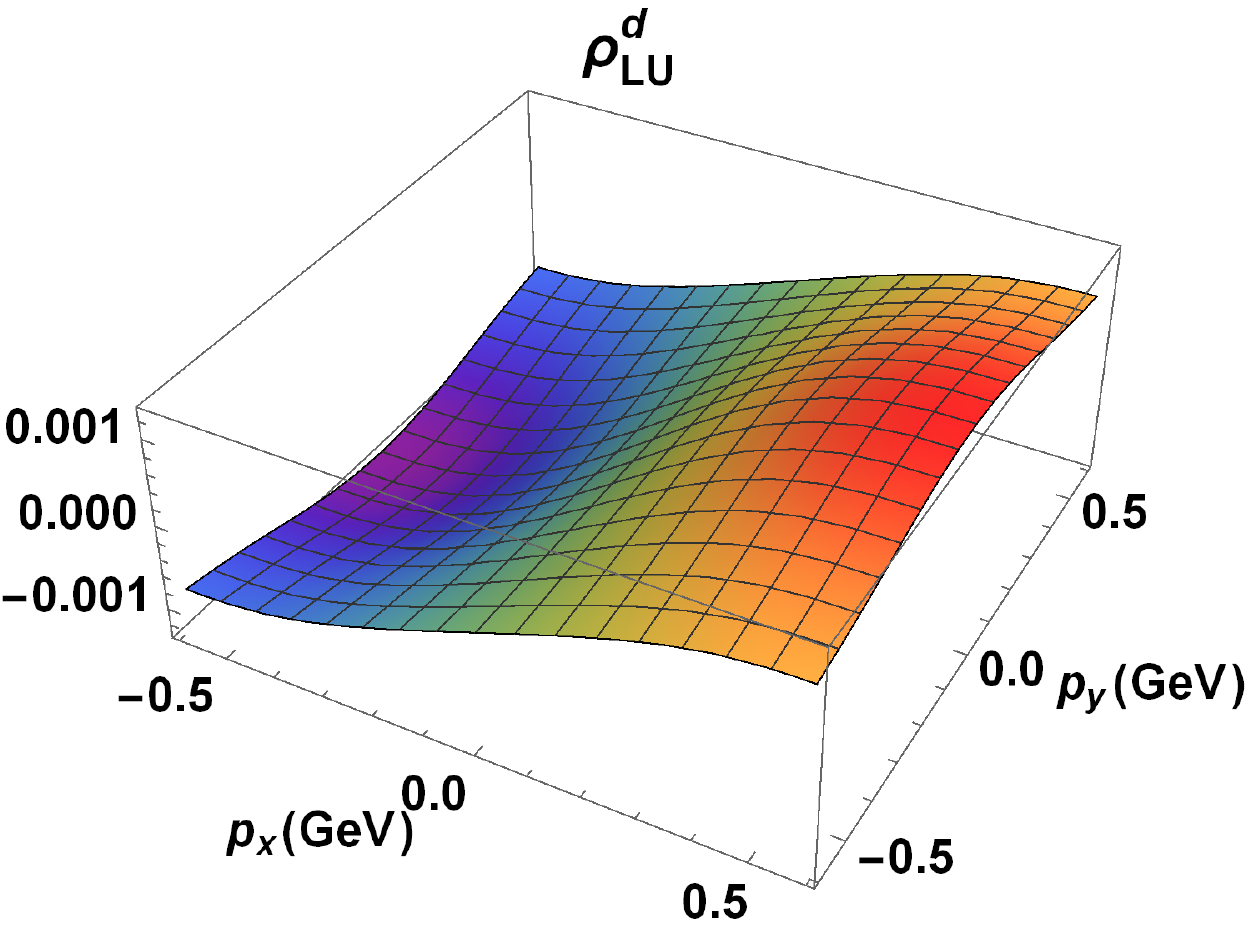}\hfill
(f)\includegraphics[width=.3\textwidth]{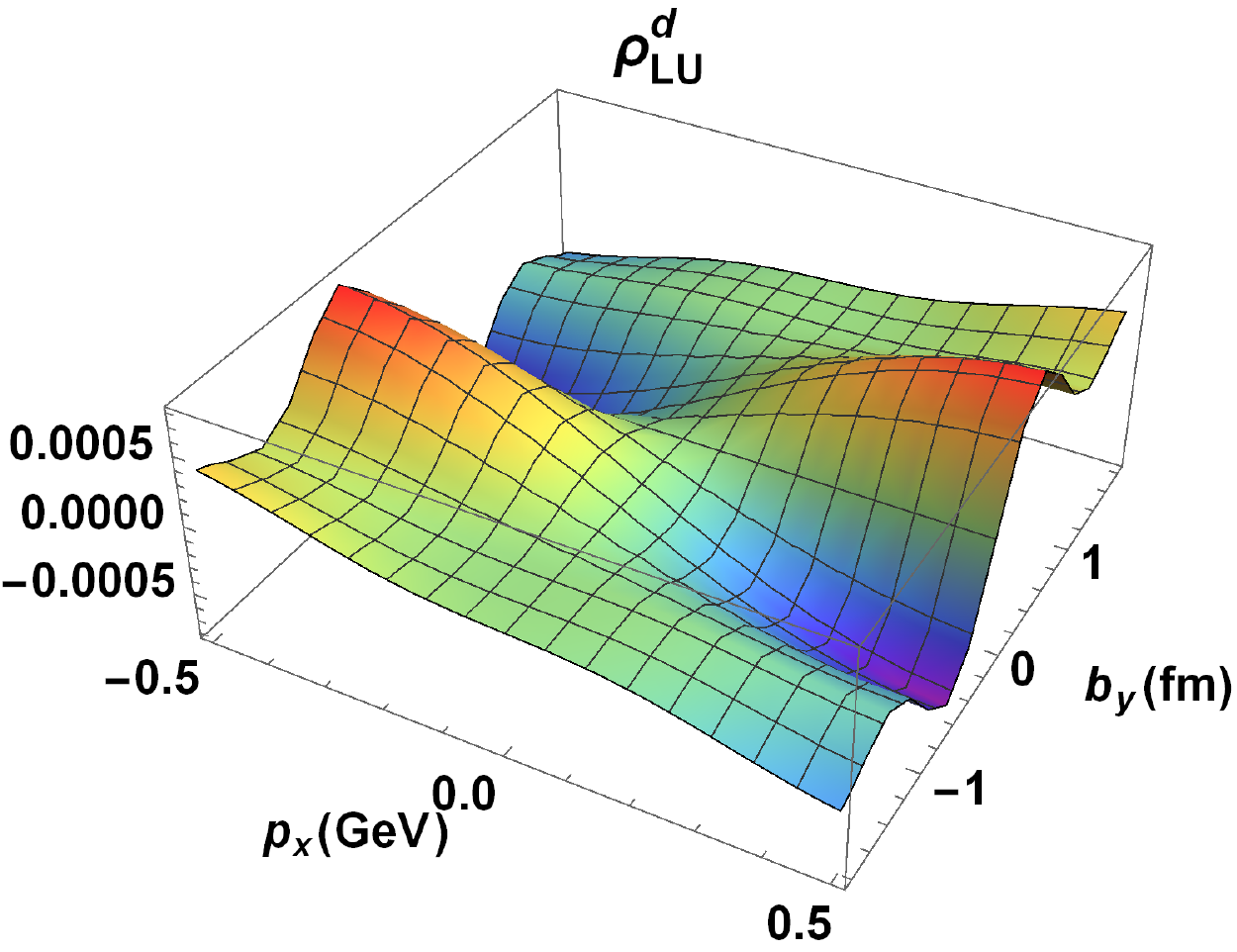}
\end{minipage}
\caption{The longitudinal-unpolarized Wigner distribution $\rho_{LU}$ in impact-parameter space $(b_x,b_y)$, in momentum space $(p_x,p_y)$ and in mixed space $(p_x,b_y)$ for $u(d)$ quark presented in upper (lower) panel.}
\label{longiu}
\end{figure}
\begin{figure}
\centering
\begin{minipage}[c]{0.98\textwidth}
(a)\includegraphics[width=.3\textwidth]{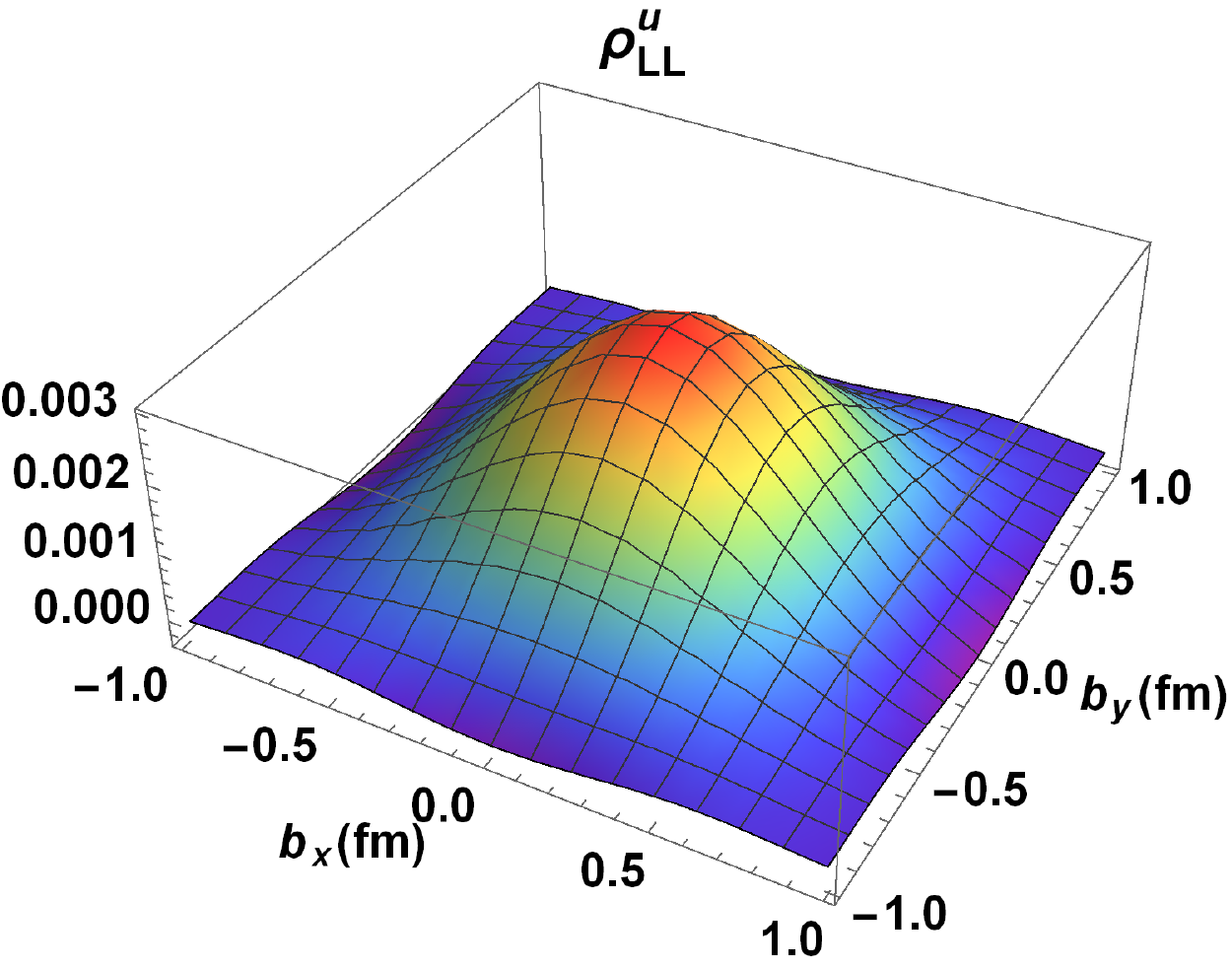}\hfill
(b)\includegraphics[width=.3\textwidth]{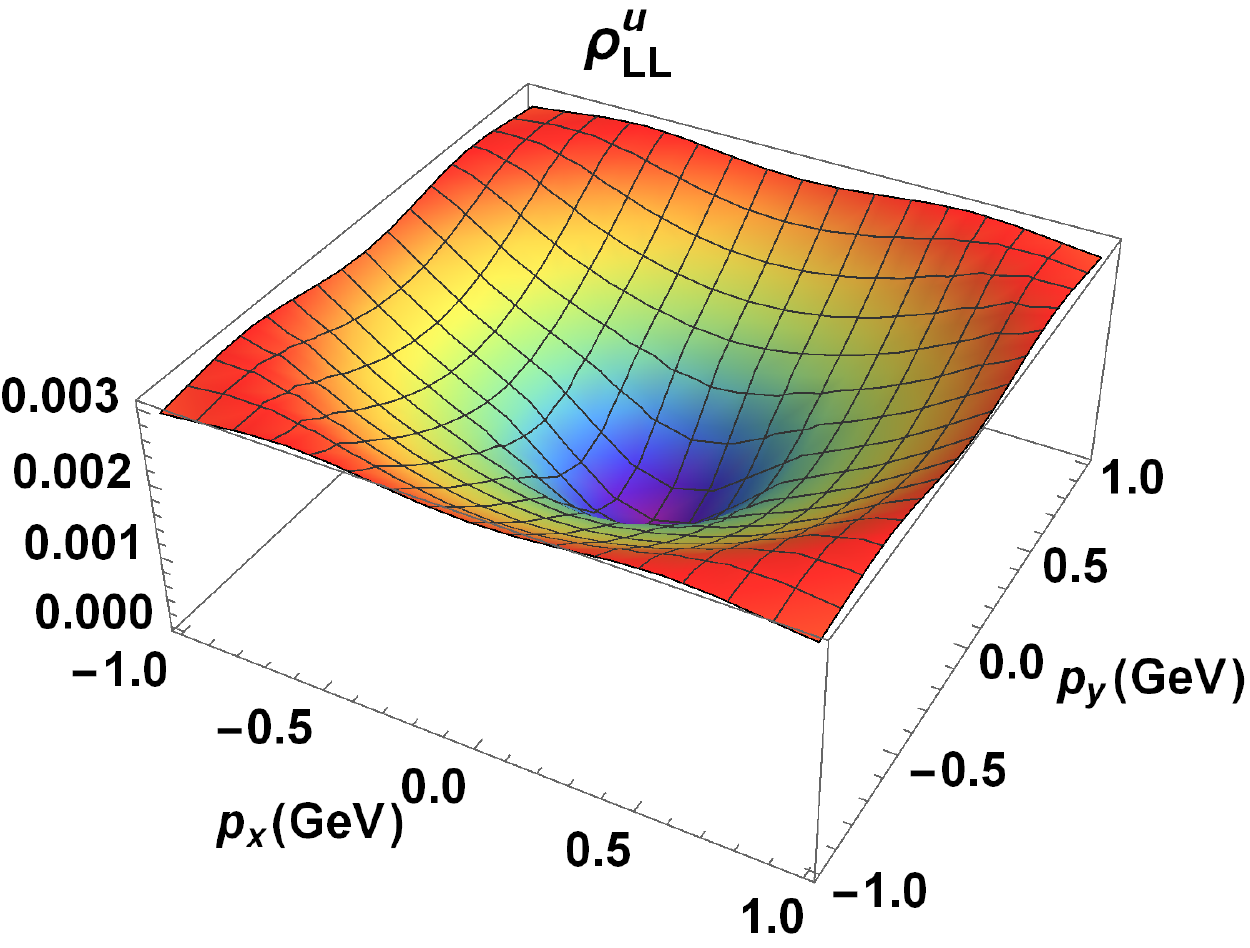}\hfill
(c)\includegraphics[width=.3\textwidth]{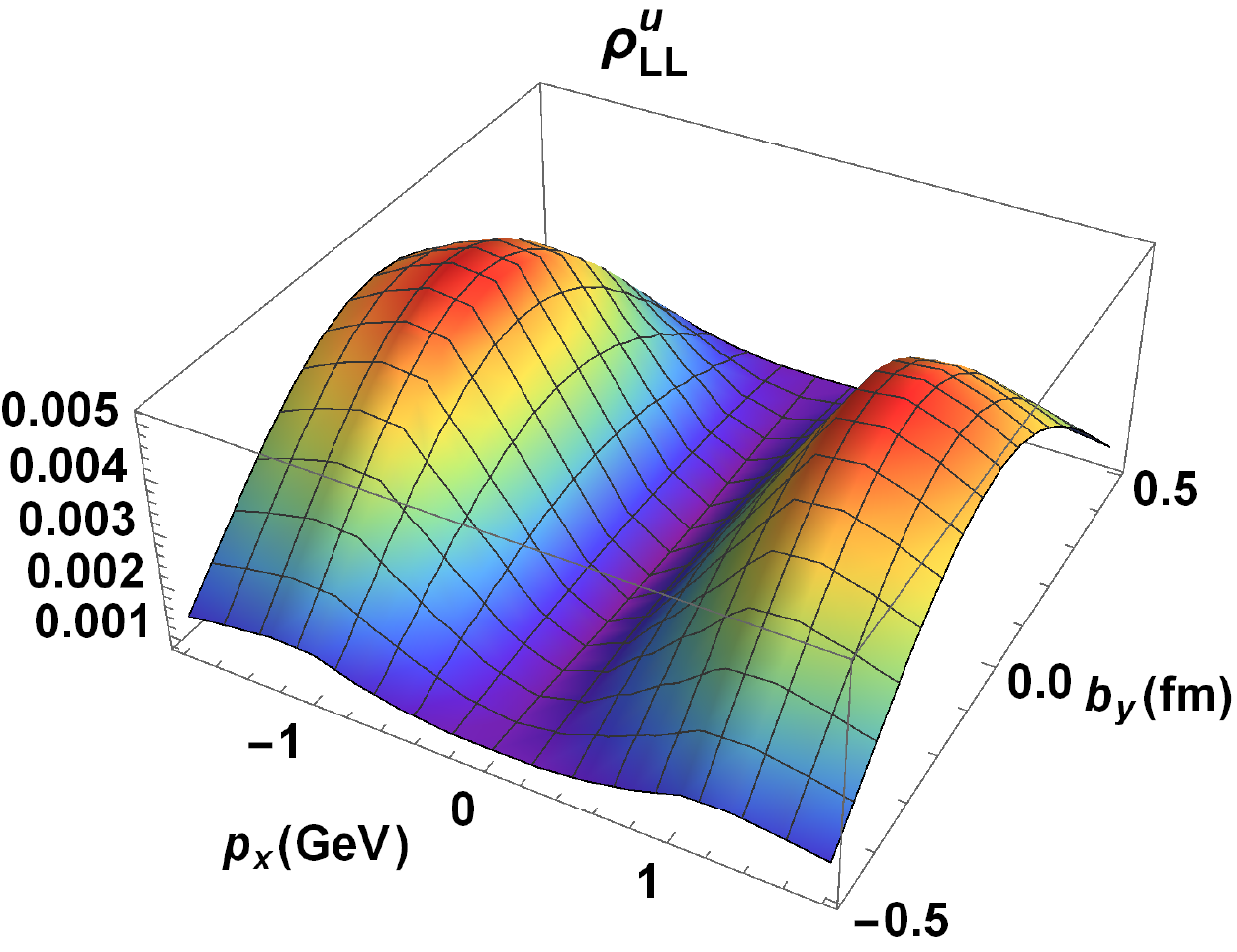}
\end{minipage}
\begin{minipage}[c]{0.98\textwidth}
(d)\includegraphics[width=.3\textwidth]{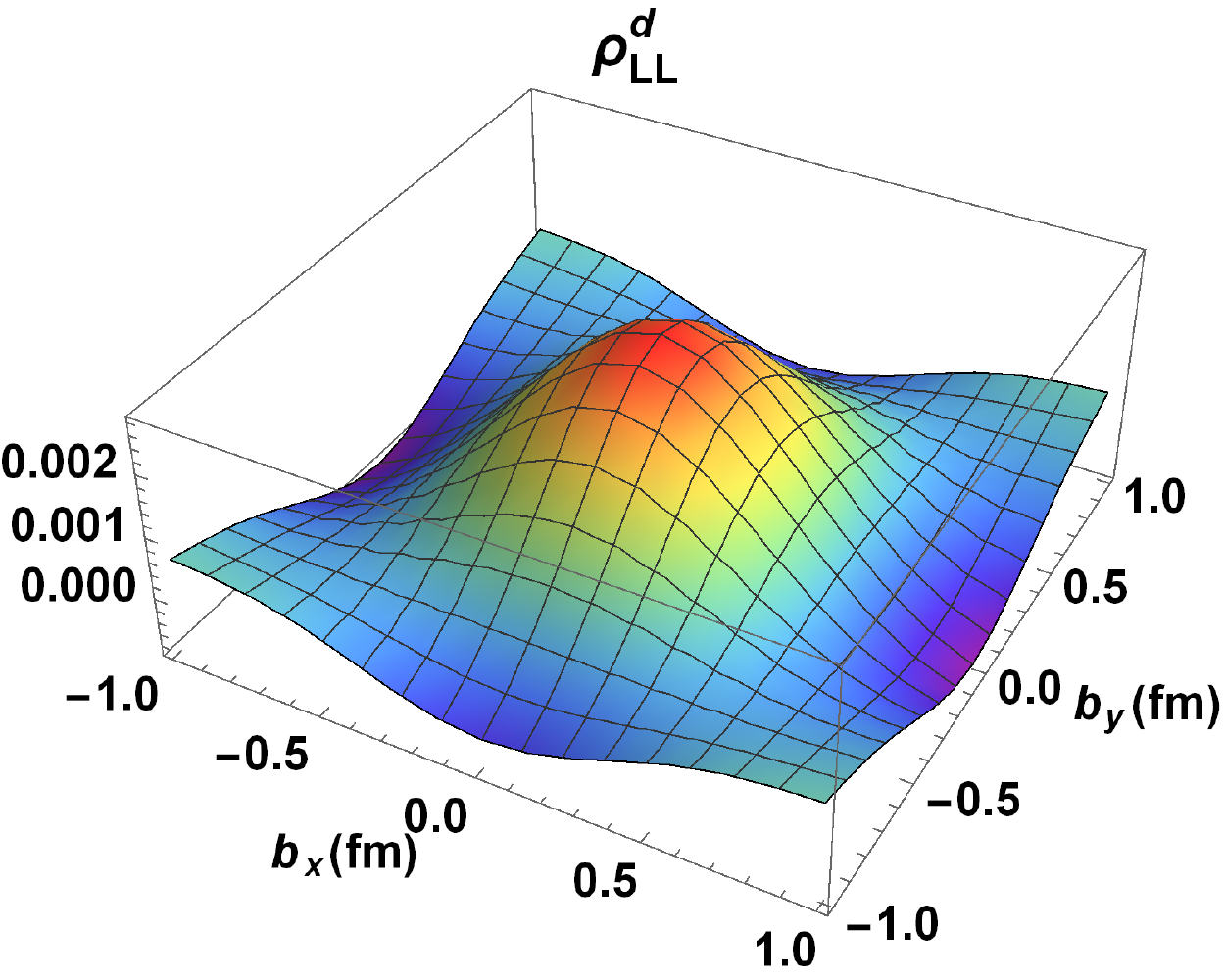}\hfill
(e)\includegraphics[width=.3\textwidth]{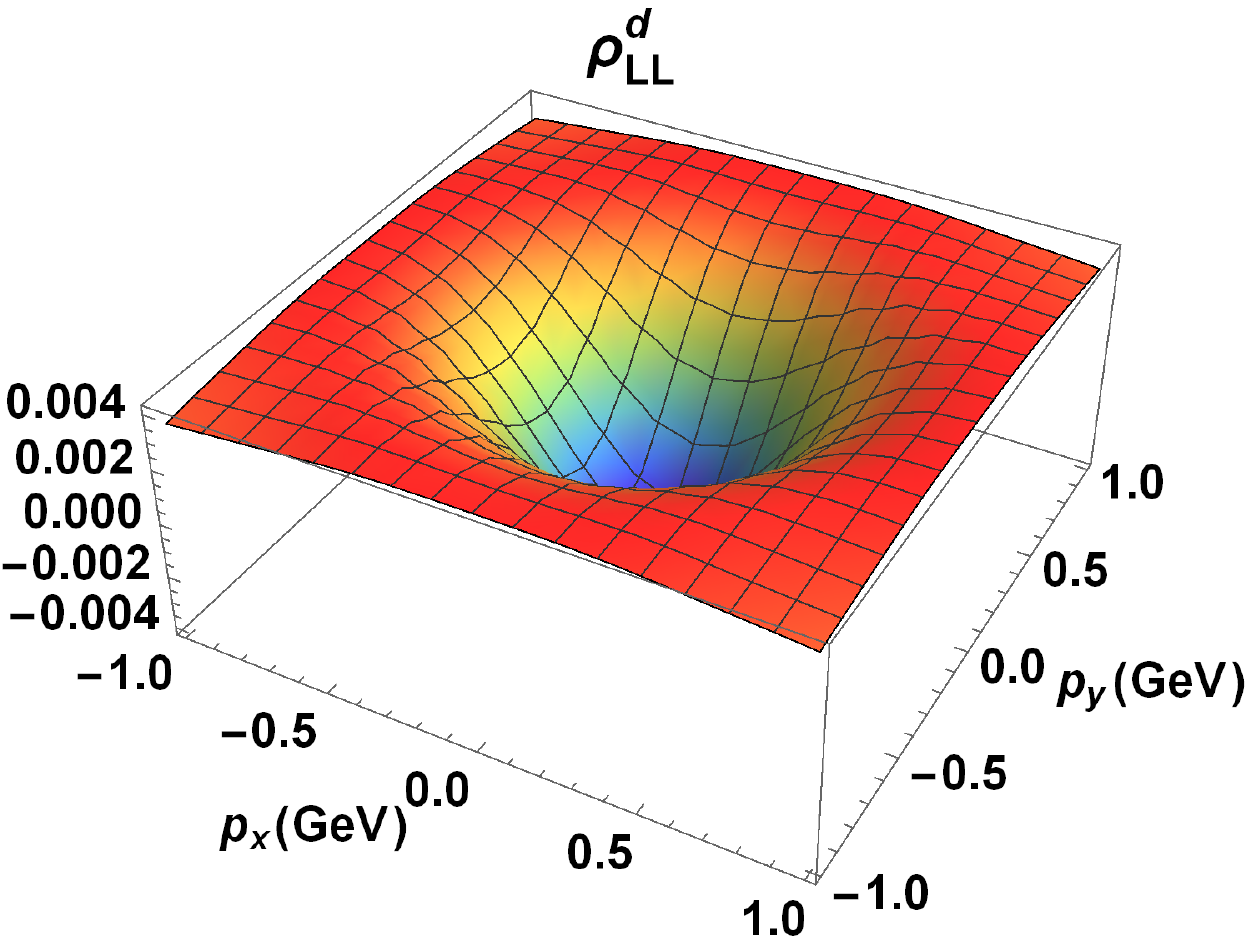}\hfill
(f)\includegraphics[width=.3\textwidth]{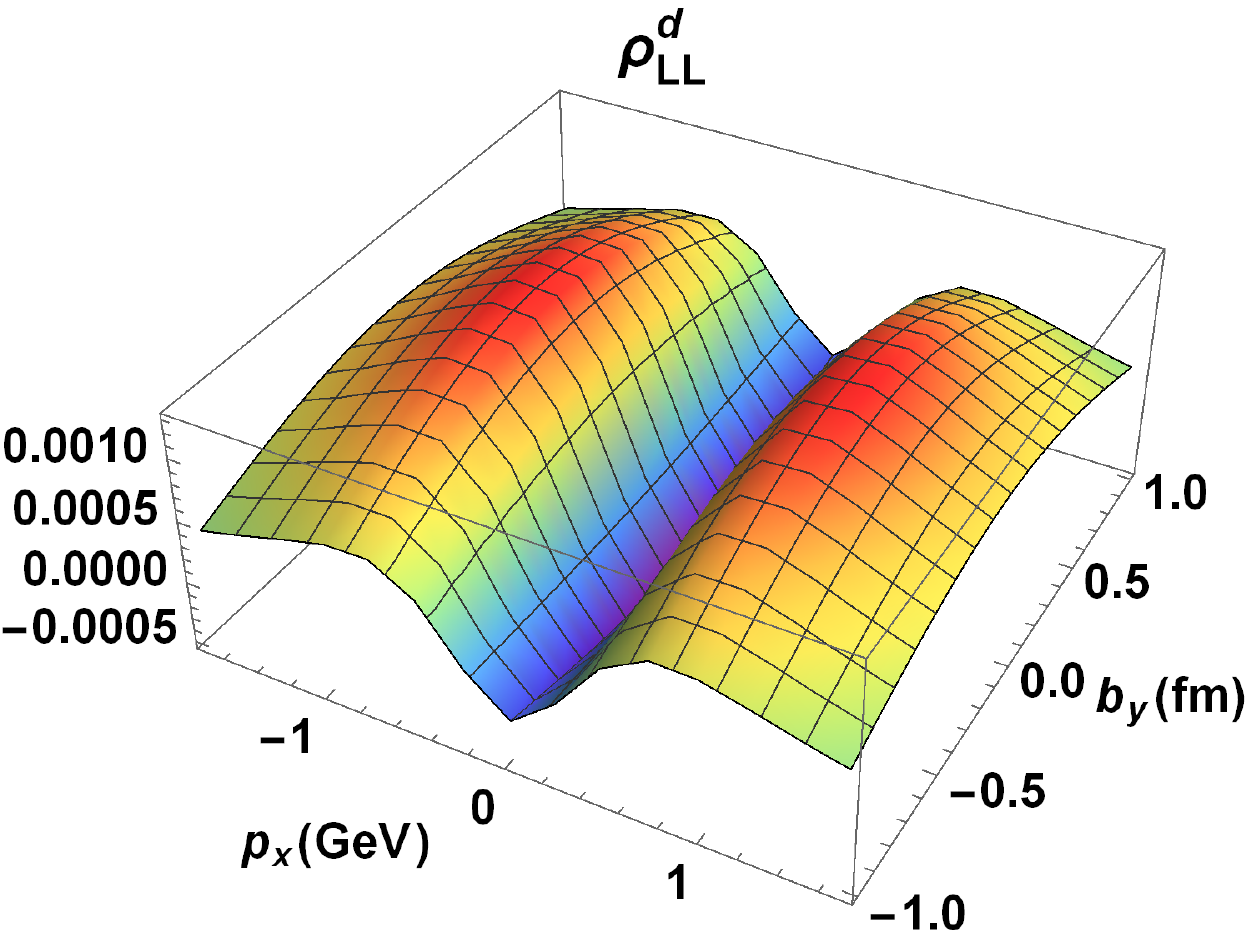}
\end{minipage}
\caption{The longitudinal Wigner distribution $\rho_{LL}$ in impact-parameter space $(b_x,b_y)$, in momentum space $(p_x,p_y)$ and in mixed space $(p_x,b_y)$ for $u(d)$ quark presented in upper (lower) panel.}
\label{longil}
\end{figure}
\begin{figure}
\centering
\begin{minipage}[c]{0.98\textwidth}
(a)\includegraphics[width=.3\textwidth]{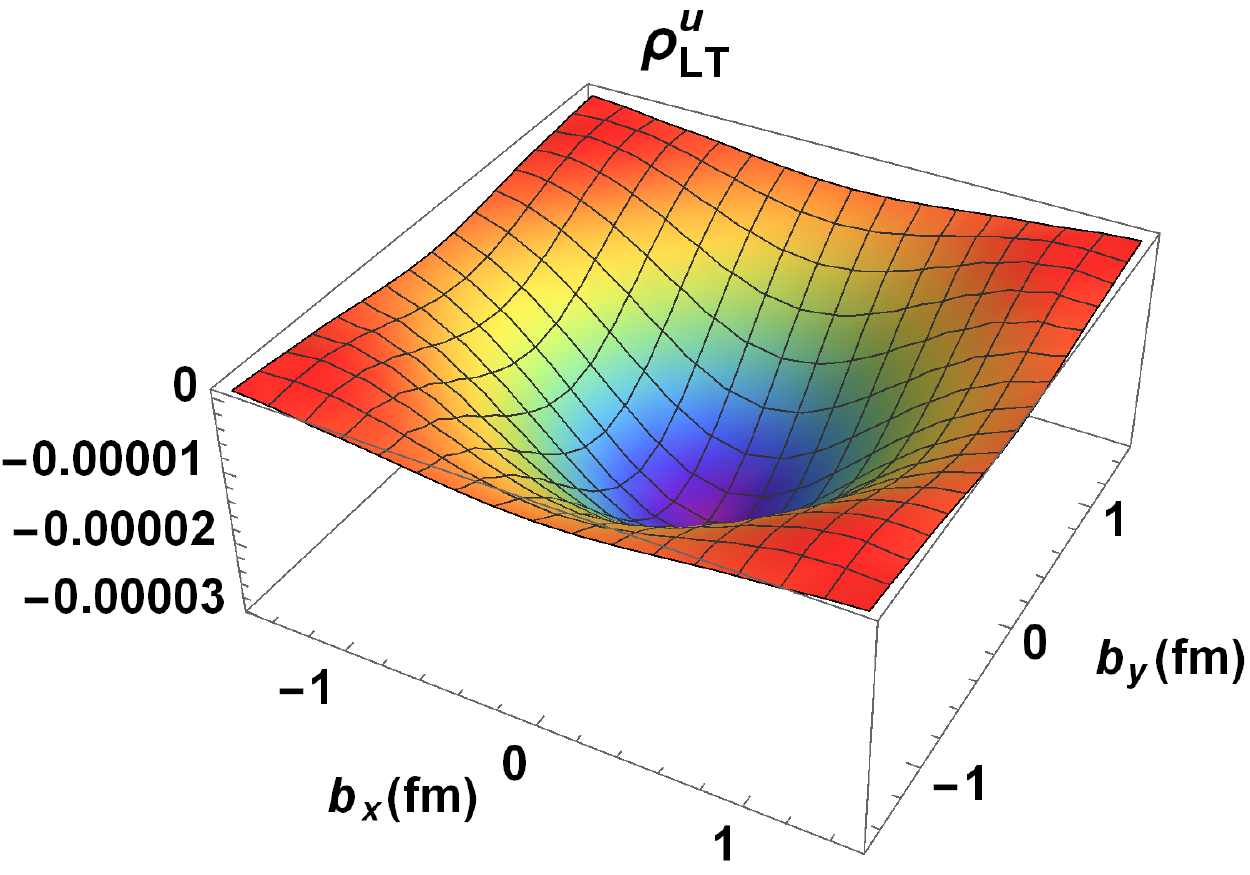}\hfill
(b)\includegraphics[width=.3\textwidth]{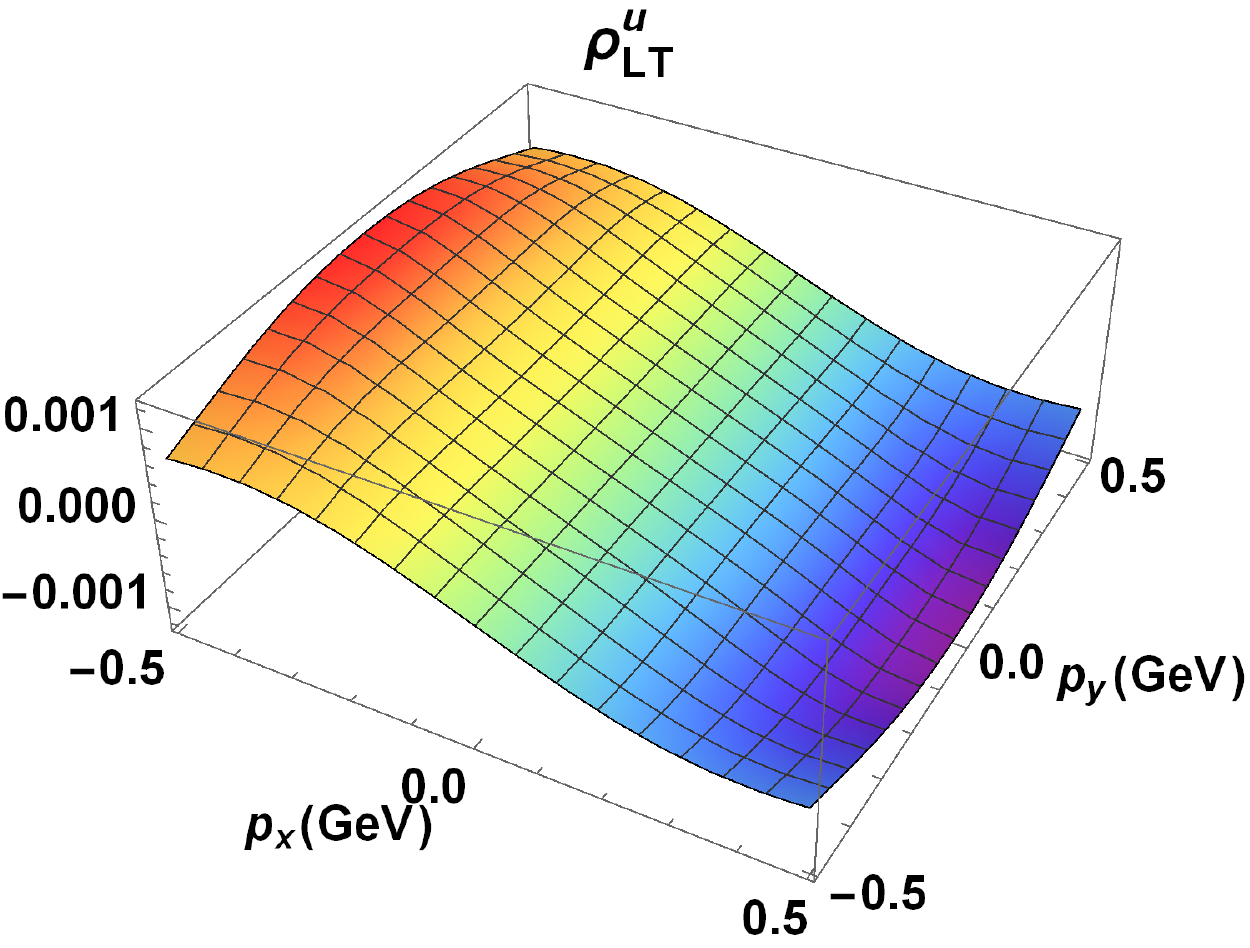}\hfill
(c)\includegraphics[width=.3\textwidth]{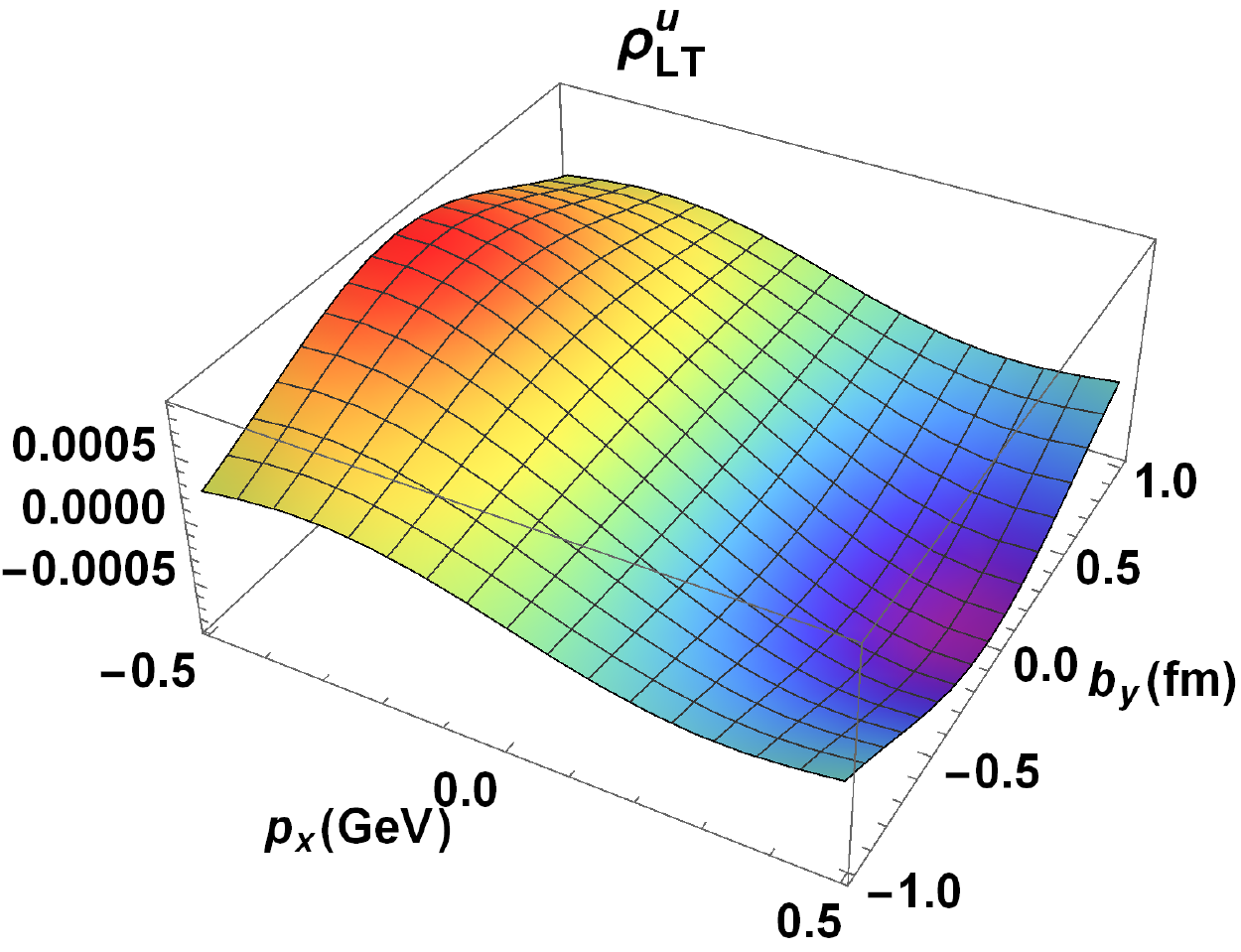}
\end{minipage}
\begin{minipage}[c]{0.98\textwidth}
(d)\includegraphics[width=.3\textwidth]{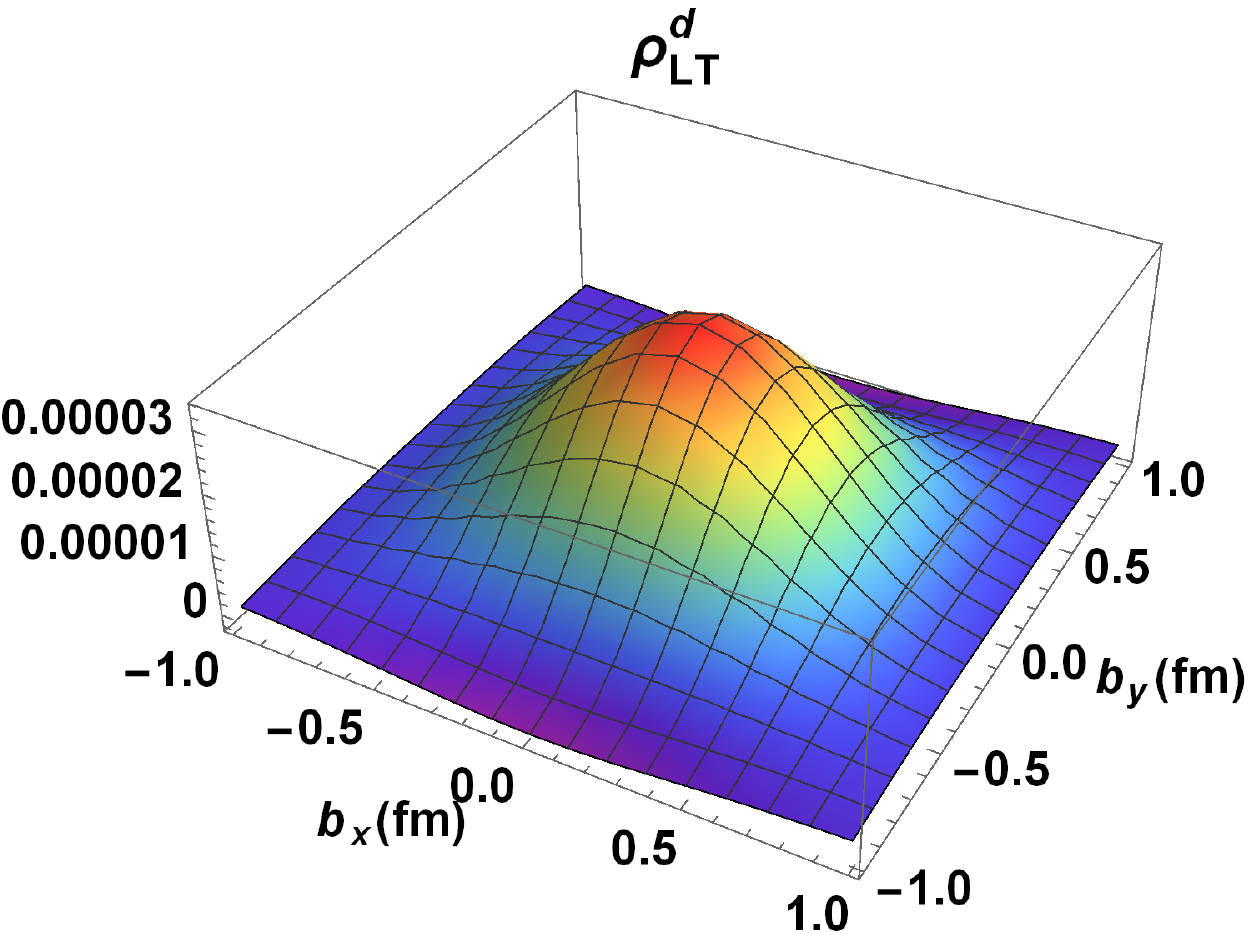}\hfill
(e)\includegraphics[width=.3\textwidth]{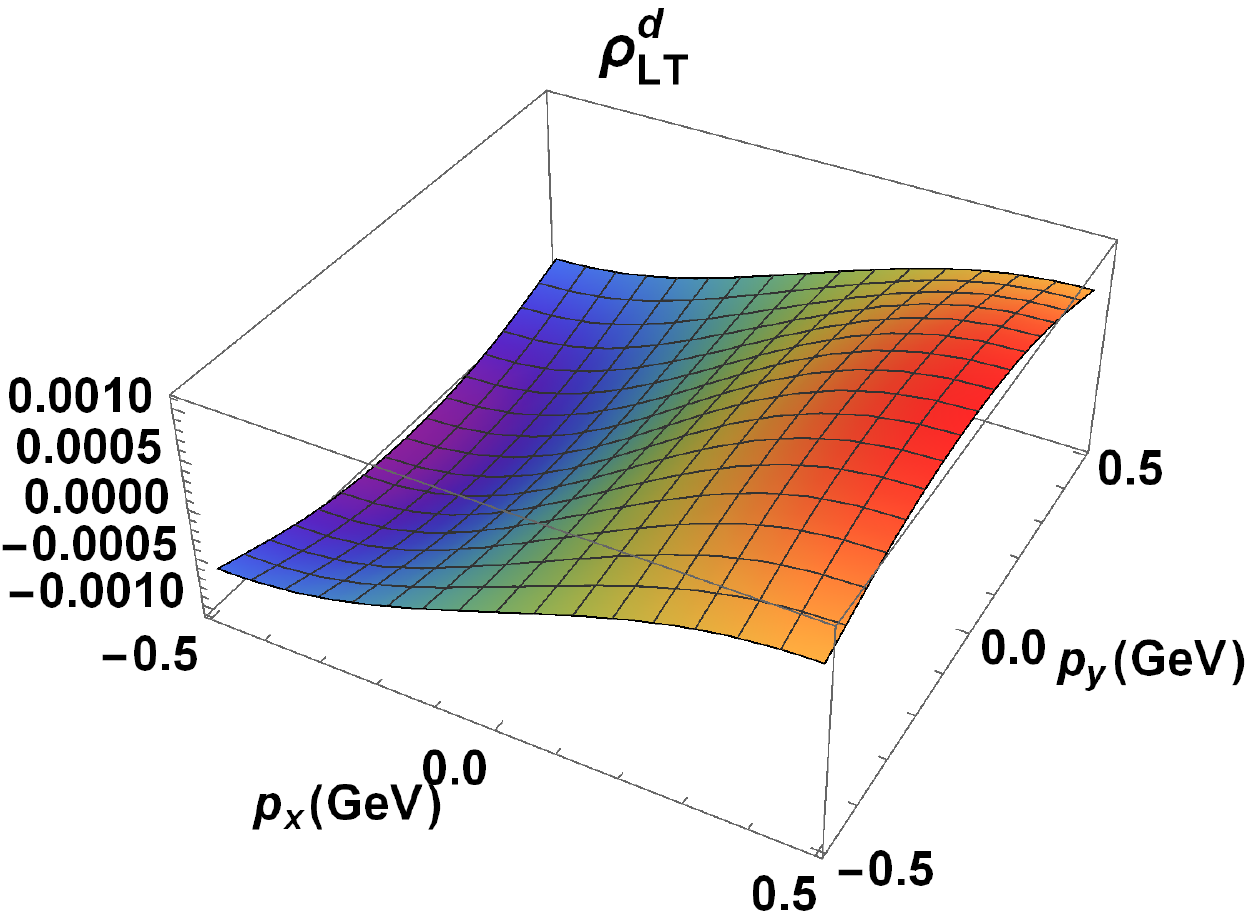}\hfill
(f)\includegraphics[width=.3\textwidth]{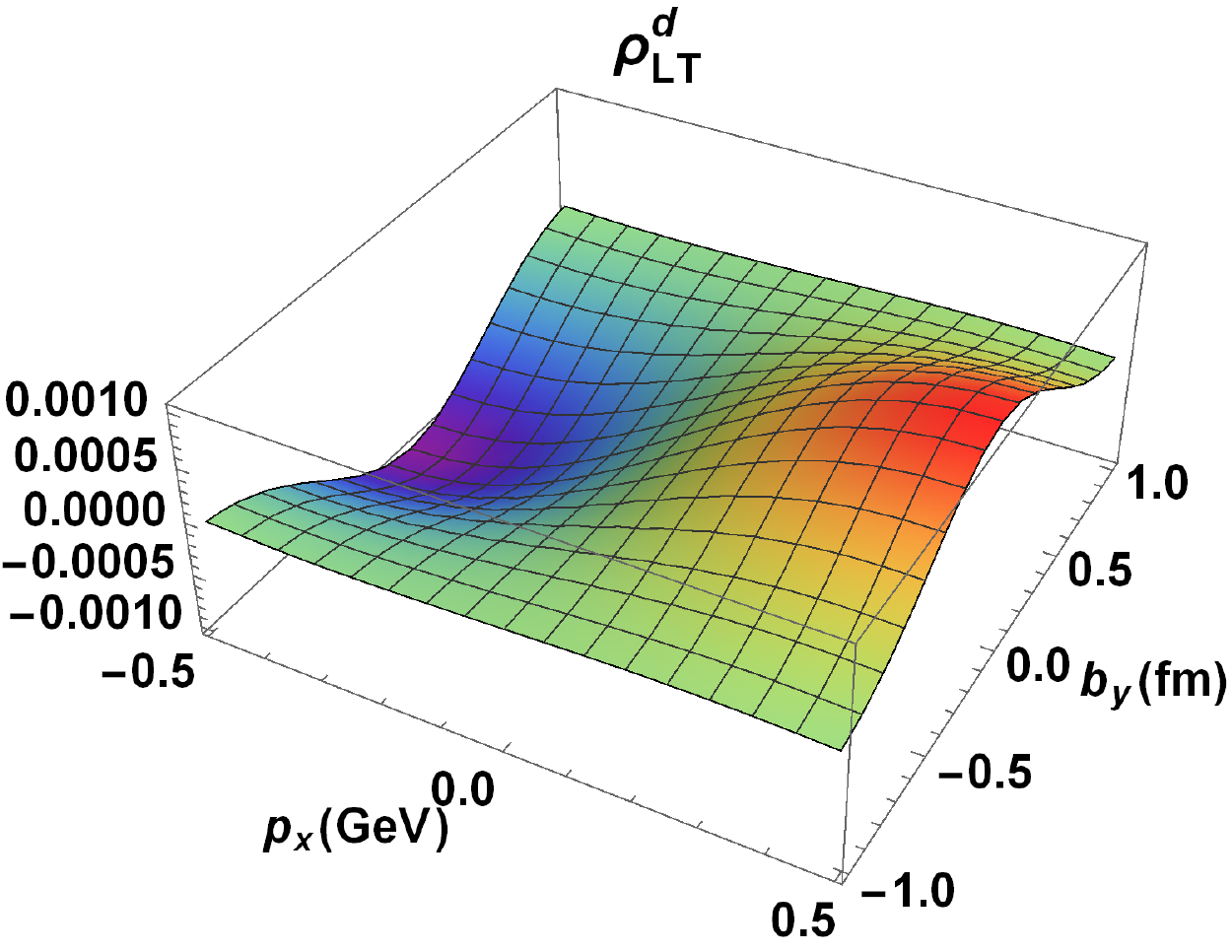}
\end{minipage}
\caption{The longitudinal-transverse Wigner distribution $\rho^j_{LT}$ in impact-parameter space $(b_x,b_y)$, in momentum space $(p_x,p_y)$ and in mixed space $(p_x,b_y)$ for $u(d)$ quark presented in upper (lower) panel.}
\label{longit}
\end{figure} 

In Fig. \ref{longiu}, we plot the Wigner distribution for a unpolarized quark in the  longitudinally-polarized proton  $\rho_{LU}$.  Figs. \ref{longiu}(a) and \ref{longiu}(d), show a dipole structure in the impact-parameter space for $\rho_{LU}$ whereas it describes the same structure with the opposite polarities in the momentum space as shown in Figs. \ref{longiu}(b) and \ref{longiu}(e). For mixed distribution $\rho_{LU}(p_x,b_y)$, Fig. \ref{longiu}(c) and \ref{longiu}(f) shows a quadrupole structure for both quarks. The polarities of this distribution are opposite for $u$ and $d$ quarks in all the three spaces. Similar to $\rho_{UL}$, in this case also, neither TMDs nor GPDs are present leading to the orbital angular momentum problem.

In Fig. \ref{longil}, the distribution $\rho_{LL}$ describes the case where both proton and quark are polarized in the longitudinal direction. $\rho_{LL}$ is plotted in the impact-parameter space, the momentum space and the mixed space. It can be clearly seen that the $\rho_{LL}$ distribution is quite similar to  $\rho_{UU}$, concentrating at the center symmetrically for both quarks in the impact-parameter space (Figs. \ref{longil}(a) and \ref{longil}(d))  while the polarity changes in momentum space (Figs. \ref{longil}(b) and \ref{longil}(e)). In mixed space, the distribution shows axially symmetric results as seen from Figs. \ref{longil}(c) and \ref{longil}(f). The TMD $g_1$ is associated with this Wigner distribution in the TMD limit. The distribution $\rho_{LL}$ is related to the combination of GPDs $H$ and $\tilde{H}$ at the IPD limit which are chiral-even GPDs.

The distribution for a transversely-polarized quark along $\hat{x}$ in the  longitudinally-polarized proton, $\rho^j_{LT}(\textbf{b}_\perp, \textbf{p}_\perp)$ is shown in Fig. \ref{longit}. In Figs. \ref{longit}(a) and \ref{longit}(d), we plot $\rho^j_{LT}$ in the  impact-parameter space where the distribution is circularly symmetric about the center, $b_x=b_y=0$ for $u$ and $d$ quarks respectively with the inverse polarities. In the momentum space (Figs. \ref{longit}(b) and \ref{longit}(e)) and in the mixed space (Fig. \ref{longit}(c) and \ref{longit}(f)), this distribution exhibits dipole structure having opposite polarities for both quarks. From Eqs. (\ref{lts}) and (\ref{lt1}), we see that if transverse momentum of quark is perpendicular to the quark polarization, this distribution will vanish. The correlation between the transverse momentum and polarization of quark is strong while the correlation between impact-parameter co-ordinates and polarization of quark is weak in this model. At TMD and IPD limits, $\rho^j_{LT}$ is connected to T-even TMD $h_{1L}^{\perp}$ and a fusion of GPDs $H_{T}$ and $\tilde{H}_{T}$ respectively.

In all the above cases for the longitudinally-polarized proton our results are in exact agreement with the light-cone spectator model \cite{spectator},  AdS/QCD quark-diquark model \cite{ads} and the light-front dressed quark model \cite{dressed1}  for the distributions in impact-parameter space, momentum space as well as mixed space as far as the type of distributions are concerned. When it comes to the signs of  explicit $u$ and $d$ quark distributions, no general comparisons can be  made between the models. This is so because the proton in these distributions is longitudinally polarized. Future measurements taking into account different polarizations of the quark and proton  will throw considerable light on the deeper implications of the interplay between the quark and proton spin.

\subsection{Wigner distributions for transversely-polarized proton}
\begin{figure}
\centering
\begin{minipage}[c]{0.98\textwidth}
(a)\includegraphics[width=.3\textwidth]{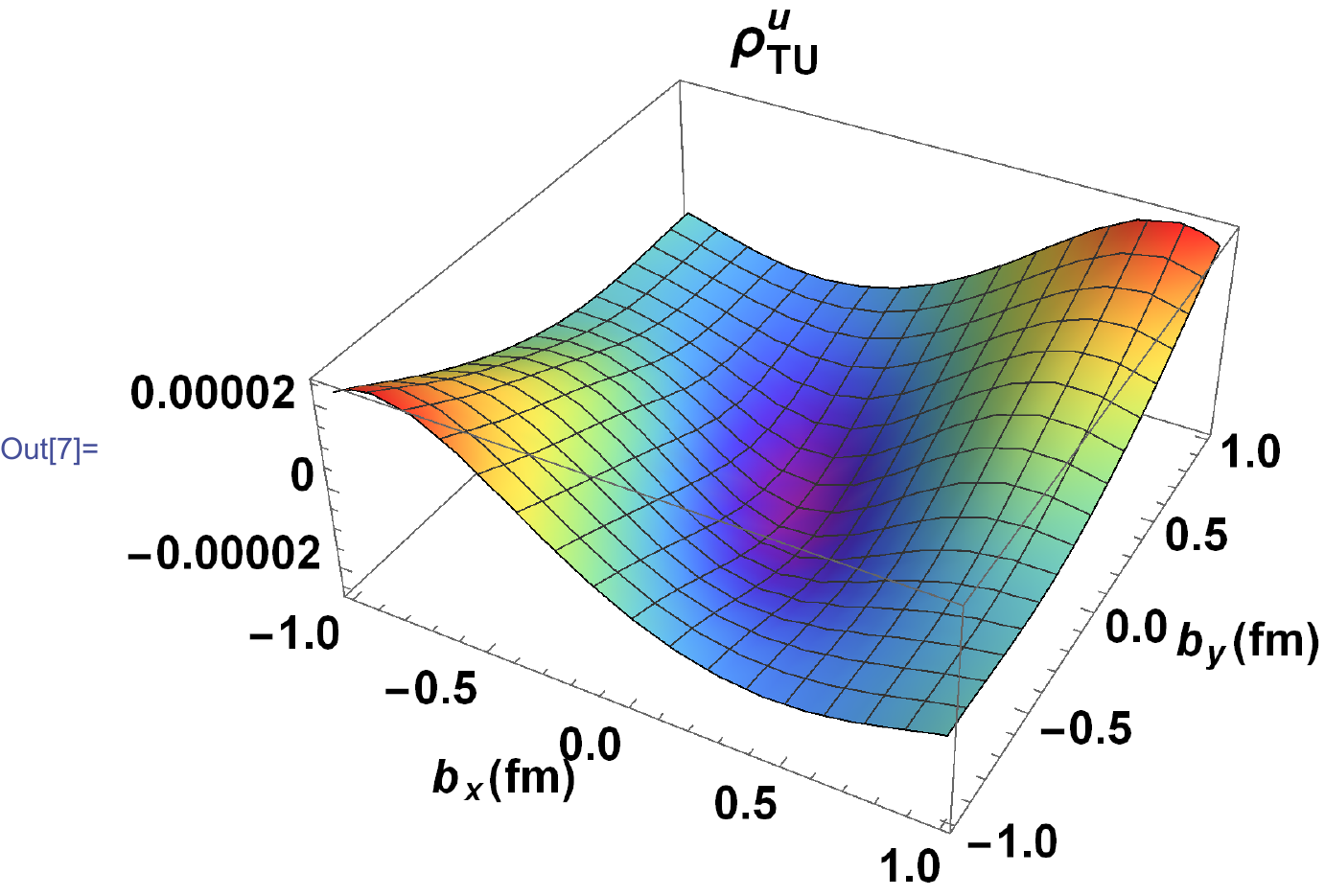}\hfill
(b)\includegraphics[width=.3\textwidth]{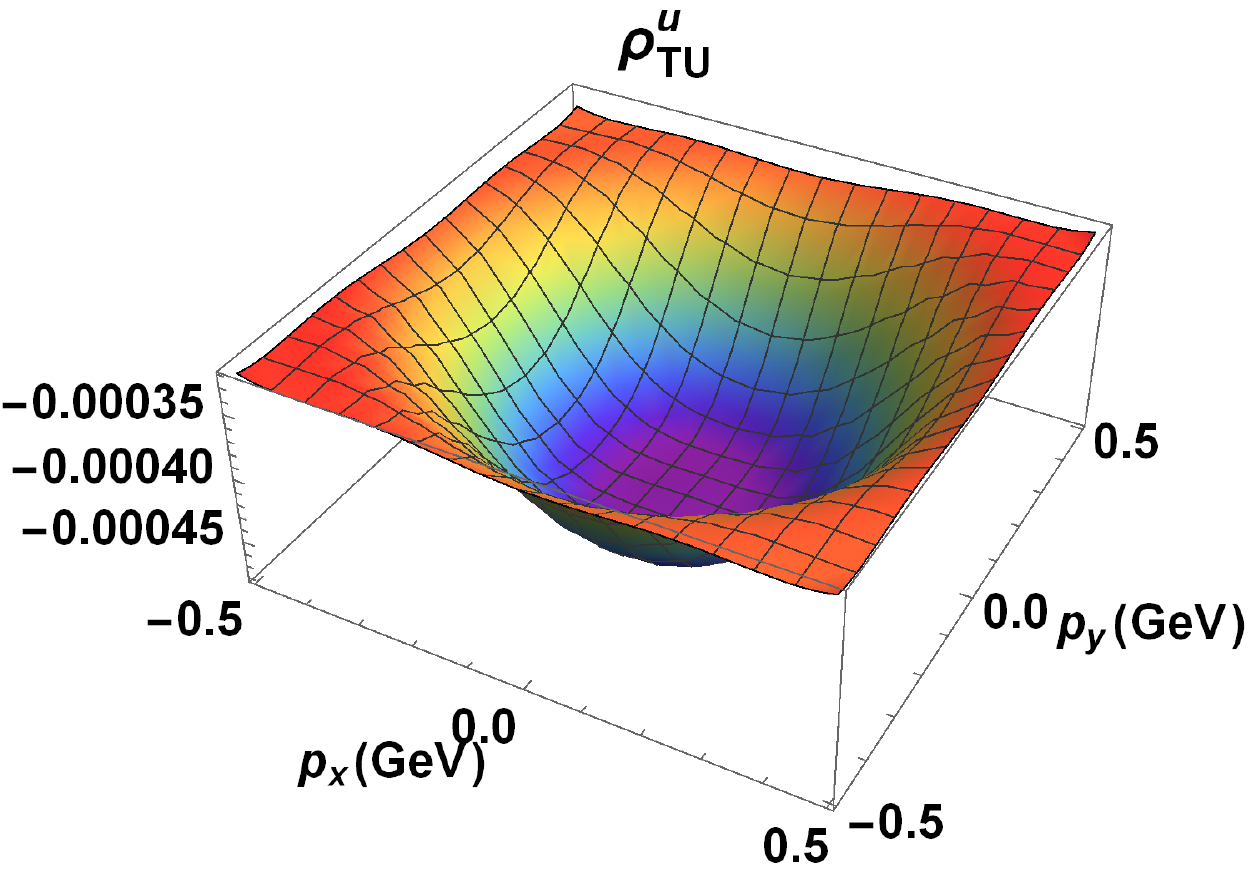}\hfill
(c)\includegraphics[width=.3\textwidth]{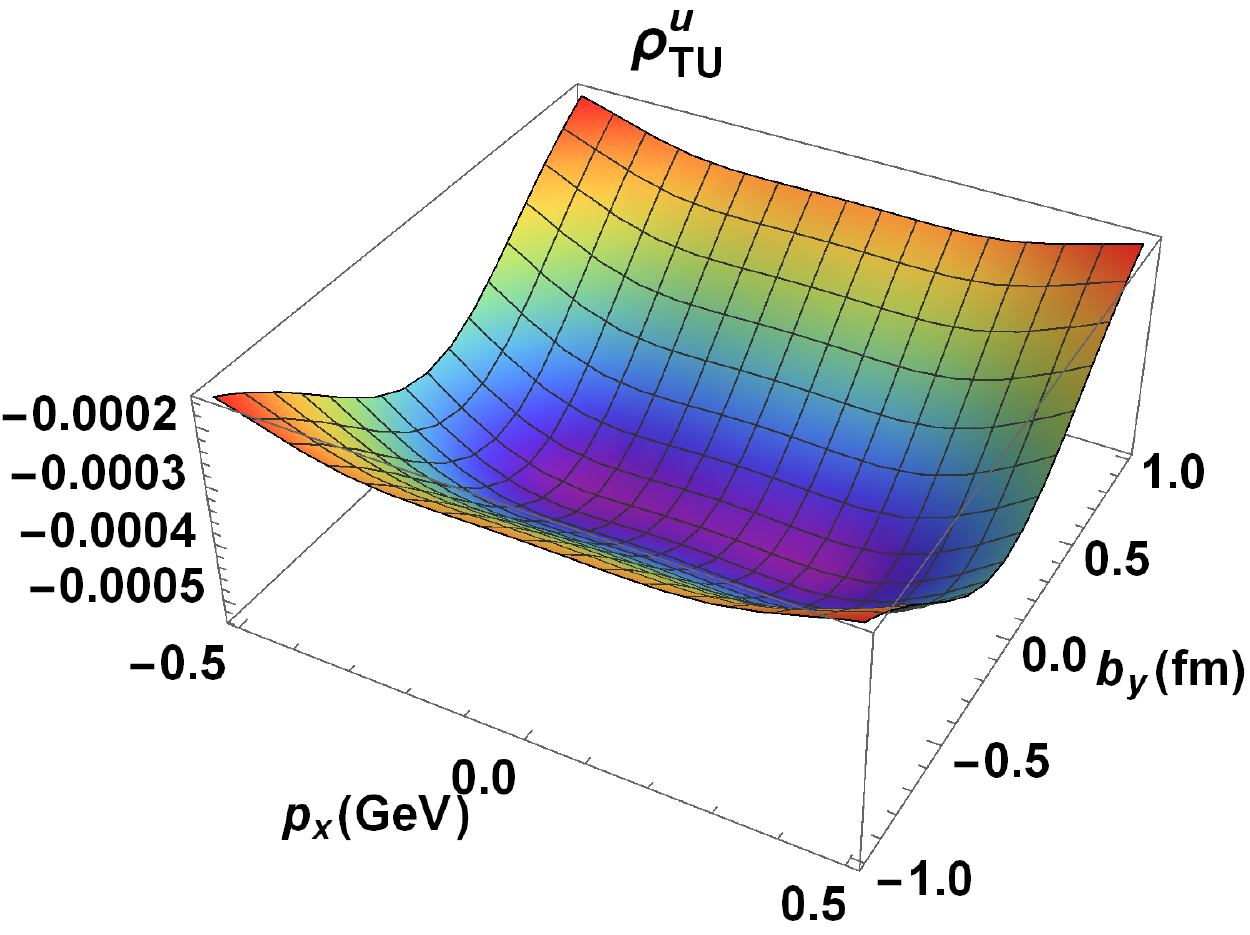}
\end{minipage}
\begin{minipage}[c]{0.98\textwidth}
(d)\includegraphics[width=.3\textwidth]{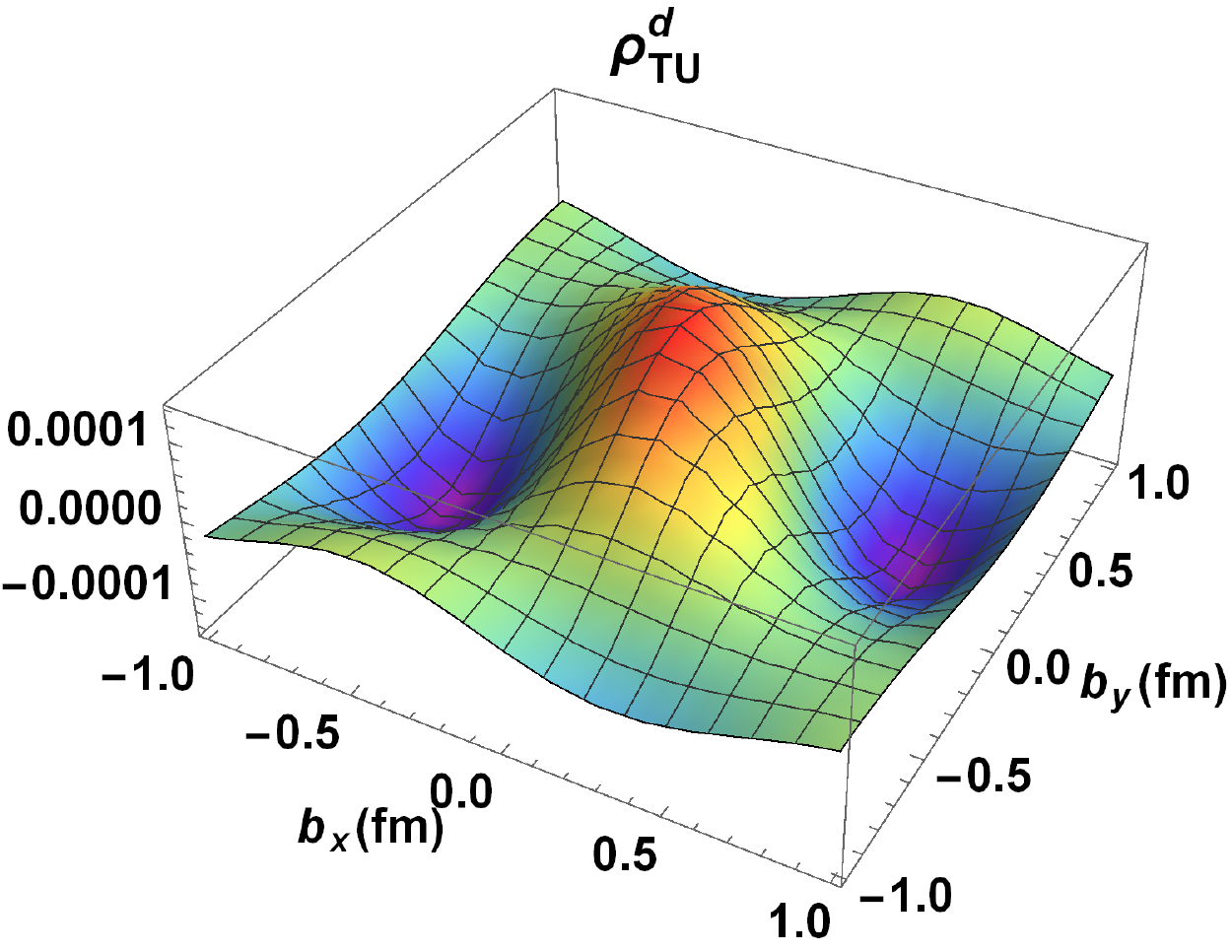}\hfill
(e)\includegraphics[width=.3\textwidth]{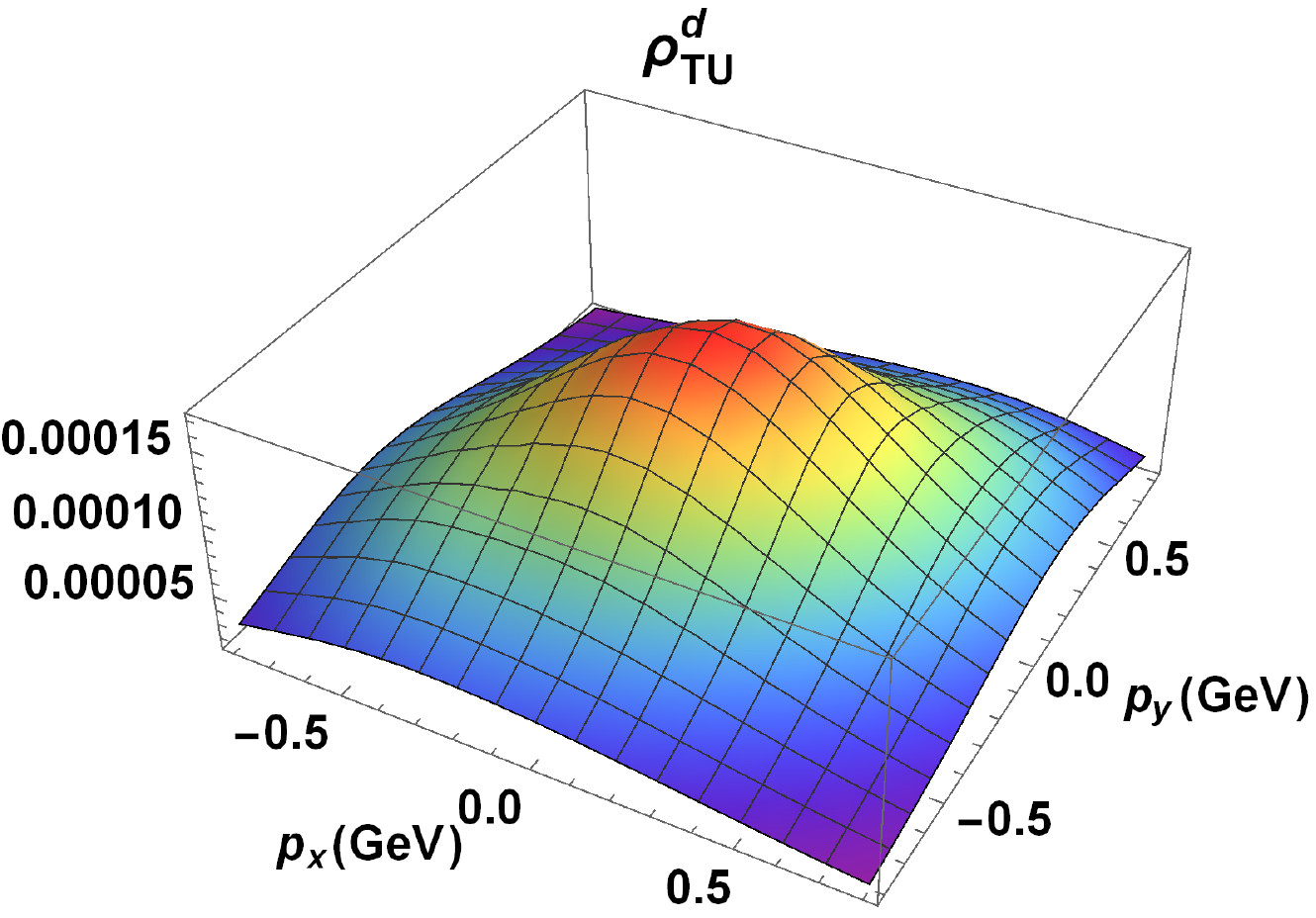}\hfill
(f)\includegraphics[width=.3\textwidth]{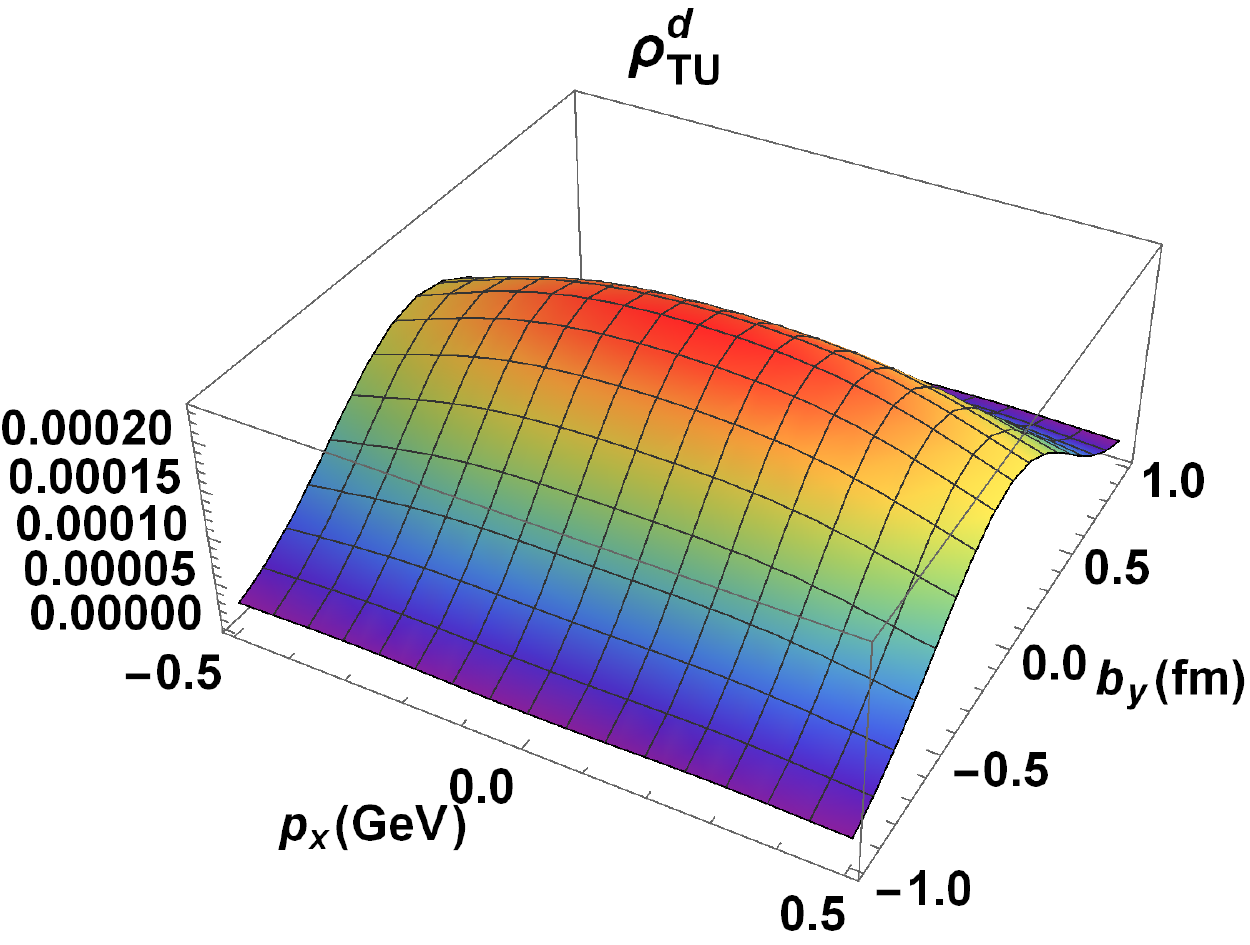}
\end{minipage}
\caption{The transverse-unpolarized Wigner distribution $\rho^i_{TU}$ in impact-parameter space $(b_x,b_y)$, in momentum space $(p_x,p_y)$ and in mixed space $(p_x,b_y)$ for $u(d)$ quark presented in upper (lower) panel.}
\label{transu}
\end{figure}
\begin{figure}
\centering
\begin{minipage}[c]{0.98\textwidth}
(a)\includegraphics[width=.3\textwidth]{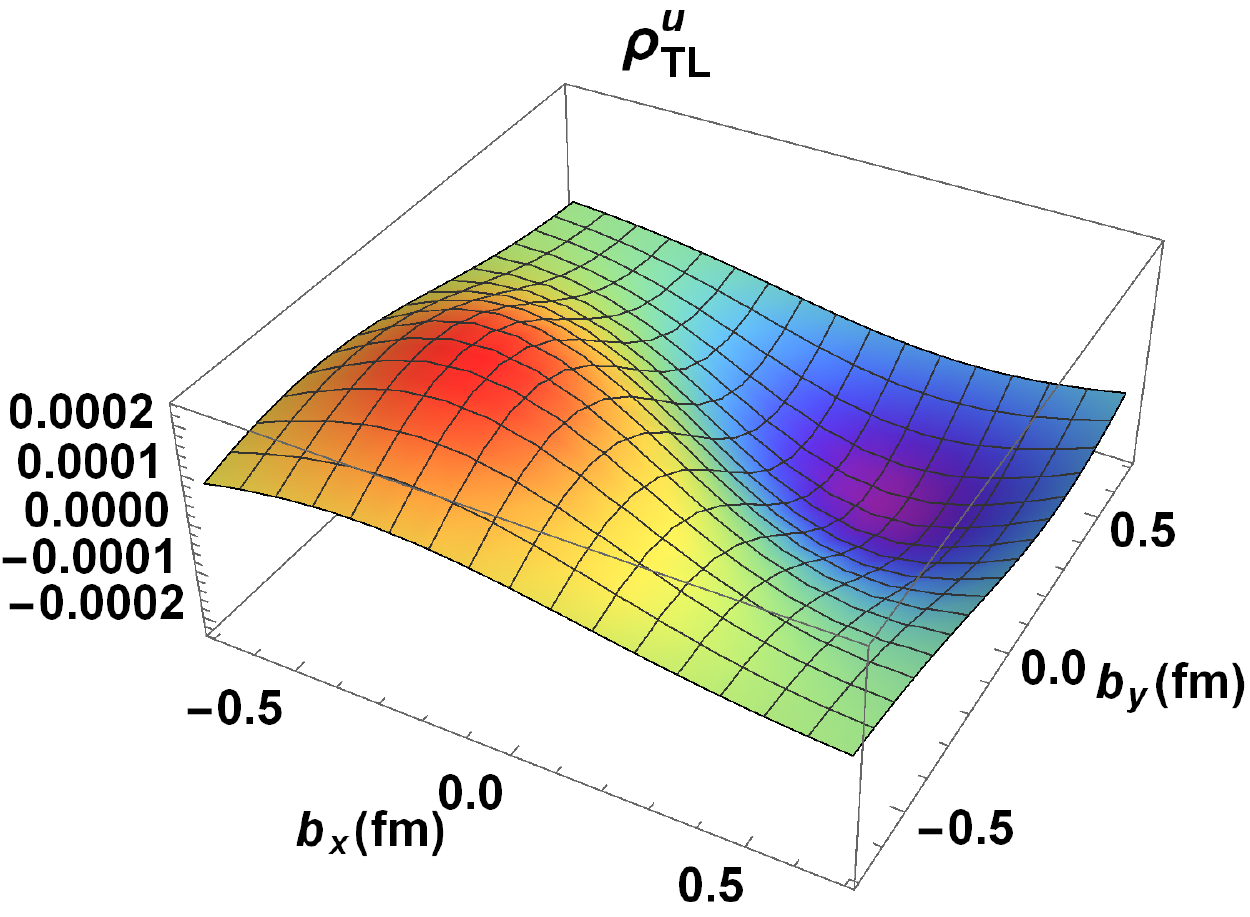}\hfill
(b)\includegraphics[width=.3\textwidth]{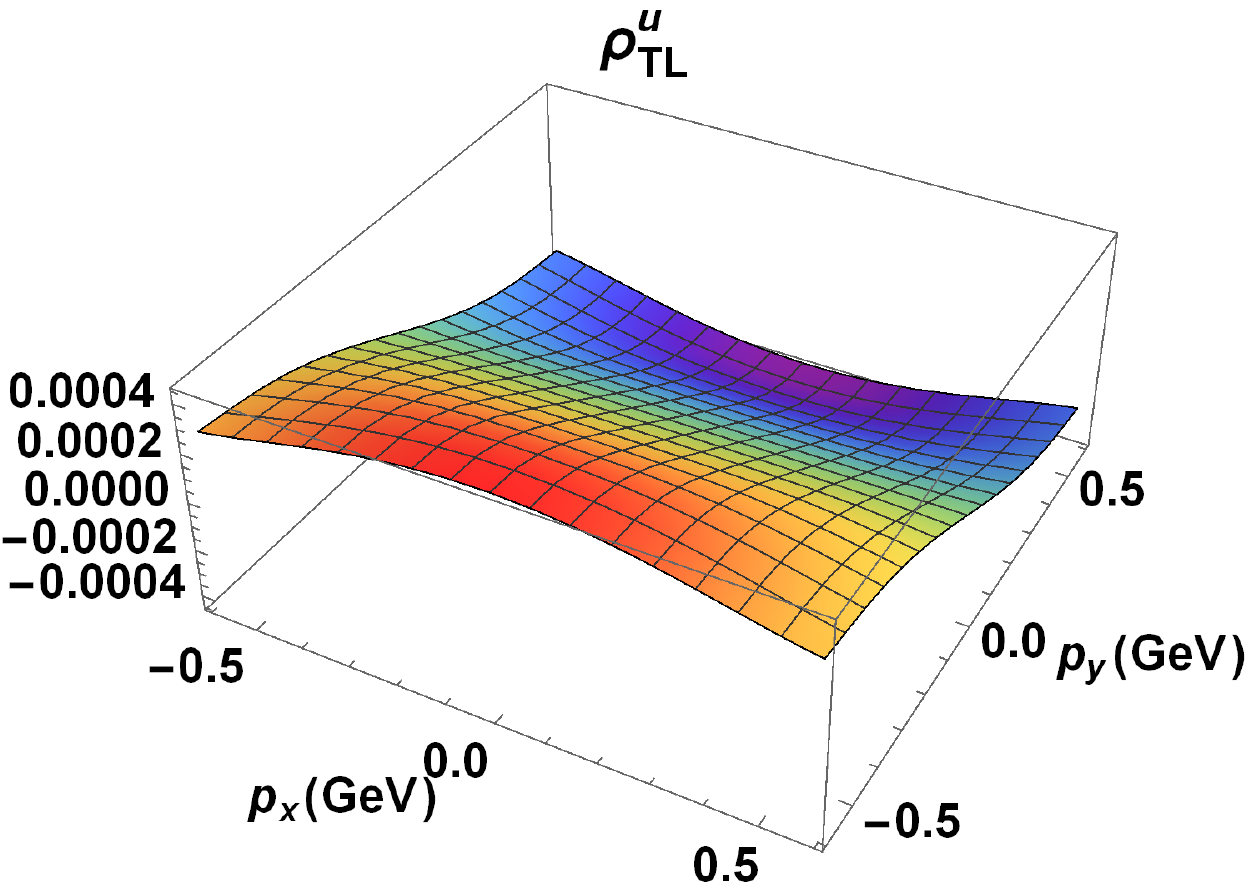}\hfill
(c)\includegraphics[width=.3\textwidth]{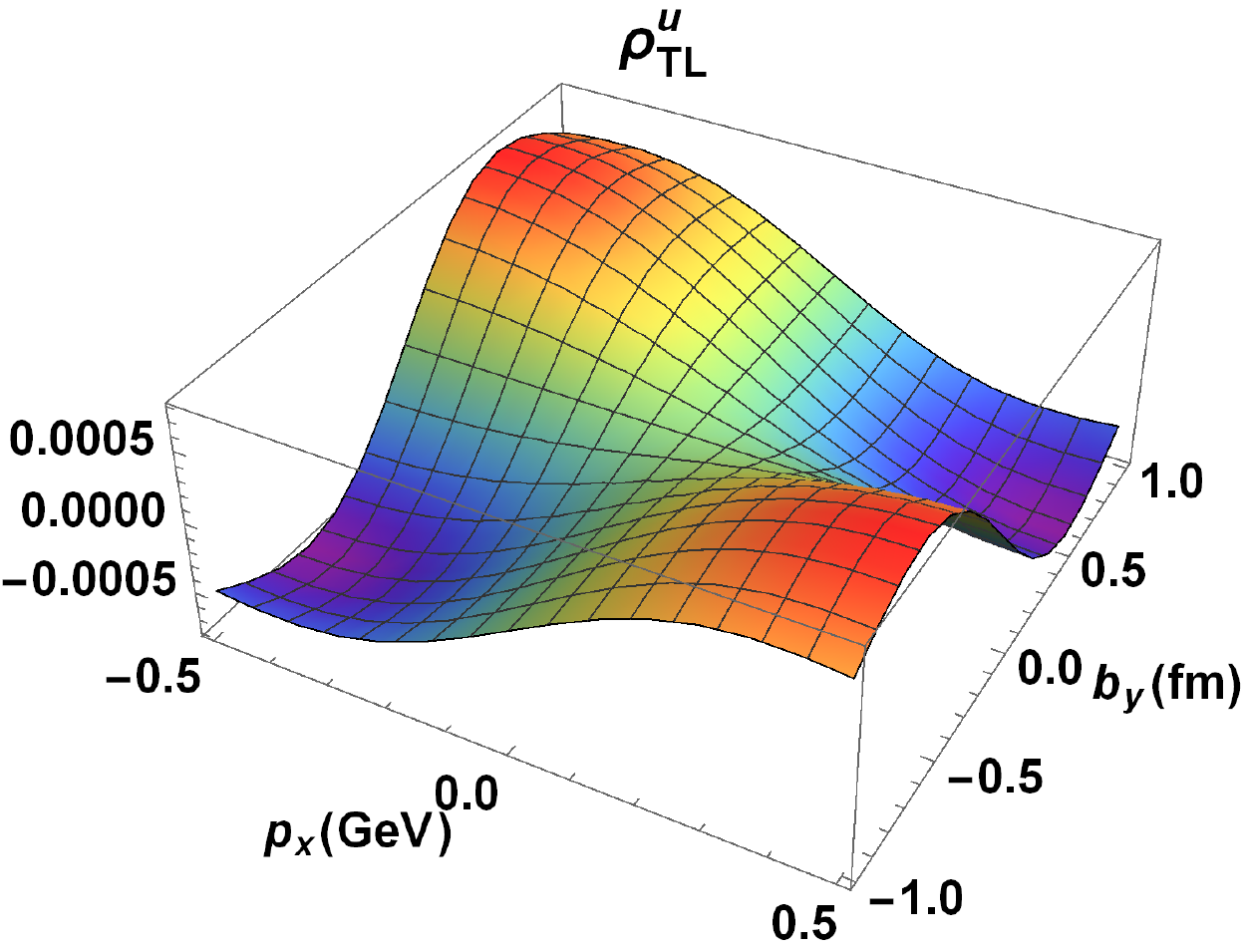}
\end{minipage}
\begin{minipage}[c]{0.98\textwidth}
(d)\includegraphics[width=.3\textwidth]{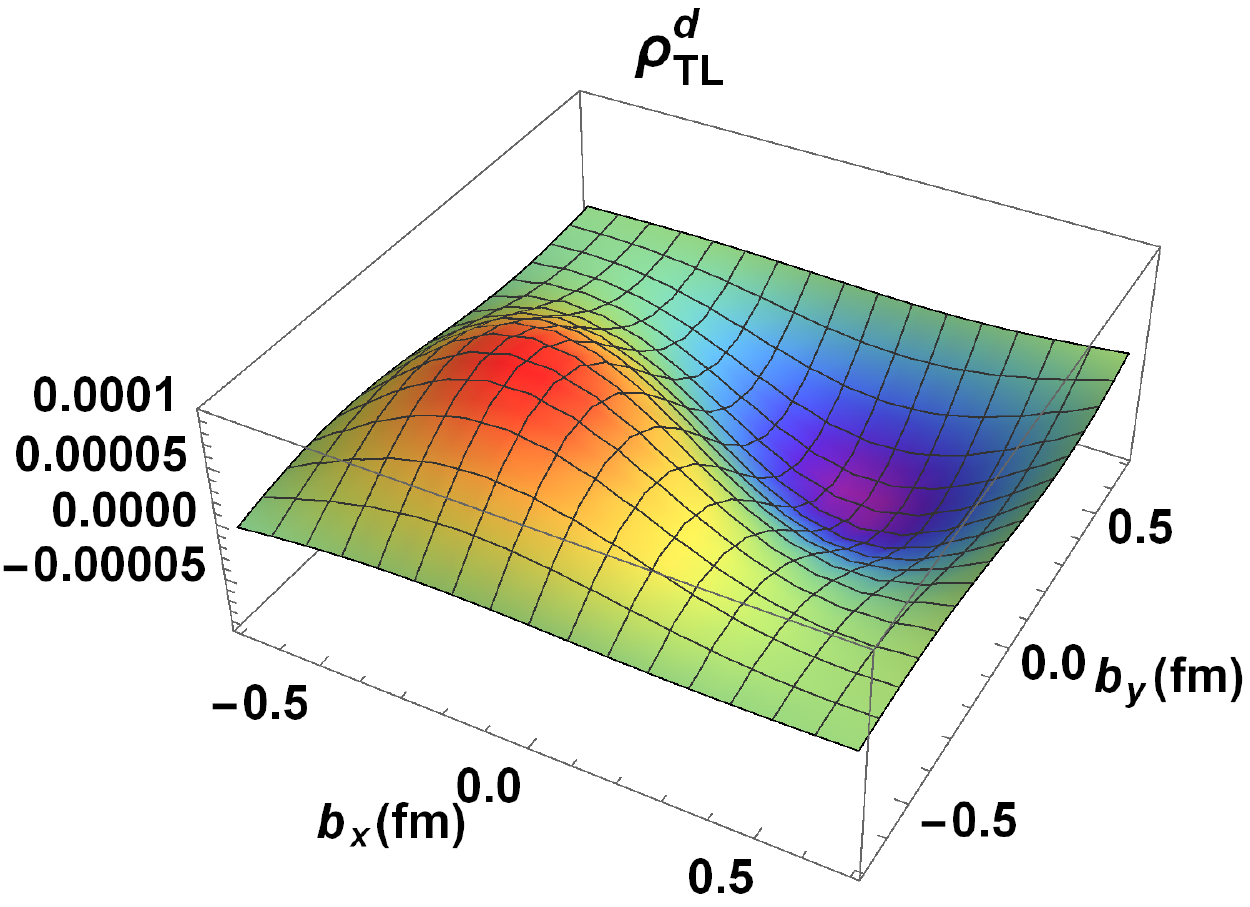}\hfill
(e)\includegraphics[width=.3\textwidth]{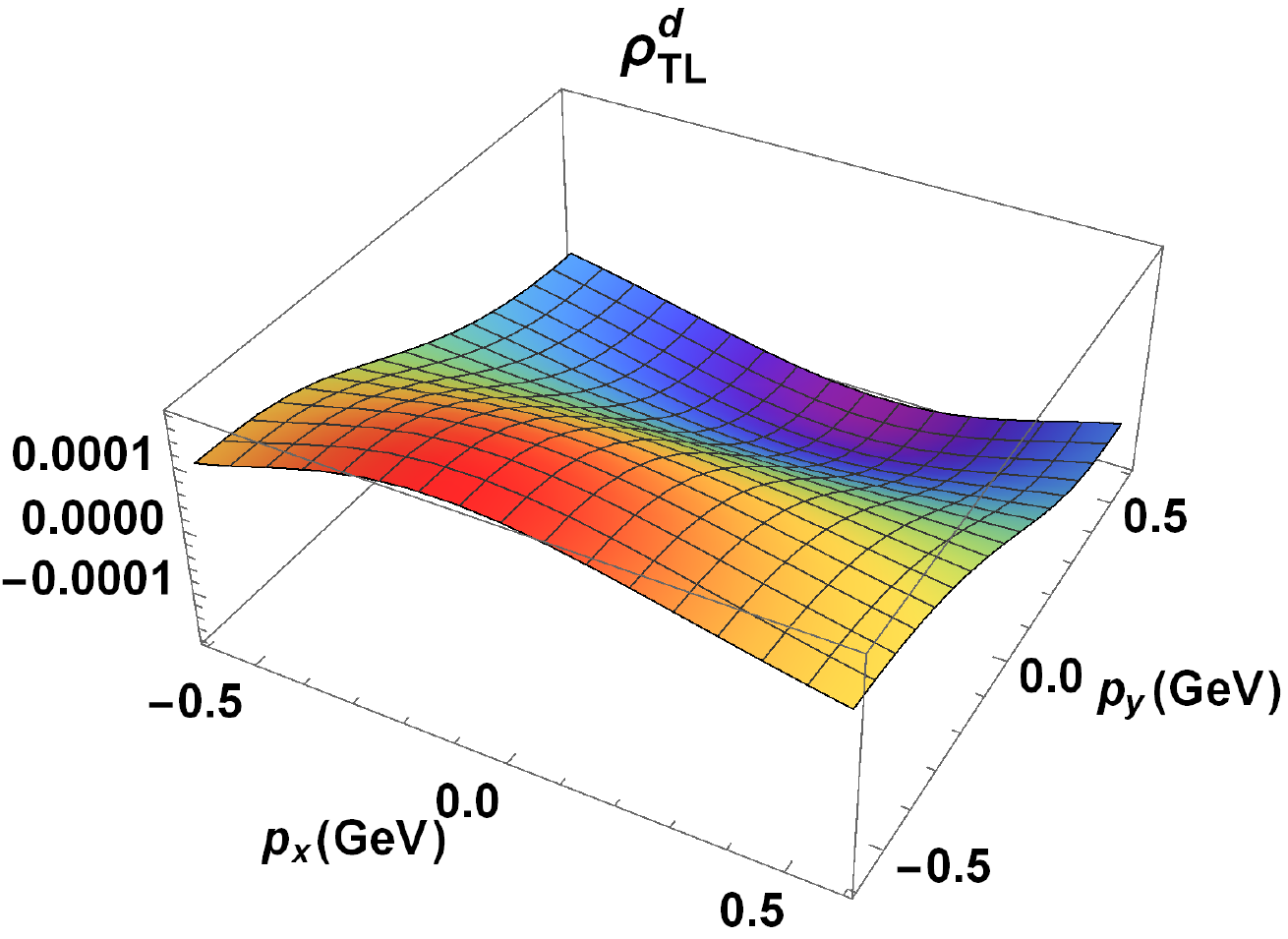}\hfill
(f)\includegraphics[width=.3\textwidth]{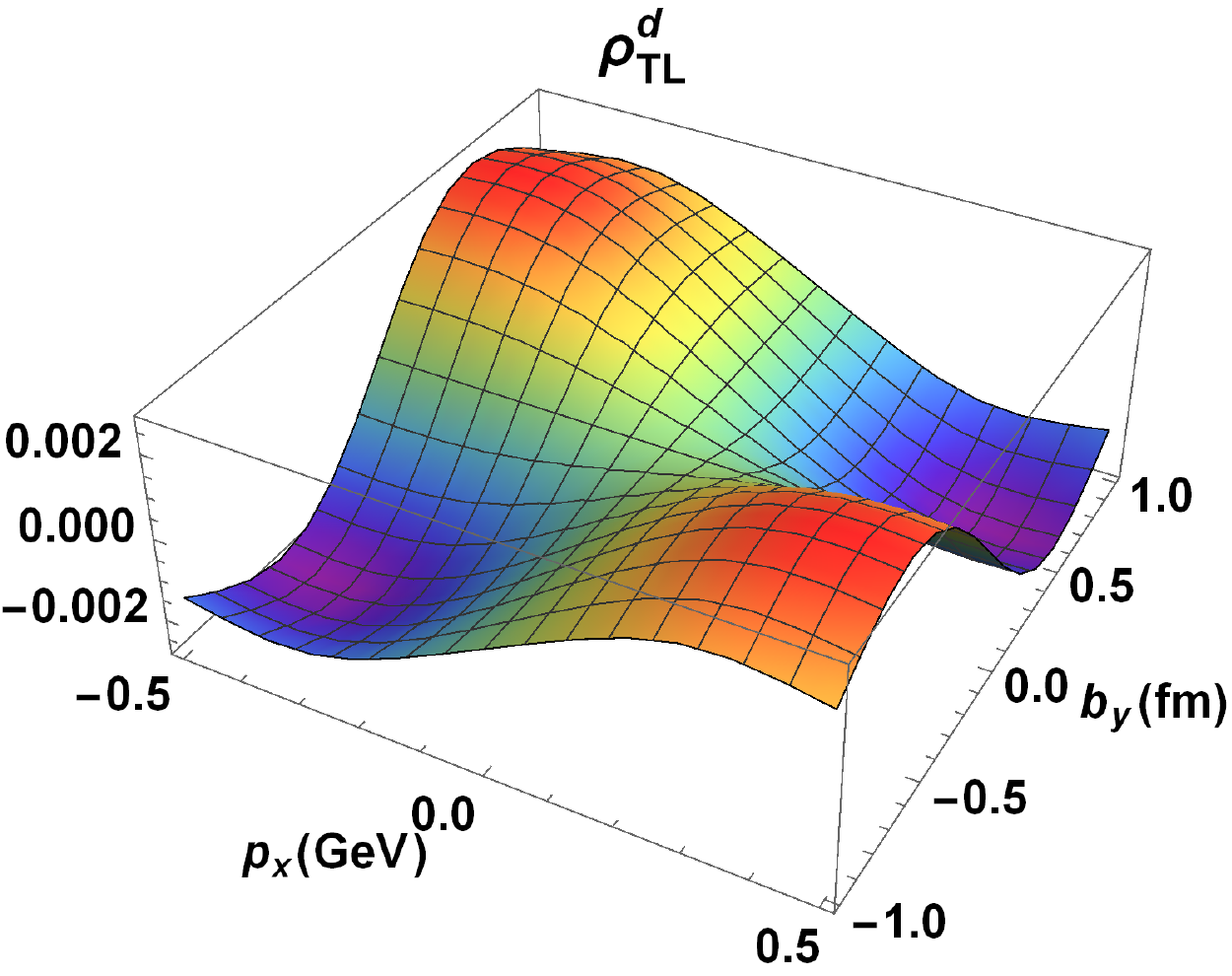}
\end{minipage}
\caption{The transverse-longitudinal Wigner distribution $\rho^i_{TL}$ in impact-parameter space $(b_x,b_y)$, in momentum space $(p_x,p_y)$ and in mixed space $(p_x,b_y)$ for $u(d)$ quark presented in upper (lower) panel.}
\label{transl}
\end{figure} 
\begin{figure}
\centering

\begin{minipage}[c]{0.98\textwidth}
(a)\includegraphics[width=.3\textwidth]{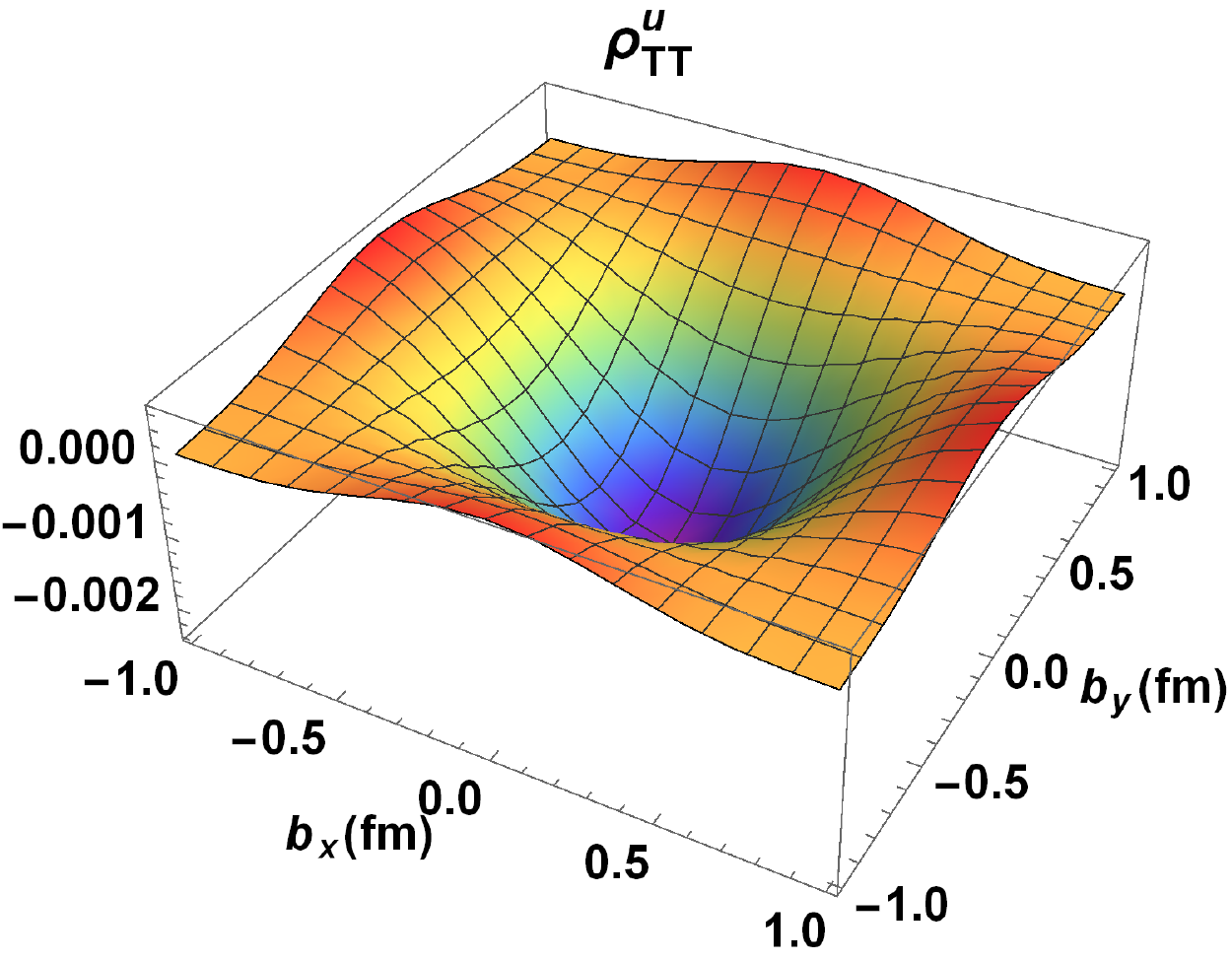}\hfill
(b)\includegraphics[width=.3\textwidth]{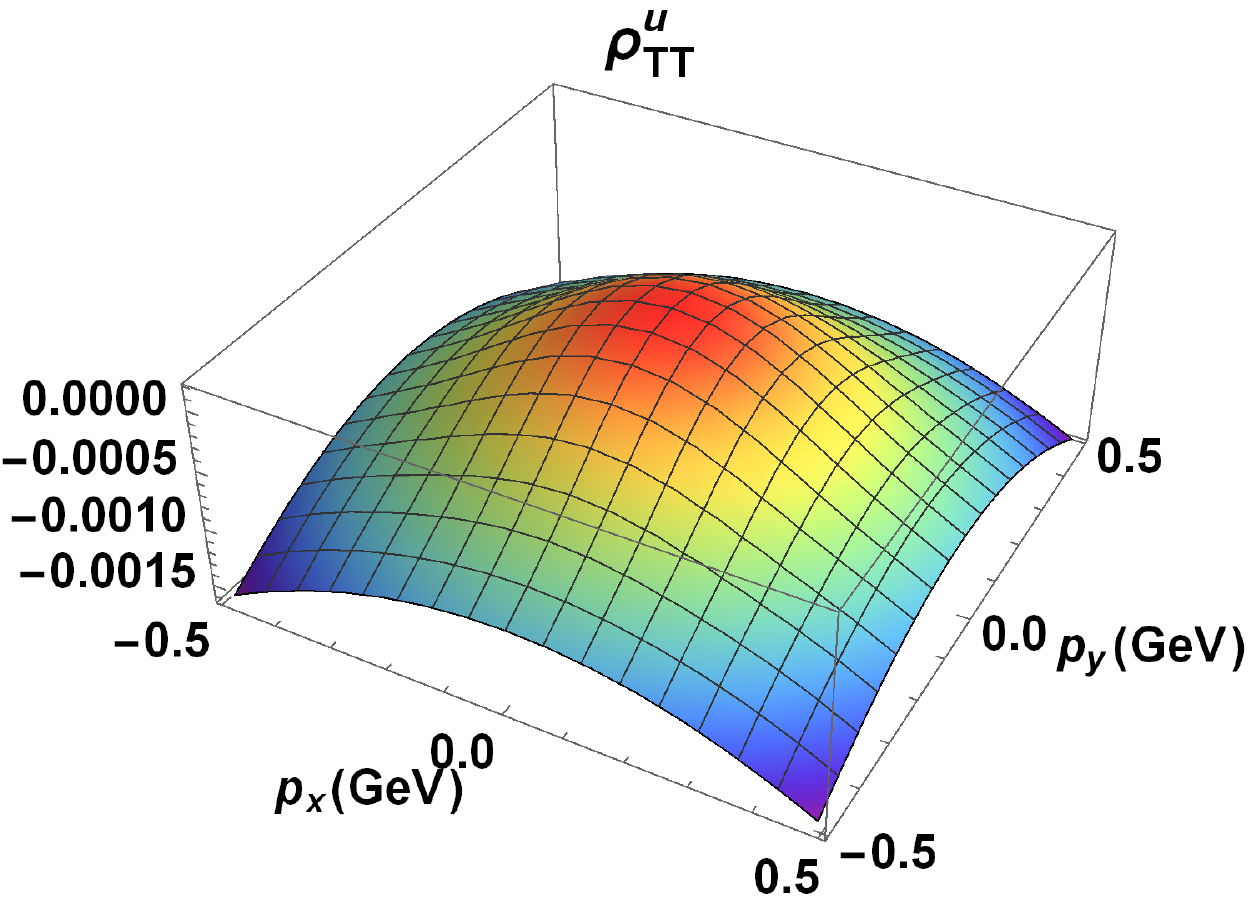}\hfill
(c)\includegraphics[width=.3\textwidth]{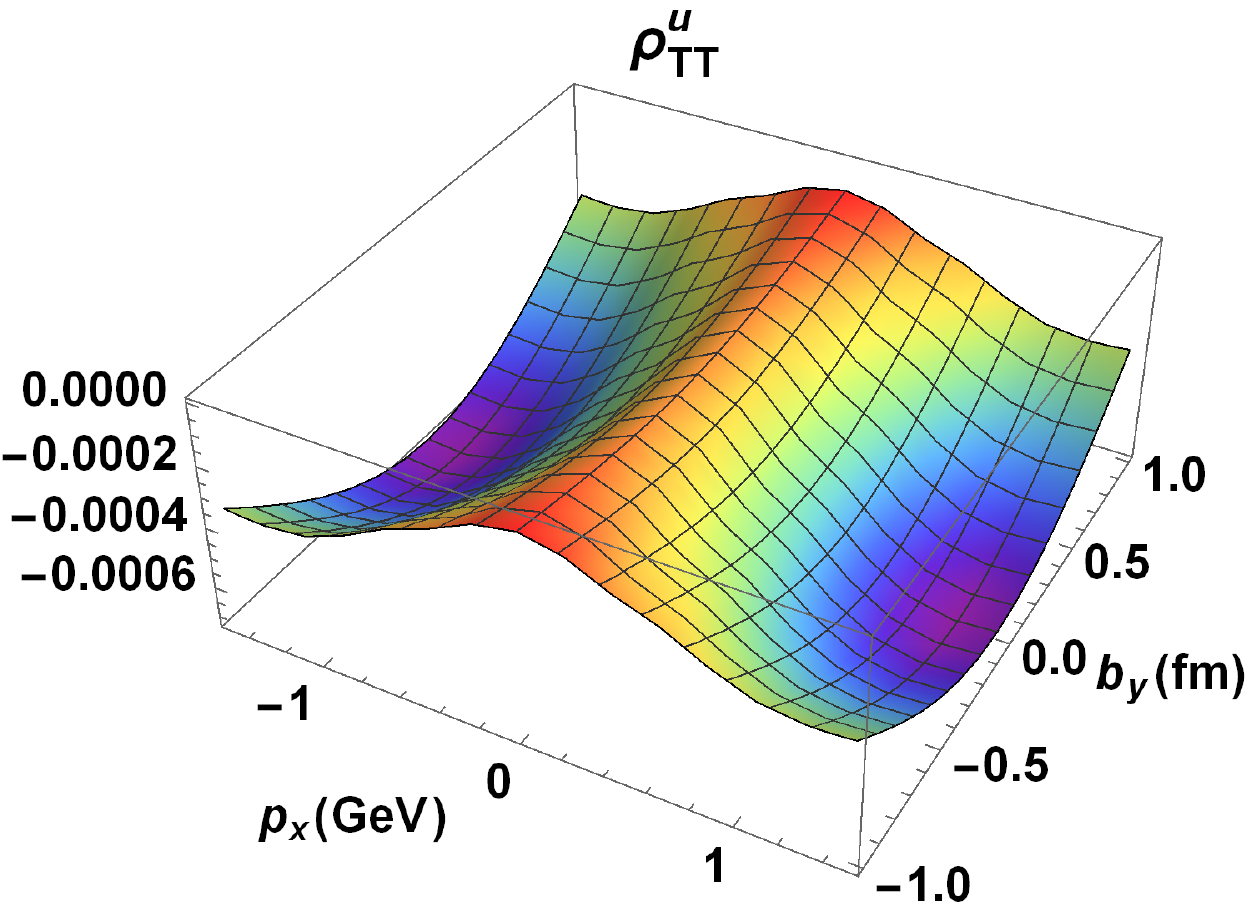}
\end{minipage}
\begin{minipage}[c]{0.98\textwidth}
(d)\includegraphics[width=.3\textwidth]{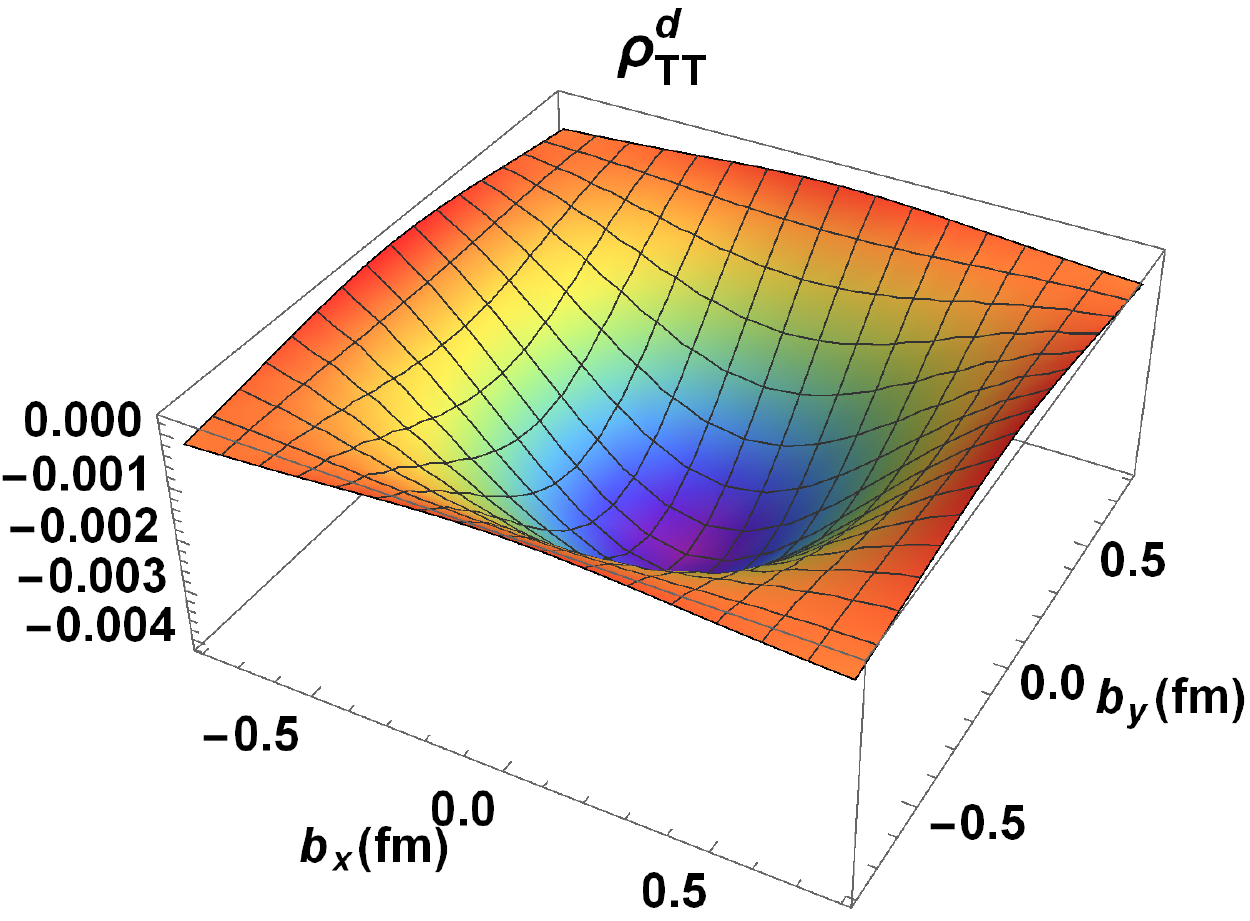}\hfill
(e)\includegraphics[width=.3\textwidth]{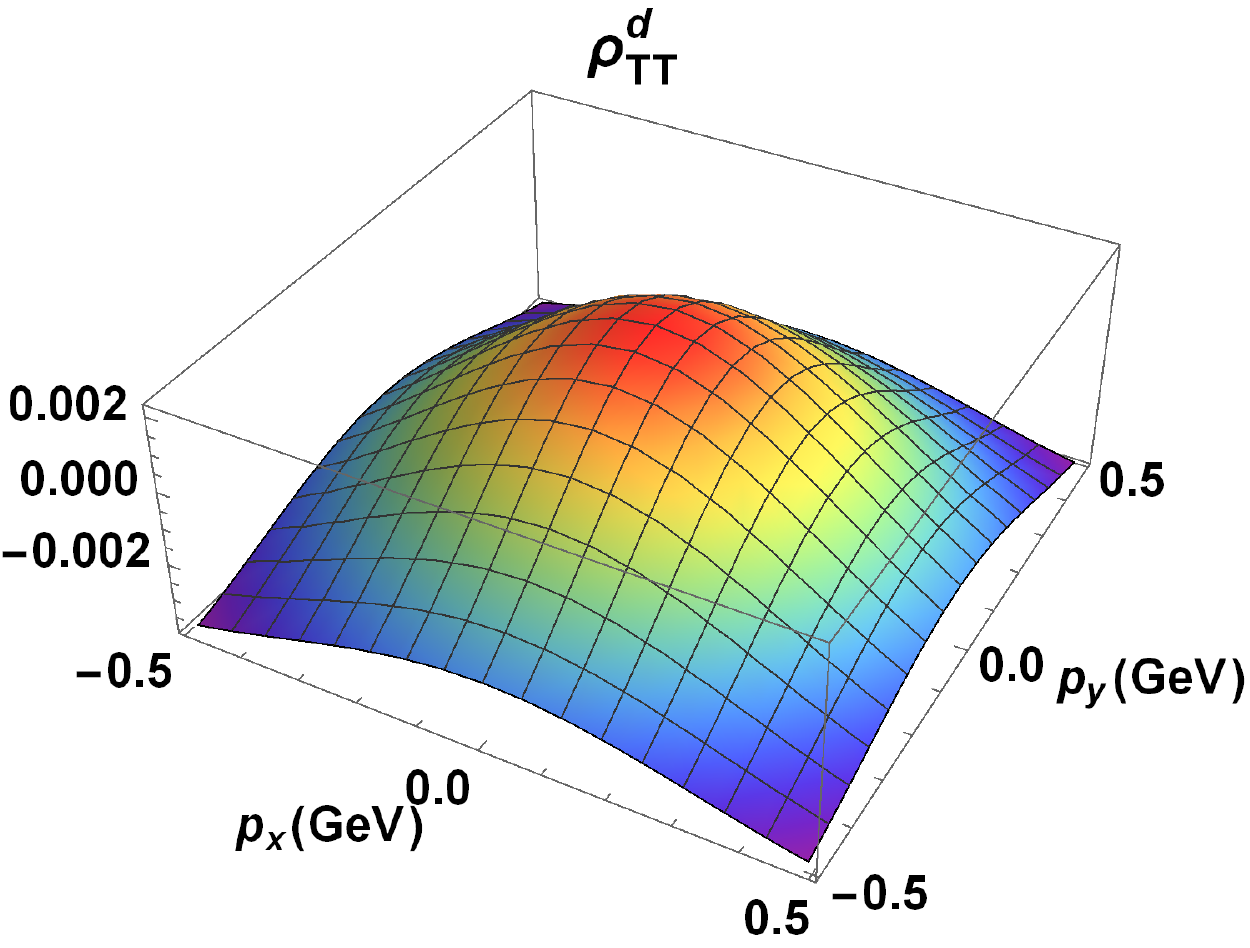}\hfill
(f)\includegraphics[width=.3\textwidth]{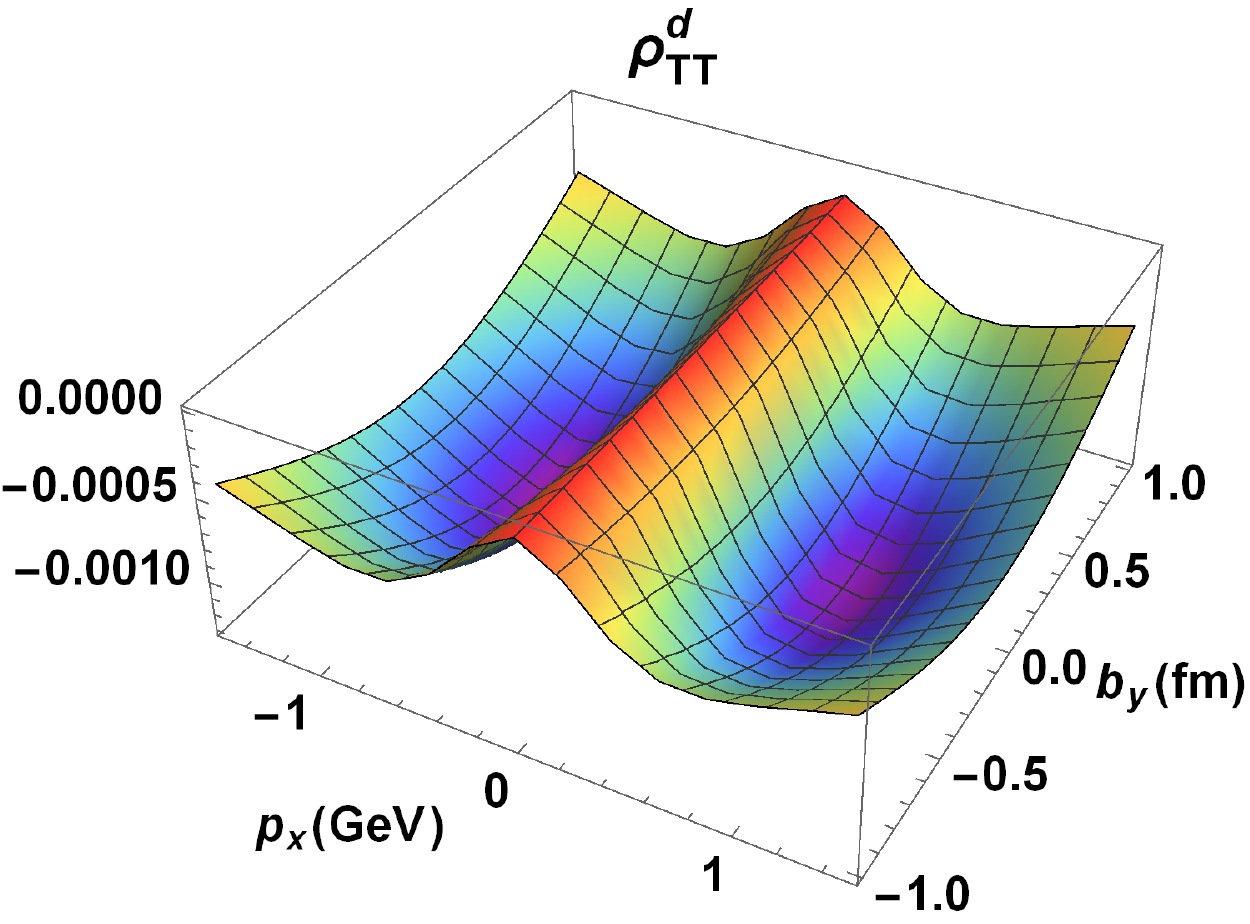}
\end{minipage}
\caption{The transverse Wigner distribution $\rho_{TT}$ in impact-parameter space $(b_x,b_y)$, in momentum space $(p_x,p_y)$ and in mixed space $(p_x,b_y)$ for $u(d)$ quark presented in upper (lower) panel.}
\label{transt}
\end{figure}
In Fig. \ref{transu}, we display the distribution of unpolarized quark in proton which is polarized along $\hat{x}$ $(i=1)$, $\rho^i_{TU}$. We observe an asymmetry in impact-parameter space as shown in Figs. \ref{transu}(a) and \ref{transu}(d). The asymmetry in this case is not located in any particular direction. But it exhibits the strong correlation of the proton polarization with the parallel transverse co-ordinate instead when spin direction is perpendicular to the quark transverse co-ordinate as can be seen from Eqs. (\ref{tus}) and (\ref{tu1}). When we see the effect of this distribution in momentum space (Figs. \ref{transu}(b) and \ref{transu}(e)), we find a circular structure. In the mixed space, $\rho^i_{TU}(p_x,b_y)$ shows axially symmetric distribution as seen from Figs. \ref{transu}(c) and \ref{transu}(f). The three spaces contribution to $\rho^i_{TU}$ show inverse polarities for $u$ and $d$ quarks. At TMD limit, transverse-unpolarized Wigner distribution is linked to Siver's function $f_{1T}^\perp$ which is a T-odd TMD. As discussed earlier since, in the present calculations, we are not taking the gluon contribution into consideration, so no T-odd TMD will be generated from  $\rho^i_{TU}$. At IPD limit, this distribution is connected to chiral-even GPDs $E$ and $H$. This leads to a very important observation that the gauge link is nontrivial and  will play a key role in building a relation between Siver's TMD and the GPDs. 

In Fig. \ref{transl}, the distribution of a longitudinally-polarized quark in the transversely-polarized proton, $\rho^i_{TL}$ is shown. We take proton spin along $\hat{x}$. The distribution in the impact-parameter space shows a dipolar structure displayed in Figs. \ref{transl}(a) and \ref{transl}(d) for $u$ and $d$ quark respectively. In momentum space, $\rho^i_{TL}$ shows a dipolar behavior locating towards $p_x$ as shown in Figs. \ref{transl}(b) and \ref{transl}(e). In mixed space, the distribution exhibits the quadrupole structure as shown in Figs. \ref{transl}(c) and \ref{transl}(f). The GPDs $\tilde{H}$ and $\tilde{E}$ are related to this distribution at IPD   limit and  the TMD $g_{1T}$ is associated with it at TMD limit.

We plot the transverse Wigner distribution $\rho_{TT}$, when both the quark and proton are transversely polarized along same direction in Fig. \ref{transt}. When the quark is polarized to the direction perpendicular to that of the proton, there is one more distribution known as pretzelous Wigner distribution denoted as $\rho_{TT}^\perp$ which comes out to be zero in this model for the case of axial-vector diquark. The transverse distribution $(\rho_{TT})$  is having an opposite polarity when compared with the unpolarized Wigner distributions $(\rho_{UU})$ and longitudinal Wigner distributions $(\rho_{LL})$ in the impact-parameter space (Figs. \ref{transt}(a) and \ref{transt}(d)), in the momentum space (Figs. \ref{transt}(b) and \ref{transt}(e)) as well as in the mixed space (Figs. \ref{transt}(c) and \ref{transt}(f)). $\rho_{TT}$ is related to the TMD $h_{1}$ and to the combination of two GPDs denoted as $H_{T}$ and $\tilde{H}_{T}$ at TMD and IPD limit respectively. 

In the different cases of transversely-polarized proton our results are mostly in agreement with the light-cone spectator model \cite{spectator} for the type of distributions with a few exceptions.   The AdS/QCD quark-diquark model \cite{ads} results differ for the type of distributions as well as for the signs of distributions. The disagreement increases when we move from the case of unpolarized quark and transversely-polarized proton to the case of transversely-polarized quark and transversely-polarized proton. Same is true in the case of the light-front dressed quark model \cite{dressed1}. This holds for the distributions in impact-parameter space, momentum space as well as mixed space. The results corresponding to the case of transversely-polarized proton are nontrivial.  The future experiments will not only provide a direct method to determine the Wigner distributions but also impose important
 constraints on the correlation between the proton polarization being parallel/perpendicular w.r.t. the quark.

\subsection{Generalized Transverse Momentum-dependent parton Distributions (GTMDs)}
\begin{figure}
\centering
(a){\includegraphics[width=7.cm,clip]{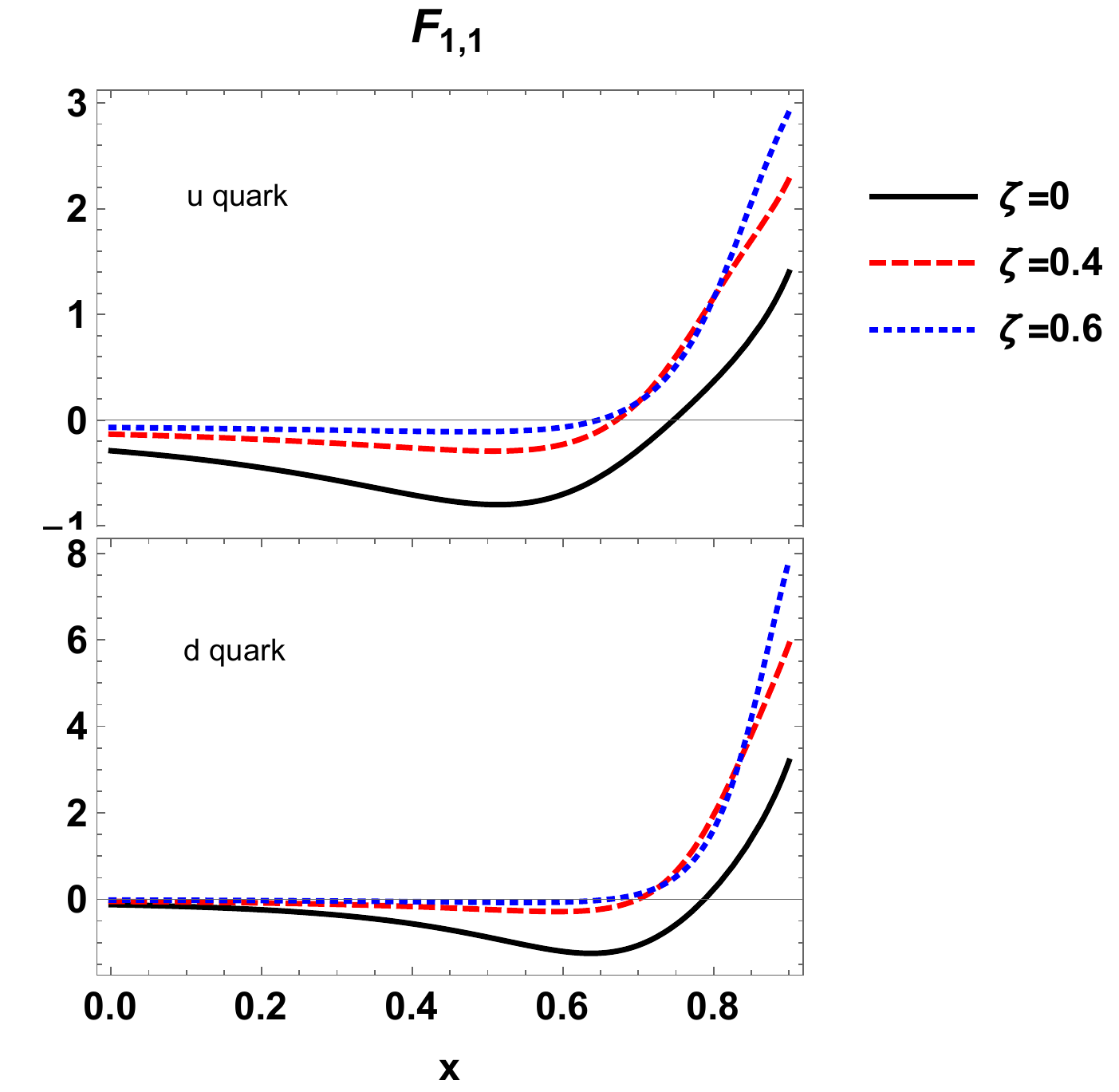}}
(b){\includegraphics[width=7.cm,clip]{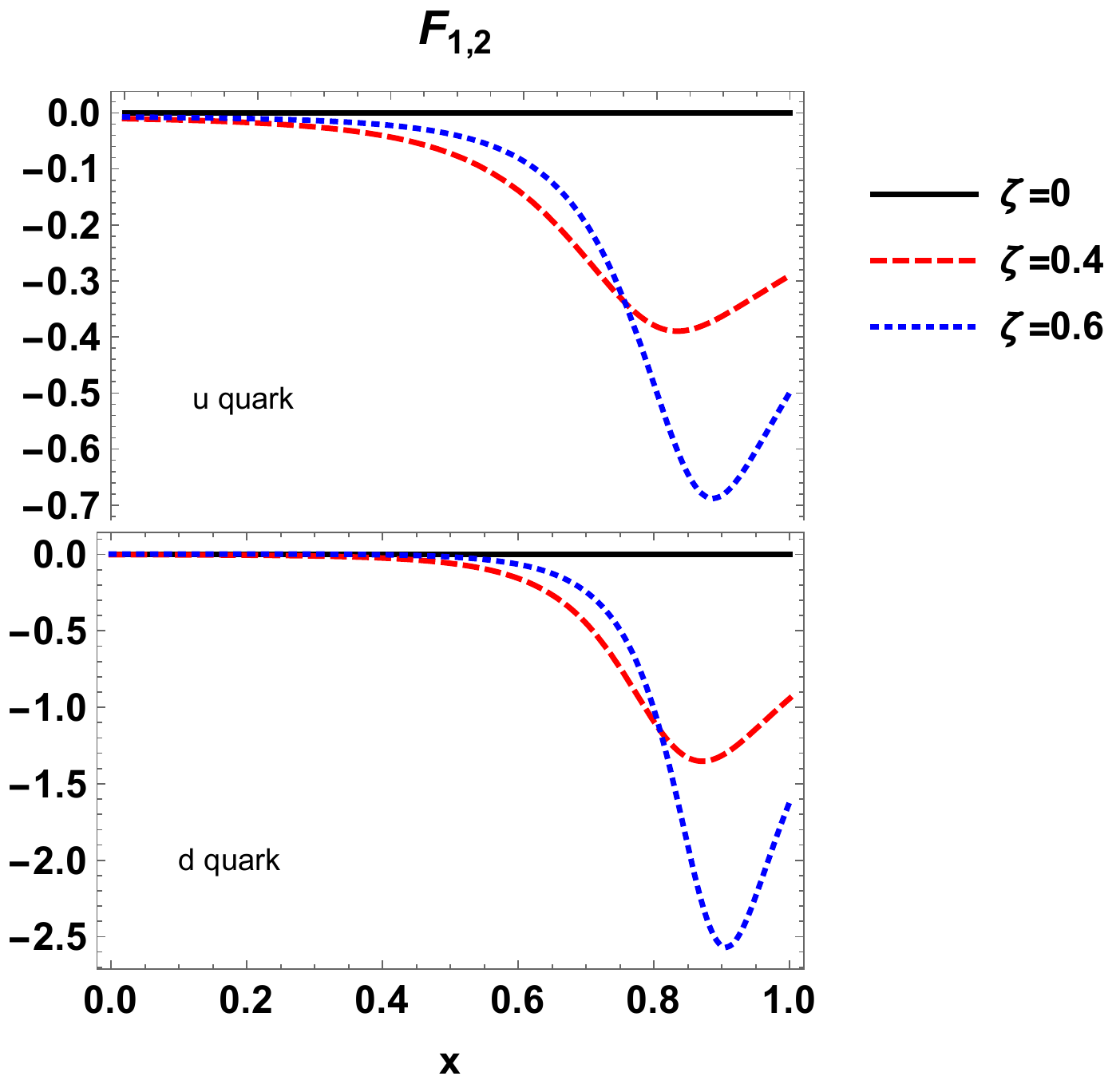}}
(c){\includegraphics[width=7.cm,clip]{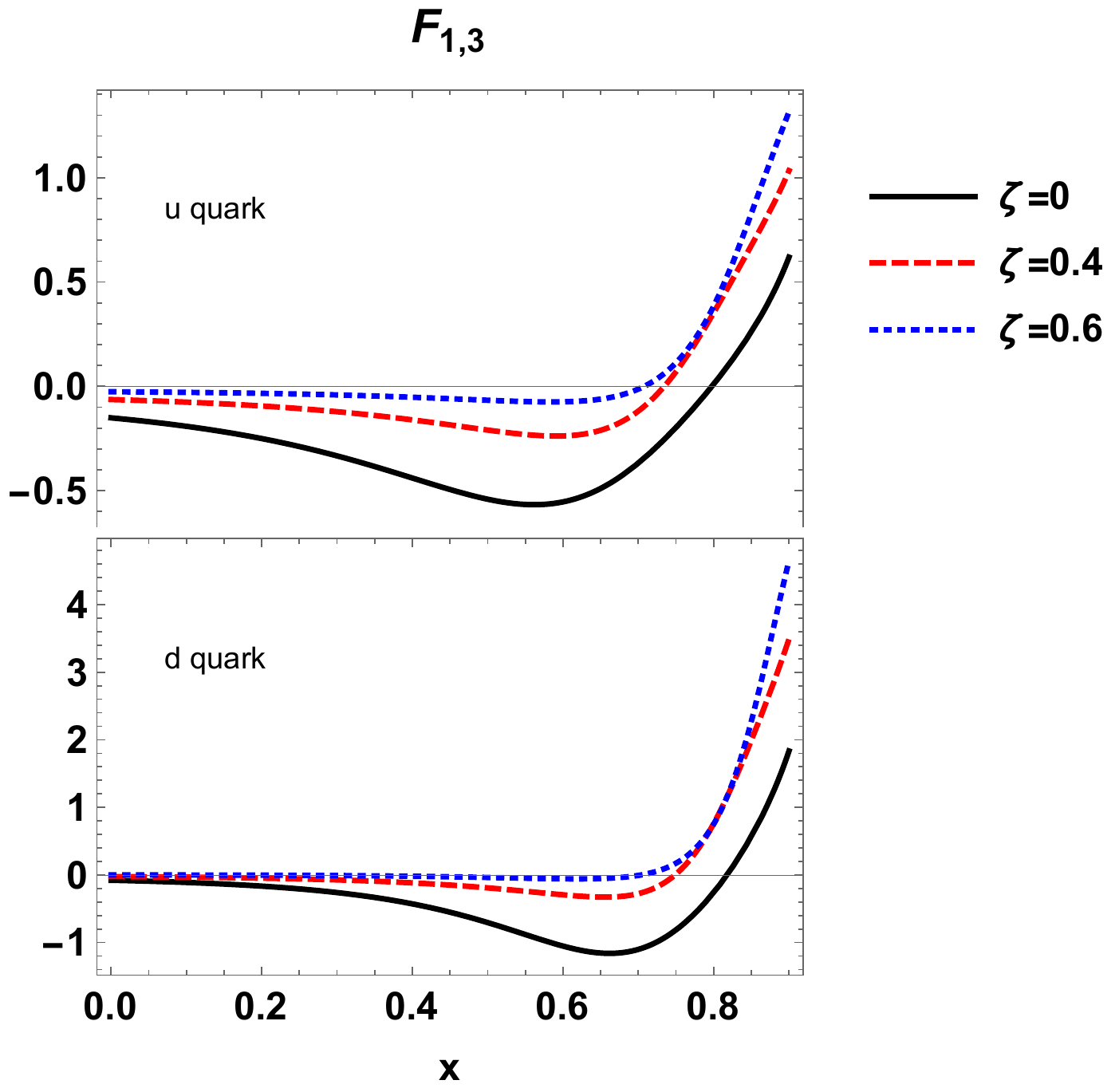}}
(d){\includegraphics[width=7.cm,clip]{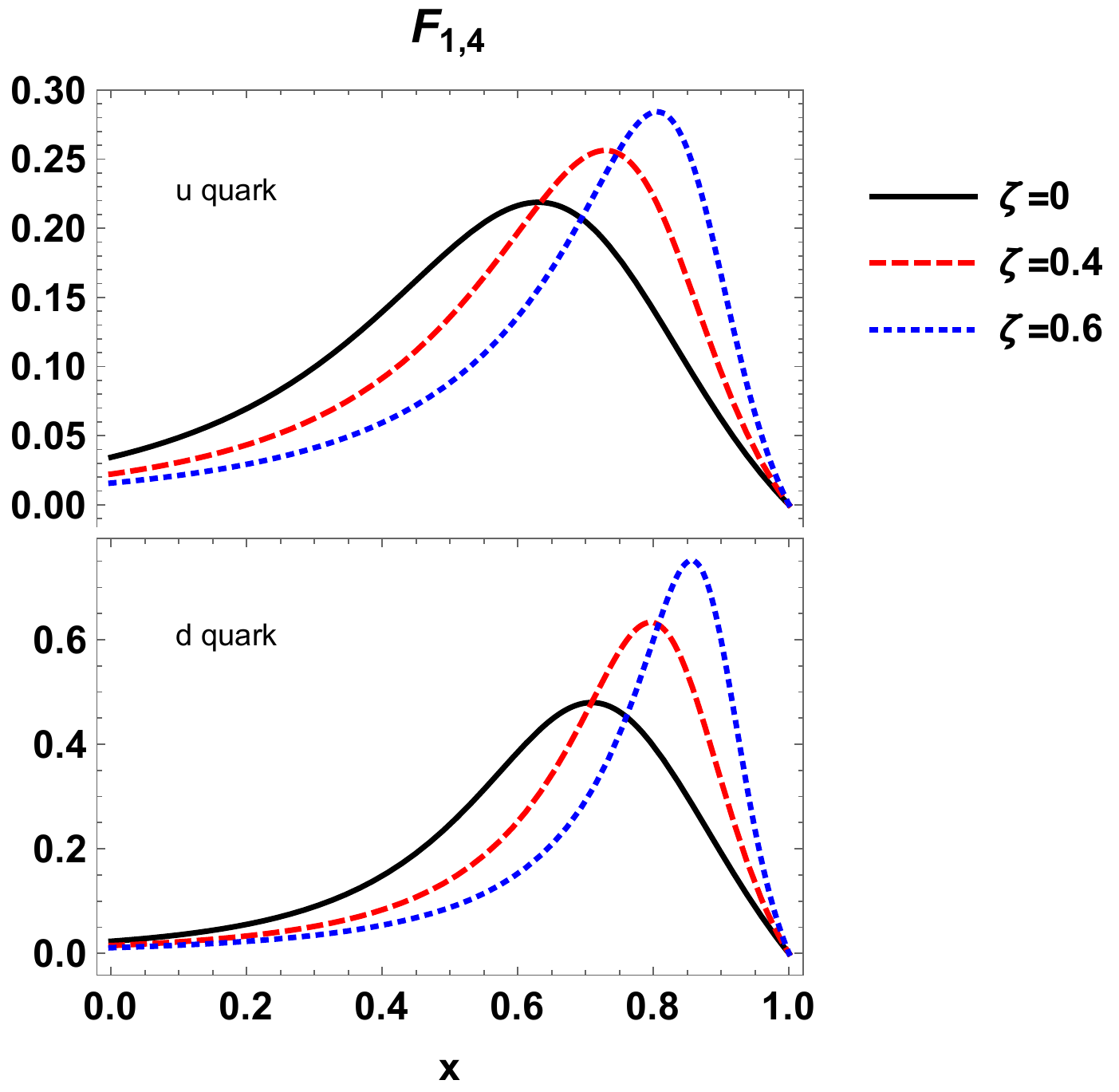}}
\caption{The variation of GTMDs $F_{1,1}(x,\zeta, {\bf \Delta}_\perp, \textbf{p}_\perp)$, $F_{1,2}(x,\zeta, {\bf \Delta}_\perp, \textbf{p}_\perp)$, $F_{1,3}(x,\zeta, {\bf \Delta}_\perp, \textbf{p}_\perp)$ and $F_{1,4}(x,\zeta, {\bf \Delta}_\perp, \textbf{p}_\perp)$ with $x$ for $u$ and $d$ quarks at $\zeta=0,0.4,0.6$.}
\label{F}
\end{figure}
\begin{figure}
\centering
(a){\includegraphics[width=7.cm,clip]{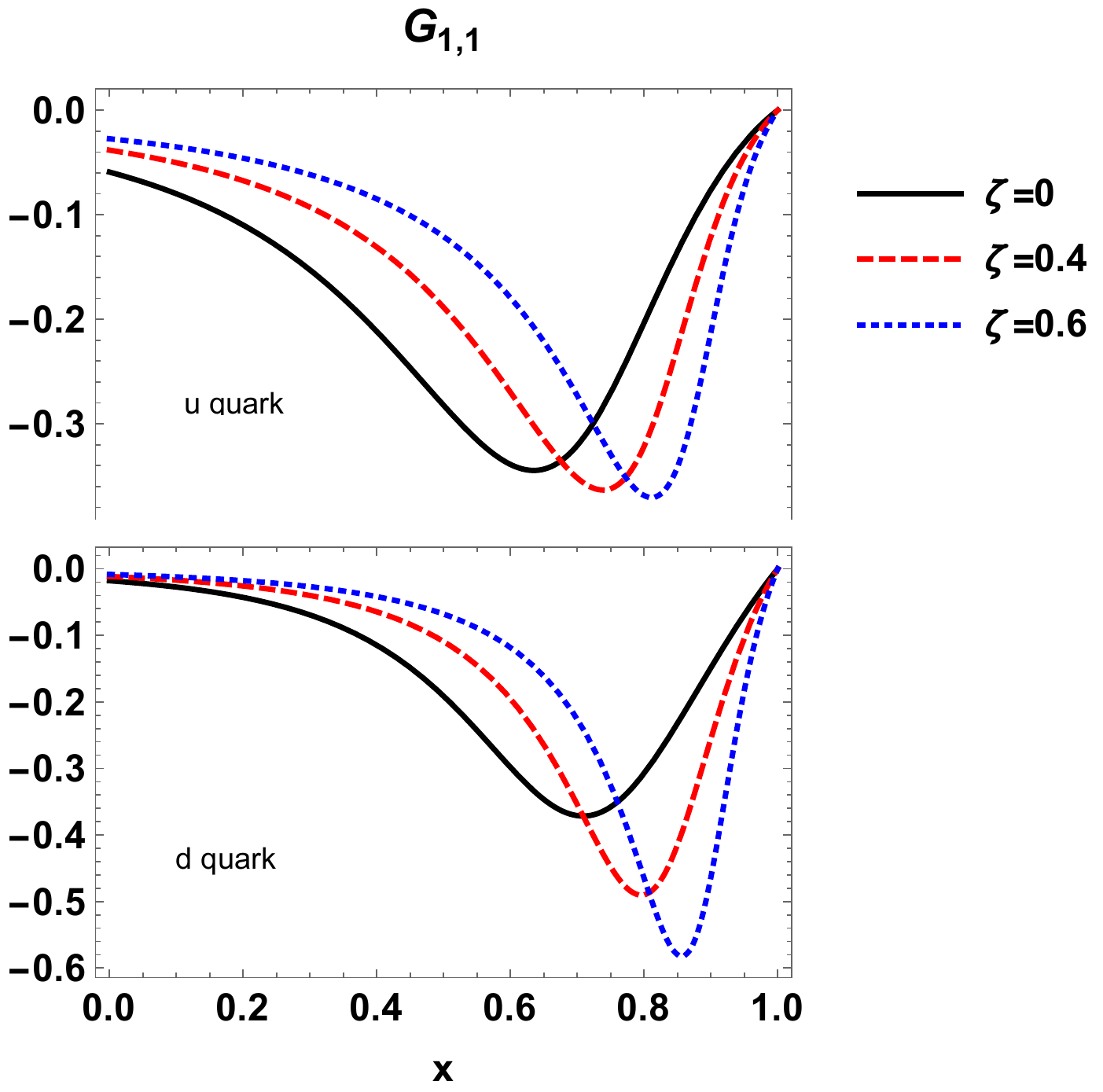}}
(b){\includegraphics[width=7.cm,clip]{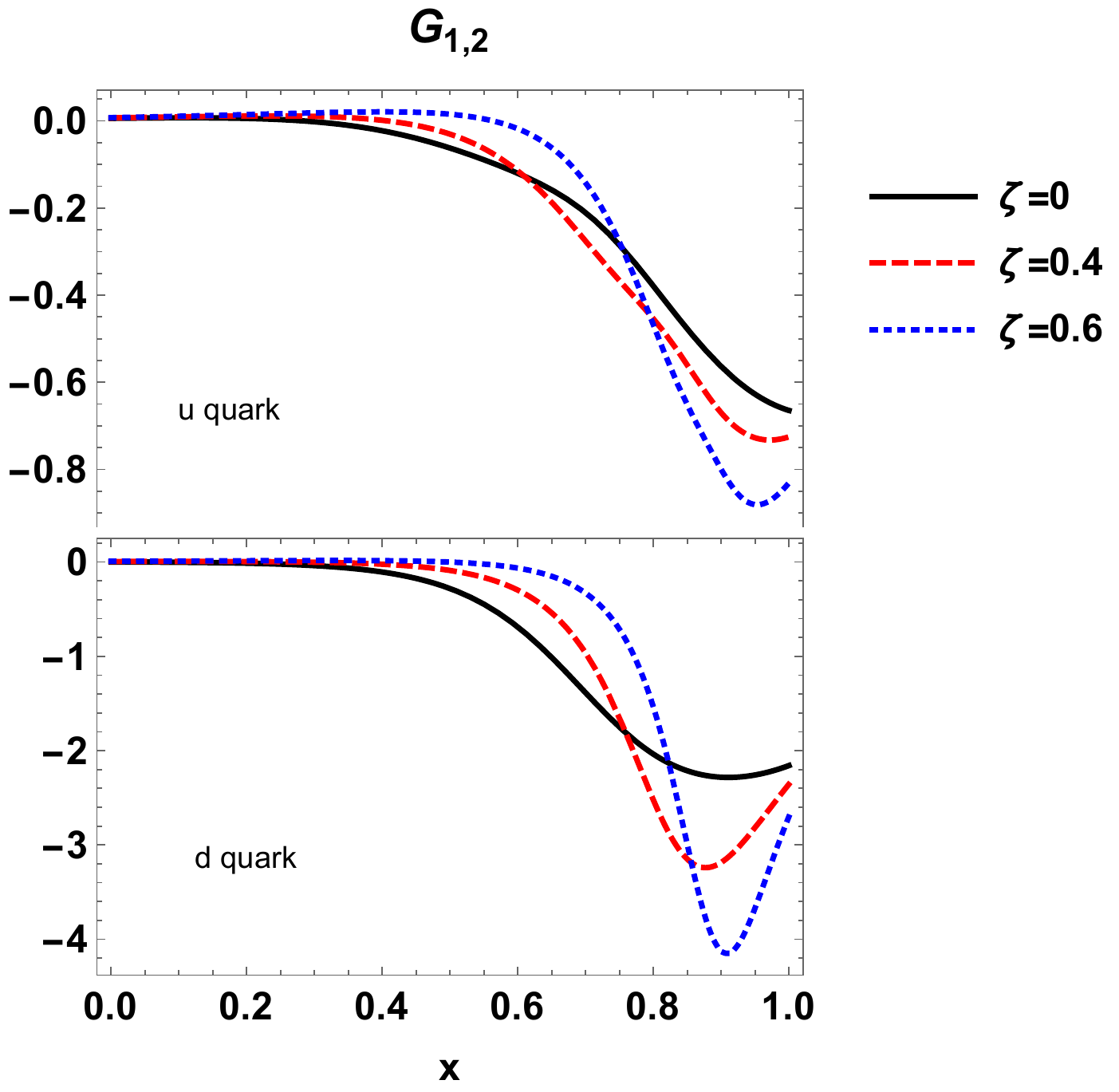}}
(c){\includegraphics[width=7.cm,clip]{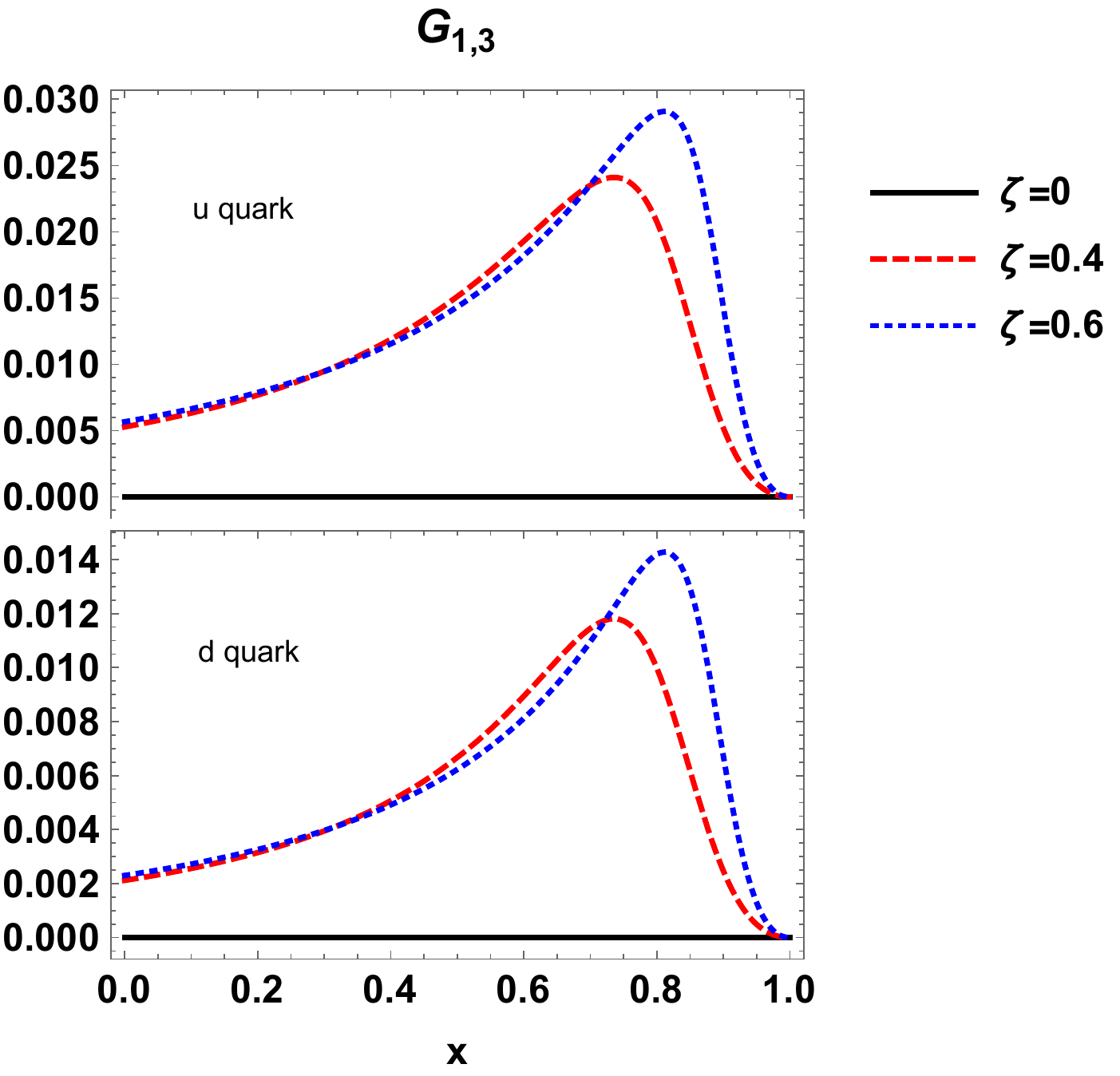}}
(d){\includegraphics[width=7.cm,clip]{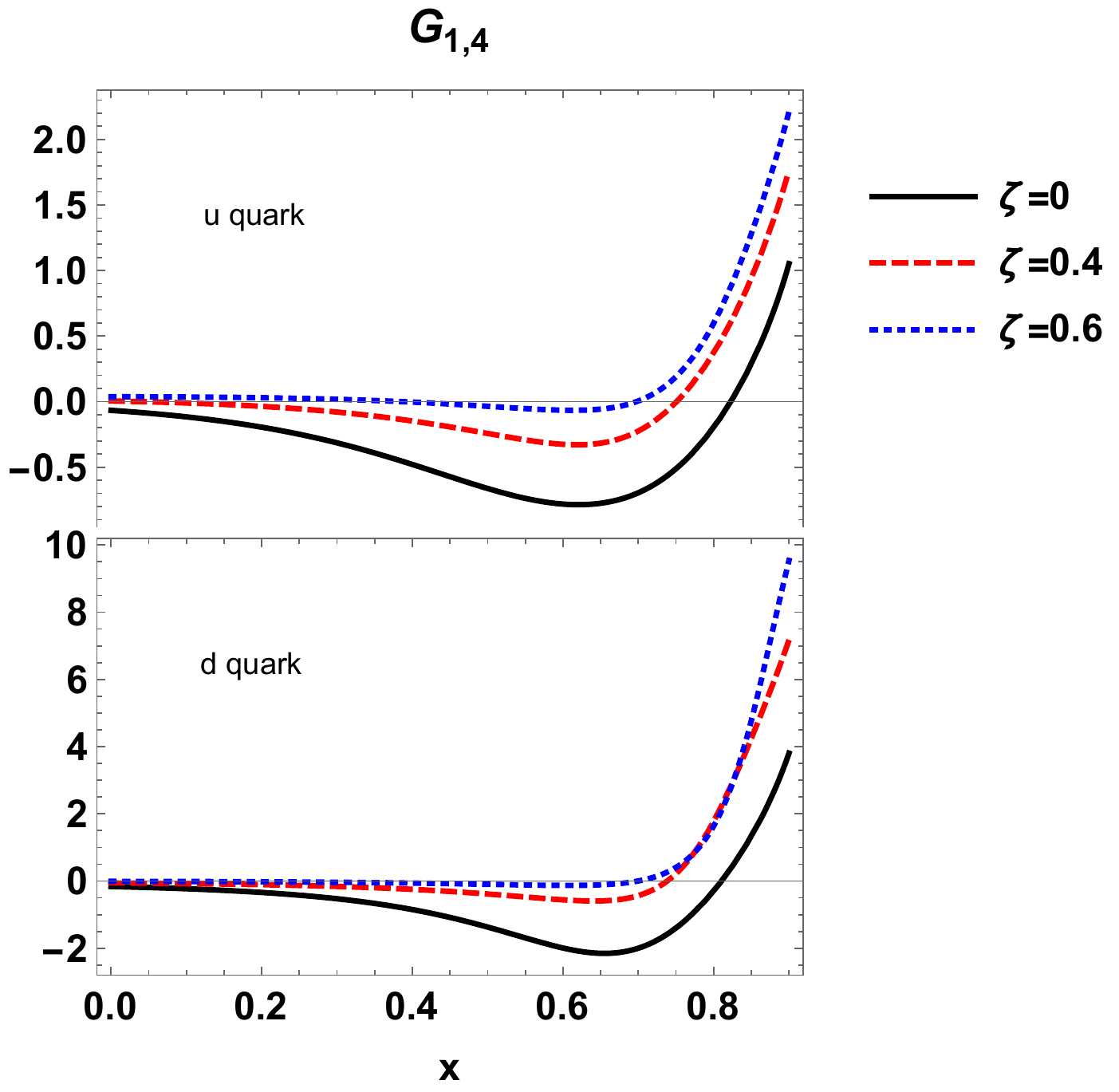}}
\caption{The variation of GTMDs $G_{1,1}(x,\zeta, {\bf \Delta}_\perp, \textbf{p}_\perp)$, $G_{1,2}(x,\zeta, {\bf \Delta}_\perp, \textbf{p}_\perp)$, $G_{1,3}(x,\zeta, {\bf \Delta}_\perp, \textbf{p}_\perp)$ and $G_{1,4}(x,\zeta, {\bf \Delta}_\perp, \textbf{p}_\perp)$ with $x$ for $u$ and $d$ quarks at $\zeta=0,0.4,0.6$.}
\label{g}
\end{figure}
\begin{figure}[hbtp]
\centering
(a){\includegraphics[width=7.cm,clip]{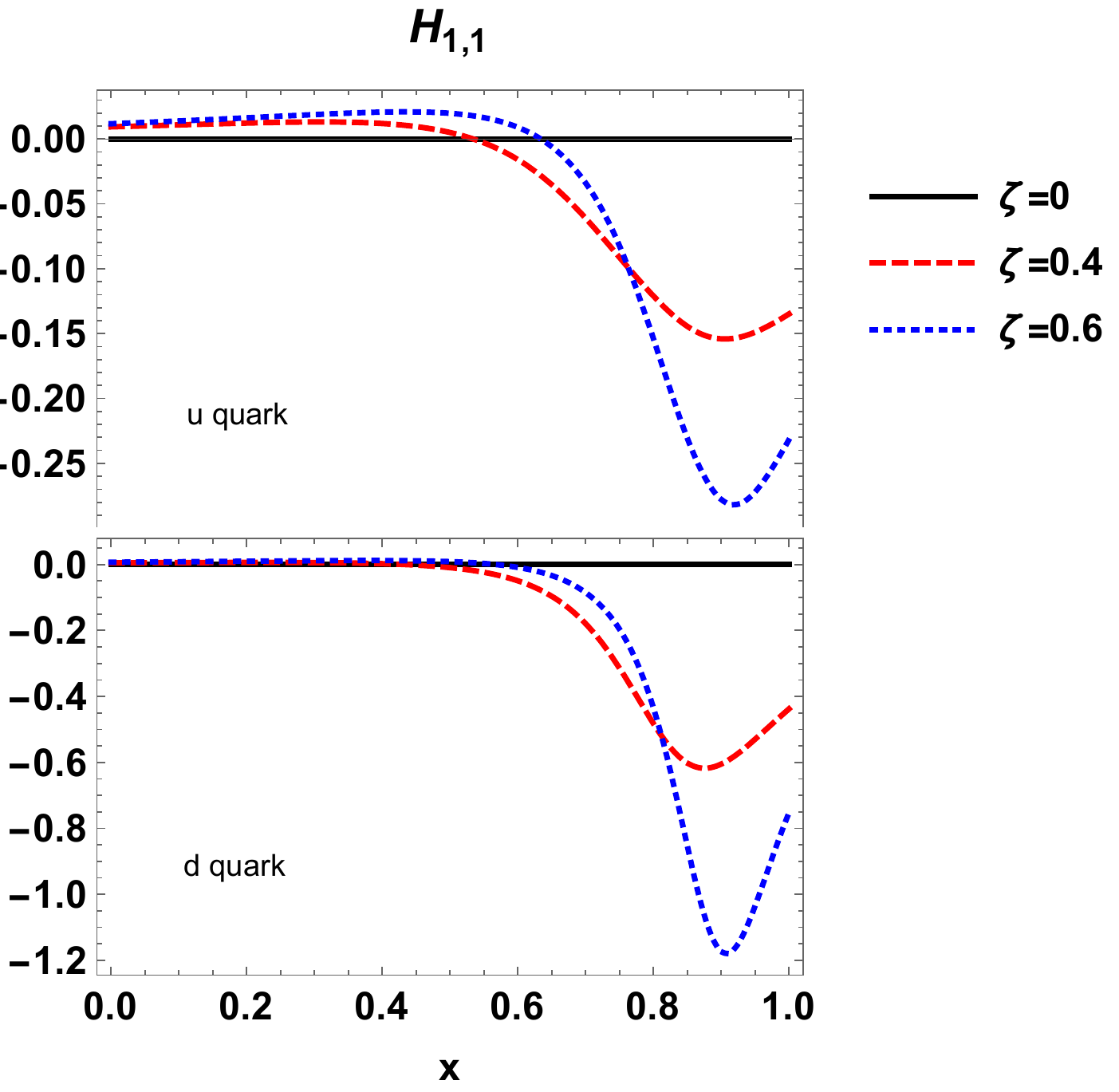}}
(b){\includegraphics[width=7.cm,clip]{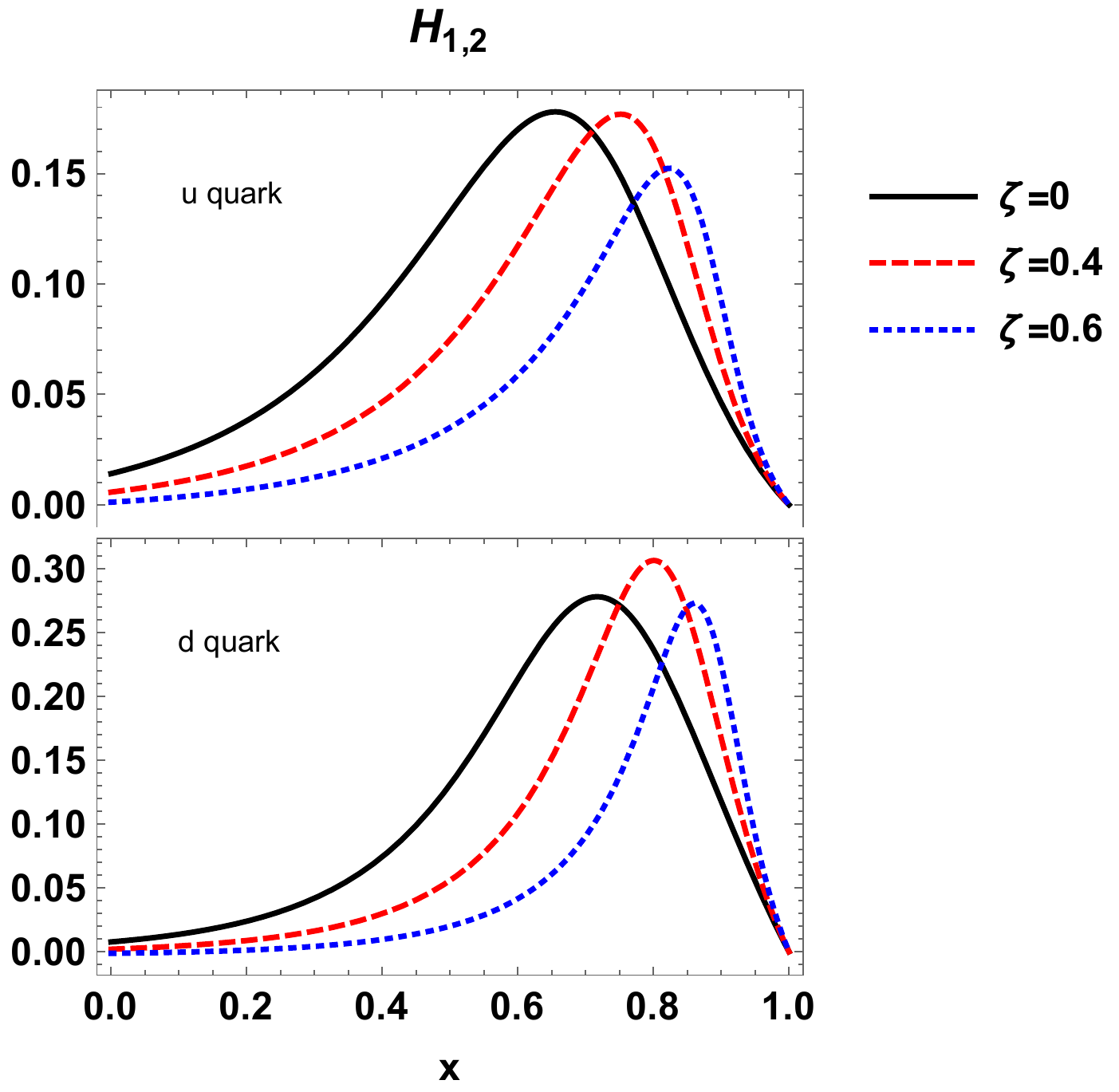}}
(c){\includegraphics[width=7.cm,clip]{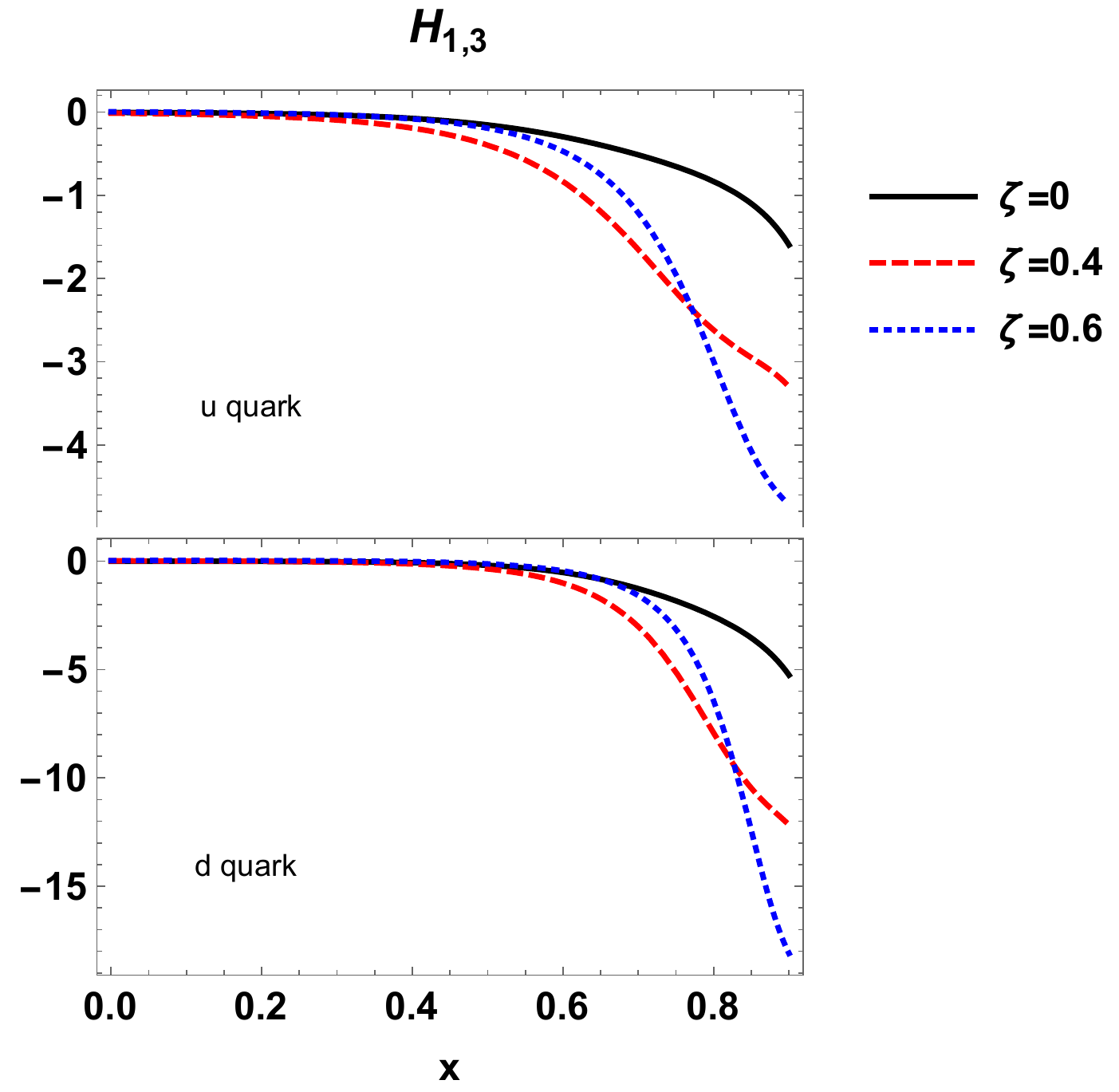}}
(d){\includegraphics[width=7.cm,clip]{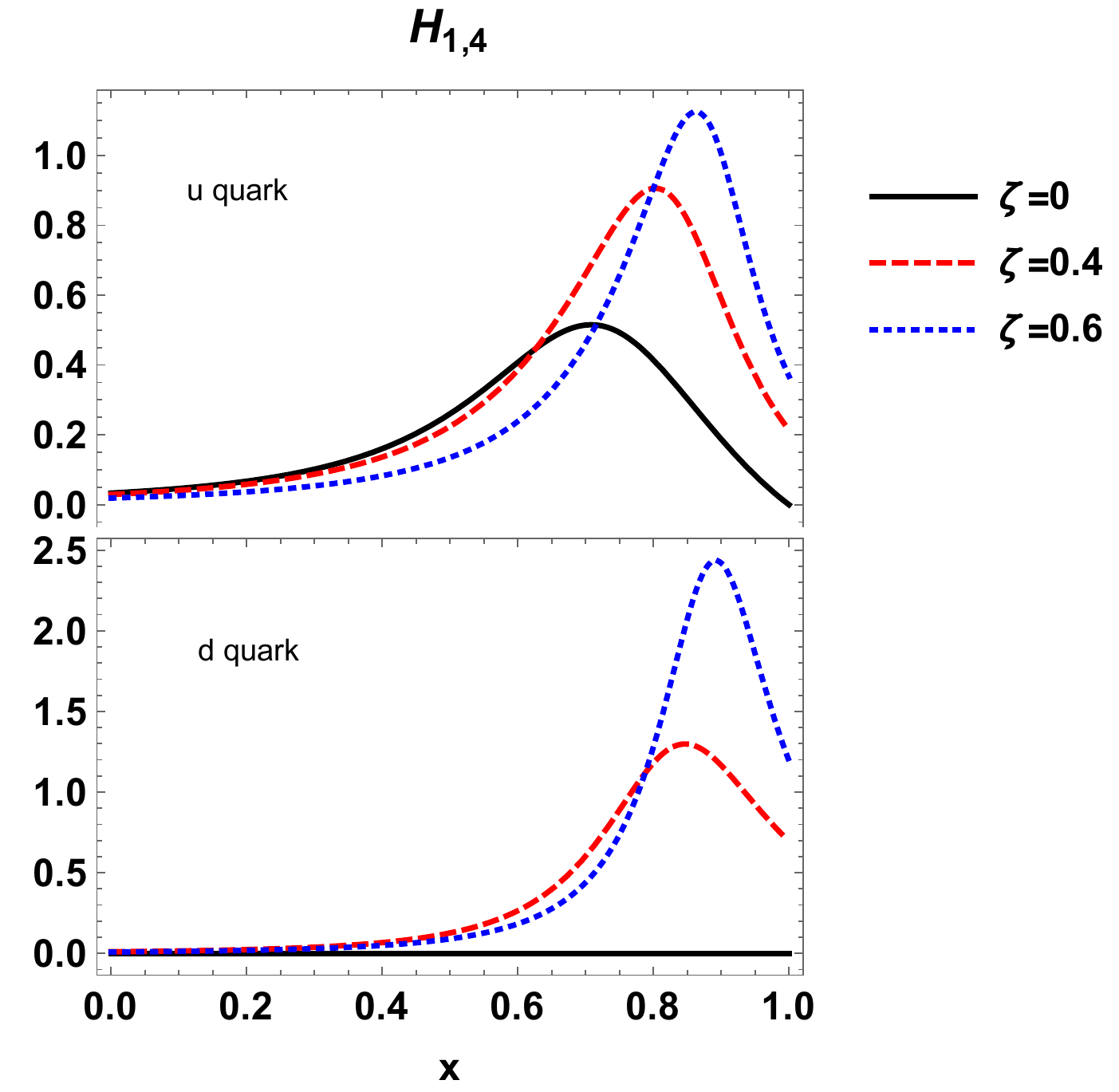}}
\caption{The variation of GTMDs $H_{1,1}(x,\zeta, {\bf \Delta}_\perp, \textbf{p}_\perp)$, $H_{1,2}(x,\zeta, {\bf \Delta}_\perp, \textbf{p}_\perp)$, $H_{1,3}(x,\zeta, {\bf \Delta}_\perp, \textbf{p}_\perp)$ and $H_{1,4}(x,\zeta, {\bf \Delta}_\perp, \textbf{p}_\perp)$ with $x$ for $u$ and $d$ quarks at $\zeta=0,0.4,0.6$.}
\label{h}
\end{figure}
\begin{figure}
(a){\includegraphics[width=7.cm,clip]{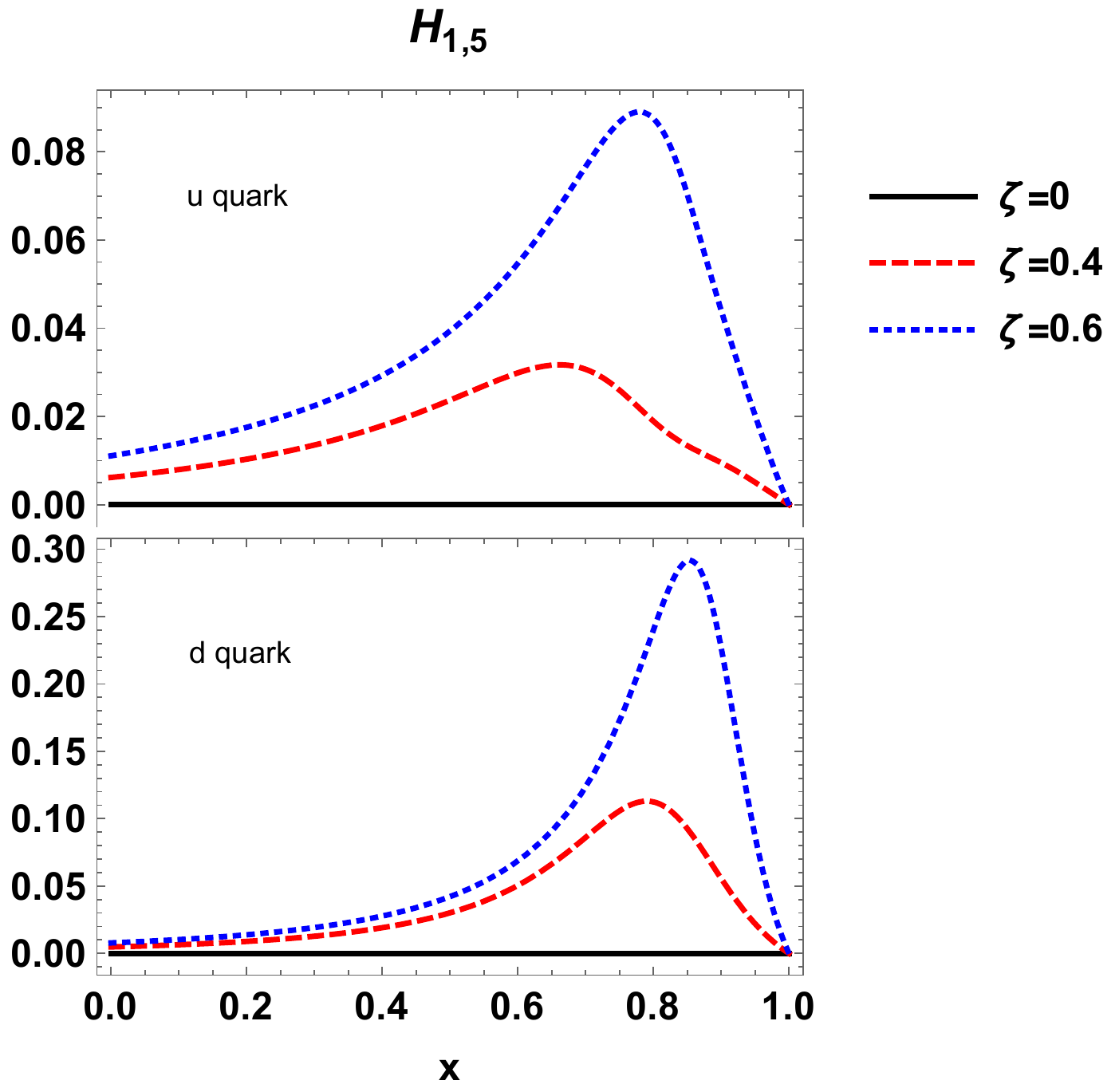}}
(b){\includegraphics[width=7.cm,clip]{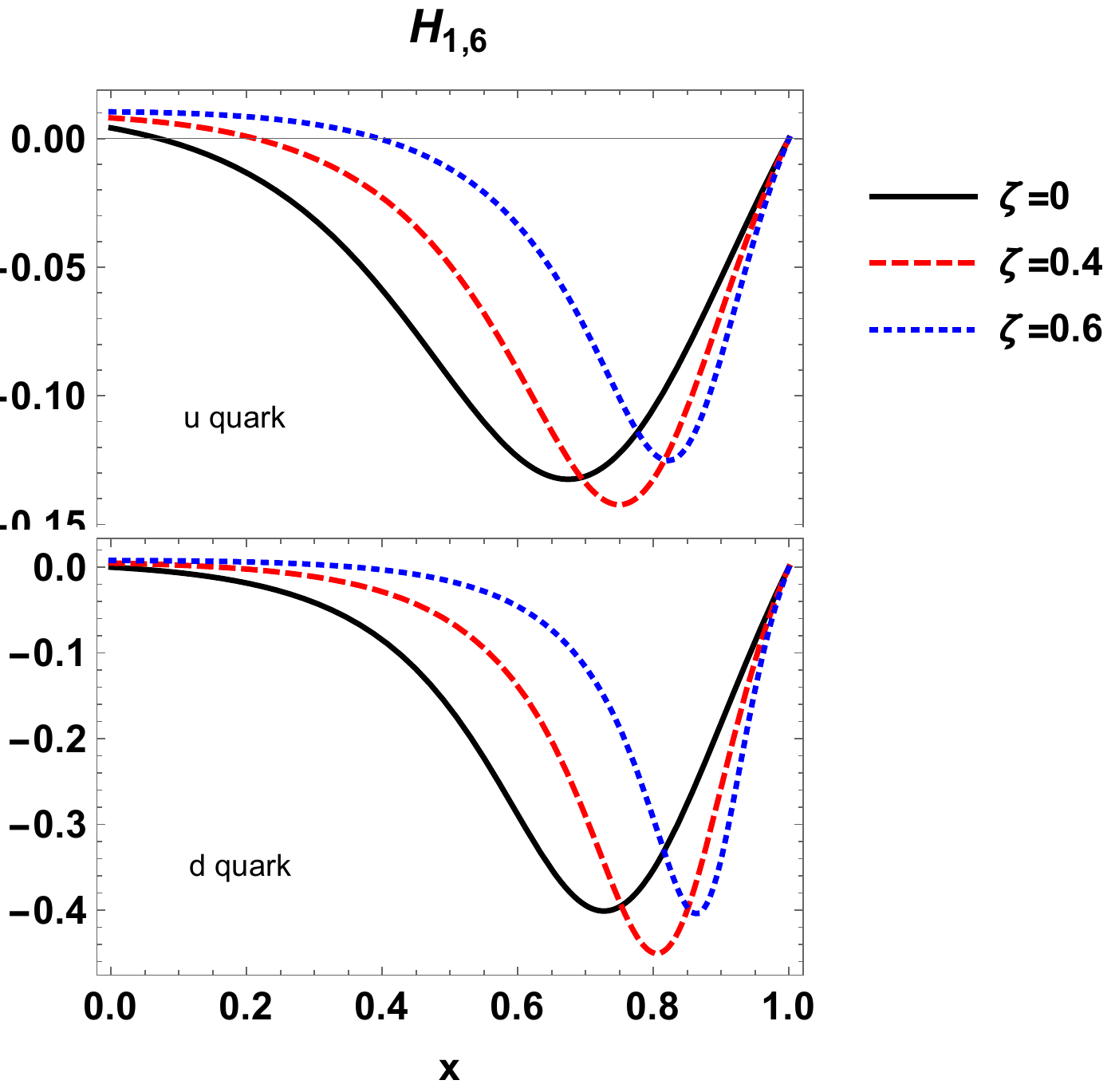}}
(c){\includegraphics[width=7.cm,clip]{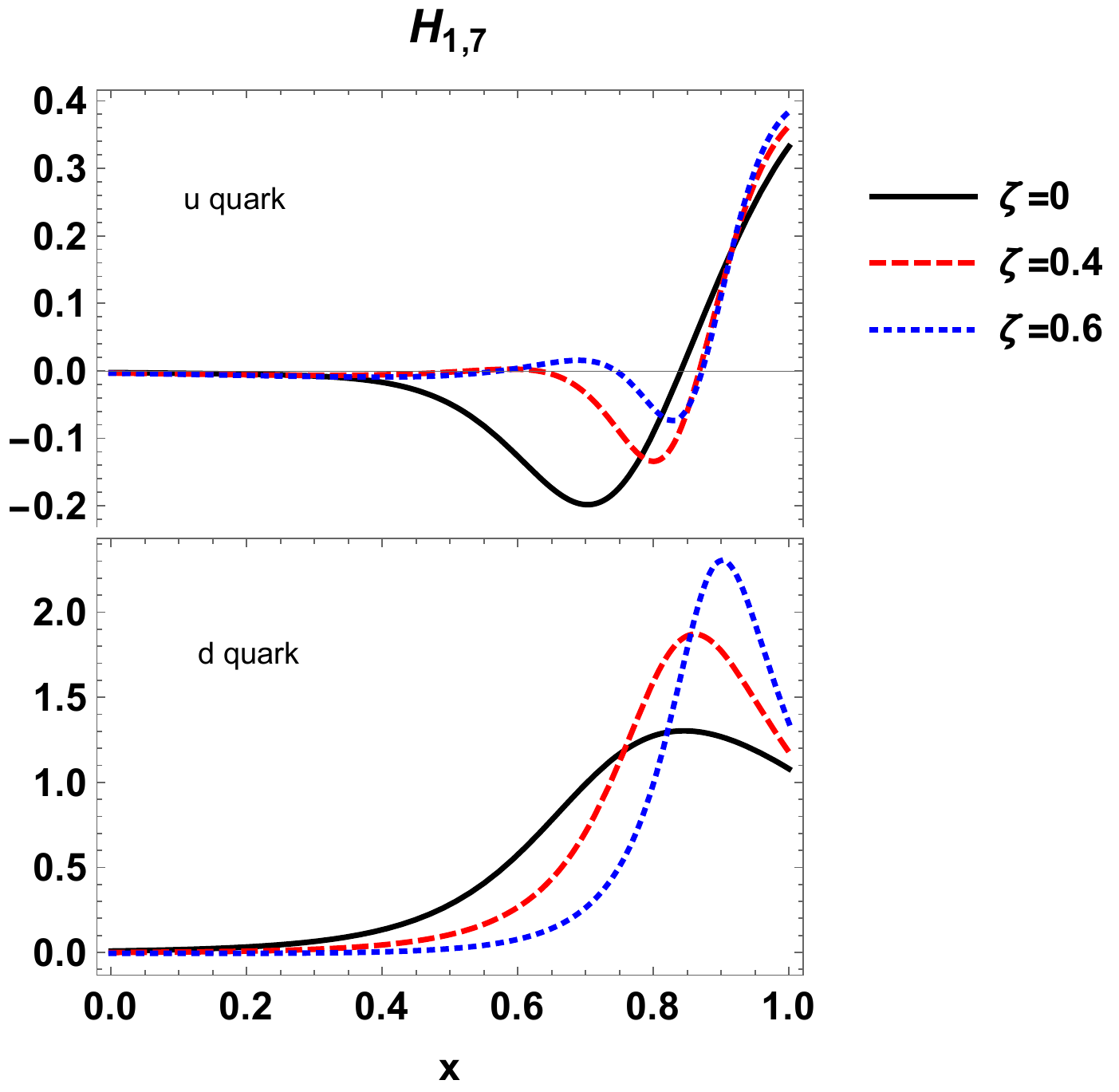}}
(d){\includegraphics[width=7.cm,clip]{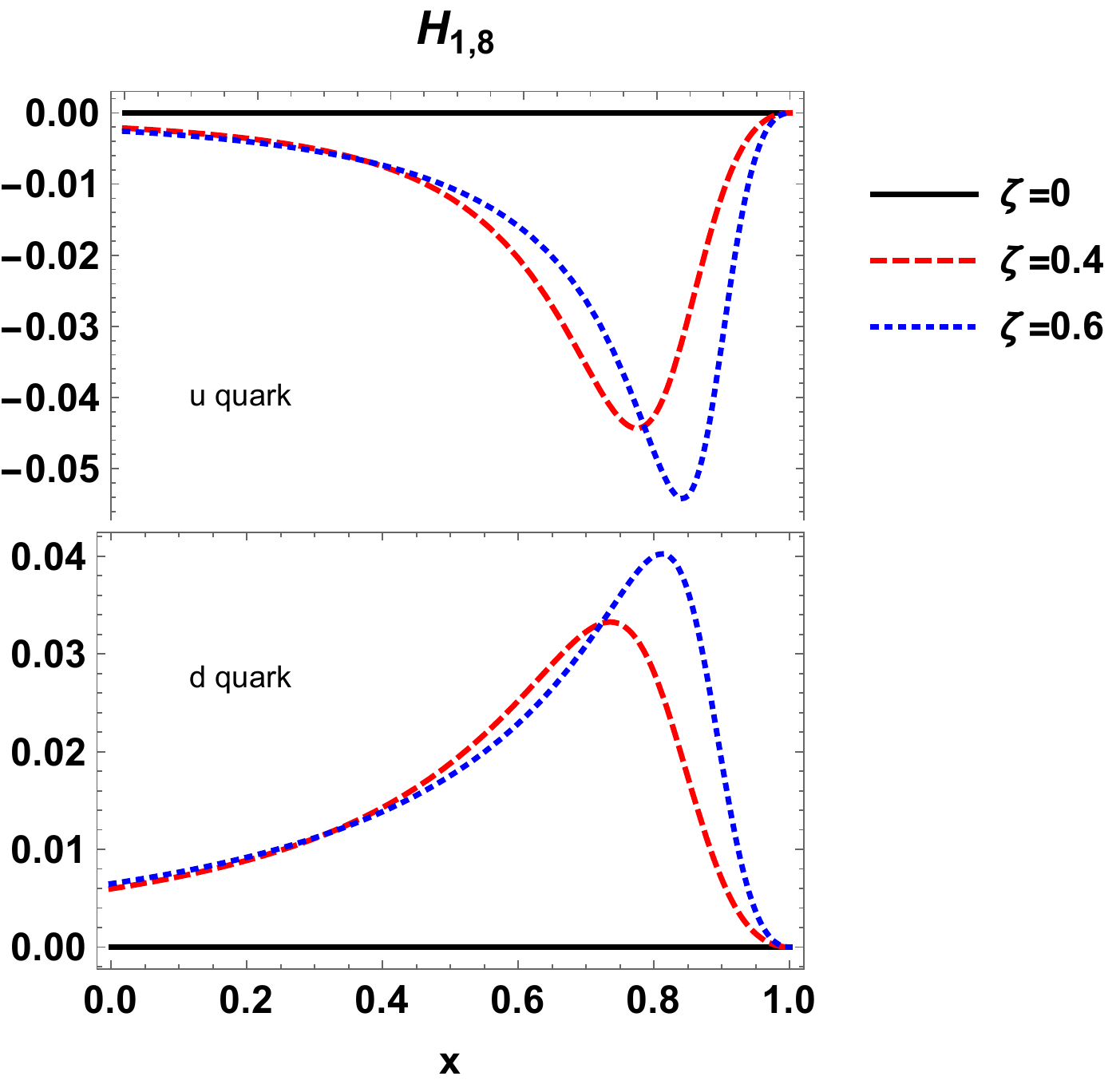}}
\caption{The variation of GTMDs $H_{1,5}(x,\zeta, {\bf \Delta}_\perp, \textbf{p}_\perp)$, $H_{1,6}(x,\zeta, {\bf \Delta}_\perp, \textbf{p}_\perp)$, $H_{1,7}(x,\zeta, {\bf \Delta}_\perp, \textbf{p}_\perp)$ and $H_{1,8}(x,\zeta, {\bf \Delta}_\perp, \textbf{p}_\perp)$ with $x$ for $u$ and $d$ quarks at $\zeta=0,0.4,0.6$.}
\label{h1}
\end{figure}
\begin{figure}
\centering
(a){\includegraphics[width=7.cm,clip]{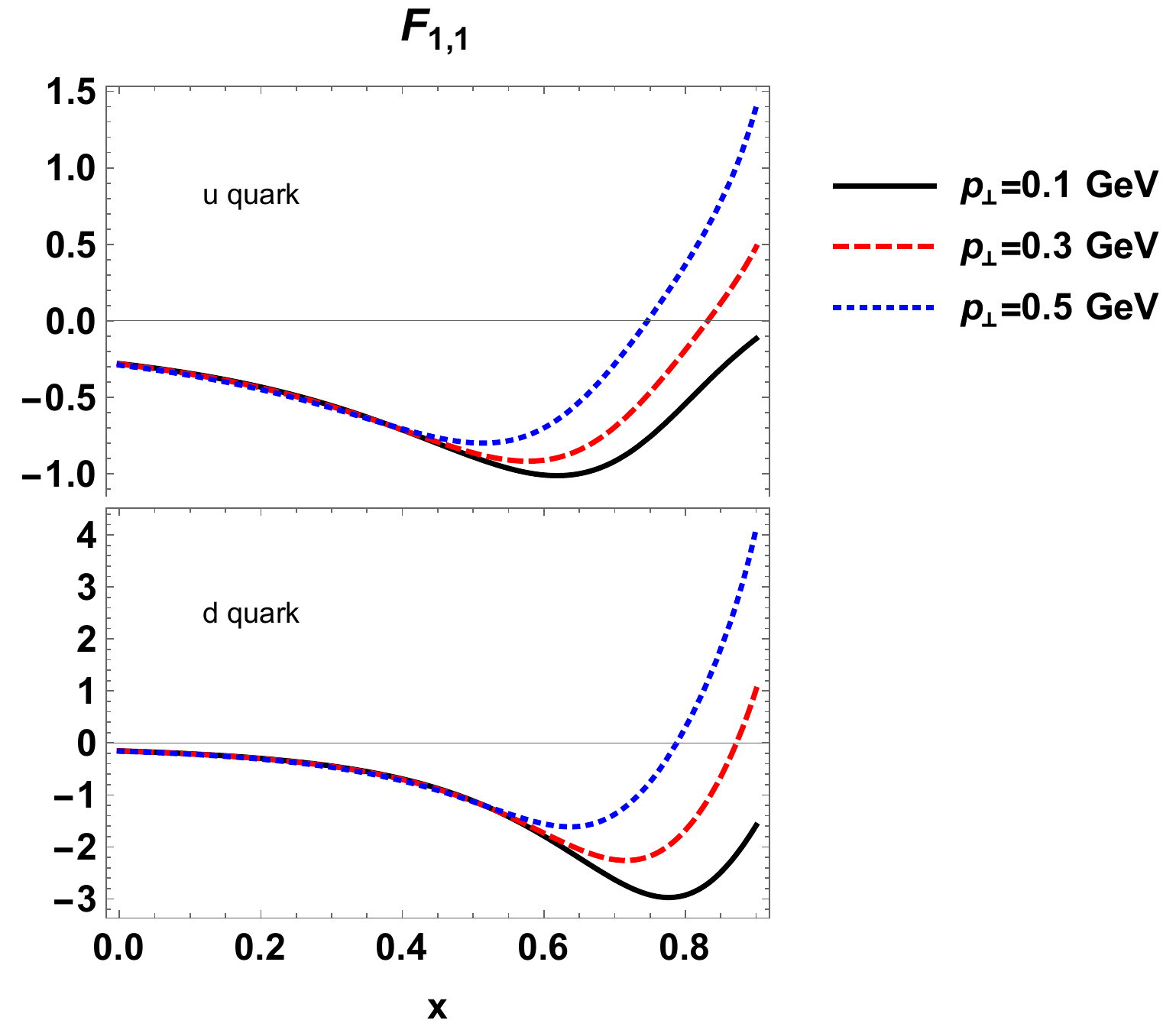}}
(b){\includegraphics[width=7.cm,clip]{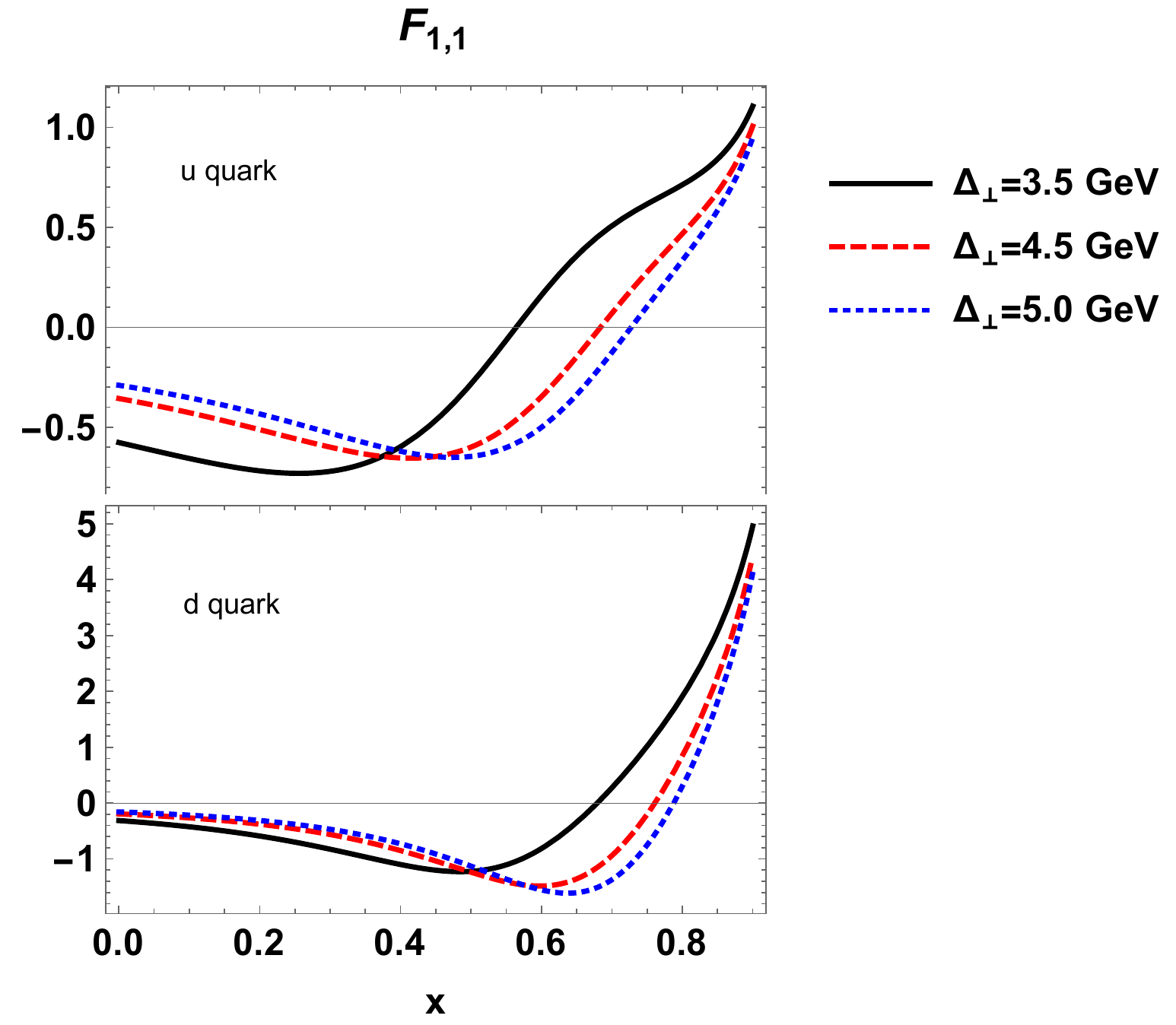}}
(c){\includegraphics[width=7.cm,clip]{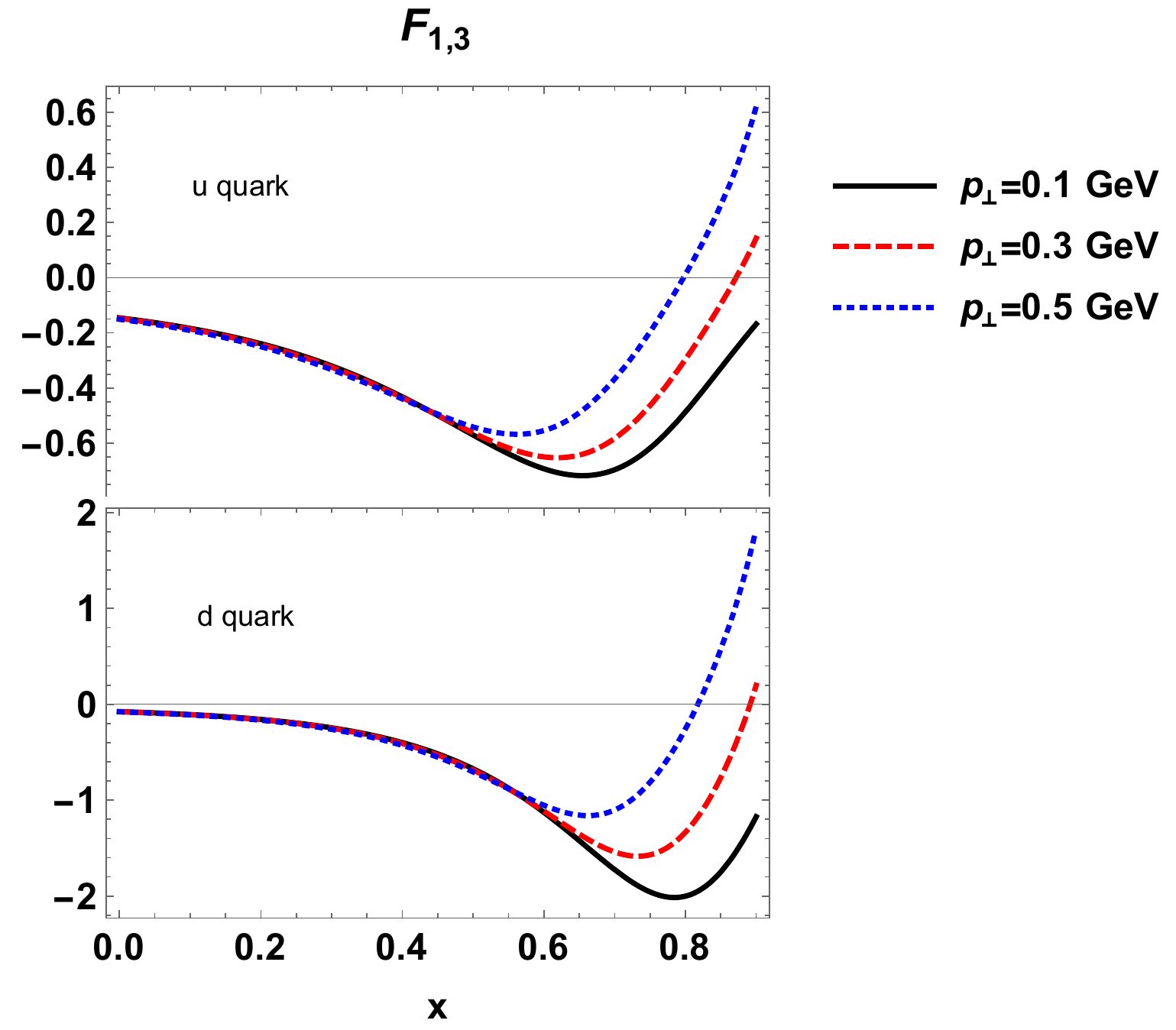}}
(d){\includegraphics[width=7.cm,clip]{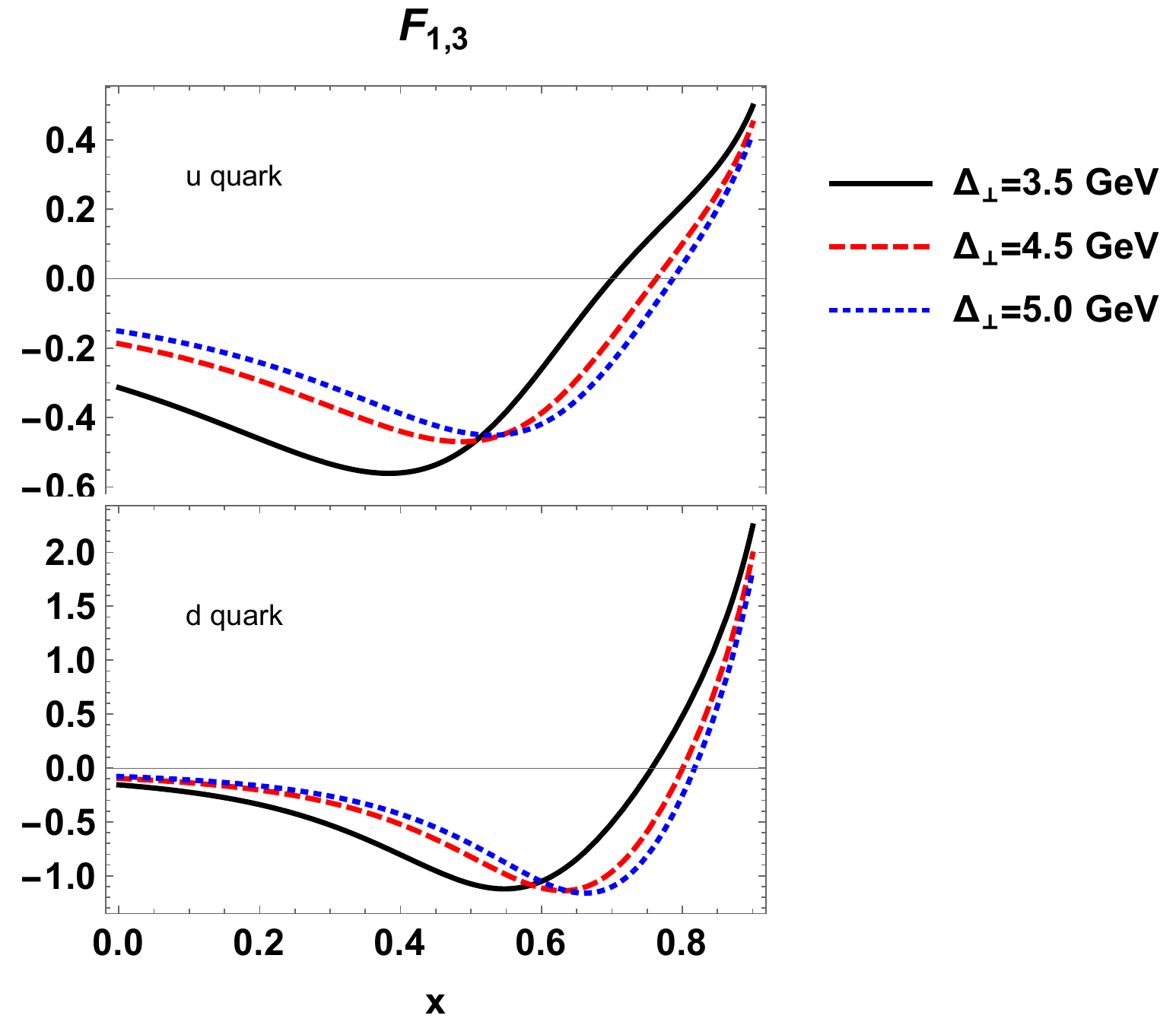}}
(e){\includegraphics[width=7.cm,clip]{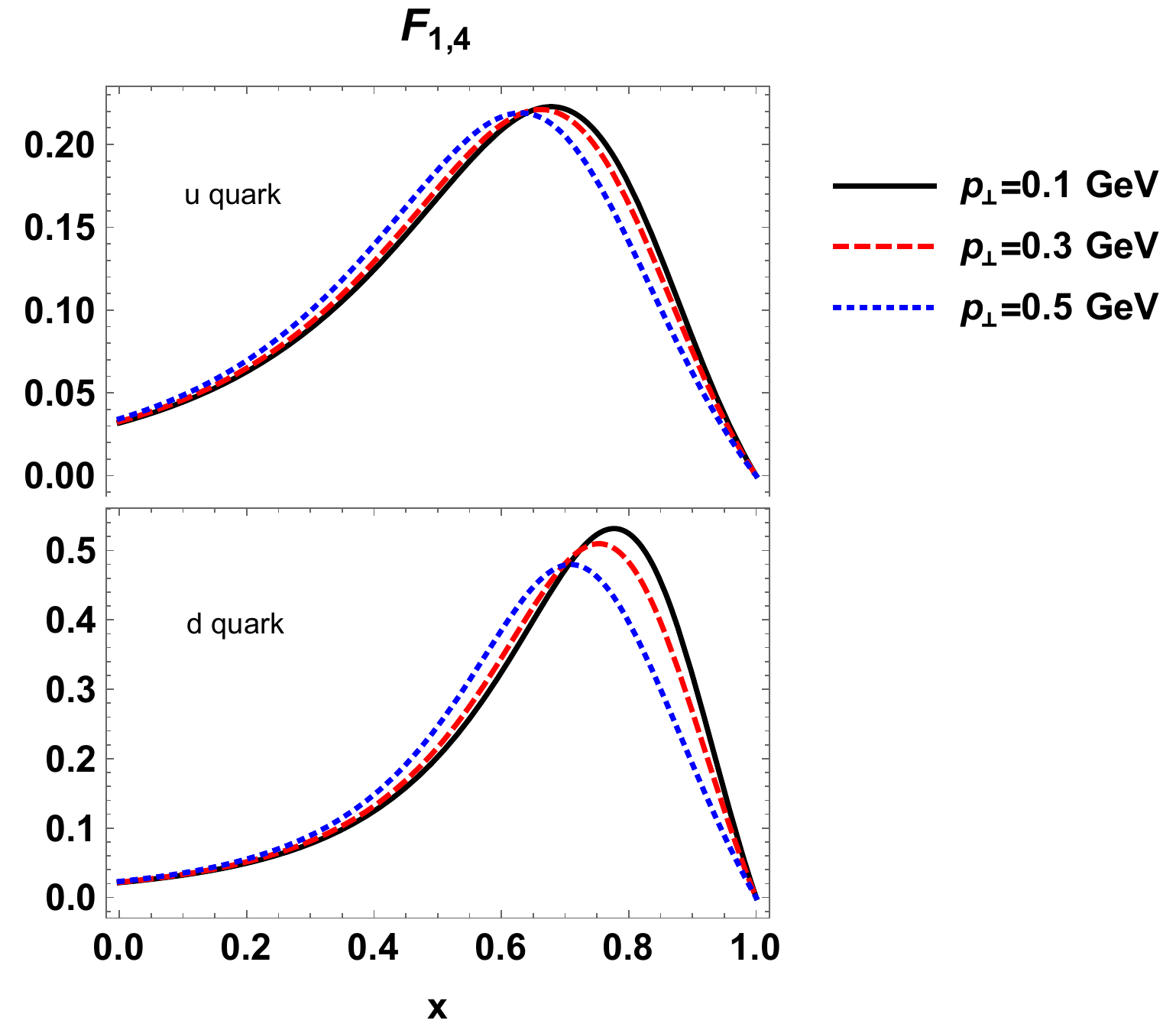}}
(f){\includegraphics[width=7.cm,clip]{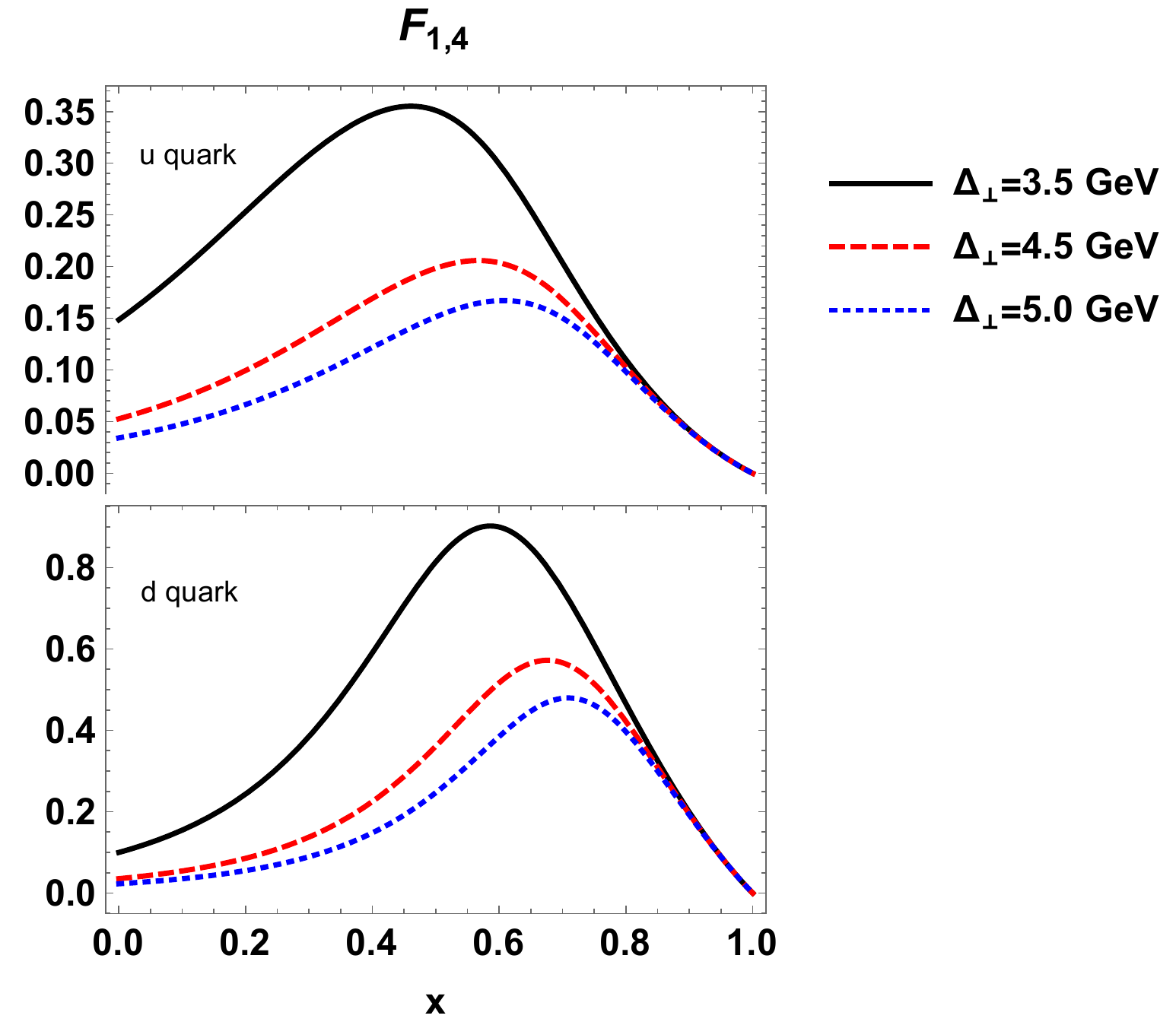}}
\caption{The variation of GTMDs $F_{1,1}(x, {\bf \Delta}_\perp, \textbf{p}_\perp)$, $F_{1,3}(x, {\bf \Delta}_\perp, \textbf{p}_\perp)$ and $F_{1,4}(x, {\bf \Delta}_\perp, \textbf{p}_\perp)$ with $x$ for $u$ and $d$ quarks at different values of $\textbf{p}_{\perp}$ with fixed ${\bf \Delta}_{\perp}$ (left panel) and at different values of ${\bf \Delta}_\perp$ with fixed $\textbf{p}_\perp$ (right panel).}
\label{fpd}
\end{figure}
\begin{figure}
\centering
(a){\includegraphics[width=7.cm,clip]{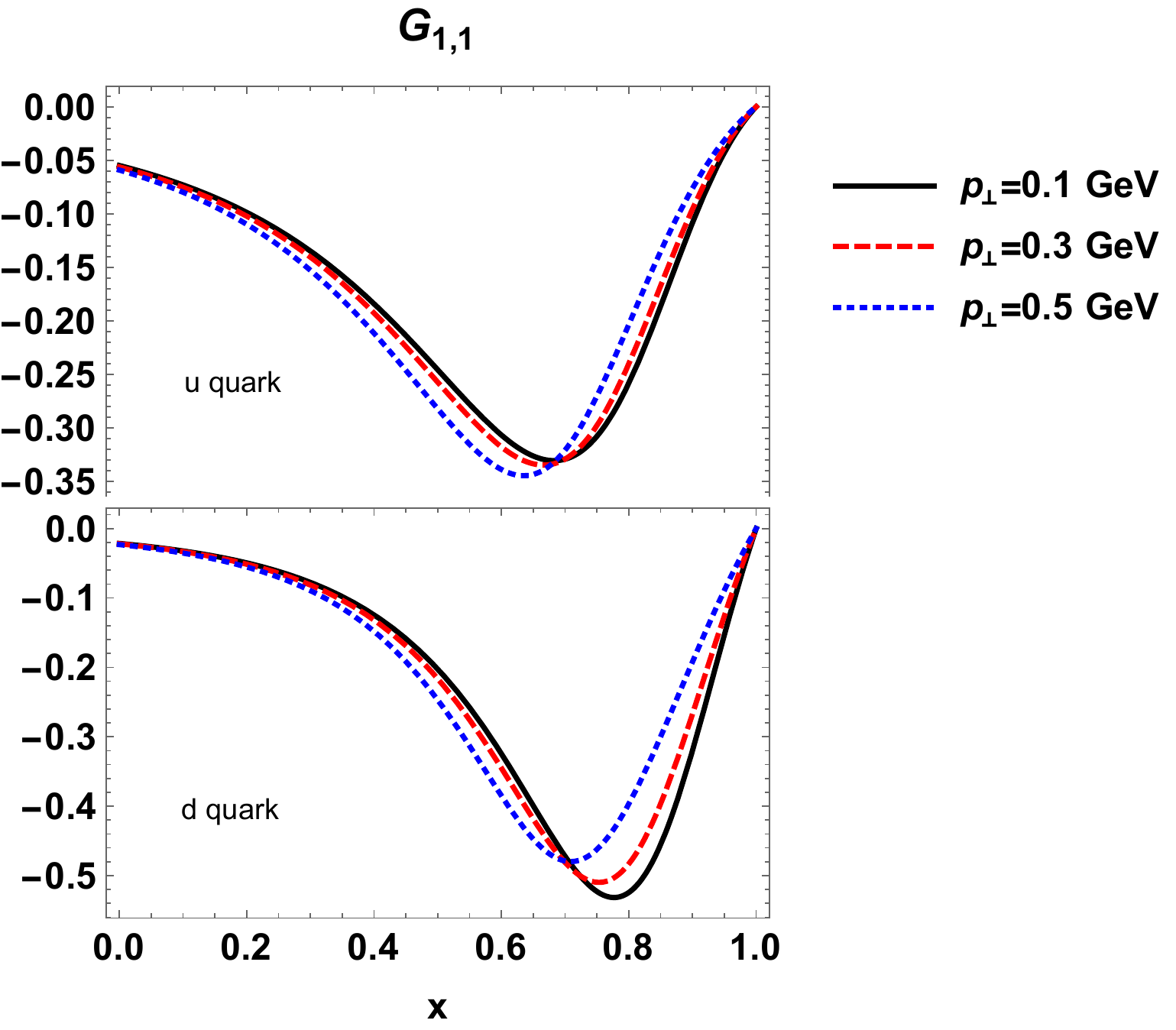}}
(b){\includegraphics[width=7.cm,clip]{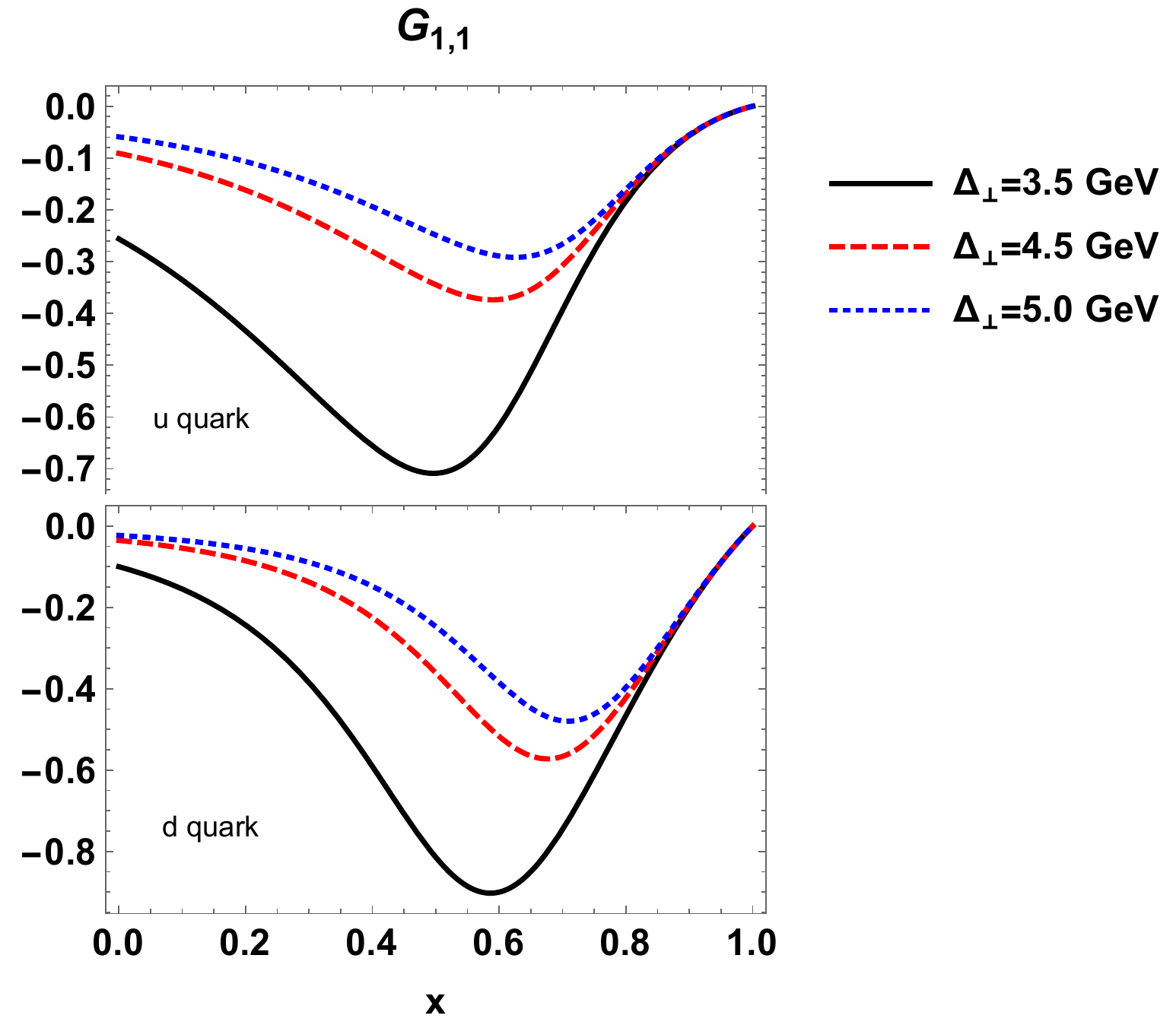}}
(c){\includegraphics[width=7.cm,clip]{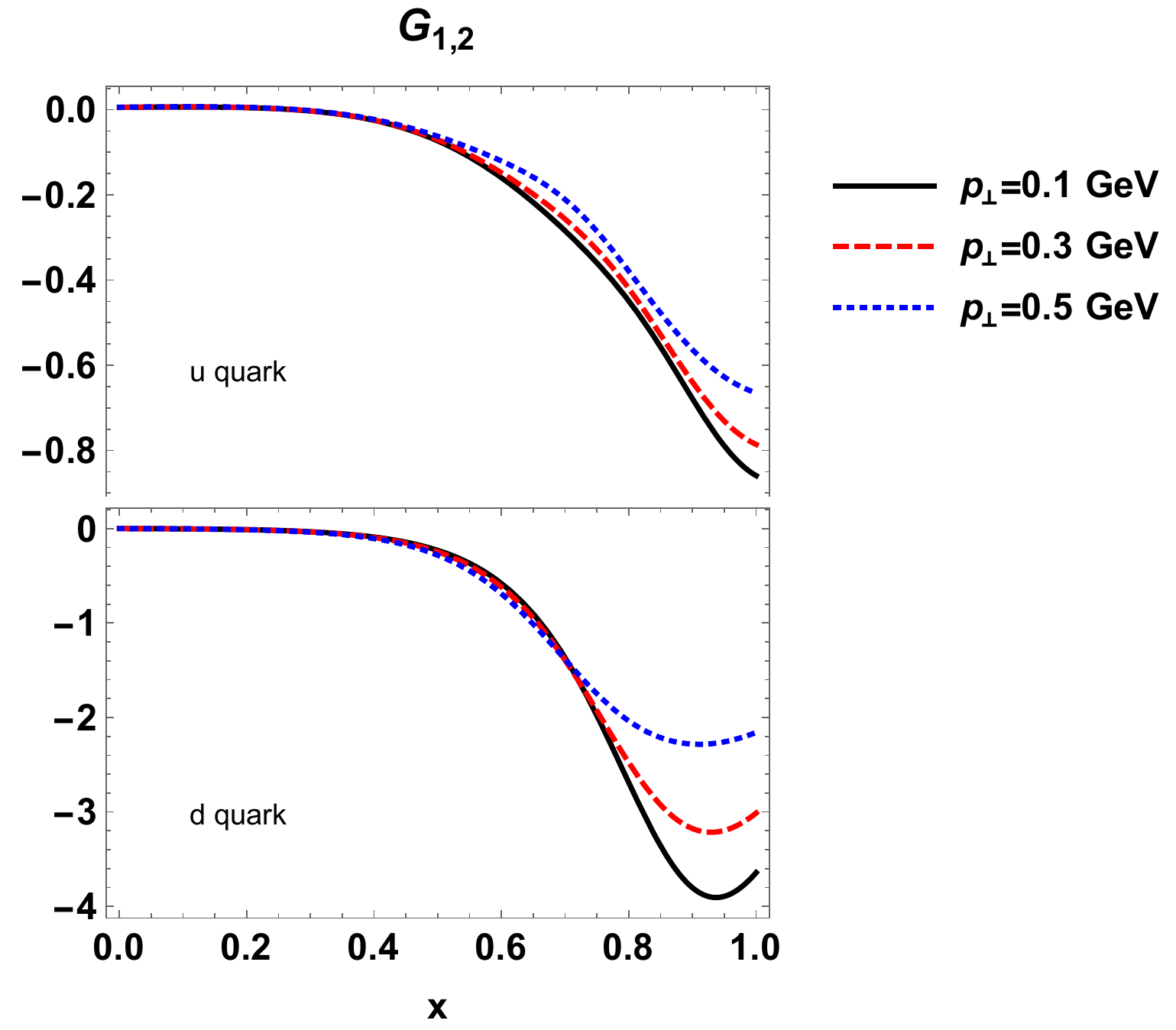}}
(d){\includegraphics[width=7.cm,clip]{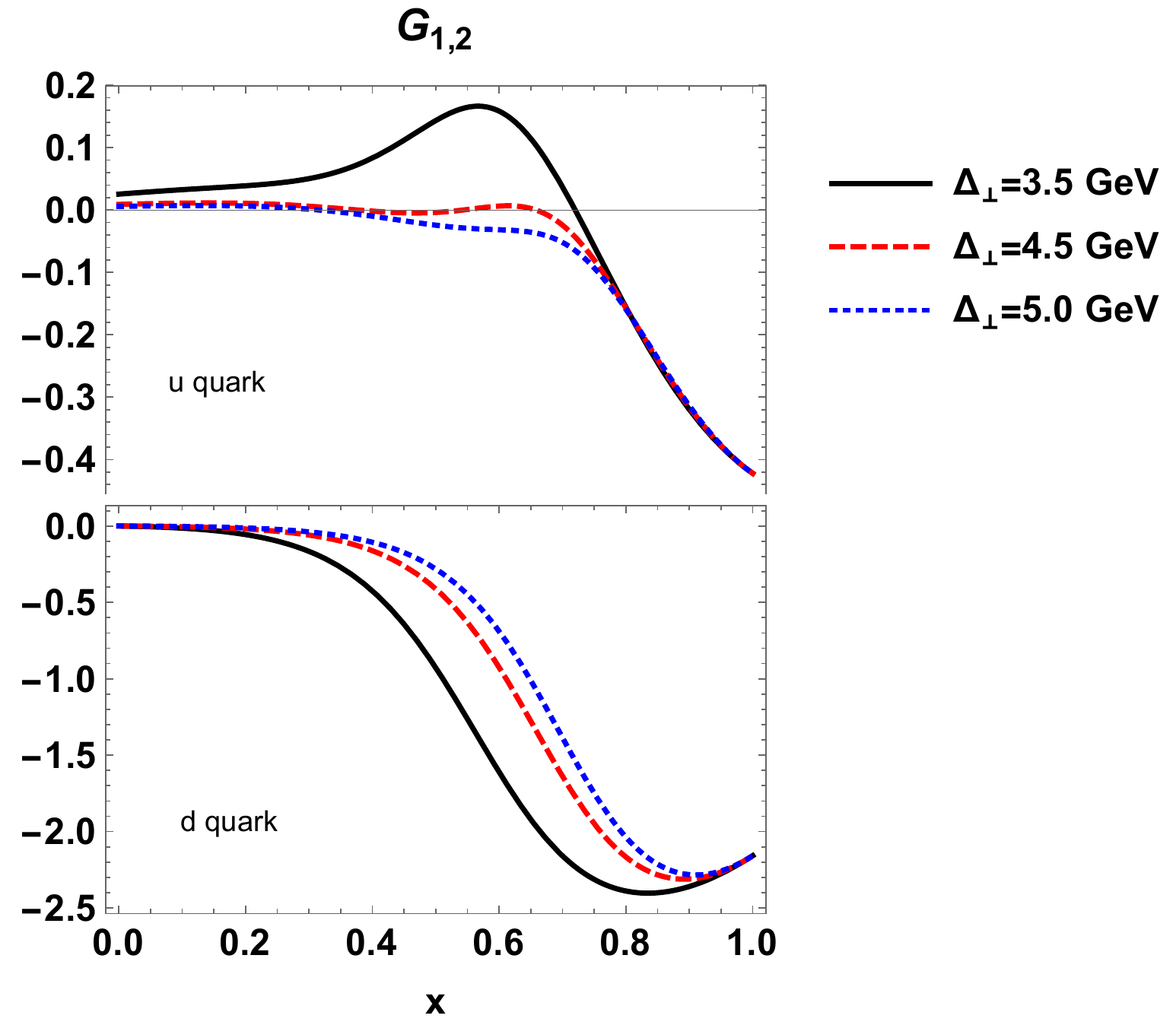}}
(e){\includegraphics[width=7.cm,clip]{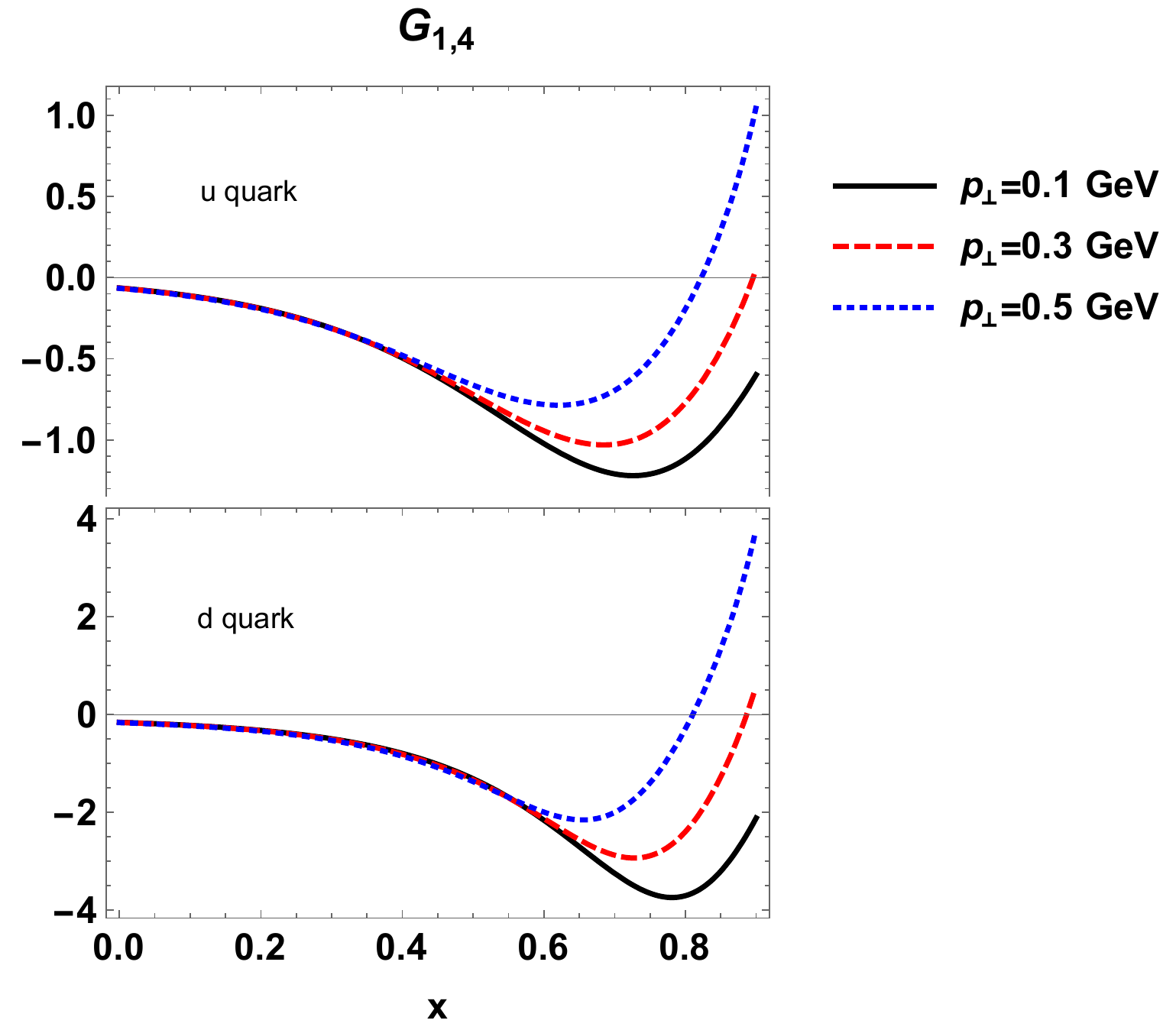}}
(f){\includegraphics[width=7.cm,clip]{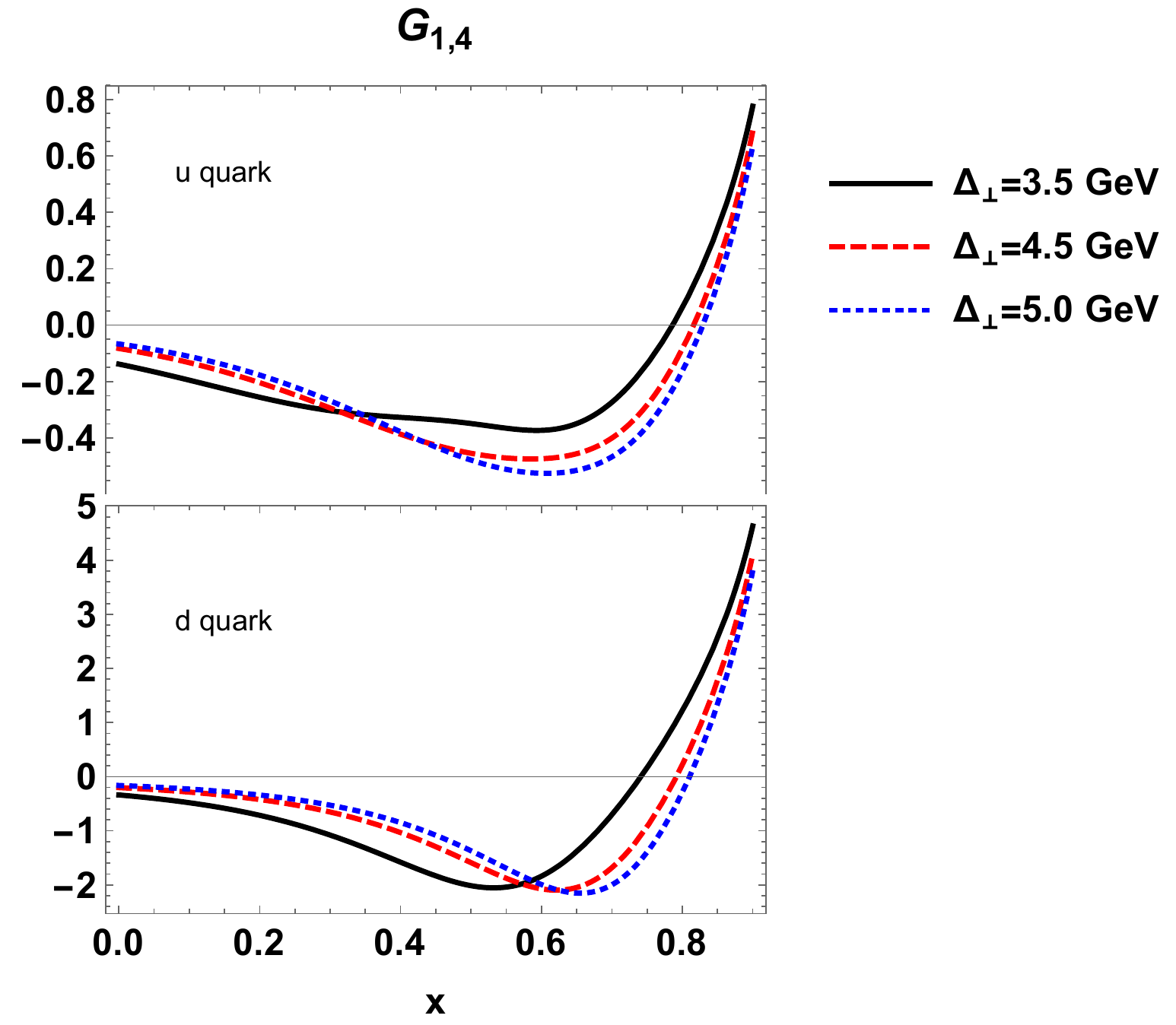}}
\caption{The variation of GTMDs $G_{1,1}(x, {\bf \Delta}_\perp, \textbf{p}_\perp)$, $G_{1,2}(x, {\bf \Delta}_\perp, \textbf{p}_\perp)$ and $G_{1,4}(x, {\bf \Delta}_\perp, \textbf{p}_\perp)$ with $x$ for $u$ and $d$ quarks at different values of $\textbf{p}_{\perp}$ with fixed ${\bf \Delta}_{\perp}$ (left panel) and at different values of ${\bf \Delta}_\perp$ with fixed $\textbf{p}_\perp$ (right panel).}
\label{gpd}
\end{figure}
\begin{figure}[hbtp]
\centering
(a){\includegraphics[width=7.cm,clip]{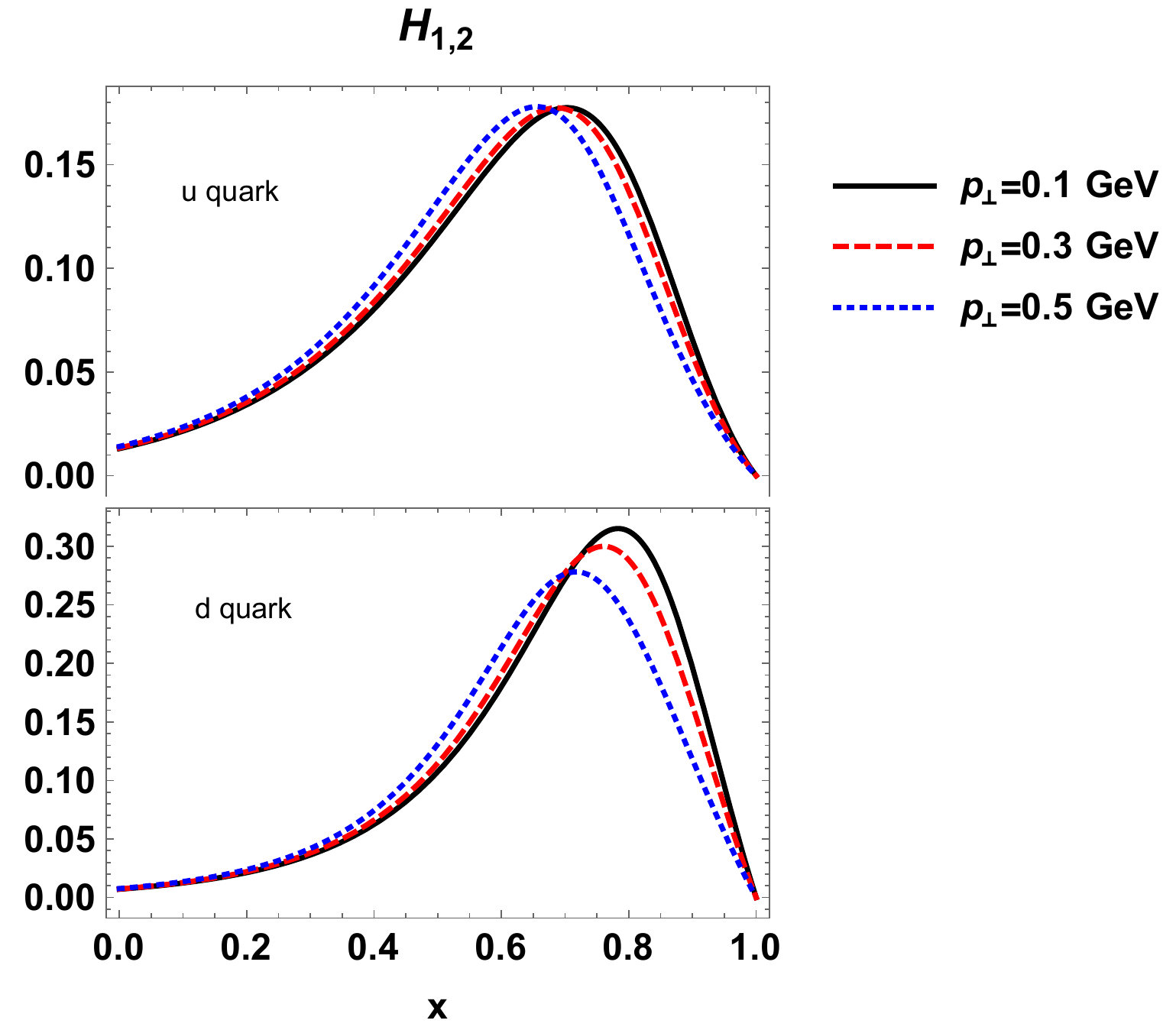}}
(b){\includegraphics[width=7.cm,clip]{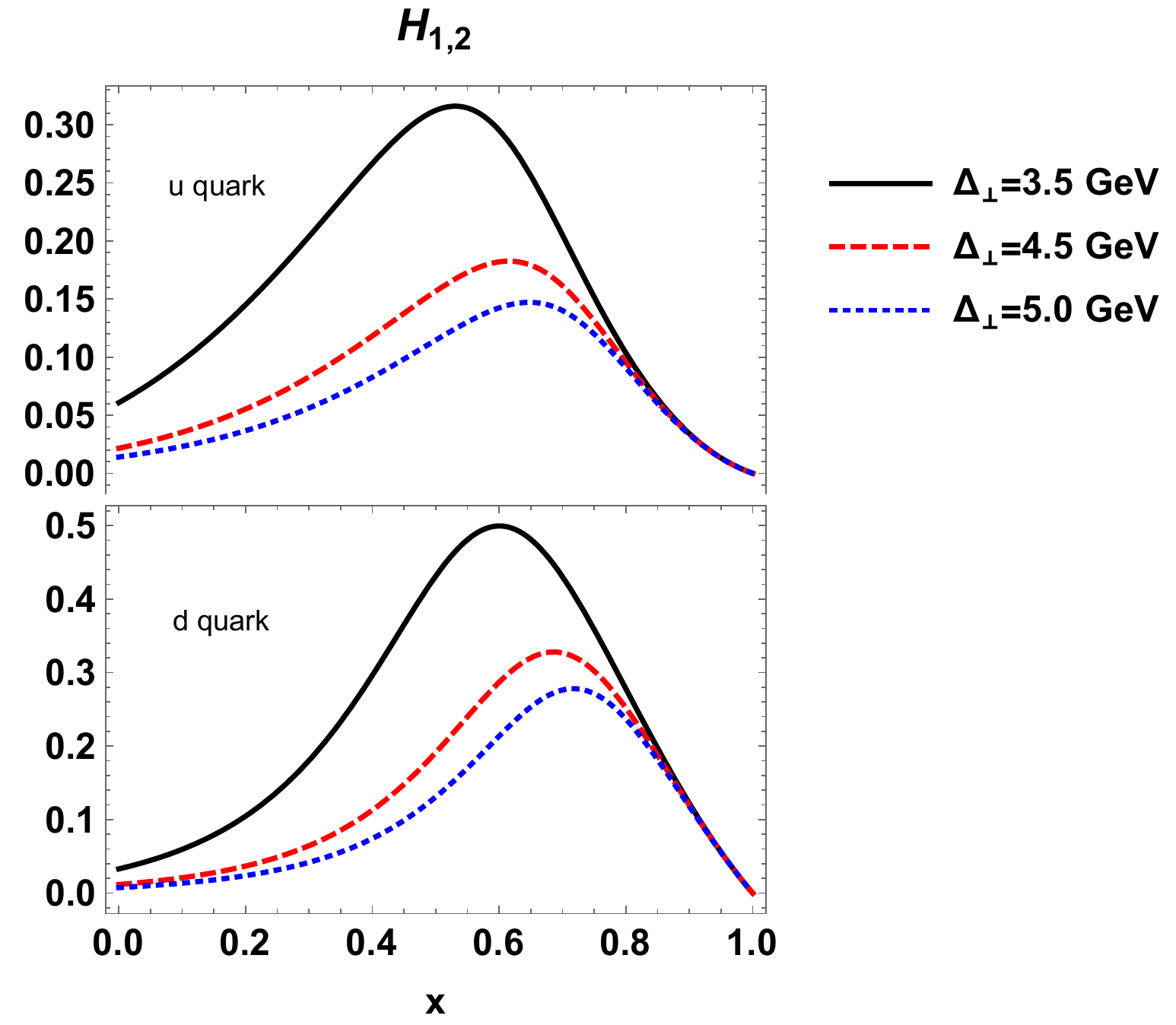}}
(c){\includegraphics[width=7.cm,clip]{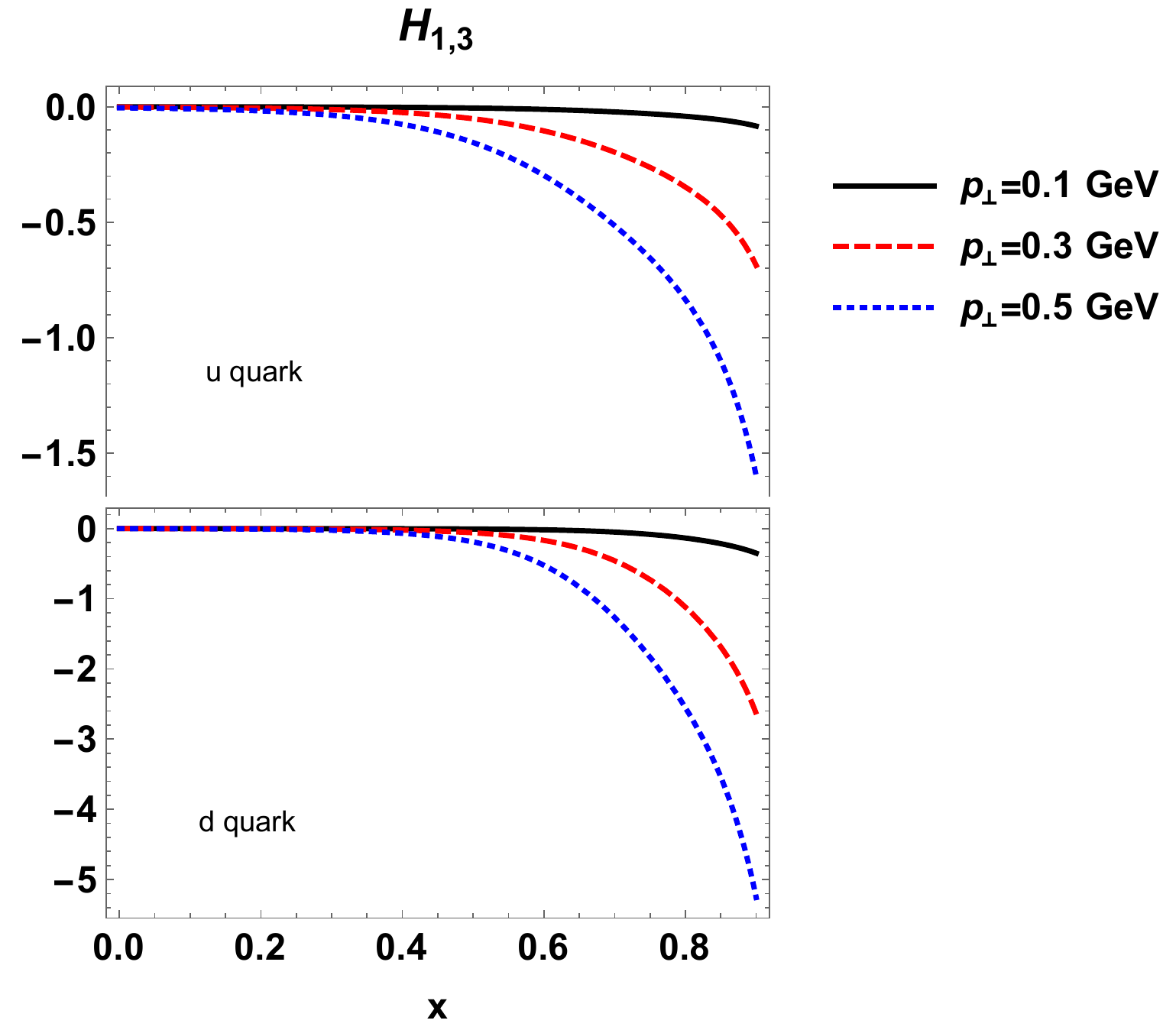}}
(d){\includegraphics[width=7.cm,clip]{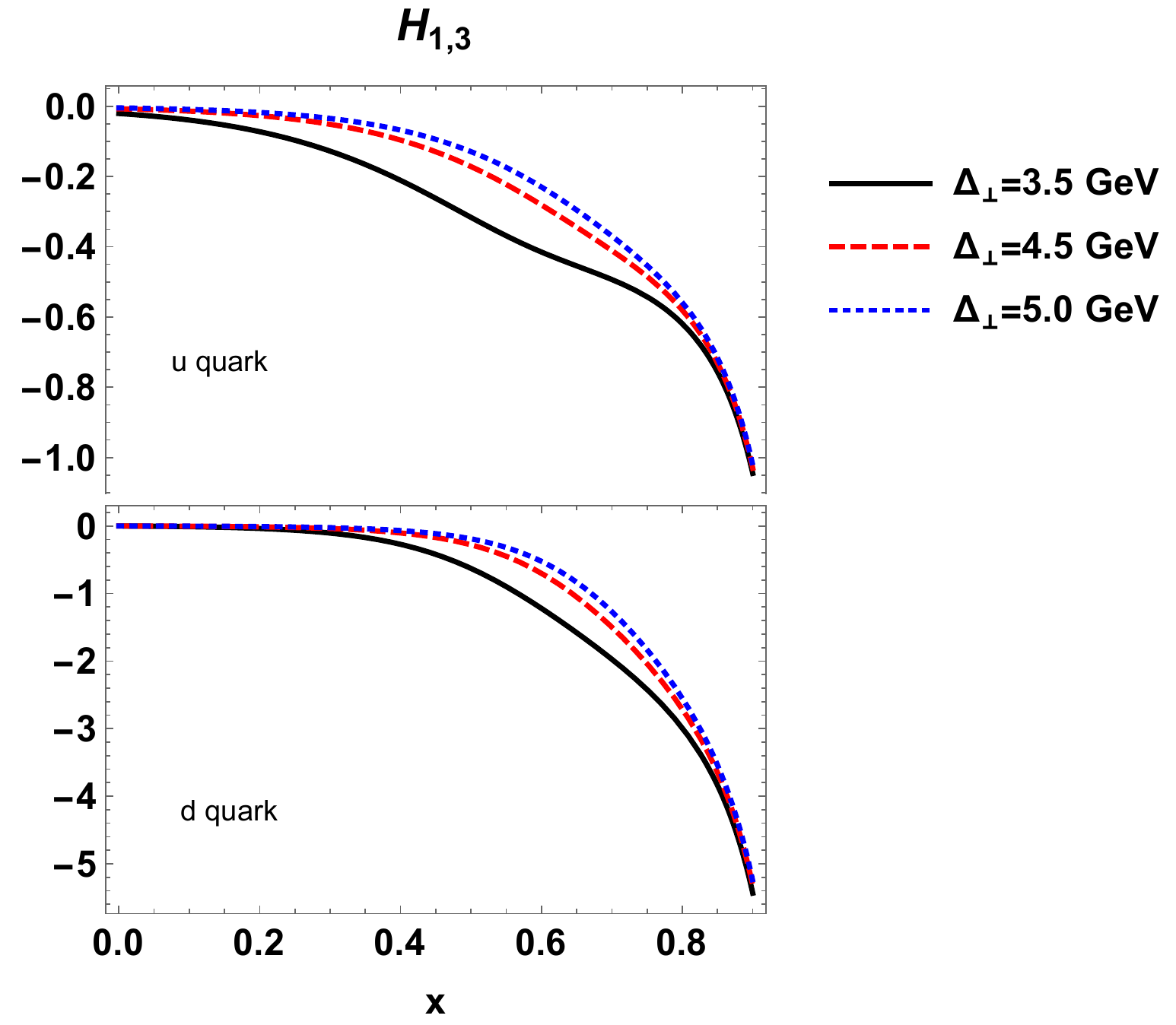}}
\caption{The variation of GTMDs $H_{1,2}(x, {\bf \Delta}_\perp, \textbf{p}_\perp)$ and $H_{1,3}(x, {\bf \Delta}_\perp, \textbf{p}_\perp)$ with $x$ for $u$ and $d$ quarks at different values of $\textbf{p}_{\perp}$ with fixed ${\bf \Delta}_{\perp}$ (left panel) and at different values of ${\bf \Delta}_\perp$ with fixed $\textbf{p}_\perp$ (right panel).}
\label{hpd1}
\end{figure}
\begin{figure}
\centering
(a){\includegraphics[width=7.cm,clip]{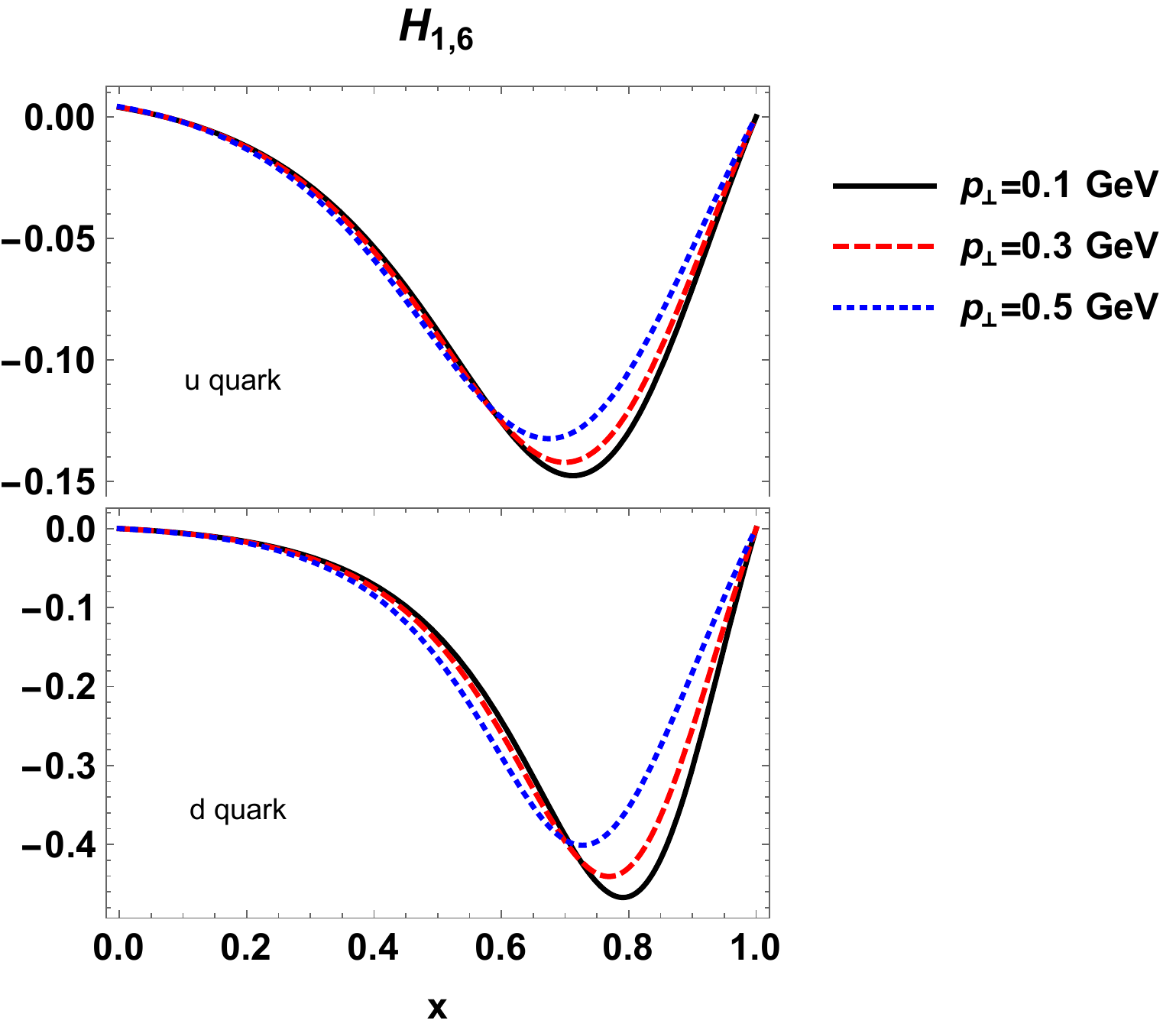}}
(b){\includegraphics[width=7.cm,clip]{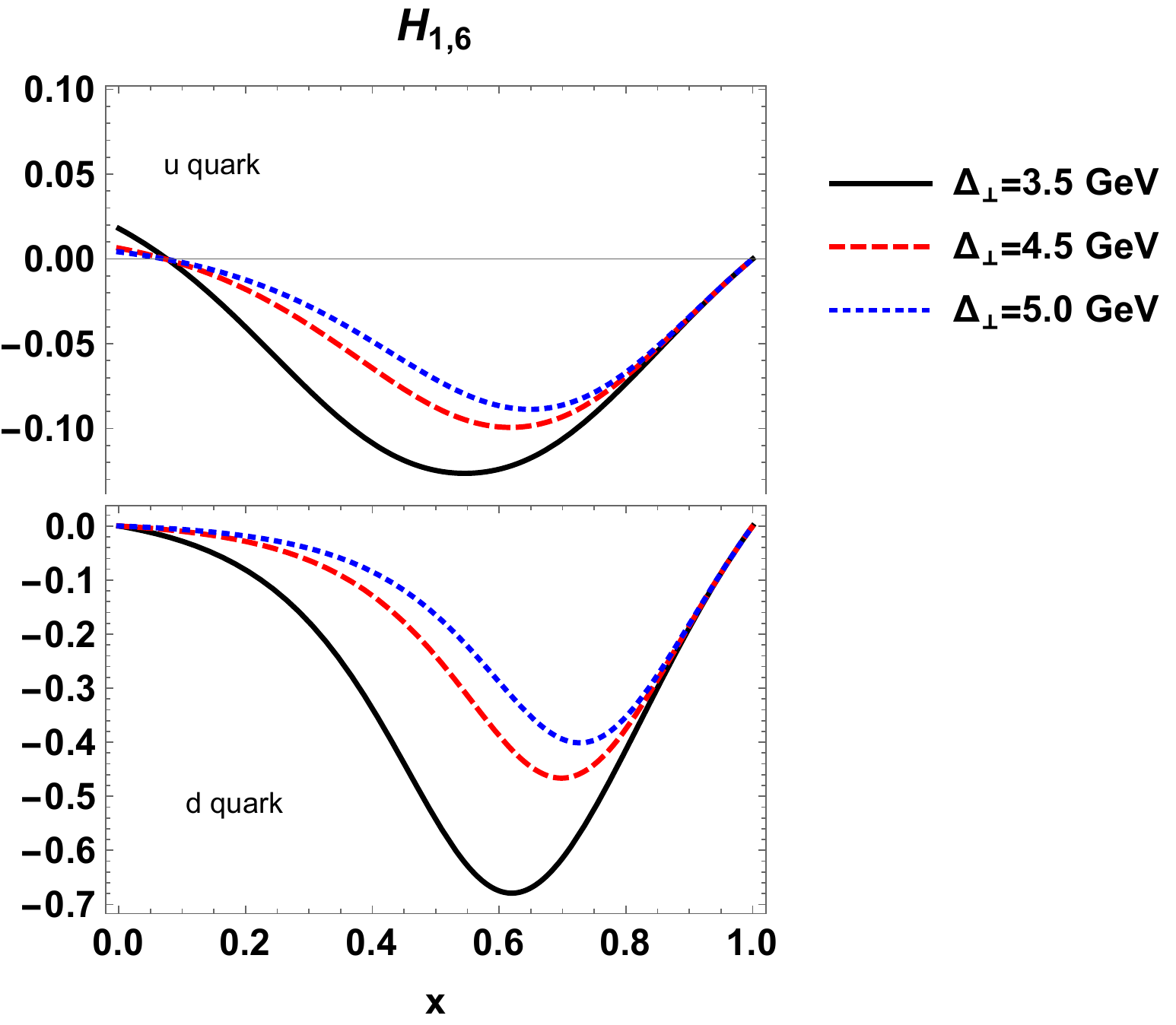}}
(c){\includegraphics[width=7.cm,clip]{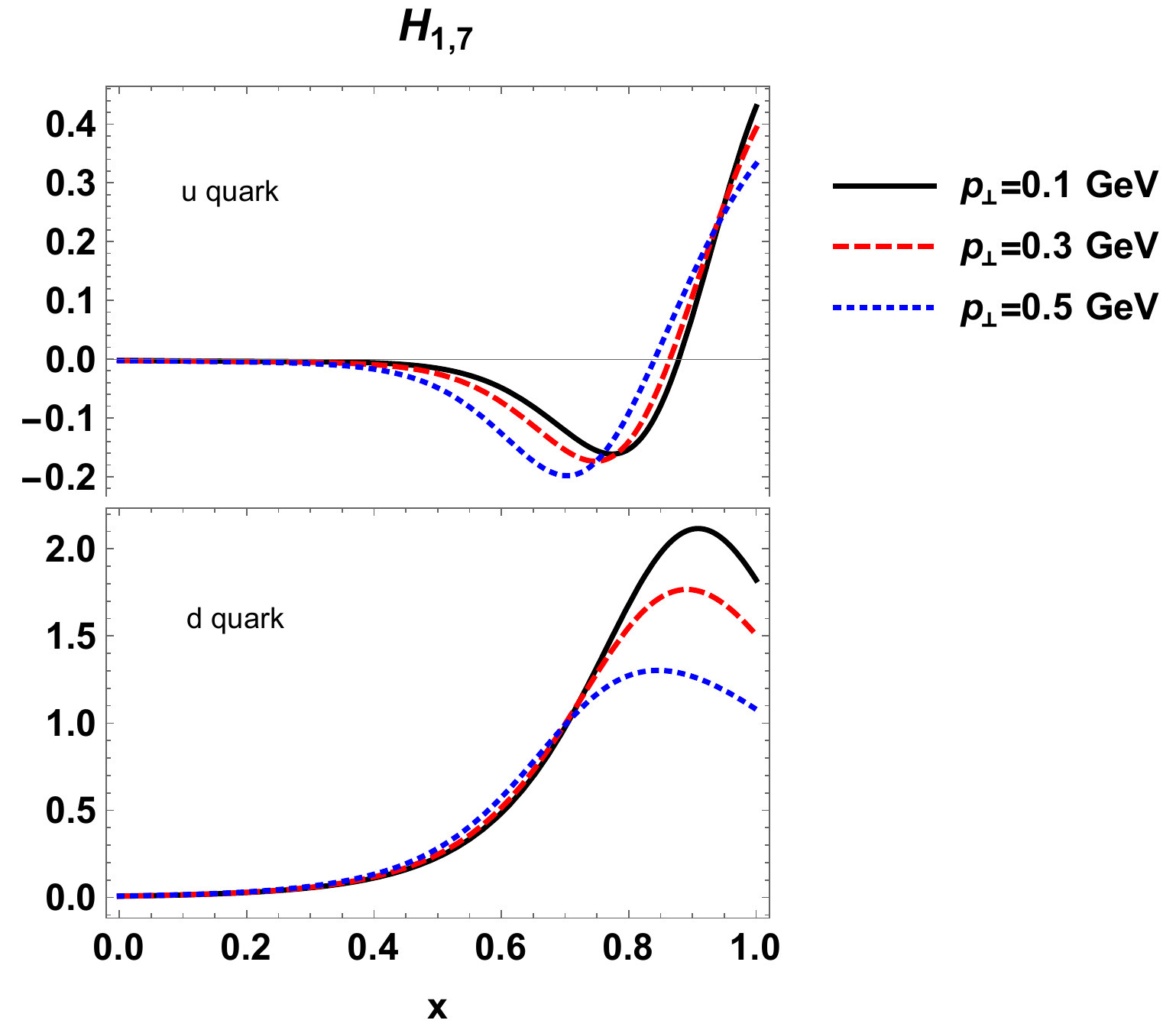}}
(d){\includegraphics[width=7.cm,clip]{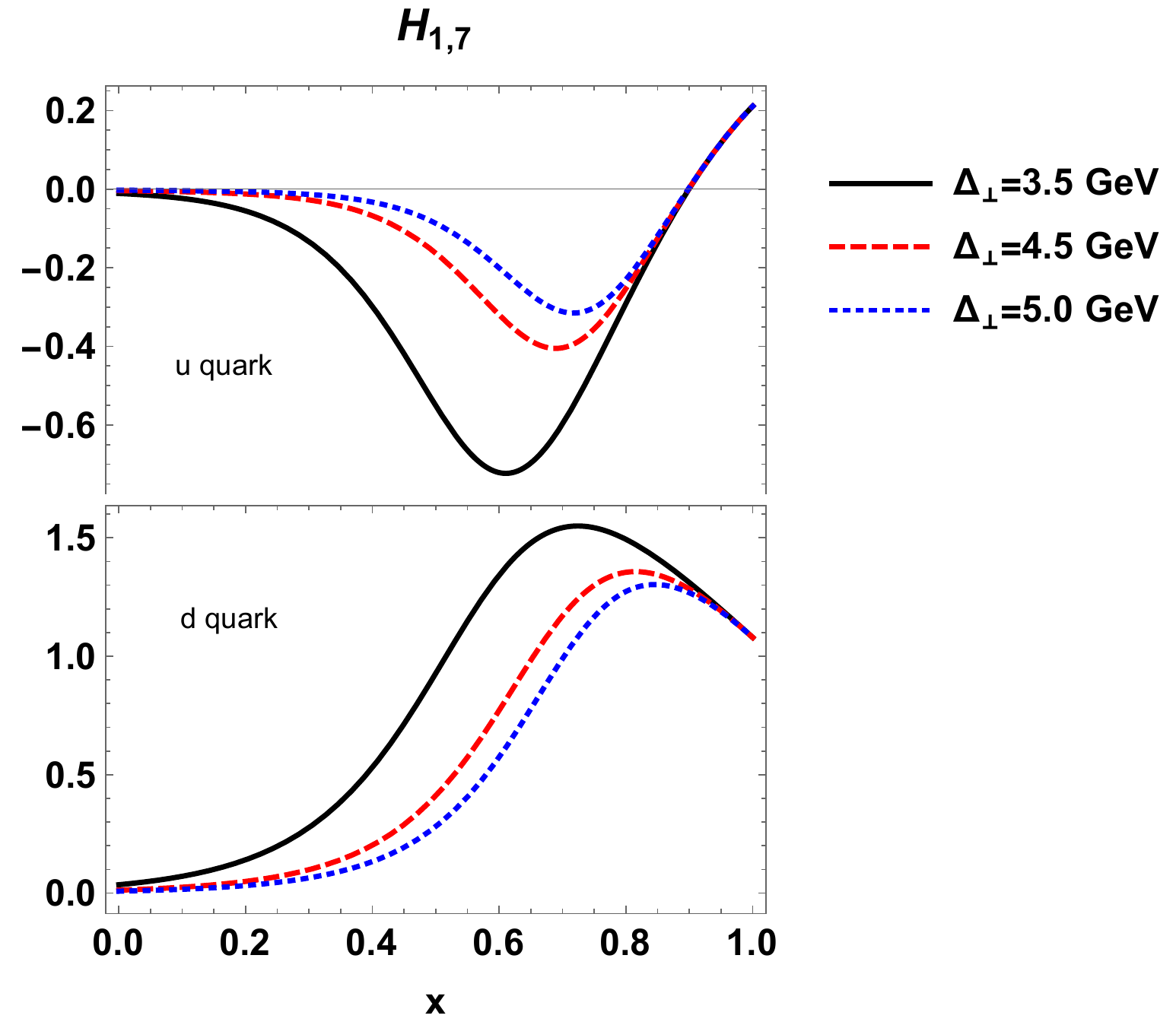}}
\caption{The variation of GTMDs $H_{1,6}(x, {\bf \Delta}_\perp, \textbf{p}_\perp)$ and $H_{1,7}(x, {\bf \Delta}_\perp, \textbf{p}_\perp)$ with $x$ for $u$ and $d$ quarks at different values of $\textbf{p}_{\perp}$ with fixed ${\bf \Delta}_{\perp}$ (left panel) and at different values of ${\bf \Delta}_\perp$ with fixed $\textbf{p}_\perp$ (right panel).}
\label{hpd}
\end{figure}

As already stated, the GTMDs of partons are entitled as mother distributions due to the reduction of these distributions to TMDs and GPDs by applying certain limits. In order to understand the significance of  $\zeta=-\frac{\Delta^+}{2P^+}$, the momentum transferred to the proton in longitudinal direction, we plot in Figs. \ref{F}-\ref{h1},  the GTMDs for different fixed values of $\zeta$. We take three different cases corresponding to  $\zeta=0,0.4$ and $0.6$.  There are total 16 complex-valued GTMDs in this model at non-zero skewness for both $u$ and $d$ quarks corresponding to diquarks $ud$ and $uu$ respectively in case of proton. It would be important to mention here that  the GTMDs can reduce to GPDs for non-zero skewness $(\zeta \neq 0)$ but not to TMDs. This is because of the fact  that TMDs do not depend on $\zeta$ but are functions of longitudinal momentum fraction $x$ and quark transverse momentum $\textbf{p}_\perp$. The transverse distance between struck quark and the center of momentum of proton differs by an amount of $\zeta b$ in the initial and final state of proton, when $\zeta \neq 0$. Further,  it would be interesting to study the GTMDs at zero skewness for two possibilities. First by fixing ${\bf \Delta}_\perp$ and taking different values of $\textbf{p}_\perp$ and second by fixing  $\textbf{p}_\perp$ and taking different values of ${\bf \Delta}_\perp$. In Figs. \ref{fpd}-\ref{hpd}, we plot the relation of GTMDs for zero skewness as a function of the longitudinal momentum fraction $x$ at different values of quark transverse momentum $(\textbf{p}_\perp)$ (${\bf \Delta}_\perp$ fixed) and at different values of momentum transferred to the proton $({\bf \Delta}_\perp)$ ($\textbf{p}_\perp$ fixed) for both $u$ quark and $d$ quarks. For the case of zero skewness, we are left with only 10 GTMDs for $u$ and $d$ quarks. We take $\textbf{p}_\perp=0.5$ $GeV$ and $\Delta_{max}=5$ $GeV$ and the quark composite mass as $m=0.3$ $GeV$ in the present calculations.

In Fig. \ref{F}, we plot twist-2 GTMDs corresponding to unpolarized quark  for $\Gamma=\gamma^+$. We have $F_{1,1}$, $F_{1,2}$, $F_{1,3}$ and $F_{1,4}$ at $\zeta=0,\ 0.4,\ 0.6$ for $u$ and $d$ quarks. From Eqs. (\ref{f11s}) and (\ref{f11}), we see that at $\zeta=0$, the second term does not contribute since $x'=x''$. The GTMD $F_{1,1}$ shown in Fig. \ref{fpd}(a), is related to unpolarized Wigner distribution $\rho_{UU}$ at $\zeta=0$. The distributions depend on the model parameters expressed in Table \ref{table1}. In Fig. \ref{F}(b), GTMD $F_{1,2}$ is shown for different values of $\zeta$ which, at $\zeta=0$, is zero in this model. For higher values of $\zeta$, the peak of $F_{1,2}$ shifts towards higher values of $x$. This clearly indicates that as the momentum transferred to the proton increases the longitudinal momentum fraction carried by the quark increases. The peak is more sharp in case of $d$ quark as compared to the case of $u$ quark. In Fig. \ref{F}(c), we see the GTMD $F_{1,3}$ shows similar variation as in the case for $F_{1,1}$ for both quarks. In Fig. \ref{F}(d), $F_{1,4}$ shows that the height of peaks distribution of both quarks  increases and shifts towards higher $x$ as $\zeta$ increases. At $\zeta=0$, $F_{1,4}$ relates to the orbital angular momentum problem. Further, its connection with orbital angular momentum would make it accessible to lattice QCD \cite{hatta}. Therefore, experimental measurements on GTMDs would have important implications for the proton structure. 

In Fig. \ref{g}, the GTMDs $G_{1,1}$, $G_{1,2}$, $G_{1,3}$ and $G_{1,4}$ describing the distributions correlated to longitudinally-polarized quark  for $\Gamma=\gamma^+ \gamma^5$ have been plotted. We observe that GTMD $G_{1,1}$ shows nearly same variation as $F_{1,4}$, however, with opposite polarities at different $\zeta$. The GTMD $G_{1,1}$ being `partner' of $F_{1,4}$ is related as $G_{1,1}=-F_{1,4}$. We observe that $G_{1,1}=-F_{1,4}$ from Eqs. (\ref{f14s}), (\ref{g11s}), (\ref{f14}) and (\ref{g11}) for both scalar and axial-vector diquarks. In case of $G_{1,2}$ shown in Fig. \ref{g}(b), the distribution becomes more concentrated at the peaks when the longitudinal momentum transferred is higher both in case of $u$ and $d$ quarks. From Fig. \ref{g}(c) it is evident that the distribution $G_{1,3}$ is zero at $\zeta=0$ while the peaks move towards higher $x$ by increasing $\zeta$. The $G_{1,4}$ distribution correlates with longitudinal Wigner distribution which shows nearly equal distribution as the case for $F_{1,1}$.

In Figs. \ref{h} and \ref{h1}, the GTMDs related to transversely-polarized quark i.e. for $\Gamma=i\sigma^{j+}\gamma^5$ are discussed, which are $H_{1,i}$ where $i=1,2,...,8$. The GTMDs $H_{1,1}$, $H_{1,5}$ and $H_{1,8}$ are zero when there is no momentum transferred in longitudinal direction $(\zeta=0)$ for both $u$ and $d$ quarks. Because of the scalar diquark contribution for $u$ quark, the distribution $H_{1,4}$ does not vanish at $\zeta=0$, while $H_{1,4}$ vanishes in case of $d$ quark as the scalar diquark contribution is absent in this case. We take the quark polarization along $\hat{x}$. In Figs. \ref{h}(a) and \ref{h}(d), the peaks of the distributions $H_{1,1}$ and $H_{1,4}$ move towards the higher values of longitudinal momentum fraction carried by struck quark. The polarities are opposite for these distributions. The distributions $H_{1,2}$ and $H_{1,6}$ are shown in Fig. \ref{h}(b) and \ref{h1}(b) respectively. The peak of these distributions for $\zeta=0.4$ is maximum for both $u$ and $d$ quarks, while for $\zeta=0.6$ $GeV$, the distribution peak is lower than the case of $\zeta=0$ for $u$ quark but for $d$ quark, the peak is almost equal for $\zeta=0$ and $\zeta=0.6$. The distribution $H_{1,6}$ exhibits negative polarities as compared to $H_{1,2}$. The distribution $H_{1,3}$ for $u$ quark and $d$ quark are shown in Fig. \ref{h}(c). For $H_{1,5}$ the distribution peaks increase with the increasing $\zeta$ values. The GTMD $H_{1,7}$ explains that the distribution peaks move towards the higher momentum fraction in longitudinal direction carried by quark with the increasing the longitudinal momentum transferred to the proton. The polarities are opposite for this distribution for $u$ and $d$ quarks. We plot $H_{1,8}$ in Fig. \ref{h1}(d), which displays that the distribution is zero for $\zeta=0$ while at non-zero skewness, the distribution peak move towards the higher $x$  taken by quark with raising $\zeta$ having opposite polarities for both quarks.


In Fig. \ref{fpd}, we have shown the GTMDs $F_{1,1}$, $F_{1,3}$ and $F_{1,4}$ for $\zeta=0$ at a fixed value of proton transverse momentum ${\bf \Delta}_\perp=5$ $GeV$ by varying the quark transverse momentum $\textbf{p}_{\perp}$ (left panel) and at fixed quark transverse momentum $\textbf{p}_{\perp}=0.5$ $GeV$ by changing the values of the proton transverse momentum ${\bf \Delta}_\perp $ (right panel). Since, the Fourier transform of $F_{1,4}$ is a function of     
$(x,0,\textbf{p}^2_{\perp}, \textbf{p}_\perp.\textbf{b}_\perp, \textbf{b}^2_\perp)$, it is similar to the Wigner distribution $\rho_{UL}(x,\textbf{p}_\perp,\textbf{b}_{\perp})$ after inclusion of a factor of $\epsilon^{ij}_{\perp}p^i_{\perp}\frac{\partial}{\partial b^j_{\perp}}$. The measurements of the distributions $F_{1,1}(x,0,{\bf \Delta}_\perp, \textbf{p}_\perp)$, $F_{1,3}(x,0,{\bf \Delta}_\perp, {\textbf{p}}_\perp)$ and $F_{1,4}(x,0,{\bf \Delta}_\perp, \textbf{p}_\perp)$ will lead to the implications of unpolarized quark in unpolarized proton, unpolarized quark in transversely-polarized proton and unpolarized proton in longitudinally polarized proton respectively. We have plotted the GTMDs corresponding to longitudinally-polarized quark at $\zeta=0$, $G_{1,1}$, $G_{1,2}$ and $G_{1,4}$ with a constant ${\bf \Delta}_\perp$ (left panel) and constant $\textbf{p}_{\perp}$ (right panel)  as  shown in Fig. \ref{gpd}. The longitudinally-polarized quark in unpolarized proton, longitudinally-polarized quark in transversely-polarized proton and longitudinally polarized quark in longitudinally-polarized proton lead to the consequences of the distributions $G_{1,1}(x,0,{\bf \Delta}_\perp, \textbf{p}_\perp)$, $G_{1,2}(x,0,{\bf \Delta}_\perp, \textbf{p}_\perp)$ and $G_{1,4}(x,0,{\bf \Delta}_\perp, \textbf{p}_\perp)$ respectively. The generalized transverse momentum-dependent distributions $H_{1,2}(x,0,{\bf \Delta}_\perp, \textbf{p}_\perp)$ and $H_{1,3}(x,0,{\bf \Delta}_\perp, \textbf{p}_\perp)$ for $u$ and $d$ quarks are shown in Fig. \ref{hpd1}, when these quarks are polarized along $\hat{x}$ at fixed ${\bf \Delta}_\perp$ (upper panel)and at fixed $\textbf{p}_{\perp}$ (lower panel). The distributions $H_{1,6}(x,0,{\bf \Delta}_\perp, \textbf{p}_\perp)$ and $H_{1,7}(x,0,{\bf \Delta}_\perp, \textbf{p}_\perp)$ at fixed ${\bf \Delta}_\perp$ and at fixed $\textbf{p}_{\perp}$ are displayed in Fig. \ref{hpd} (upper panel) and \ref{hpd} (lower panel) respectively.

\section{Conclusion}

We have studied the quark Wigner distributions and GTMDs in the light-front quark-diquark model, where the proton can be formed by the coupling of a quark and a diquark. The diquark can be considered as either scalar (spin-0) or axial-vector diquark (spin-1).
The axial-vector diquark can further be distinguished as isoscalar and isovector diquarks based on the realistic flavor analysis.

The Wigner distributions for $u$ and $d$ quarks in the proton have been discussed for different polarization configurations and they provide the abundant information about the multi-dimensional structure of proton. The TMDs and GPDs can be calculated from Wigner distributions and GTMDs under TMD limit and IPD limit respectively. We have used the overlap representation of light-front wavefunctions (LFWFs) to evaluate the Wigner distributions.  The input parameters containing diquark masses and coupling constants for scalar-isoscalar, vector-isoscalar and vector-isovector diquarks are shown in Table \ref{table1} and the constituent quark mass is taken to be $m=0.33$ $GeV$.

We have presented the quark Wigner distributions in impact-parameter space $(\textbf{b}_\perp)$ and momentum space $(\textbf{p}_\perp)$ by fixing the quark transverse momentum $(p_y=0.5$ $GeV)$ and impact-parameter co-ordinate $(b_y=0.4$ $GeV)$ respectively. In impact-parameter space, the distributions named $\rho_{UU}$, $\rho_{LL}$, $\rho_{LT}$ and $\rho_{TT}$ show a circularly symmetric behavior while $\rho_{UL}$, $\rho^j_{UT}$, $\rho_{LU}$ and $\rho^i_{TL}$ describe a dipolar structure. In this model, the distributions $\rho_{UL}$ and $\rho_{LU}$ are similar and have important implications for the orbital angular momentum problem. In momentum space, the distributions $\rho_{UU}$ and $\rho_{LL}$ are circularly symmetric in negative direction while in case of $\rho_{TT}$ and $\rho^j_{UT}$ the polarity changes. The Wigner distributions $\rho_{UL}$, $\rho_{LU}$, $\rho^j_{LT}$ and $\rho_{TT}$ describe the dipolar behavior of both $u$ and $d$ quarks in the proton. The mixed distributions $(p_x,b_y)$ for both $u$ and $d$ quarks have also been presented. We find  quadrupole distributions in the case of $\rho_{UL}$, $\rho_{LU}$ and $\rho^i_{TL}$ whereas  axially symmetric behavior is found in the case of $\rho_{UU}$, $\rho_{LL}$, $\rho^i_{TU}$ and $\rho_{TT}$. The distributions $\rho^j_{UT}$ and $\rho^j_{LT}$ show the dipolar behavior for both $u$ and $d$ quarks. 

The Wigner distributions in general indicate a strong interplay between the polarization of the quark as well as proton. Further, the correlation to the direction of quark as well as proton spin is nontrivial. Even though all TMDs and GPDs can be obtained from the Wigner distributions at certain limits, the Wigner distributions contain detailed information in the content of longitudinal as well as transverse distributions in the phase space as well as the momentum space. The general dipole and quadrupole structure of Wigner distributions depend entirely on the spin orientations of the quark and proton. Further studies on Wigner distributions will surely provide as deeper understanding of the spin-orbit correlation and orbital angular momentum.

We have also discussed the twist-2 generalized transverse momentum-dependent parton distributions (GTMDs) of $u$ and $d$ quarks related to the longitudinal momentum fraction $x$ at different values of momentum transferred in the longitudinal direction $\zeta$ which provide the maximum information about the proton structure. The GTMDs at zero skewness i.e. at $\zeta=0$, related to the longitudinal momentum fraction at different quark transverse momentum and at different momentum transferred to the proton, have been studied. The measurements of the distributions $F_{1,1}(x,0,{\bf \Delta}_\perp, \textbf{p}_\perp)$, $F_{1,3}(x,0,{\bf \Delta}_\perp, \textbf{p}_\perp)$, $F_{1,4}(x,0,{\bf \Delta}_\perp, \textbf{p}_\perp)$ would imply the unpolarized quark in unpolarized proton, in transversely-polarized proton and in longitudinally-polarized proton while the GTMDs corresponding to the longitudinally-polarized quark i.e. $G_{1,1}(x,0,{\bf \Delta}_\perp, \textbf{p}_\perp)$, $G_{1,2}(x,0,{\bf \Delta}_\perp, \textbf{p}_\perp)$ and $G_{1,4}(x,0,{\bf \Delta}_\perp, \textbf{p}_\perp)$ measure the longitudinally-polarized quark in unpolarized proton, in transversely-polarized proton and longitudinally-polarized proton respectively. We have calculated  16 GTMDs for $u$ and $d$ quarks in the case of proton at $\zeta \neq 0$. In this model, there are 10 GTMDs at $\zeta=0$ corresponding to the unpolarized, longitudinally polarized and transversely polarized quark. The GTMDs quantify the correlation between the quark/proton spin and the orbital angular momentum of the quarks. Several polarization observables have been proposed  which will give access to GTMDs, either directly or indirectly.

To conclude, the results can perhaps be substantiated by measurements of Wigner distributions and  GTMDs by future experiments. The experiments would not only restrict the model parameters and provide better knowledge of the distributions but will also impose important constraints on the correlation between the proton polarization being parallel/perpendicular w.r.t. the quark. Further, the possibility of inclusion of appropriate gauge link will play a very important role in building general relations between TMDs and the GPDs as the T-odd TMDs require explicit gauge degrees of freedom.

\section{Acknowledgements}
H.D. acknowledges financial support received from Science and Engineering Research Board a statutory board under Department of Science and Technology, Government of India (Grant No. EMR/2017/001549).

\end{document}